\documentclass[11pt,a4paper,oneside]{book}

\usepackage[pdftex]{graphicx}
\usepackage{tree-dvips}
\usepackage{fancyheadings}
\usepackage{url}
\usepackage{dropping}
\usepackage[cmex10]{amsmath}
\usepackage{amssymb}
\usepackage{amsbsy}
\usepackage{algorithm2e}
\usepackage[Glenn]{fncychap}
\usepackage[hang,sl]{caption}
\usepackage{setspace}
\usepackage{thesispage}
\usepackage{tabulary}
\usepackage{longtable}
\usepackage{algorithmic}
\usepackage{url}
\usepackage{xspace}
\usepackage[caption=false]{subfig}
\usepackage{caption}
\usepackage{array}
\usepackage{lscape}
\usepackage{algorithmic}
\usepackage{url}
\usepackage{xspace}
\usepackage{multirow}
\usepackage{float}
\usepackage{footnote}
\usepackage{accents}
\usepackage{geometry}
\usepackage{color}
\usepackage{spverbatim}
\usepackage{listings}
\usepackage{kbordermatrix}

\definecolor{pblue}{rgb}{0.13,0.13,1}
\definecolor{pgreen}{rgb}{0,0.5,0}
\definecolor{pred}{rgb}{0.9,0,0}
\definecolor{pgrey}{rgb}{0.46,0.45,0.48}

\graphicspath{{pics/}}
\DeclareGraphicsExtensions{.pdf,.jpeg,.png}
\newcommand{\etal}{\emph{et al.}\xspace}
\newcommand{\cf}{\emph{cf.}\xspace}
\makeatletter
\newcommand*{\etc}{%
    \@ifnextchar{.}%
        {etc}%
        {etc.\@\xspace}%
}

\newcommand{\MsgCharSet}{\mbox{$\mathcal{C}$}}
\newcommand{\Message}{\mbox{$m$}}
\newcommand{\MessageSet}{\mbox{$\mathcal{M}$}}

\newcommand{\MessageCharacter}{\mbox{$c$}}

\newcommand{\Request}{\mbox{\it $m_{req}$}}
\newcommand{\Response}{\mbox{\it $m_{res}$}}
\newcommand{\Interaction}{\mbox{$i$}}

\newcommand{\TraceSet}{\mbox{$\mathcal{T}$}}
\newcommand{\TraceLibrary}{\mbox{$\mathcal{L}$}}
\newcommand{\InteractionSet}{\mbox{$\mathcal{I}$}}
\newcommand{\InteractionCluster}{\mbox{$\mathcal{C}$}}
\newcommand{\InteractionTrace}{\mbox{$t$}}

\newcommand{\Pair}[2]{\mbox{$(#1, #2) $}}
\newcommand{\ConsensusPrototype}{\mbox{$p$}}

\newcommand{\ClusterCentre}{\mbox{${\Interaction}^{\star}$}}

\newcommand{\DistFunc}{\mbox{$dist$}}
\newcommand{\SimiCalc}{\mbox{$simi$}}
\newcommand{\TransFunc}{\mbox{$trans$}}
\newcommand{\MatchingFunc}{\mbox{$matching$}}
\newcommand{\AnalysisFunc}{\mbox{$analysis$}}
\newcommand{\ClusteringStep}{\mbox{$cluster$}}
\newcommand{\FormulatingStep}{\mbox{$formulate$}}
\newcommand{\SelectingStep}{\mbox{$select$}}
\newcommand{\SymmetricField}{\mbox{$\mathcal{S}$}}
\newcommand{\IdentifyingStep}{\mbox{$identify$}}
\newcommand{\SubstitutingStep}{\mbox{$substitute$}}

\newcommand{\wildcard}{{\boldsymbol ?}}
\newcommand{\gap}{{\boldsymbol *}}
\newlength{\aligncharw}
\newcommand{\agap}{\makebox[\aligncharw]{$\gap$}}

\newcommand{\myspace}{1.5}
\begin{document}

\nocite{*}

\settoheight{\parskip}{K}
\setlength{\parindent}{0cm}
\pagestyle{fancy}
\ChTitleVar{\Huge\sf}

\lhead{}

%%%%%%%%%%%%%%%%%%
%% Title macros %%
%%%%%%%%%%%%%%%%%%

\newcommand{\titleVal}{opaque response generation}

\newcommand{\subtitleVal}{Enabling Automatic Creation of Virtual Services for Service Virtualisation}

\newcommand{\authorVal}{Miao Du}

\newcommand{\factVal}{Faculty of Science, Engineering and Technology}
\newcommand{\instVal}{Swinburne University of Technology}
\newcommand{\yearVal}{2016}

%%%%%%%%%%%%%%%%%%%%%%%%%
%% Title page geometry %%
%%%%%%%%%%%%%%%%%%%%%%%%%
%\newgeometry{top=50mm, left=40mm, bottom=40mm, right=20mm}

\begin{titlepage}
  \hrule
  \begin{center}
    \Huge\sc \titleVal
  \end{center}

  \vspace{4mm}

  \begin{center}
    \Large \subtitleVal
  \end{center}

  \begin{center}
    \LARGE \sc By
  \end{center}

  \begin{center}
    \huge \sc \authorVal
  \end{center}

  \hrule

  \vfill
  \begin{center}
    A thesis submitted in total fulfilment of the requirements of the
    degree of Doctor of Philosophy.
  \end{center}

  \vfill
  \begin{center}
    \Large\yearVal
  \end{center}

  \vspace{1cm}
  \begin{center}
    \Large \factVal\\\instVal
  \end{center}
\end{titlepage}

% LocalWords:  Ph

\thispagestyle{empty}
Nowadays, an enterprise system interacts with many other systems to perform functionalities. With the trend of ever increasing inter-connectedness between software systems, one of the primary obstacles facing today's software development teams is the ability to develop and test a software system independent of the other systems on which it depends. Replicating a production environment for testing is expensive, time-consuming and error-prone. Moreover, software developers will usually have only limited access to it. However, continuous access to production-like conditions to test their application is an important part of the modern development method em Development-Operations (known as "DevOps"). Service virtualisation technique can make continuous access into a reality.

Service virtualisation is a method to create virtual service models that can mimic interaction behaviors between a system under test and the target system. With service virtualisation, the development team can get access to the production-like conditions whenever and however many times they need, enabling frequent and comprehensive testing. Previous techniques for service virtualisation have relied on explicitly modelling the target services by a service expert and require detailed knowledge of message protocol and structure. However, neither of these are necessarily available.

In this thesis, we introduce our novel opaque response generation approach. This approach enables services to be virtualised automatically without any expert knowledge or documentation of system protocol and interaction behaviours. Given a collection of interactions exchanged between a system under test and a target real service, our approach can 1) organise the same type of interactions into the same cluster and derive a cluster prototype for each cluster; 2) search a given incoming request for its the most similar request in the interaction library; 3) learn knowledge from the incoming request and the recorded interaction; and 4) generate a response. A framework and proof-of-concept implementation of our opaque response generation approach is described. Experimental results show our opaque response generation approach is able to automatically generate accurate responses in real time with an accuracy rate over 99\% on average.

\newpage

\thispagestyle{empty}
\cleardoublepage

\Large \textbf{Acknowledgements}
\normalsize

When I am about to wrap up my work and life in Swinburne, I look back at these years and realise that I was really fortunate to receive help and support from many people. Without them, I would not be able to finish my Ph.D dissertation.

First and foremost, I sincerely express my deepest gratitude to my supervision team, Prof. John Grundy, A/Prof. Jean-Guy Schneider and Dr. Steve Versteeg, for their seasoned supervision and continuous encouragement throughout my PhD study. Their advice and supports were critical for my Ph.D study. Their technical advice taught me how to conduct cutting-edge research;
their patience and trust gave me confidence to overcome the problems in my research.

I thank Swinburne University of Technology and the Faculty of Science, Engineering and Technology for providing the opportunity to undertake
my PhD. Additionally, thanks to CA Technologies, in particular CA Labs, for offering a practical research topic to explore and funding with which
to explore it.

My thanks also go to Prof. Jun Han, to my review panel members Prof. Richard Sadus, Dr. Man Lau, Dr. Clinton Woodward, and to staff members, research students and research assistants at SUCCESS for their help, suggestions, friendship and encouragement, in particular, Dr. Cameron Hine, Dr. Mohamed Abdelrazek, Dr. Iman Avazpour, Dr. Feifei Chen.

I am deeply grateful to my parents Tao Du and Yingchun Qiu for raising me up, teaching me to be a good person, and supporting me to study abroad. Last but not the least, I thank my husband, Dong Yuan, for his love, understanding, encouragement, sacrifice and help. His love, care and support have been and will be the source of my strength.

\clearpage

\newpage

\thispagestyle{empty}
\cleardoublepage

\Large \textbf{Declaration}
\normalsize
 
  I, the candidate Miao Du, hereby declare that the examinable
  outcome embodied by this dissertation:
  \begin{itemize}
  \item contains no material which has been accepted for the award to
    the candidate of any other degree or diploma, except where due
    reference is made in the text of the examinable outcome;
  \item to the best of our knowledge contains no material
    previously published or written by another person except where due
    reference is made in the text of the examinable outcome; and
  \item where the work is based on joint research or publications,
    discloses the relative contributions of the respective workers or
    authors.
  \end{itemize}
  \vspace{2cm}
  \rule{6cm}{.1pt}\\
  \large{\emph{Miao Du}}

\clearpage 
\newpage

\thispagestyle{empty}
\cleardoublepage

\Large \textbf{Publications}
\normalsize

During the course of this project a number of peer-reviewed publications were produced. They are presented here for reference and also to highlight the corresponding content in the thesis.

\begin{itemize}
    
    \item Du, M., Automatic Generation of Interaction Models for Enterprise Software Environment Emulation, published in doctoral symposium allocated at the {\em 22nd Australasian Software Engineering Conference (ASWEC 2013)}, June 2013, Melbourne, Australia.
        \begin{itemize}
            \item
                Corresponds to the framework for opaque response generation approach, which is presented in Chapter~\ref{chap4:framework}.
        \end{itemize}
    
    \item Du, M., Schneider, J.-G., Hine, C., Grundy, J. and Versteeg, S., Generating Service Models by Trace Subsequence Substitution, published in proceedings of the {\em 9th International ACM Sigsoft Conference on Quality of Software Architectures (QoSA 2013)}, 2013, Vancouver, British Columbia, Canada.
        \begin{itemize}
            \item
                This is the whole library approach, the first implementation of the opaque response generation framework, to generating responses from network traffic presented in Chapter~\ref{chap5:qosa}.
        \end{itemize}
    
    \item Du, M., Versteeg, S., Schneider, J-G, Han, J. and Grundy, J.C., Interaction Traces Mining for Efficient System Responses Generation, published in proceedings of the \textit{2nd International Workshop on Software Mining (SoftMine 2013)}, 11th Nov 2013, Palo Alto, CA, USA. (Best paper award)
        
        \begin{itemize}
            \item
                Presents the cluster centroid approach for improving efficient virtual service response performance described in Chapter~\ref{chap6:SoftMine}.
        \end{itemize}
        
    \item Du, M., Versteeg, S., Schneider, J-G, Han, J. and Grundy, J.C., From Network Traces to System Responses: Opaquely Emulating Software Services. (Pre-printed)      

        \begin{itemize}
            \item
                Presents the consensus prototype approach that is efficient and robust to generate accurate responses, discussed in chapter~\ref{chap7:consensus}.
        \end{itemize}
\end{itemize}

\clearpage

\newpage

\thispagestyle{empty}
\cleardoublepage

\Large \textbf{Patents}
\normalsize

\begin{itemize}
    
    \item M. Du, J.-G. Schneider, C. Hine, J. Grundy, J. Han, and S. Versteeg, “Message matching for opaque service virtualization,” US Patent Application No. 14/223,607, March, 2014.
    \item S. Versteeg, J. Bird, N. Hastings, M. Du, and J. Dahan, “Entropy weighted message matching for opaque service virtualization,” US Patent Application No. 14/211,933, March, 2014.
    \item M. Du, S. Versteeg, J.-G. Schneider, J. Han, and J. Grundy, “System and methods for clustering trace messages for efficient opaque response generation,” US Patent Application No. 14/305,322, June, 2014.
    \item S. Versteeg, M. Du, J.-G. Schneider, J. Grundy and J. Han, “Systems and Methods For Automatically Generating Message Prototypes for Accurate and Efficient Opaque Service Emulation,” US Patent Application No. 14/535,950, November, 2014.
    \item M. Du, S. Versteeg, “Response Prototypes with Robust Substitution Rules for Service Virtualisation,” US Patent Application No. 14/661,514, March, 2015.
\end{itemize} 

\clearpage
\newpage

\setcounter{page}{1}
\pagenumbering{roman}
\tableofcontents
\listoftables
\listoffigures
\newpage
%\frontheadings
\setcounter{page}{1}
\pagenumbering{arabic}
\chapter{Introduction}
\label{chap1:introduction}

Modern enterprise software environments integrate a large number of software systems to facilitate complicated business processes. These systems are critical to the operation of the organisation. Many of these software systems need to interact with services provided by other systems to fulfil their responsibility. Upgrading such a software system or replacing a system or installing an additional system poses a high risk, to not only the system itself (the system-under-test), but also to the other systems in the environment (the dependency systems) that the system-under-test interacts with. This creates significant engineering challenges, namely 1) how to assure, before live deployment, the quality of the {\em system-under-test} interoperating across the often heterogeneous services provided by large-scale production environments; and 2) how to investigate the effects of different environment configurations on the system-under-test's operational behaviour.

\begin{figure}
  \centering
  % Requires \usepackage{graphicx}
  \includegraphics[width=0.8\textwidth]{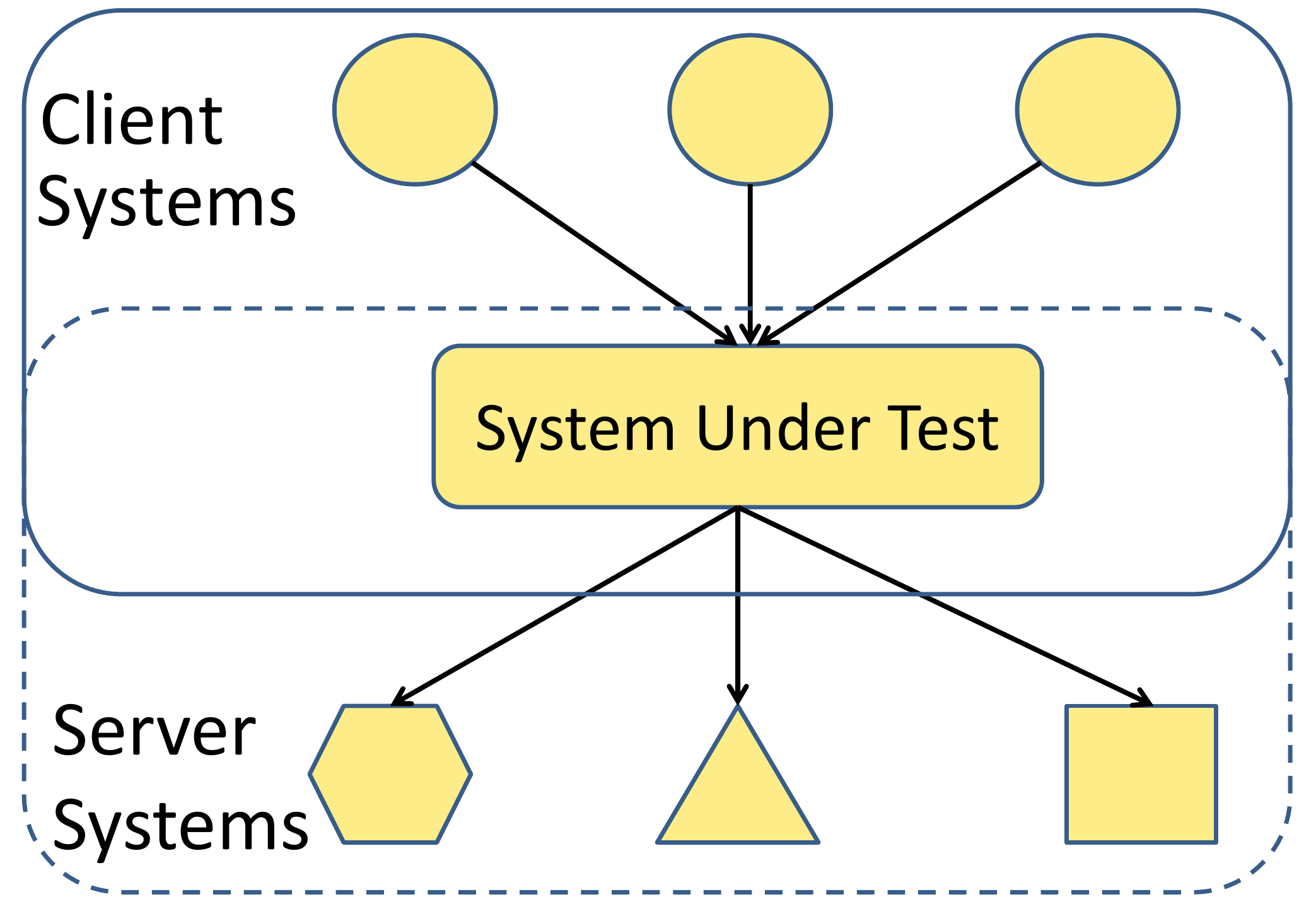}\\
  \caption{Traditional test-bed environment versus interactive test-bed environment}
  \label{chap1fig:test-bedenvironment}
\end{figure}

In enterprise environments, a system-under-test may act as both a client and a server. It may offer a service responding to client requests, as well as making requests to other services in the environment. These client and server interactions with other systems are depicted in Figure~\ref{chap1fig:test-bedenvironment}. There have been many methods and tools developed for testing functional and non-functional properties of a system-under-test which focus on mimicking the clients in environment - tools that generate requests to which the system-under-test responds~\cite{HPLoadRunner}~\cite{JMeter} (the scenario shown with the solid line rectangle in Figure~\ref{chap1fig:test-bedenvironment}). However, there is a gap in the current generation of testing tools for reactive situations, to deal with requests made by the system-under-test to other systems in the environment. This is the scenario shown with the dash line rectangle in Figure~\ref{chap1fig:test-bedenvironment}.

{\em Service virtualisation} has been postulated as a promising approach to providing a reactive production-like environment. By creating virtual services capable of emulating interaction behaviours of server-side system in the environment and simultaneously executing a number of those virtualised services, the service virtualisation technique provides an interactive representation of an enterprise software environment which, from the perspective of a system-under-test, appears to be a real operating environment.

This thesis proposes a novel approach to service virtualisation by utilising a genome-sequencing algorithm to help realise the virtualisation of services, without requiring any expert knowledge. The novel research reported in this thesis eliminates the need for detailed protocol knowledge. Instead, it matches requests based on byte-level patterns to provide accurate responses based on a network traffic library. Experimental evaluation demonstrates that our approach is an automated, accurate, efficient and robust way to support service virtualisation. %Furthermore, the larger the library, the more data our approach can learn from, and thus the more accurate the response it produces.

This chapter introduces the background, motivations, key issues and major contributions of this research. It is organised as follows. Section~\ref{chap1:background} gives a brief introduction to service virtualisation and discusses some major challenges. Section~\ref{chap1:motivation} presents a motivating example for this work. Section~\ref{chap1:questions} outlines key research questions. Section~\ref{chap1:contributions} describes major contributions of this research. Section~\ref{chap1:method} presents methodology for conducting this research. Finally, Section~\ref{chap1:overview} presents an overview of this thesis.

%One popular approach that developers use to test their application's dependence on other systems is to install the other systems on virtual machines (such as VMware) \cite{Sugerman:01}.  However virtual machines are time consuming to configure and maintain.  Furthermore the configuration of the systems running on the virtual machine is likely to be different to the production environment. An alternative that is gaining increasing traction is service emulation, where models of services are emulated - sometimes into the many thousands of service emulations - to provide more realistic scale and less complicated configuration \cite{hine:09a}. However, existing approaches to service emulation have relied on explicitly modelling the target services by a system expert and require detailed knowledge of message protocols and structure.  This is often infeasible if the required knowledge is unavailable for the wide range of services in a real deployment environment \cite{Ghosh99issuesin}.
%
%Our aim is to develop an automated approach to service emulation which uses no explicit knowledge of the services, their message protocols and structures and yet can simulate, to a high degree of accuracy and scale, a realistic enterprise system deployment environment.  To achieve this we need an automated, accurate, efficient and robust method for service emulation that derives service responses from collected actual network traces between an enterprise system and a service -- the focus of this thesis.
\section{Service Virtualisation and its challenges}
\label{chap1:background}

% why do we need service virutalisation and what is service virtualisation
% balance software quality and time to market
%Software innovation drives today's enterprises. Business leaders expect IT department to constantly drive innovation
%across the enterprise, while reducing costs, increasing operational efficiency, and maximizing competitiveness. The extent to which IT
%departments meet these expectations is often determined by their ability to deliver software development
%projects on time at the desired quality.   

With the trend of ever increasing inter-connectedness between software systems, one of the primary obstacles facing today's enterprise software development teams is the ability to develop and test a software system, independent of the other systems on which it depends. In a typical deployment scenario, an enterprise system might interact with many other systems, such as a mainframe system, directory servers, databases, third-party systems and many other types of software services. When you upgrade or replace software systems, it is risky if you do not know what the cascade effects will be. A failure in a single software system can cause a cascade of failures across multiple systems bring a catastrophic failure across multiple IT services. An example of this was during 2012 when Citylink, the company which operates the road tunnel going under central Melbourne, upgraded their billing system. An unexpected interdependency between systems triggered a failure in the tunnel safety systems, causing the tunnel to close and traffic chaos~\cite{citylinkchoas}.

A development team wants to check whether the services will function correctly in terms of its interactions with other services prior to their actual deployment. Getting access to the actual production environment for testing is not possible due to the risk of disruption. Large organisations often have a test environment, which is a close replication of their production environment, but this is very expensive. Furthermore the test environment is in high demand, so software developers will have only limited access to it.  Enabling developers to have continuous access to production-like conditions to test their application is an important part of Development-Operations (known as `DevOps'). Service virtualisation technique can make this a reality.

%why current service depend on other services
{\em Service virtualisation} is a method to create virtual services that are able to emulate the interactive behaviour of specific components in heterogeneous component-based applications, such as API-driven applications, cloud-based applications and service-oriented architectures~\cite{michelsen2012servicevirtualisation}. It has been postulated as a promising approach to providing a production-like environment for testing purpose. Specifically, when developing and testing a software system, software development and quality assurance teams often need to get access to certain services which are not constantly available or difficult to access. They create virtual service models for emulating interactive behaviours of those services. By configuring a test environment with service models, software development and quality assurance teams can get continuous access to continue their development task.

In a nutshell, service virtualisation works in three steps: the {\em capture} step, the {\em process} step and the {\em model} step.  {\em Capture} is a step where a set of the interactions between a system-under-test and other services they rely on can be gathered. Interactions can be captured from log files or recording the traffic occurring on a live system. In the {\em process} step, a {\em decoder} is required to detect the protocol of the live interactions and to process the syntax and semantics of the data. The {\em model} step is the final stage where we specify how virtual services respond to the system-under-test.

However, in a realistic environment, there are cases where you want to virtualise a service for an application that has to talk to legacy or proprietary systems and the documentation is not available. Defining a decoder becomes extremely difficult. It can either take a lot of time to reverse engineer the message structure, or the project can fail if the team decides that it is too hard. Additionally, the large number of services and service interactions make setting up the virtual test environment very hard, time consuming and error prone.

\section{Motivating Example}
\label{chap1:motivation}

Assume an enterprise is going to upgrade its enterprise resource planning (ERP) system, which is our {\em system under test}. In its production environment, the ERP system is to be integrated with many other systems for managing and interpreting data from many business activities. The other systems (called services) include a legacy mainframe program, a directory server and a web service.  The behaviour of the example ERP system depends on the responses it receives from these dependency services.

The directory server uses a proprietary protocol which is poorly documented. It has operation types add, search, modify and remove, each with various payloads and response messages. The operation type is encoded in a single character. The ERP system needs to interact with this service using this protocol, to search for user identities, register new identities, de-register identities and so on. Other systems in the environment share the same service.

Enterprise software environment emulation is a promising approach to providing an executable, interactive testing environment for enterprise systems using such a messaging protocol~\cite{hine:thesis}. To realise emulated testing environments for enterprise systems, virtual service models are used. When sent messages by the enterprise system under test, these would receive messages, responding with approximations of ``real'' service response messages. %\cite{servicemodel}.

The most common approach to developing service models is to manually define interaction models with the use of available knowledge about the underlying interaction protocol(s) and system behaviour(s), respectively. Challenges with these approaches include the time-consuming and sometimes error-prone processes of developing the models, lack of precision in the models especially for complex protocols, and ensuring robustness of the models under diverse loading conditions. % \cite{sun2012usefulness}.
To assist developing reusable service models, approaches either reverse engineer message structures %\cite{messageformatinference}
\cite{cui:07a}, or discover % \cite{processmining} \cite{processdiagnostics}
processes for building behavioral models. % \cite{behavioralmodel}.
While these allow engineers to develop more precise models, none of them can automate interaction between enterprise system under test and the emulated dependency services.

\begin{table}[t]
\footnotesize
\begin{center}
\begin{tabular}{|c||l|l|}
\hline
Index & \multicolumn{2}{c|}{Request/Response Messages} \\ \hline\hline
\multirow{2}{0.8cm}{\centering 1} & Req. & \{id:1,op:B\} \\ \cline{2-3}
& Res. & \{id:1,op:BindRsp,result:Ok\} \\ \hline
\multirow{2}{0.8cm}{\centering 2} & Req. & \{id:2,op:S,sn:Du\} \\ \cline{2-3}
& Res. & \{id:2,op:SearchRsp,result:Ok,gn:Miao,sn:Du,mobile:5362634\} \\ \hline
\multirow{2}{0.8cm}{\centering 13} & Req. & \{id:13,op:S,sn:Versteeg\} \\\cline{2-3}
& Res. & \{id:13,op:SearchRsp,result:Ok,gn:Steve,sn:Versteeg,mobile:9374723\} \\ \hline
\multirow{2}{0.8cm}{\centering 24} & Req. & \{id:24,op:A,sn:Schneider,mobile:123456\} \\ \cline{2-3}
& Res. & \{id:24,op:AddRsp,result:Ok\} \\ \hline
\multirow{2}{0.8cm}{\centering 275} & Req. & \{id:275,op:S,sn:Han\} \\ \cline{2-3}
 & Res. & \{id:275,op:SearchRsp,result:Ok,gn:Jun,sn:Han,mobile:33333333\} \\ \hline
\multirow{2}{0.8cm}{\centering 490} & Req. & \{id:490,op:S,sn:Grundy\}\\ \cline{2-3}
 & Res. &
\{id:490,op:SearchRsp,result:Ok,gn:John,sn:Grundy,mobile:44444444\} \\ \hline
\multirow{2}{0.8cm}{\centering 2273} & Req. & \{id:2273,op:S,sn:Schneider\} \\ \cline{2-3}
& Res. & \{id:2273,op:SearchRsp,result:Ok,sn:Schneider,mobile:123456\} \\ \hline
\multirow{2}{0.8cm}{\centering 2487} & Req. & \{id:2487,op:A,sn:Will\} \\ \cline{2-3}
& Res. & \{id:2487,op:AddRsp,result:Ok\} \\ \hline
\multirow{2}{0.8cm}{\centering 3106} & Req. & \{id:3106,op:A,sn:Hine,gn:Cam,Postcode:33589\} \\ \cline{2-3}
& Res. & \{id:3106,op:AddRsp,result:Ok\} \\
\hline
\multirow{2}{0.8cm}{\centering 3211} & Req. & \{id:3211,op:U\} \\ \cline{2-3}
& Res. & \{id:3211,op:UnbindRsp,result:Ok\} \\
\hline\hline
\multirow{2}{0.8cm}{\centering 1} & Req. & \{id:1,op:B\} \\ \cline{2-3}
& Res. & \{id:1,op:BindRsp,result:Ok\} \\ \hline
\multirow{2}{0.8cm}{\centering 12} & Req. & \{id:12,op:S,sn:Hine\} \\ \cline{2-3}
& Res. & \{id:12,op:SearchRsp,result:Ok,gn:Cam,sn:Hine,Postcode:33589\} \\ \hline
\multirow{2}{0.8cm}{\centering 34} & Req. & \{id:34,op:A,sn:Lindsey,gn:Vanessa,PostalAddress1:83 Venton Road\} \\ \cline{2-3}
& Res. & \{id:34,op:AddRsp,result:Ok\} \\ \hline
\multirow{2}{0.8cm}{\centering 145} & Req. & \{id:145,op:S,sn:Will\} \\ \cline{2-3}
& Res. & \{id:145,op:SearchRsp,result:Ok,sn:Will,gn:Wendy,mobile:54547\} \\ \hline
\multirow{2}{0.8cm}{\centering 1334} & Req. & \{id:1334,op:S,sn:Lindsey,gn:Vanessa,PostalAddress1:83\ Venton\ Road\} \\ \cline{2-3}
& Res. & \{id:1334,op:SearchRsp,result:Ok,gn:Vanessa,PostalAddress1:83\ Venton\ Road\} \\ \hline
\multirow{2}{0.8cm}{\centering 1500} & Req. & \{id:1500,op:U\} \\ \cline{2-3}
& Res. & \{id:1500,op:UnbindRsp,result:Ok\} \\ \hline
\end{tabular}
\end{center}
\caption{Directory Service Interaction Library Example}
\label{Chap1tab:tl}
\end{table}

Recording and replaying message traces is an alternative approach. This involves recording request messages sent by the enterprise system under test to real services and the response messages sent back from these services, and then using these message traces to `mimic' the real service response messages in the emulation environment~\cite{cui:06}. Table~\ref{Chap1tab:tl} shows a small sample set of request and response messages (a \emph{transaction library}) for the directory service. It is from a fictional protocol that has some similarities to the widely used LDAP protocol~\cite{Sermersheim2006}. Such message traces are captured by either network monitoring tools, such as Wireshark, or proxies. Some approaches combine record-and-replay with reverse-engineered service models. Although they can mimic interactions automatically, they still require some knowledge of the services which is not necessarily available. Hence, there is a need for a new approach to
creating virtual service models which uses no explicit knowledge of the services, their message protocols and structures and yet can simulate - to a high degree of accuracy and scale - a realistic enterprise system deployment environment.

\section{Research Questions}
\label{chap1:questions}

The main research question tackled in this thesis is:

\hspace{0.5cm} \textbf{\em ``Can virtual services be created automatically in the absence of experts or expert knowledge, which are able to return accurate approximations of the real responses at runtime?''}

In order to address this major research question, we have divided it into three sub-questions as follows.

\begin{enumerate}
\item {\em ``Is it feasible to create virtual services automatically?''}

    Virtual services can emulate real target services by responding to a system under test with responses that enable the system under test to keep on running. This question focuses on how to create a virtual service that can automatically generate responses to respond to a system under test. To answer this question, we propose a framework that targets to generate responses directly from network traffic. This is addressed in Chapter \ref{chap4:framework} of this thesis.

\item {\em ``Can the automatic creation of a virtual service provide responses to a system under test accurately?''}

    To enable an operation to continue, what a system under test needs is not just a response but a valid response that conforms to both the message structure specification and the system current state. This question focuses on evaluating whether the automatic creation of a virtual service can generate accurate, protocol-conformant responses. To answer this question, we propose and implement a machine learning approach for automatically creating virtual service. In addition, we define several criteria to check the validity of created virtual services. This is addressed in Chapter~\ref{chap5:qosa}, Chapter~\ref{chap7:consensus}.

\item {\em ``Can the automatic creation of a virtual service respond to a system under test in a timely fashion?''}

    Response time is the total time that a target service/virtual service takes to respond to a request from a system under test. A slow response time may lead to a service failure, which will terminate the rest of the operation. This question focuses on measuring if the automatic creation of a virtual service can respond to a request as fast as a real service can. To answer this question, we propose and implement a data mining approach to assist with virtual service creation. In addition, we evaluate the efficiency and accuracy of our new approach to creating virtual services. These aspects are addressed in Chapter~\ref{chap6:SoftMine}, Chapter~\ref{chap7:consensus}.

\end{enumerate}

In short, we have invented an opaque response generation approach, enabling virtual services to be created from network traffic without the existence of experts or expert knowledge. In order to answer these research questions, we evaluate the feasibility and efficiency of this approach to creating accurate virtual service models.

\section{Research Contributions}
This thesis demonstrates that:

\hspace{0.5cm}
{\em Opaque response generation} is an effective and efficient approach to automatically creating virtual services capable of acting as and appearing to be real target service, enabling real services to be emulated and replaced in the absence of experts or expert knowledge. This approach is accurate, efficient and robust at producing responsive behaviours of target services so that the software development team can get continuous access to required services for testing purpose.

More specifically, this thesis contributes:

\label{chap1:contributions}
\begin{enumerate}

    \item A framework for the opaque response generation approach described in Chapter~\ref{chap4:framework}, which simulates the creation of virtual services at the raw bits and bytes level. We define some notations that are needed to formalise our research problem and describe our framework. We also present an overview of the architecture for our opaque response generation approach. Following paper describes this work in detail:
    \begin{itemize}
        \item
        ``Automatic Generation of Interaction Models for Enterprise Software Environment Emulation,'' in the Doctoral Symposium allocated at the 22nd Australasian Software Engineering Conference. (ASWEC 2013)
     \end{itemize}

    \item An opaque message matching approach described in Chapter~\ref{chap5:qosa} which implements the automatic creation of virtual service models. By employing a genome-sequencing alignment algorithm and a common field substitution algorithm, the opaque message matching approach generates responses directly from network traffic, enabling service to be virtualised without requiring knowledge of the underlying protocol or message structure. Following paper describes this work in detail:
    \begin{itemize}
        \item
        ``Generating service models by trace subsequence substitution,'' in Proceedings of
        the 9th International ACM Sigsoft Conference on Quality of Software Architectures. (QoSA 2013)
    \end{itemize}

    \item A method for clustering a large collection of network traffic for efficient opaque response generation described in Chapter~\ref{chap6:SoftMine}, which allows responses to be generated in real time. This approach extends the initial opaque response generation approach by applying clustering techniques to organise the same type of messages into the same group before the response generation process, thereby making the response generation much more efficient at runtime. A conference paper was published describing this approach:
        \begin{itemize}
            \item
            ``Interaction traces mining for efficient system responses generation,” in Proceedings of the 2nd International Workshop on Software Mining. (SoftMine 2013)
        \end{itemize}

    \item A method of producing a cluster prototype expression to describe the cluster for efficient, accurate and robust opaque response generation described in Chapter~\ref{chap7:consensus}. This approach uses a bioinformatics-inspired multiple sequence alignment algorithm to derive message prototypes from the network traffic collection and a scheme for response generation. The following paper describes this work:

        \begin{itemize}
            \item
            ``From network traces to system responses: opaquely emulating software services”, Technical Report, Swinburne University of Technology, 2015
        \end{itemize}

\end{enumerate}

\section{Research Method}
\label{chap1:method}

To perform our research, some key steps were undertaken as outlined as follows.

\begin{itemize}
    \item We conducted a literature review of approaches to providing production-like environments for testing purpose, and identified that service virtualisation is a promising technique for the provision of an interactive approximation of the enterprise software environment. However, we also found that creating virtual service models is a big challenge as it relies on the availability of protocol experts and expert knowledge. The target of our work is to address these problems with existing service virtualisation techniques.

    \item We investigated widely used application-level protocols to identify requirements for virtual service models interacting with a system-under-test.

    \item We invented an approach to automatically create virtual service models by generating responses by the analysis of network traffic.

    \item We collected a wide variety of network traffic using network sniffer tools.

    \item We developed a prototype for automatic response generation.

    \item We refined this automatic response generation approach by clustering network traffic using offline analysis.

    \item We further refined this new automatic response generation approach by producing a message consensus prototype from each cluster of network traffic, enabling the runtime processing to be more accurate and efficient.

    \item We then evaluated the refined approach with four enterprise messaging protocols.

    \item Finally, we drew some conclusions from our review, prototypes, refinement and evaluations.

\end{itemize}

\section{Thesis Overview}
\label{chap1:overview}

The following chapters are organised as follows.

Chapter~\ref{chap2:relatedwork}: Related Work

This chapter introduces the related work to this research. We start with introducing major issues of existing approaches to producing interactive assessment environments in testing distributed software systems, as well as covering the major issues in testing distributed software systems, and then we move to service virtualisation technique, which is proposed to address those issues. By introducing major steps that the service virtualisation works for creating interactive virtual service models, we raise the issue of creating interactive virtual service models in the absence of experts or expert knowledge. At last, we introduce research in network message structure reverse engineering, which is an important reference to our work.

Chapter~\ref{chap3:requirements}: Protocol Investigation and Identified Requirements

This chapter compares some popular application-layer protocols which are used by software services to transmit messages, and then identify some requirements for producing valid interactive service models.

Chapter~\ref{chap4:framework}: Framework for Automatic Opaque Response Generation

This chapter presents a conceptual framework for our automatic opaque response generation approach. The framework includes three important components, which are the {\em analysis function}, the {\em matching function} and the {\em substitution function}. By introducing each component in detail, it gives a high-level overview of the opaque response generation approach to creating interactive service models from network traffic.

Chapter~\ref{chap5:qosa}: Opaque Response Generation Approach

This chapter describes an implementation of the opaque response generation approach, creating interactive service models directly from network traffic, recorded between a system-under-test and a software system service it depends upon.
%The network traffic . By utilising a genome sequencing alignment algorithm and a field substitution algorithm, which is to automatically create interactive virtual service models.

Chapter~\ref{chap6:SoftMine}: Interaction Trace Analysis

In this chapter, we develop an approach to processing and analysing network traffic during the offline processing, utilising data mining techniques. It can dramatically reduce the number of interactions to be searched, from the whole network traffic library to the number of clusters, thereby speeding up the matching procedure. It can enable responses to be generated efficiently at runtime.

Chapter~\ref{chap7:consensus}: Cluster Consensus Prototype Inference

In this chapter, we refine the interaction trace analysis approach by producing a representative consensus prototype which is able to robustly captures most of the important information contained in the cluster messages. In addition, a modified Needleman-Wunsch algorithm is introduced for distance calculation during message matching for increased accuracy, as well as high efficiency.

%; 3) apply the refined distance function to cluster either request messages or response messages to achieve homogeneous clustering.

Chapter~\ref{chap8:conclusion}: Conclusion and Future Work

This chapter summaries the new ideas presented in this thesis, and discusses consequent further research works. 
\chapter{Related Work}
\label{chap2:relatedwork}

% service virtualisation and DevOps
% the purpose of this chapter is to 1) compare enterprise software environment emulation against tradition approach to providing a testing environment; 2) discuss current approach to creating interactive service models; 3) introduce existing approaches that
% Define devops, challenges that devops teams have to meet Why devops
``DevOps'', short for Development-Operations, is a new software development method which enables development and operations teams to communicate and collaborate more effectively~\cite{bass2015devops}. It represents the most recent industry efforts aimed at providing realistic operating environments for system development/testing to achieve continuous delivery of software~\cite{chen2015continuous}. %High performance teams which embrace DevOps have been demonstrating that they can deliver quality systems with amazing speed and agility. Many DevOps teams focus on scripting the application build, package and deployment process, resulting in an automated framework, known as the deployment pipeline. Deployment automation enables the organization to enjoy the benefits of continuous integration and continuous delivery, significantly enhancing productivity and agility.
%
%The time available for delivery of such an integration is often so short that building a new system is not an option and therefore existing system components have to be integrated into a distributed system that appears as an integrating computing facility. Secondly, the time available for providing new services are decreasing.
%
%
%More and more {\em DevOps} organisations are adopting {\em service virtualisation} to produce an interactive test environment for software quality assurance.
The service virtualisation technique is a novel DevOps approach where the target services of a system under test are replaced by virtual service models~\cite{servicevirtualisation}, allowing engineers have continuous access to the services needed to test a system. However, it is still a new research topic in computer science and software engineering. To the best of our knowledge, there is no directly relevant work. To define the scope of our research work, the approach we have adopted is to review literature from different areas, selecting work from each which appears most relevant to our topic. A number of existing tools are capable of achieving some of our goals, and hence are reviewed as part of this chapter.

The structure of this chapter is as follows. In Section~\ref{chap2sec:motivation}, we describe a concrete industry scenario where the software quality team requires a production-like interactive testbed environment for assessing run-time properties of the software-under-test. In Section~\ref{chap2sec:interactiveenv}, we discuss some existing approaches and tools for producing an interactive testbed environment, and identify their major limitations. In Section~\ref{chap2sec:envemulation}, we introduce a latest technique, namely {\em service virtualisation}, which is to provide a production-like interactive testbed environment. This testbed environment is also known as an emulation environment where requests sent by an external system-under-test are received and processed by virtual service models to generate responses. In Section~\ref{chap2sec:modelcreation}, we analyse some approaches to creating virtual service models and raise the major issues of these approaches. In Section~\ref{chap2sec:reverseengineering}, we present the reverse engineering technology which is the most important reference to our work.

\section{Industry Scenario}
\label{chap2sec:motivation}

\begin{figure}[tb]
  \centering
  \includegraphics[width=\columnwidth]{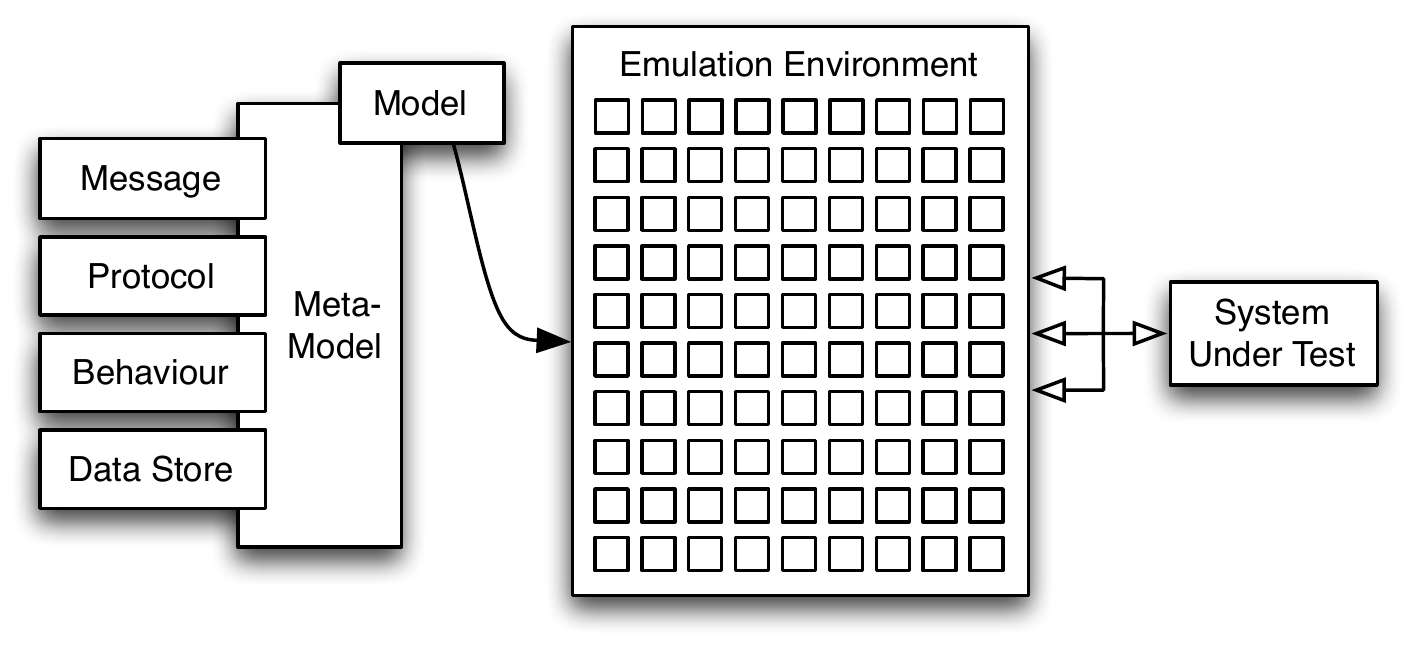}
  \caption{Service Emulation Approach.}
  \label{Chap2fig:service-emulation}
\end{figure}

CA Identity Minder (IM) enterprise system~\cite{IdentityManager} is an enterprise-grade identity management suite which helps large organisations to manage users' identity and control access of computational resources in the enterprise's environment. It is used by large corporations to manage identities and access permissions for users. These identities and permissions are then applied to the enterprise's environment by IM interacting with required computational resources within the enterprise environment using whichever application-level protocol is supported by each resource.

Consider IM, our
\emph{system-under-test}, in a proposed deployment environment, as shown in
Figure~\ref{Chap2fig:service-emulation}. Engineers want to evaluate whether IM
can scale to handling the number of likely computational resources (or called \emph{endpoints}) in the deployment
scenario, given (i) likely maximum number of endpoints; (ii) likely maximum
number of messages between endpoint and system; (iii) likely frequency of
message sends/receives needed for the system to respond in acceptable
timeframe; (iv) likely size of message payloads given deployment network
latency and bandwidth; and (v) the system's robustness in the presence of
invalid messages, too-slow response from end-points, or no-response from
end-points. Messages being exchanged between the system and endpoints adhere
to various protocols. For example, an LDAP message sent by an endpoint to IM needs to be responded to with a suitable response message sent in reply by IM, in an acceptable timeframe and with acceptable message payload.
Subsequent messages sent by the endpoint to IM using the LDAP response
message payload need to utilise the previous response information. A major challenge when attempting to assure the quality of highly interconnected enterprise software systems is to produce a suitable testbed environment.

\section{Traditional approaches for providing a testbed environment}
\label{chap2sec:interactiveenv}

Over the years, a number of tools and approaches have been proposed to provide testing environments suitable for quality assurance
activities required to test enterprise software systems. These tools and approaches fall into the following areas: 1) physical replication of the deployment environment, 2) hardware virtualisation tools, 3) mock testing, 4) software system performance and load testing tools and 5) network emulation.

\begin{enumerate}
  \item {\textbf{\em Physical replication of a deployment environment}}

  {\em Physical replication of a deployment environment} is a way of producing a ``typical'' physical deployment environment, using appropriately configured real environment systems. This process involves construction and configuring every aspect of a real software environment from bottom to top. Specifically, software quality assurance teams manually construct and configure an alternative environment with all realistic settings of the production environment where a system under test is intended to operate in. Then, operation of the system under test within the replicated environment can be observed and assessed.

  Physical replication is the most straightforward way of providing an interactive test-bed environment. Even if this approach is significantly labour intensive, it can be well-suited in certain situations. For example, production environments are of modest scale and complexity, or the necessary resources and expertise are readily available. As the replicated environment copies all configurations of the real environment,  interactions between the system under test and the replicated environment are similar to those which occur a real environment. This guarantees accuracy and consistency of the results of any test undertaken in such environments.

  In many situations, however, physical replication is not applicable, especially for large enterprise environments. As a host machine is required by every software system present in the real environment, the scale and complexity of many enterprise environment make physical replication extremely expensive and time consuming. Furthermore, the replicated environment is very rigid. Once configurations of the real environment have been modified, the replicated environment has to be reconfigured accordingly, which requires significant efforts. Additionally, it is difficult to reuse replicated host machines. Although replicated machines are reusable, they may require physical relocation and manual reconfiguration. Real system data may not be available due to privacy issues.

   \item {\textbf{\em Hardware virtualisation tools}}

  Hardware virtualisation tools, such as VMWare Workstation~\cite{Sugerman:01} and VirtualBox~\cite{website:VirtualBox}, are popularly used in industry to create virtual machines for providing interactive representations of enterprise software environments. A {\em virtual machine (VM)} is a software implement of a computer system where programs operate the same as on a physical machine. Each virtual machine image has all interactive features of a full physical machine, requiring fewer resources.

  Virtual machine technology has numerous advantages over the physical replication approach to providing interactive test environments. Firstly, virtual machines are easier to manage and maintain than physical replication approach. Once the virtual machine is properly configured, a snapshot is taken to preserve its initial proper sate. If anything goes wrong while operating, the initial state with proper configuration can be easily recovered. In addition, as virtual machines require less resources of their host physical machine, they are much scalable than their physical counterparts. A single physical host is able to execute multiple virtual machines simultaneously. For high-end server systems, one may be able to fit as many as hundreds of virtual machines on a single host. Although the scalability of the virtual machine technology is superseding the physical replication approach, it is still insufficient for representing larger enterprise environment, consisting of thousands or even tens-of-thousands of individual machines. The same issues of complex setup and data availability exist.

  \item {\textbf {\em Mock testing}}

  In object-oriented programming, mock objects are simulated objects that mimic the behavior of real objects in controlled ways~\cite{freeman2004mock}. Mock testing is a way that a programmer creates a mock object to test the behavior of some other object. It is becoming a common and standard integrated programmatic technique to mimic the interactive behaviour of real remote systems~\cite{mostafa2014empirical}. There are currently mocking frameworks available for practically all popular programming languages, such as Mockito~\cite{acharya2015mockito} for Java, RSpec~\cite{rspec} for Ruby, and Mockery~\cite{Mockery} for PHP etc. By applying this integrated programmatic techniques developers can have representations of distributed software environments which are highly controllable and immensely useful for performing functional unit testing.

  There are however, a couple of key limitations when interactive representations of distributed environments are provided through mock testing approaches. Firstly, the code of mock objects are typically coupled with the code of the software under investigation. A mock object, for example, is generally a local object in the object-oriented programming language sense, which the software under (test) investigation interacts with in place of another object that is used in deployment that provides an interface for interaction with the real distributed system. Coupling the interactive environment with the code under investigation is only possible if the code itself is available, which is not necessarily the case.

  Secondly, mock objects are usually written by developers, occasionally shared with QA teams and very rarely shared with other development teams due to the inter-operability problems. The lack of sufficient knowledge of objects' interactive behaviour becomes a big issue for those people to do mock testing. Furthermore, the results of an analysis of the run-time properties of a software system intended to interact with distributed software systems, while it interacts with local code in lieu of real distributed systems, are likely to be considerably inaccurate. %All tasks, establishing and terminating connections, sending and receiving messages, are more complex and can take unpredictable lengths of time longer when being carried out on a distributed computer network.

  \item {\textbf {\em Software application performance testing tools}}

  Software application performance testing is the process of determining how an software is performing and its impact on the other software~\cite{performanceengineering}. There are a number of aspects to be measured when undertaking performance testing, which can be divided into two types: {\em service-oriented} and {\em efficiency-oriented}. Service oriented aspects include {\em availability} and {\em response time}. They measure how well a software system is performing and providing services to other users. Efficiency-oriented aspects contain {\em throughput} and {\em utilization}. They measure how well a software system utilizes resources of its deployment environment.

  Performance testing tools provide a means to evaluate the performance and scalability of server software applications. Tools such as HP’s LoadRunner~\cite{HPLoadRunner}, the Apache
  Software Foundation’s JMeter~\cite{JMeter} and  Appvance~\cite{appvance} operate to emulate many thousands of concurrent client systems, sending requests to some server systems under test, and then measure if responses sent back from server systems under test meet performance goals. Some more advanced performance testing tools are able to provide in-deep performance diagnosis of the large-scale system-under-test target, such as AgileLoad~\cite{AgileLoad} and Selenium~\cite{Selenium}. AgileLoad~\cite{budai2013performance} features functions like automatic modeling, real time data correlation, anomaly diagnostic and recommendations, enabling automatic identification of performance bottlenecks in the systems. Selenium~\cite{monier2015evaluation} is an automated testing suite for web applications across different browsers. It provides a record/playback tool for authoring tests without learning a test scripting language (Selenium IDE). It also provides a test domain-specific language (Selenese) to write tests in a number of popular programming languages, including Java, C\#, Groovy, Perl, PHP, Python and Ruby so that the tests can then be run against most modern web browsers. These performance testing tools enable testing automation, and particularly useful in generating scalable client {\em load} towards some server system under test. They are not, however, well-suited to provide a suitable test environment with system-under-test to environment load scaling {\em client-to-server load}. They require detailed knowledge of target protocols and suitable datasets.

  There are also a lot approaches to evaluating whether the software system satisfies the user performance goal~\cite{performanceengineering}. Modelling is the key technique applied in this field. Some of works are able to generate executable distributed test-beds from high-level models of the system. The SoftArch/MTE testbed~\cite{softarch}, for example, is intended to help evaluate the performance of different architecture, design and deployment choices for a system during system development. MaramaMTE~\cite{marama2008} is another tool with similar capabilities. The architecture view of MaramaMTE allows specification of relationships between proposed services and a database. This information is used in the synthesis of code which mimics proposed service behaviour replacing the real service implementation for testing purposes. These tools provide a means to emulate many thousands of software system clients with limited resources. The limitation of these tools is they are designed to generate scalable client load towards a target system, rather than the opposite direction needed for our problem situation.

  \item {\textbf {\em Network Emulation}}

  In the field of computer network, simulation and emulation techniques are popularly used for performance evaluation of real distributed protocol implementations~\cite{networkemulation}. Network emulation is used to assess the performance of a network under test or to predict the impact of possible changes under controlled and repeatable conditions which mimic those found in real deployments~\cite{emulatinsurvey}. A network emulator, therefore, is able to interact with real network traffic. A network simulator, on the other hand, tries to model all the details of the real world networks so as to allow researchers to assess performance of the network under test~\cite{simulationsurvey}. It is closed and does not interact with any real external network devices.

  Researches used WAN emulator~\cite{TCPVegas} to perform part of an evaluation of the TCP Vegas congestion control algorithm. This emulator required a lot of manual effort and many physical machines. Later on, Dummynet~\cite{dummynet} was introduced, enabling the emulation of multi-hop network topologies on a single physical machine. In 2002, ModelNet~\cite{modelnet} was proposed, which addressed issues of scalable network emulation. ModelNet runs at the top of a cluster infrastructure and allows researchers to deploy unmodified distributed application implementations into large-scale network environments. In the following year, NIST Net~\cite{NISTNet} features emulation of diverse performance dynamics over high levels of system load using a single commodity class Linux machine.

  Although NIST Net only requires limited resources for the provision of a network emulation, the network emulator techniques are not directly applicable to tackle our problem. This is because network emulator techniques are focused on emulating underlying communication infrastructure of networks, which is lower that which we are interested. The aim of our research is to emulate interactions at the service or application-layer of the OSI reference model~\cite{osi}. The different levels of abstraction lead to some different major problems. In particular, the actual payload of packet content handled by network emulators is not a big issue, so long as it is consistent, i.e. has a valid checksum. At the application-layer however, some payload information need to consider. The message identifier of a response for instance, may need to match the message identifier of the corresponding request.

  \end{enumerate}

  In conclusion, although these approaches have addressed some specific aspects of providing a production-like testing environment, neither of them is capable of generating an interactive test-bed environment, representing large-scale heterogeneous production-like environment.

\section{Enterprise software environment emulation to provide a large-scale interactive test-bed environment}
\label{chap2sec:envemulation}

To overcome shortcomings of these existing approaches and tools in providing a production-like test environment, some researchers have proposed to create ``virtual'', or emulated, deployment environments. In~\cite{Ghosh99issuesin}, Ghosh and Mathur state that ``an
emulation environment, which is able to provision representations of diverse
components, enables tests of components that interact with a heterogeneous
environment, while scalable models put scalability and performance testing for
components into practice''. Enterprise software environment emulation is a promising approach to providing an executable, interactive testing environment where quality assurance teams monitor and assess run-time properties of a system under test. Some work have been done for An enterprise software
environment emulator, called Kaluta, was proposed in~\cite{hine:thesis,hine:09a,hine:10a}. It provides a large-scale and
heterogeneous emulation environment capable of simultaneously emulating
thousands of endpoint systems on one or a few physical machines. They have shown its scalability meets the needs of enterprise system QA as outlined in Section~\ref{chap2sec:motivation}. However, the
creation of executable endpoint models relies on the availability of a precise
specification of the interaction protocols used. It is also time consuming, error-prone and subject to considerable maintenance effort in heterogeneous deployment environments.

CA Service Virualization~\cite{Michelsen:11} is a commercial software tool which aims to
emulate the behaviour of services which a system under test interacts with in
its deployment environment. It does this by `mimicking' responses that a real
service would produce when sent a request by the enterprise system under test.
One of the key features of LISA is that, after recording a set of real
interactions between an enterprise system and an endpoint, it uses these to
produce responses to further requests, thus behaving like a `virtual' service.
Service Virtualisation is able to consider the interaction state when sending a response, and
uses field substitution (called \emph{magic strings}) in the responses for
fields it detects are identical in the request and response. However, it requires
the transport protocol and the service protocol to be known in detail and in advance of the
recording for the modelling to be effective.

\section{Creating interaction models for enterprise software environment emulation}
\label{chap2sec:modelcreation}

There are two common ways of setting up an interaction model for service virtualisation. The first is to use the communication contract ({\eg} a WSDL or other protocol-specific description) and create interaction models by defining responses. This can be done manually by using normal programming languages, such as Java or .NET. An advantage of this method is that the software test team can conduct test in parallel with the software development team, without waiting for the real version of the service to be completed. However, the disadvantage is that the communication contract is not always publicly available. Even if the communication contract is available, challenges with this include the time-consuming and sometimes error-prone processes of developing the models, lack of precision in the models especially for complex protocols, and ensuring robustness of the models under diverse loading conditions.

Recording and replaying network traffic is the other way of creating interaction models. A tool sits between a system under test and target services. Basically, This tool acts as a proxy, recording request messages sent by the system under test to the target services and response messages sent in the opposite direction. Then, these recordings are used to `mimic' the real service response messages in the emulation environment. There have been fully automatic approaches like the RolePlayer system~\cite{cui:06}, in which the authors automatically modelled the server responses for various protocols to fool attackers into attempting to exploit vulnerabilities. Similarly, Ma {\etal}~\cite{ma2006} proposed an approach that attempts to classify packet streams into protocols based on learned Markov models. While these more
automatic methods are able to effectively fool attackers and identify protocols, the models they learn are not sufficient to interact with external system under test.

\section{Protocol Reverse Engineering}
\label{chap2sec:reverseengineering}

Our research is mainly inspired by the work in the protocol reverse engineering area.
To assist developing reusable service models, approaches target to either reverse engineer message structures %\cite{messageformatinference}
\cite{cui:07a}, or discover/mine processes~\cite{processmining} for building behavioural models. While these allow engineers to develop more precise models, none of them can automate interaction between enterprise system under test and the emulated dependency services.

%%!!%% This paragraph was incomplete - need to fix quickly...
Research in protocol reverse engineering is an important reference to our
work. %% It has been attempted in various domains.
Early effort in reverse engineering was for protocol determination. By
analysing a large amount of packets and traces captured on networks,
researchers are able to obtain structure information of the target protocol
for network analysis % \cite{beddoe:2004}
and even automatically reverse engineering the state-machine model of network
protocols \cite{Comparetti:2009}. Cui {\etal} \cite{cui:06}, for example,
proposed an emulator aiming to mimic both client and server side
behaviours. With the emulator, they can record/replay the interactions of web
applications for checking conformance of web server behaviours. Although the
proposed approach deals with the emulation of interaction process, they
essentially intend to test conformance of the client-side systems rather than
work in the opposite direction as we do.

\section{Summary}
\label{chap2sec:summary}

In this chapter, we reviewed some literatures related to service virtualisation approaches for providing enterprise environment emulation. We start with introducing an industry scenario for scoping our research problem. By discussing some existing approaches and tools, we introduce that environment emulation is the most promising approach to address the issue in that scenario. Service virtualisation is a widely used approach for the provision of an emulated environment. We analyse major issues of current service virtualisation approaches, which are what we aim to deal with. At last, we introduce some works about protocol reverse engineering which is an important reference to our work. 
\chapter{Application-layer Protocol Study}
% Main LaTeX file for ASWEC 2013 doctoral symposium submission
%Requirement for service emulation
\label{chap3:requirements}

In this chapter, we survey some widely-used application-layer protocols, in particular, their message structures and interactive behaviours, enabling us to define clear requirements that virtual services have to meet for producing valid responses. These protocols were chosen because: 1) they are widely used in the enterprise environment; 2) they are representative protocols and have distinguishing features; and 3) their documents are publicly available so that we can use this knowledge to define criteria for validation.

The rest of this chapter is organised as follows: In Section \ref{sec:overview}, we give an overview of the application layer services and protocols. Then, we introduce features of some popular protocols in detail in Section~\ref{sec:protocols}. This is followed by a discussion of protocol features in Section~\ref{sec:taxonomy}. %In Section~\ref{chap3:traceexample}, we introduce our way of getting Twitter (a social networking application) traffic.
Finally, we summarise this chapter in Section~\ref{chap3sec:conclusion}.

\section{Overview of Application Layer Functionality and Protocols}
\label{sec:overview}

In the realm of computer networks, a networked application refers to programs that run on different systems (host systems) and communicate with each other over the wire. When developing a new network application, the application developer should first plan the application architecture that dictates how programs of the application are structured over the various end systems. There are three predominant architectural paradigms used in modern network applications: the client-server architecture, the peer-to-peer (P2P) architecture and the message-oriented middleware.

In client-server architecture~\cite{CSmodel}, there is a host, called the server system, which serves requests from many other hosts, called client systems. A classic example is the Web application for which an Web server services requests form browsers running on client hosts. There are two characteristics for the client-server architecture: 1) clients do not directly communicate with each other; 2) the server has a fixed, well-known address so that a client can always contact the server by sending a packet to the server's address.

In P2P architecture~\cite{P2Parch}, the application exploits direct communication between pairs of intermittently connected hosts, called \textit{peers}. There are several characteristics for this architecture: 1) all nodes are both clients and servers; 2) nodes are autonomous and able to collaborate directly with each other; 3) network is dynamic, that is, nodes enter and leave the network frequently.

%The message-oriented middleware is software or hardware infrastructure supporting messages are sent and received between distributed systems~\cite{curry2004message}.
A message bus~\cite{curry2004message} is a logical component that allows different systems to communicate through a shared set of interfaces. It enables separate applications to work together in a decoupled fashion so that applications can be easily added or removed without affecting the others. A message bus contains three key elements: 1) a set of message schemas; 2) a set of common command messages; 3) a shared infrastructure for bus messages to recipients.

% To describe the protocol specification
For our work in this thesis, we only consider protocols using the client/server paradigm. The P2P architecture and the message bus are tasks we intend to cover in future work. A client/server protocol consists of a number of messages which are passed between respective peers and establish the rules for dialogue exchanges when a communication requirement between any two end users is initiated. Any messages, which are being directed to their respective peer entities, are then passed between end systems using the connected physical medium and actioned according to the relevant procedures and semantics of each protocol concerned. Entities (processes or modules) local to each user communicate with one another via the services of the lower layer. The interaction among entities in providing services of the layers constitutes the actual protocol. Hence a protocol specification must describe the operation of each entity within a layer in response to commands from its users, messages from the other entities (via the lower layer service, also called Protocol Data Units (PDU)), exchange of interactions (via service primitives) with service users at service access points, and internally initiated actions ({\eg}timeouts).

Thus, a protocol specification typically includes: a list of the types and formats of messages exchanged between the entities (the static aspect); and the rules governing the reaction of each entity to user commands, messages from other entities, and internal events (the dynamic aspect).

After choosing the application architecture, the developer also needs an application-layer protocol that is able to regulate communication among two or multiple end systems. An application-layer protocol defines message formats that are composed of a sequence of fields organized with certain rules and that can take values from a given domain. It also identified the order in which the message can be transmitted while programs execute. In particular, an application-layer protocol provides:

\begin{itemize}
  \item %The types of messages exchanged, for example, request messages and response messages
      {\em Knowledge of protocol syntax}

      Protocol syntax is a mechanism for participants to identify and response with each other. Moreover, it provides a set of rules for formatting messages that specify how data is transformed into messages sent and received.

  \item {\em Knowledge of Protocol State-machine}

    A protocol state-machine characterises all possible legitimate states and sequences of messages. A state machine of a protocol can be considered as:
    \begin{itemize}
        \item {a set of input messages,}
        \item {a set of output messages,}
        \item {a set of states,}
        \item {a function that maps states and input to output,}
        \item {a function that maps states and inputs to states (which is called a state transition function), and}
        \item a description of the initial state.

    \end{itemize}

\end{itemize}

Based on the message format definition, an application-layer protocol can be either textual or binary. Textual protocols use ASCII text to
encode both the structure and the content of messages in the human-readable form, while binary protocols use machine structures to encode
data, which is intended or expected to be read by a machine rather than a human being.

\section{Protocol Study}
\label{sec:protocols}

\begin{figure}
  \centering
  % Requires \usepackage{graphicx}
  \includegraphics[width=0.75\textheight,height=0.8\textwidth,angle=90]{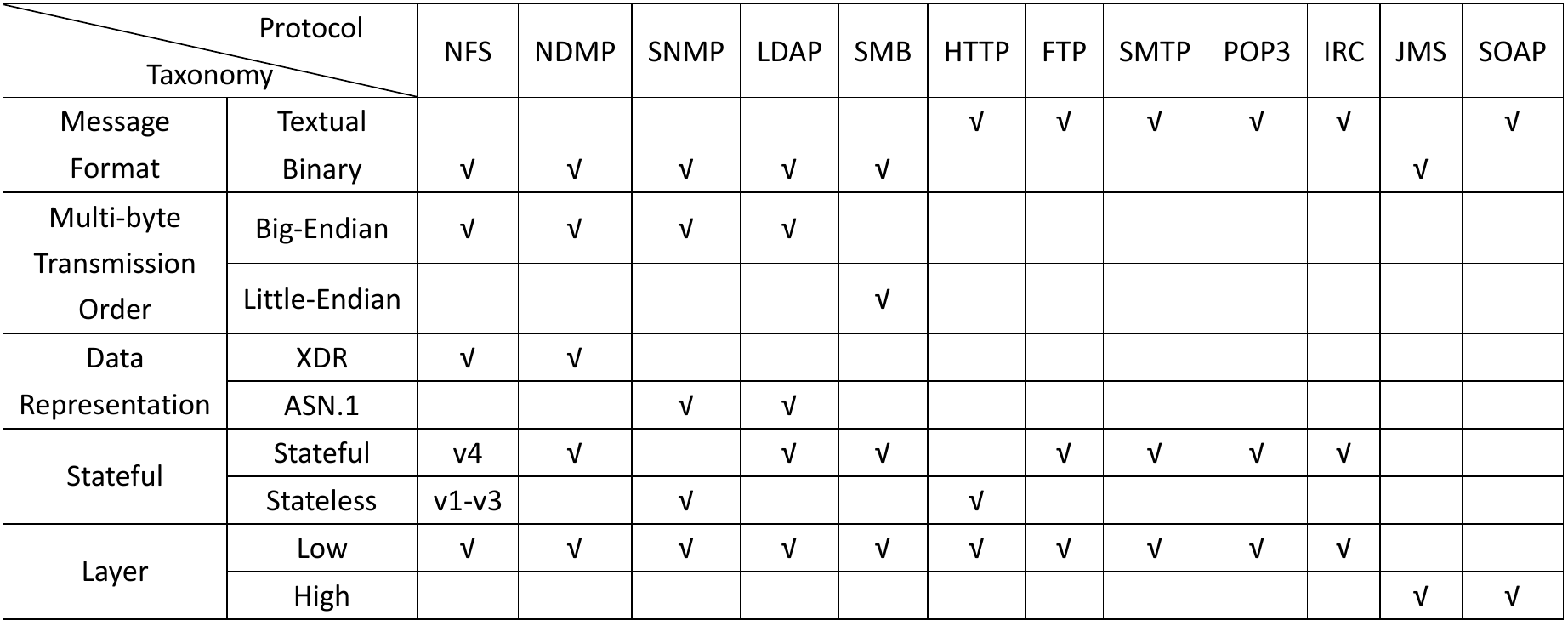}\\
  \caption{Protocol Summary Table}\label{chap3:protocolsummary}
\end{figure}

We investigate twelve popularly used application protocols, including five binary protocols, five textual protocols, and two protocols that need to be transmitted over the other protocols. Following introduce each protocol with their key features, which are summarised in Figure~\ref{chap3:protocolsummary}. %Besides these, we introduce 2 high-layer protocols, which are Java Message Service (JMS) and Simple Object Access Protocol (SOAP), respectively. The Summary of protocols based on previous taxonomy can be
%seen in Appendix.

\subsection{Network File System (NFS)}

Network File System (NFS) \cite{NFS} is a distributed file system protocol, allowing a user on a client system to access file over a network in a manner
similar to how local storage is accessed. NFS builds on the Open Network Computing Remote Procedure Call (ONC RPC) system for this protocol.

\begin{itemize}
  \item NFS version 1-3 were intended to be as stateless as possible. That is, a server should not need to maintain any protocol state
   information about any of its clients in order to function correctly.
  \item NFS version 4 became stateful as it introduces new operations, such as opening, locking, reading, and writing. Therefore, the
  usage information of an object by an NFSv4 client is maintained by the server.
  \item XDR and RPC are used for the NFS protocol.
  \item RPC authentication protocols are allowed, among which the RPCSEC\_GSS security flavour must be used to
   enable the mandatory security mechanism.
\end{itemize}

\subsection{Network Data Management Protocol (NDMP)}

Network Data Management Protocol (NDMP) \cite{NDMP} is used to transport data between network-attached storage (NAS) devices and backup services. This protocol:

\begin{itemize}
  \item Uses client-Server paradigm
  \item The message protocol is specified with the XDR language
  \item Stateful protocol
  \item Uses asynchronous messages and a message does not require a reply, however, many of the messages may result in a reply message
  \item A message consists of a message header that is optionally followed by a message body
  \item Message header contains a message identifier
\end{itemize}

\subsection{Simple Network Manage Protocol (SNMP)}

Simple Network Manage Protocol (SNMP) \cite{SNMP} is the dominant protocol in network management. It allows an SNMP manager (the controller) to control
an SNMP agent (the controlee) by exchanging SNMP messages. The purpose of an SNMP message is to get and set parameters on an SNMP agent.
A parameter is an instance of a more generic object. This is:

\begin{itemize}
  \item A request/response protocol
  \item Uses ASN.1 to define the data types used to build an SNMP message
  \item Stateless protocol
  \item Has a \emph{msg}\_flags field in the variable-length header to indicate whether the encryption/authentication is used to protect the privacy/authenticity of
  the message

\end{itemize}

\subsection{Light Weight Directory Protocol (LDAP)}

Light Weight Directory Protocol (LDAP) \cite{Sermersheim2006} is used for accessing and maintaining distributed directory information services over an Internet Protocol (IP)
network. This protocol:

\begin{itemize}
  \item Uses single-request/multiple-response paradigm
  \item Uses ASN.1 to describe the protocol
  \item Stateful protocol
  \item The function of the Bind operation is to allow authentication
   information to be exchanged between the client and server. Anonymous binds or SASL authentication is allowed.
\end{itemize}

\subsection{Server Message Block (SMB)}

Server Message Block (SMB) \cite{SMB}, also known as Common Internet File System (CIFS) operates as an application-layer network protocol mainly used for providing
shared access to files, printers, serial ports, and miscellaneous communications between nodes on a network. This is:

\begin{itemize}
  \item A single-request/single-response protocol
  \item Stateful protocol
  \item A SMB message can be broken down into three parts:
  \begin{enumerate}
    \item A fixed-length \textit{header}, which has 32-byte length, contains a command field that indicates the operation code that the client requests to or
    to which the server responds
    \item A variable-length \textit{SMB Parameters Block} contain 1-byte \textit{length} field that indicates the size of the following content
    \item A variable-length \textit{SMB Data} contain 2-byte \textit{length} field that indicates the size of the following content
  \end{enumerate}
  \item Multibyte fields in an SMB message must be transmitted in little-endian byte order

\end{itemize}

\subsection{Hypertext Transfer Protocol (HTTP)}

The Web clients and Web servers, executing on different systems, cooperate with each other by exchanging HTTP messages \cite{HTTP}.
\begin{itemize}
  \item A request/response protocol
  \item Stateless protocol
  \item Uses text-based ``Field:Value'' pairs formatted according to RFC 2822
  \item Request message
  \begin{itemize}
    \item Begin with a single Start\_Line specifies the operation, the web page and the HTTP version
    \item One/multiple subsequent Header lines
    \item Each line ends with a carriage return and a line feed
    \item Request operations: get(most), head, post, put, delete, trace, connect
    \item Blank line indicates the end of request headers
  \end{itemize}
  \item Response message
  \begin{itemize}
    \item First line specifies HTTP version, three-digit code and a text string
    \item Blank line indicates the end of response head
    \item Unstructured body data follows, with specified size
  \end{itemize}
\end{itemize}

\subsection{File Transfer Protocol (FTP)}

File Transfer Protocol (FTP) \cite{FTP} is a standard network protocol used to transfer files from one host to another host over a TCP-based network, such as the Internet. This is:

\begin{itemize}
  \item Built on a client-server architecture and uses separate control and data connections between the client and the server.
  \item Stateful protocol
  \item Text-based ``Field:Value'' pairs formatted according to RFC 2822
\end{itemize}

\subsection{Simple Mail Transfer Protocol (SMTP)}

Simple Mail Transfer Protocol (SMTP) \cite{SMTP} is an Internet standard for electronic mail (e-mail) transmission across Internet Protocol (IP) networks. Email is submitted
by a mail client to a mail server using SMTP. This protocol:

\begin{itemize}
  \item Uses line-by-line request-response pattern
  \item Stateful protocol
  \item Uses text-based ``Field:Value'' pairs formatted according to RFC 2822
  \item All requests are four characters followed by data
  \item All responses are numeric followed by explanatory text
  \item The message structure, that is, submitted E-mail, is similar to HTTP message structure, which consists of header, blank line, then body
\end{itemize}

\subsection{Post Office Protocol (POP3)}

The Post Office Protocol (POP) \cite{POP3} is an application-layer Internet standard protocol used by local e-mail clients to retrieve e-mail from a remote server over a TCP/IP connection. This protocol:
\begin{itemize}
  \item Uses line-by-line request-response pattern
  \item Stateful protocol
  \item Uses text-based ``Field:Value'' pairs formatted according to RFC 2822
  \item All requests consist of a case-insensitive keyword, possibly followed by one or more arguments
  \item All responses consist of a status indicator and a keyword
   possibly followed by additional information
  \item The message structure, that is, retrieval E-mail, is similar to HTTP message structure, which consists of header, blank line, then body
\end{itemize}

\subsection{Internet Relay Chat (IRC)}

Internet Relay Chat (IRC) \cite{IRC} is a protocol for live interactive Internet text messaging (chat) or synchronous conferencing. This protocol is:

\begin{itemize}
    \item Servers and clients send each other messages, which may or may not generate a reply
    \item Stateful protocol
    \item IRC messages are always lines of characters terminated with a CR-LF(Carriage Return - Line Feed) pair
    \item Each IRC message may consist of up to three main parts: the prefix
   (OPTIONAL), the command, and the command parameters. The prefix, command, and all parameters are separated
   by one ASCII space character each
   \item Messages are formatted with Augmented BNF
\end{itemize}

\subsection{Java Message Service (JMS)}
\label{sec:jms}

The Java Message Service (JMS) \cite{JMS} provides a Java Message Oriented Middleware (MOM) API specification for sending messages between two or more clients. Apache
ActiveMQ is an open source message broker which fully implements the Java Message Service (JMS). OpenWire is a cross language \textit{wire Protocol} to allow
native access to ActiveMQ from a number of different languages and platforms. The OpenWire transport is the default transport in ActiveMQ. The JMS characteristics include:

\begin{itemize}
  \item OpenWire is used to marshal objects to byte arrays and back. The marshalled objects are referred to as commands.
  \item Every command is encoded a triple of the form \textless size, type, value\textgreater, where \textit{size} (4 bytes) holds how many subsequent bytes are in the
  command, \textit{type} (1 byte) is the command type identifier, and \textit{value} represents data for the command. The most significant difference between this form and
  ASN.1 on-the-wire data representation is the order of size and type's fields.
  \item Every command is encoded in big-endian byte order.
  \item Every command is able to nest other commands in its fields.
\end{itemize}

\subsection{Simple Object Access Protocol (SOAP)}

SOAP is a lightweight protocol intended for exchanging structured information in a decentralized, distributed environment. SOAP uses XML technologies to define an extensible messaging framework, which provides a message construct that can be exchanged over a variety of underlying protocols. The framework has been designed to be independent of any particular programming model and other implementation-specific semantics. SOAP protocol:

\begin{itemize}
  \item Is a way for a program running in one operating system to communicate with a program running in either the same or a different operating system, using HTTP (or any other transport protocol) and XML.
  \item Messages consist of three parts: \em {SOAP Envelope}, \em{SOAP Header (optional)}, \em{SOAP Body}
\end{itemize}

\section{Protocol Taxonomy}
\label{sec:taxonomy}

Based on what we have learnt from above protocols, there are two main encoding methods, binary and textual methods. The binary encoding method targets to a standardized data structure, while the textual method is for text strings. According to the encoding method being employed, an application-layer protocol is either a binary protocol or a textual protocol.

%Each protocol defines a particular message format for exchanging data. These formats can roughly be classified into
%\textit{text} and \textit{binary}. According to the category of a message format, an application-layer protocol
%can be referred to as either the \textit{textual protocol} or the \textit{binary protocol}.

\textit{\textbf{Binary protocols}} rely on specific data structure; and hence, transmitted messages usually resort to fixed-length fields or to a special notation to indicate the length of variable fields.

\textit{\textbf{Textual protocols}} are built around the notion of message fields encoded with text data and separated by known characters, which
are often formatted in the form of ``Field:Value'' pairs. ASCII text is often used to encode both the structure and the content of text messages, following some sort of grammar that is often formalized in Backus-Naur form (BNF).

\subsection{Binary protocols}

Binary protocols include {\em primitive protocols} and {\em advanced protocols}. For most {\em primitive protocols}, they can be further divided based on the multi-byte order and data representations, which are illustrated as follows:

\begin{enumerate}
\item{\textbf{Big-endian order and little-endian order}

Big-endian order and little-endian order describe the order in which the bytes of a multi-byte number are transmitted over the wire.

\textit{\textbf{Big-endian order}} represents the most significant byte is at the highest address. TCP/IP defines big-endian as the standard
network byte order, and thus many higher level protocols over TCP/IP also define its network byte order to be big-endian.

\textit{\textbf{Little-endian order}} means the most significant byte is at the lowest address.
}

\item{\textbf{Network data representations}

One of the most common transformation of network data is from the representation used by the application program into a form that is suitable for transmission over
a network and vice versa. As illustrated in Figure \ref{fig:representation}, the sending program translate the data it wants to transmit from representation
it uses internally into a message that can be transmitted over the network; that is, the data is encoded into a message. On the receiving side, the communicating program
translates this arriving message into a representation that it can the process; that is, the message is decoded.

\begin{figure}[ht]
\centering
  \includegraphics[width=4in]{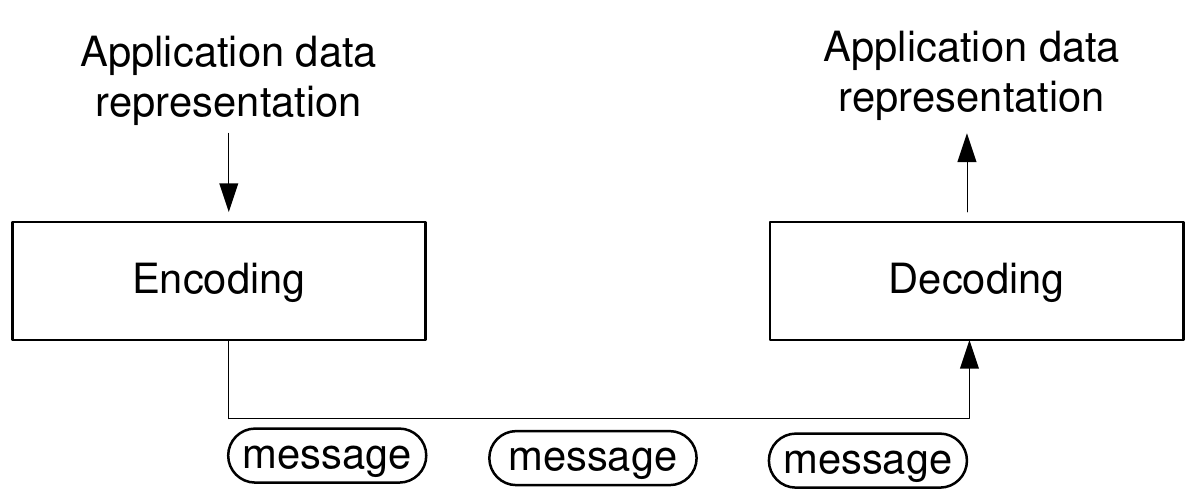}
  \caption{Encoding and decoding application data}\label{fig:representation}
\end{figure}
}

There are two popular network data representations (\ie {\em External Data Representation (XDR)} and {\em Abstract Syntax Notation One (ASN.1)}), proposed to encode primitive data types and compound data structures. Both {\emph XDR} and {\emph ASN.1} specify how primitive data types and compound data structure are encoded so they can be sent from a client program to a server program over the wire.

\begin{itemize}
\item{\textbf{External Data Representation (XDR)}

External Data Representation (XDR) \cite{XDR} is the network format used to transfer data between different kinds of computer systems.

 \begin{itemize}
   \item Supports C-type system without function pointers
   \item 4-bytes (32 bits) base unit
   \item Uses big-endian order for serialisation
   \item Dose not use tags (except to indicate array lengths)
   \item Encodes the components of a structure in the order of their declaration in the structure
 \end{itemize}

 Figure \ref{fig:xdr} shows a structure example and depicts XDR's on-the-wire representation of this structure when the field \textit{name} is four characters long
 and the array \textit{list} has two values in it.

\begin{figure}[ht]
\centering
  \includegraphics[width=4in]{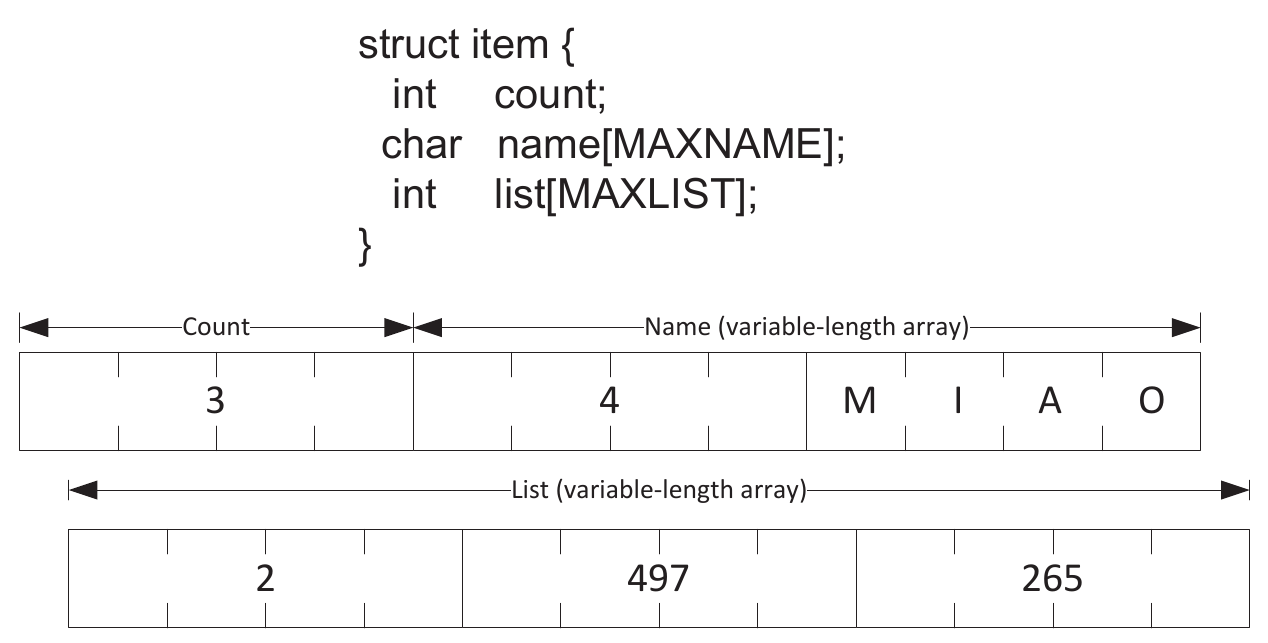}
  \caption{Example encoding of a structure in XDR}\label{fig:xdr}
\end{figure}
}

\item{\textbf{Abstract Syntax Notation One (ASN.1)}

Abstract Syntax Notation One (ASN.1) \cite{ASN.1} is an standard that defines a representation for data sent over a network.
 \begin{itemize}
   \item Supports C-type system without function pointers
   \item Represents each data item with a triple of the form \textless tag, length, value\textgreater, where both the \textit{tag} and \textit{length}
   are typically 8-bit fields, and the \textit{value} is represented in exactly the same way as XDR
   \item Uses big-endian order for serialisation
   \item Use type tags
   \item Compound data types can be constructed by nesting primitive types, shown in Figure \ref{fig:compoundtypeASN1}
 \end{itemize}

\begin{figure}[ht]
\centering
  \includegraphics[width=4in]{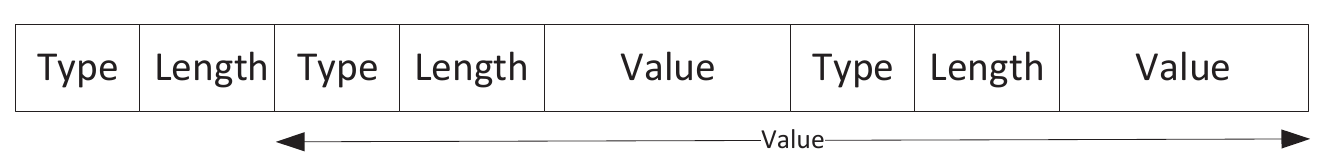}
  \caption{Compound types created by means of nesting in ASN.1}\label{fig:compoundtypeASN1}
\end{figure}

}
\end{itemize}
\end{enumerate}
{\em High-layer protocols} often refer to middleware service protocols that enable communication and management of data in distributed applications. The middleware service protocol is defined as the software layer that lies between the operating system and the applications on each site of the system \cite{middleware}. Services that can be regarded as middleware include enterprise application integration, data integration, message oriented middleware (MOM), object request brokers (ORBs), and the enterprise service bus (ESB). An example of message oriented middleware service (\ie JMS) is introduced in Section \ref{sec:jms}.

\subsection{Textual Protocols}

The classic {\em Open Systems Interconnection model} (OSI) characterizes and standardizes the internal functions of a networked system by partitioning it into logic layers. According to if one protocol relies on another protocol for message transmission, we further classify textual application-layer protocols into two groups: {\em low-layer protocols} and {\em high-layer protocols}. Specifically, \emph {low-layer protocols} refer to protocols using ``Field:Value'' format, while \emph {high-layer protocols} use structured standards to define message format and rely on \emph {lower-layer protocols}, most notably Hypertext Transfer Protocol (HTTP) or Simple Mail Transfer Protocol (SMTP), for message negotiation and transmission. Some commonly used structural standards are introduced as follows:
%we which embed in the ``value'' part textual protocol also includes some HTTP-based However, the availability of applications and data on the Web create a new situation in which Web applications need to communicate with each other and
%understand each other's data. The approach taken in the Web today to enable communication
%among web servers is based on the \textit{Markup language}, of which \textit{HyperText Markup Language} (HTML) and \textit{Extensible Markup language} (XML)
%are both examples.
\begin{enumerate}

\item{\textbf{Extensible Markup Language (XML)}

XML \cite{XML} is a framework for defining different markup languages for different kinds of data.
\begin{itemize}
  \item XML defines a basic syntax for mixing markup with data text, but
the designer of a specific markup language has to name and define its markup
  \item XML syntax provides for a nested structure of tag/value pairs, which is equivalent to a tree structure for the represented data. This is similar to XDR and
ASN.1's ability to represent compound types, but in a format that can be both processed by programs and read by humans
  \item The formal definition of a specific XML-based language is given by a \textit{schema}, which is a database term for a specification of how to interpret a
collection of data
  \item The XML namespace enable a schema to be reused as part of another schema
\end{itemize}
}

\item{\textbf{Javascript Object Notation (JSON)}

JavaScript Object Notation (JSON)~\cite{JSON} is a text-based data interchange format. The JSON format is derived from the JavaScript scripting language for serialising and transmitting structured data over a network connection. It is used primarily to transmit
data between a server and web applications, serving as an alternative of XML.
\begin{itemize}
  \item JSON can represent four primitive types (strings, numbers, boolean,
   and null) and two structured types (objects and arrays)
  \item A string is a sequence of zero or more Unicode characters
  \item An object is an unordered collection of zero or more name/value
   pairs, where a name is a string and a value is a string, number,
   boolean, null, object, or array
  \item An array is an ordered sequence of zero or more values
  \item The formal definition of a specific JSON-based language is given by a JSON schema, which is a specification for what JSON data is
  required for a given application and how it can be modified, much like the XML schema provides for XML.
\end{itemize}
}
\end{enumerate}

\section{Discussion}
\label{chap3sec:conclusion}

In a distributed system environment, messages are exchanged between systems according to the various interaction protocols. In general, interaction protocols define two
types of messages, that is, the request messages and the response messages. The request messages are sent by a requestor system to a replier system which receives and processes the request, ultimately returning a message in response. The observable interaction request messages and response messages communicated between a system under test and a target system contain two types of information: (i) protocol structure information (such as the operation name, field names and field delimiters), used to describe the type and format of a message, and (ii) payload information, which includes attribute values of a records or objects and metadata. In general, given a collection of message interactions conforming to a specific interaction protocol, the repeated occurrence of protocol structure information may be common, as only a limited number of operations are defined in the protocol specification. In contrast, payload information is typically quite diverse according to the various objects and records exposed by the service. %Our research interest it to infer the interaction behaviour among enterprise system elements
%based on mining interaction traces.

Our research goal is to automatically learn required structural or interactive knowledge from recorded interactions, and generate responses for incoming request. By studying these widely used application-layer protocols, we have following findings:
\begin{enumerate}
    \item Both binary protocols and textual protocols can be further classified into 2 groups, the high layer protocols and low layer protocols. High layer protocols rely on lower layer application protocols for message transmission
    \item We find common features of both text and binary protocols, based on which we can design a generic approach to abstract and present protocol information with a
    hierarchical data structure, for instance, a tree structure.
    \item With regard to binary protocols, we find that different protocols use different network data representations, where {\em ASN.1} and {\em XDR} are the most popular ones. For a specific data representation, we need to design specific rules to obtain the embedded protocol information.
\end{enumerate}

\section{Summary}

In this chapter, we discussed twelve widely-used application-layer protocols, NFS, NDMP, SNMP, LDAP, SMB, HTTP, FTP, SMTP, POP3, IRC, JMS and SOAP. Based on the protocols we investigated, we summarised general characteristics of these protocols. Moreover, based on different methods used for processing the protocols, we classified all protocols into two categories: the binary protocol and the textual protocol. For each category, we analysed their message structures and interactive behaviours, enabling us to evaluate the validity of responses being produced by using our opaque response generation approach.

%%%%%%%%%%%%%%%%%%%%%%%%%%%%%%%%%%%%%%%%%%%%%%%%%%%%%%%%%%%%%%%%%%%%%%%%%%%%%

\chapter{Framework for Automatic Opaque Response Generation}
\label{chap4:framework}

Service virtualisation is an emerging technique for creating realistic executable models of server-side behaviour and
is particularly useful in producing an interactive testing environment for quality assurance purposes: replicating production-like conditions for large-scale enterprise software systems. This allows performance engineers to mimic very large numbers of servers and/or provide a means of controlling dependencies on diverse third-party systems. However, previous approaches to service virtualisation rely on manual definition of interaction models requiring significant human effort. They also rely on either a system expert or documentation of a system protocol and behaviour, neither of which are necessarily available. In this chapter, we propose a novel opaque response generation approach to creating executable service models for automatic service virtualisation. This approach can address the limitations of conventional service virtualisation methods~\cite{hine:09a}~\cite{hine:10a}. Moreover, this approach create service models to meet all executable requirements: 1) requiring created service models are able to generate responses conforming to message format specification defined by services to be virtualised; 2) requiring interaction behaviours between system under test and created service models conform to temporal property defined by the protocol behavioural specification.
.% for virtual services  identified in Chapter \ref{chap3:requirements}.  %Its three major components with technical details are described in the subsequent chapters.

In this chapter, we present a framework of our opaque response generation approach. %Specifically, Section \ref{chap4sec:motivation} introduces the key motivation for our technique using a concrete enterprise system emulation example.
Section \ref{chap4sec:frameworkdesign} gives the framework overview. Section \ref{chap4sec:preliminaries} defines some preliminaries that are needed to describe our framework. Section \ref{chap4sec:analysis} introduces the design of the first component, {\ie} the {\em analysis function}. Section \ref{chap4sec:distance} introduces the design of the second component, {\ie} the {\em matching function}. The third component, {\ie}{\em the translation function} is described in Section \ref{chap4sec:translation}. Finally, Section \ref{chap4sec:summary} summaries the main contributions of this chapter.

The framework introduces in this chapter is based on our work present in~\cite{Du:2013}.

\section{Framework Overview}
\label{chap4sec:frameworkdesign}

The aim of our framework is to provide a high-level overview of the automatic service virtualisation technique that uses message exchange recordings collected a priori to {\em produce} a response on behalf of a service when invoked by a system-under-test at runtime. To this end, there are three requirements that will be addressed in this work, which are listed as follows.

\begin{itemize}
  \item processing and analysing network traffic between a {\em system-under-test} and a target server system for extracting inherent protocol structure information;
  \item creating interaction models on the basis of the extracted information;
  \item using virtual service models to communicate with the system under test, thereby replacing the real target service for quality assurance purpose.
\end{itemize}

Communicating pairs are able to continuously exchange messages over the network channel by agreeing to and faithfully implementing an appropriate protocol specification. A protocol specification contains two essential aspects: message format and protocol state machine~\cite{cui:07a}. Message format characterises the syntax and semantics for all possible legitimate messages, while the protocol state machine defines rules for exchanging messages between entities. As long as the communication among communicating peers is still continuing, each message must conform to message format rules, and the sequence of exchanged messages must conform to protocol state machine. Hence, we can assume that message exchange recordings are conformant with the above two aspects. 

\begin{figure}[t]
\centering
  \includegraphics[width=\textwidth]{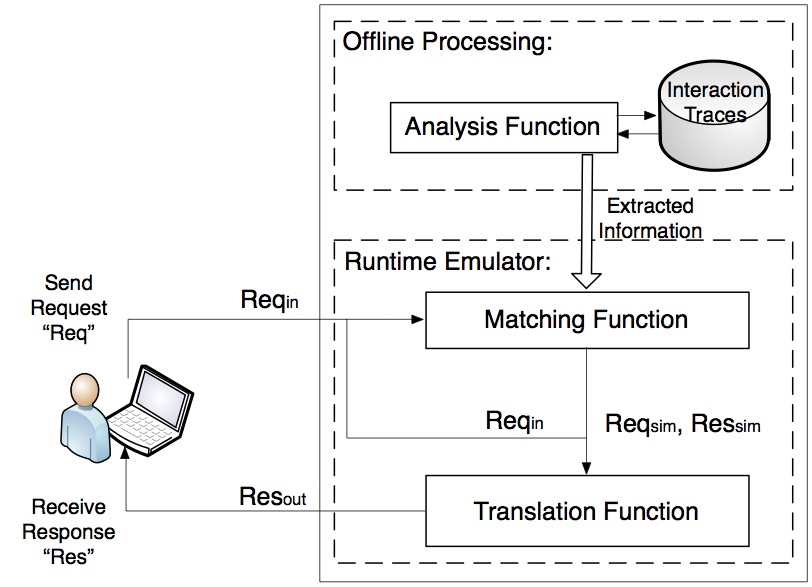}\\
  \caption{An Overview of the Framework}
  \label{fig:approchoverview}
\end{figure}

For the purpose of the proposed framework, we assume that for a given service to be virtualised, we are able to record a sufficiently large number of interactions between this service and the external system under test. Tools like Wireshark~\cite{website:wireshark} have the functionality to filter network traffic and record messages of interest in a suitable format for further processing. We also assume that these recordings are``valid'', that is, that the sequence of recorded interactions are (i) correct with regards to the temporal properties of the underlying protocol and that (ii) each request and response message is well-formed, conforming to message format rules. 
Based on the above assumptions, we propose a framework to deal with these identified requirements. This framework includes three components, that is, the {\em analysis function}, the {\em matching function}, and the {\em translation function}, as shown in Figure \ref{fig:approchoverview}. Specifically, the {\em analysis function} is performed offline. It takes as input a collection of messages, divides these messages into groups of similar messages, and infers {\em cluster prototypes} capable of summarising contents of the groups of messages. %is performed offline to process and analyze the {\em interaction traces}, which is to uncover protocol information ({\ie} message structural information defined by a particular protocol specification), versus payload information ({\ie} variables that are produced/consumed by application programs).
At runtime, given an incoming request from a client under test, the {\em matching function} compares it with every cluster prototype for the purpose of locating this incoming request into an appropriate message group, enabling the detection of its suitably “similar” recorded interaction. And then, the {\em translation function} is performed to (i) identify the commonalities and accommodates variations between the incoming request and the recorded request, and (ii) alter the recorded responses by using the identified commonalties to produce a new response, respectively.

\section{Preliminaries}
\label{chap4sec:preliminaries}
%summarise framework

This section defines a number of basic notations that are needed to express our framework more formally.

\textbf{Definition 1} (Message Character). The set of message characters, denoted as {\MsgCharSet}, is the most basic building block. For the purpose of our study, {\MsgCharSet} comprises of the set of valid {\em Bytes} that can be transmitted over a network or the set of printable {\em Characters} as a dedicated subset. We require equality and inequality are defined for the elements ${\MessageCharacter} \in {\MsgCharSet}$.  %As network communication occurs in streams of {\em bytes}, each valid {\em message character}  a valid {\em byte} that can be transmitted over a network,  {\MsgCharSet} denotes the set of all valid message
%Let {\MsgCharSet} denote the set of {\em message characters}. {\MessageSet} is the set of all possible {\em messages} (see Def. 2). Any element in ${\MessageSet}$ is a finite sequence of message characters from {\MsgCharSet}.

\textbf{Definition 2} (Message). {\MessageSet} denotes the set of all (possibly empty) {\em messages} that can be defined using the set of message characters {\MsgCharSet}. A message ${\Message} \in {\MessageSet}$ is a non-empty, finite sequence of {\em message characters} ${\MessageCharacter}_1{\MessageCharacter}_2\ldots{\MessageCharacter}_n$ with ${\MessageCharacter}_i \in {\MsgCharSet}$, $1 \leq i \leq n$. We require equality and inequality to be defined. Given two messages ${\Message}_1 = {\MessageCharacter}_{1,1}{\MessageCharacter}_{1,2}\ldots {\MessageCharacter}_{1,l}$
and ${\Message}_2= {\MessageCharacter}_{2,1}{\MessageCharacter}_{1,2}\ldots {\MessageCharacter}_{2,n} $ to be equal if $l=n$ and $c_{1,i} =
c_{2,i}, 1 \leq i \leq n$. Otherwise, these two messages are considered to be inequal. In real enterprise software environment, there could be cases where the same message is encoded differently using different encoding rules. Our approach is to operate directly on the collection of raw bits transmitted over the wire. Therefore, if messages are encoded into different raw bits, they are considered to be different.
\textbf{Definition 3} (Interaction). An interaction {\Interaction}, denoted as ${\Interaction} = {\Pair{\Request}{\Response}}$, is a request-response pair between a client system and a target service system. {\InteractionSet} is the set of interactions. Both {\Request} and {\Response} are elements of {\MessageSet}, where {\Request} denotes a message sent from a client system to the target service system, and {\Response} denotes zero or one or many returning message(s) sent back from the target service system to the client system. %consists of a request, denoted by {\Requeset}, as well as the corresponding response, denoted by {\Response}. system.  It in which a client sends a request message to a target service which receives and processes the request, ultimately returning zero/one/many response message(s).
Without loss of generality, we assume that each request is always followed by a single response. If a request does not generate a response, we insert a dedicated ``no-response'' message (\textit{NO RESPONSE}) into the response part of this interaction. If, on the other hand, a request leads to multiple responses, these are concatenated into a single response. Network sniffer tools like Wireshark have the functionality to record message of interest in a suitable format by setting the source ip address and the destination ip address. If a message from the source ip address is followed by another message from the source ip address rather than from the destination ip address, we insert the dedicated {\textit {NO RESPONSE}}. If a message sent from the source ip address is followed by several message from the destination ip address, we concatenate them into a single response.

Considering the directory service interaction library example given in Table~\ref{Chap1tab:tl} in Chapter~\ref{chap1:introduction}, \{id:1,op:S,sn:Du\} is considered to be a request message, and its following response message is \{id:1,op:SearchRsp,result:Ok,gn:Miao, sn:Du,mobile:5362634\}. Their interaction can be written as below:

\hspace{0.48cm} (\{id:1,op:S,sn:Du\}, \{id:1,op:SearchRsp,result:Ok,gn:Miao,sn:Du,mobile:5362634\})

%Each interaction $I_k$ has to conform to following rules,
%
% \begin{itemize}
%   \item {\Request} denotes a non-empty request message ${\Request} \in \MessageSet$. A {\em request message} is a triple ${\Request} = <\OperationType,i,k>$;
%   \item {\Response} is a set of returning response messages ${\Response} \subset \MessageSet \cup \{no-response\}$. A {\em response message} is a triple ${\Response} = <\OperationType,i,k>$,
% \end{itemize}
%
% where $\OperationType$ represents operation type of a request/response message, and $k \leq 1$ is an numeric index that uniquely indicates the order (position) of the current interaction {\Interaction} within an interaction trace ${\InteractionTrace}_i$ ({\cf} Definition 4).

% Not sure if I should explain that for the purpose of our work, we require each interaction trace to be complete and sequential.

\textbf{Definition 4} (Interaction Trace). {\TraceSet} denotes the set of interaction traces. An interaction trace {\InteractionTrace} is defined as a finite, non-empty sequence of interactions ${\InteractionTrace}= {\Interaction}_1{\Interaction}_2\ldots{\Interaction}_n$ with ${\Interaction}_i \in {\InteractionSet}, 1 \leq i \leq n$. An interaction trace describes the message changes between two host for accomplishing specific tasks. Figure~\ref{chap4fig:trace} shows two interaction trace examples captured at different times, where the column {\em index} show a time order that interactions take place.
%%An interaction trace is denoted by ${\InteractionTrace}=<\prec, {\Interactions}>$, where $\prec$ denotes the completes-before relation among interactions. The completes-before relation contains interactions in two aspects, that is, the communication completenessThe target of our work is to generate responses by , we require each communication between a client system and the target service to be complete to This relation must satisfy both communication completeness constraint and communication consistence constraint: (i) communication completeness only allows a client to establish a new communication with the target service until their current communication is complete; (ii) communication consistence constrains a client to suspend sending the next request message (see Definition 2) until it receives all response message(s) of this request message from the target service.

%   We express above constrains as follows,
%    \begin{enumerate}
%       \item $\forall I_i \in \InteractionSet$, $\exists I_i \in \InteractionTrace$ with $i \in [1,|{\InteractionTrace}|]$, and $I_j \in \InteractionTrace'$ with $j \in [1, |{\InteractionTrace}'|]$, $I_i \prec I_j \iff {\InteractionTrace} \prec {\InteractionTrace}'$,
%       \item $\forall I_i, I_{i+1} \in \InteractionTrace$ with $i \in [1,|{\InteractionTrace}|-1]$, $\Response_i \in I_i, \Request_{i+1} \in I_{i+1} \Rightarrow \Response_i \prec \Request_{i+1}$
%    \end{enumerate}
%\end{itemize}
\begin{figure}[!tbh]
\centering
\footnotesize
\subfloat[Interaction Trace 1\label{Chap4tab:t1}]{
    \begin{tabular}{|c||l|l|}
    \hline
    Index & \multicolumn{2}{c|}{Request/Response Messages} \\ \hline\hline
    \multirow{2}{0.8cm}{\centering 1} & Req. & \{id:1,op:B\} \\ \cline{2-3}
& Res. & \{id:1,op:BindRsp,result:Ok\} \\ \hline
    \multirow{2}{0.8cm}{\centering 2} & Req. & \{id:2,op:S,sn:Du\} \\ \cline{2-3}
    & Res. & \{id:2,op:SearchRsp,result:Ok,gn:Miao,sn:Du,mobile:5362634\} \\ \hline
    \multirow{2}{0.8cm}{\centering 13} & Req. & \{id:13,op:S,sn:Versteeg\} \\\cline{2-3}
    & Res. & \{id:13,op:SearchRsp,result:Ok,gn:Steve,sn:Versteeg,mobile:9374723\} \\ \hline
    \multirow{2}{0.8cm}{\centering 24} & Req. & \{id:24,op:A,sn:Schneider,mobile:123456\} \\ \cline{2-3}
    & Res. & \{id:24,op:AddRsp,result:Ok\} \\ \hline
    \multirow{2}{0.8cm}{\centering 275} & Req. & \{id:275,op:S,sn:Han\} \\ \cline{2-3}
     & Res. & \{id:275,op:SearchRsp,result:Ok,gn:Jun,sn:Han,mobile:33333333\} \\ \hline
    \multirow{2}{0.8cm}{\centering 490} & Req. & \{id:490,op:S,sn:Grundy\}\\ \cline{2-3}
     & Res. &
    \{id:490,op:SearchRsp,result:Ok,gn:John,sn:Grundy,mobile:44444444\} \\ \hline
    \multirow{2}{0.8cm}{\centering 2273} & Req. & \{id:2273,op:S,sn:Schneider\} \\ \cline{2-3}
    & Res. & \{id:2273,op:SearchRsp,result:Ok,sn:Schneider,mobile:123456\} \\ \hline
    \multirow{2}{0.8cm}{\centering 2487} & Req. & \{id:2487,op:A,sn:Will\} \\ \cline{2-3}
    & Res. & \{id:2487,op:AddRsp,result:Ok\} \\ \hline
    \multirow{2}{0.8cm}{\centering 3106} & Req. & \{id:3106,op:A,sn:Hine,gn:Cam,Postcode:33589\} \\ \cline{2-3}
    & Res. & \{id:3106,op:AddRsp,result:Ok\} \\
    \hline
    \multirow{2}{0.8cm}{\centering 3211} & Req. & \{id:3211,op:U\} \\ \cline{2-3}
    & Res. & \{id:3211,op:UnbindRsp,result:Ok\} \\
    \hline
    \end{tabular}
    }\\

    \subfloat[Interaction Trace 2\label{Chap4tab:t2}]{
    \begin{tabular}{|c||l|l|}
    \hline
    Index & \multicolumn{2}{c|}{Request/Response Messages} \\ \hline\hline
    \multirow{2}{0.8cm}{\centering 1} & Req. & \{id:1,op:B\} \\ \cline{2-3}
    & Res. & \{id:1,op:BindRsp,result:Ok\} \\ \hline
    \multirow{2}{0.8cm}{\centering 12} & Req. & \{id:12,op:S,sn:Hine\} \\ \cline{2-3}
    & Res. & \{id:12,op:SearchRsp,result:Ok,gn:Cam,sn:Hine,Postcode:33589\} \\ \hline
    \multirow{2}{0.8cm}{\centering 34} & Req. & \{id:34,op:A,sn:Lindsey,gn:Vanessa,PostalAddress1:83 Venton Road\} \\ \cline{2-3}
    & Res. & \{id:34,op:AddRsp,result:Ok\} \\ \hline
    \multirow{2}{0.8cm}{\centering 145} & Req. & \{id:145,op:S,sn:Will\} \\ \cline{2-3}
    & Res. & \{id:145,op:SearchRsp,result:Ok,sn:Will,gn:Wendy,mobile:54547\} \\ \hline
    \multirow{2}{0.8cm}{\centering 1334} & Req. & \{id:1334,op:S,sn:Lindsey,gn:Vanessa,PostalAddress1:83\ Venton\ Road\} \\ \cline{2-3}
    & Res. & \{id:1334,op:SearchRsp,result:Ok,gn:Vanessa,PostalAddress1:83\ Venton\ Road\} \\ \hline
    \multirow{2}{0.8cm}{\centering 1500} & Req. & \{id:1500,op:U\} \\ \cline{2-3}
    & Res. & \{id:1500,op:UnbindRsp,result:Ok\} \\ \hline
    \end{tabular}
    }
\caption{Interaction Trace Examples. The index column indicates a time order that interactions take place.}
\label{chap4fig:trace}
\end{figure}

%Considering the interaction trace library example in Table~\ref{Chap1tab:tl} in Chapter~\ref{chap1:introduction}. 
 % show the interaction traces of this example. Each interaction trace consists of a number of interactions over a particular communication channel, which is established between a client system and a target service system before normal interactions over the channel begins.

\textbf{Definition 5} (Trace Library). A trace library {\TraceLibrary} is denoted by {\TraceLibrary} = $<${\TraceSet}, {\InteractionSet}$>$, where {\TraceSet} (Definition 4) denotes the set of interaction traces and  {\InteractionSet} (Definition 3) denotes the set of interactions that compose every interaction trace in {\TraceSet}. Table~\ref{Chap4tab:t4} shows the example of an interaction trace library.

\begin{table}[!tbh]
\footnotesize
\begin{center}
\begin{tabular}{|c||l|l|}
\hline
Index & \multicolumn{2}{c|}{Request/Response Messages} \\ \hline\hline
\multirow{2}{0.8cm}{\centering 1} & Req. & \{id:1,op:B\} \\ \cline{2-3}
& Res. & \{id:1,op:BindRsp,result:Ok\} \\ \hline
\multirow{2}{0.8cm}{\centering 2} & Req. & \{id:2,op:S,sn:Du\} \\ \cline{2-3}
& Res. & \{id:2,op:SearchRsp,result:Ok,gn:Miao,sn:Du,mobile:5362634\} \\ \hline
\multirow{2}{0.8cm}{\centering 13} & Req. & \{id:13,op:S,sn:Versteeg\} \\\cline{2-3}
& Res. & \{id:13,op:SearchRsp,result:Ok,gn:Steve,sn:Versteeg,mobile:9374723\} \\ \hline
\multirow{2}{0.8cm}{\centering 24} & Req. & \{id:24,op:A,sn:Schneider,mobile:123456\} \\ \cline{2-3}
& Res. & \{id:24,op:AddRsp,result:Ok\} \\ \hline
\multirow{2}{0.8cm}{\centering 275} & Req. & \{id:275,op:S,sn:Han\} \\ \cline{2-3}
 & Res. & \{id:275,op:SearchRsp,result:Ok,gn:Jun,sn:Han,mobile:33333333\} \\ \hline
\multirow{2}{0.8cm}{\centering 490} & Req. & \{id:490,op:S,sn:Grundy\}\\ \cline{2-3}
 & Res. &
\{id:490,op:SearchRsp,result:Ok,gn:John,sn:Grundy,mobile:44444444\} \\ \hline
\multirow{2}{0.8cm}{\centering 2273} & Req. & \{id:2273,op:S,sn:Schneider\} \\ \cline{2-3}
& Res. & \{id:2273,op:SearchRsp,result:Ok,sn:Schneider,mobile:123456\} \\ \hline
\multirow{2}{0.8cm}{\centering 2487} & Req. & \{id:2487,op:A,sn:Will\} \\ \cline{2-3}
& Res. & \{id:2487,op:AddRsp,result:Ok\} \\ \hline
\multirow{2}{0.8cm}{\centering 3106} & Req. & \{id:3106,op:A,sn:Hine,gn:Cam,Postcode:33589\} \\ \cline{2-3}
& Res. & \{id:3106,op:AddRsp,result:Ok\} \\
\hline
\multirow{2}{0.8cm}{\centering 3211} & Req. & \{id:3211,op:U\} \\ \cline{2-3}
& Res. & \{id:3211,op:UnbindRsp,result:Ok\} \\
\hline\hline
\multirow{2}{0.8cm}{\centering 1} & Req. & \{id:1,op:B\} \\ \cline{2-3}
& Res. & \{id:1,op:BindRsp,result:Ok\} \\ \hline
\multirow{2}{0.8cm}{\centering 12} & Req. & \{id:12,op:S,sn:Hine\} \\ \cline{2-3}
& Res. & \{id:12,op:SearchRsp,result:Ok,gn:Cam,sn:Hine,Postcode:33589\} \\ \hline
\multirow{2}{0.8cm}{\centering 34} & Req. & \{id:34,op:A,sn:Lindsey,gn:Vanessa,PostalAddress1:83 Venton Road\} \\ \cline{2-3}
& Res. & \{id:34,op:AddRsp,result:Ok\} \\ \hline
\multirow{2}{0.8cm}{\centering 145} & Req. & \{id:145,op:S,sn:Will\} \\ \cline{2-3}
& Res. & \{id:145,op:SearchRsp,result:Ok,sn:Will,gn:Wendy,mobile:54547\} \\ \hline
\multirow{2}{0.8cm}{\centering 1334} & Req. & \{id:1334,op:S,sn:Lindsey,gn:Vanessa,PostalAddress1:83\ Venton\ Road\} \\ \cline{2-3}
& Res. & \{id:1334,op:SearchRsp,result:Ok,gn:Vanessa,PostalAddress1:83\ Venton\ Road\} \\ \hline
\multirow{2}{0.8cm}{\centering 1500} & Req. & \{id:1500,op:U\} \\ \cline{2-3}
& Res. & \{id:1500,op:UnbindRsp,result:Ok\} \\ \hline
\end{tabular}
\end{center}
\caption{Trace Library Example}
\label{Chap4tab:t4}
\end{table}

%A {\em trace library} can be denoted as {\TraceLibrary} = $<\Gamma, {\InteractionSet}> $,
%\begin{itemize}
%    \item $\Gamma$ is a set of {\em interaction traces} $\Gamma$ = {\InteractionTraces} (see Def.4);
%    \item $|\Gamma|=n$ represents the number of interaction traces in the trace library {\TraceLibrary};
%    \item {\InteractionSet} is the set of distinguishing interactions in the interaction traces $\Gamma$ of the trace library {\TraceLibrary}, and $\forall I \in \InteractionSet$, $\exists I \in T_i$ with $i \in [1,|\Gamma|]$.
%\end{itemize}

%process interaction traces, cluster interactions into clusters, synthesize a representative for each cluster
\section{Analysis Function}
\label{chap4sec:analysis}

Given a trace library ${\TraceLibrary} = <{\TraceSet}, {\InteractionSet}>$, we can obtain a set of interaction traces that comprise of interactions in the interaction set {\InteractionSet}. These interactions are of different message types. Each message type is defined by a certain {\em message format specification}. A message format specification specifies static and dynamic parts of common format of messages~\cite{SemanticRE:2014}. The value at a static part is stable, that is, it does not change across messages, while values at a dynamic part may have high variability. Some of values at static parts serve to distinguish different types of messages. %The motivation behind our analysis function is if messages are very similar to each other (having a suitable notion of ``similarity''), their common features should give us an indication where static
%have the same message format defined by re very similar to each other, then they should have several static parts; if the same static parts are found in another  then so that they should be of the same type. Hence, if messages can be found the same static parts, they should have the same message format. Hence, %without req                   uiring knowledge of the underlying protocol or message structure,
Hence, extracting the commonality and accommodating variations in a group of similar recorded messages would assist with revealing message common format defined by the message format specification. %Hence, identifying static parts of messages of the same type without relying on expert knowledge regarding the underlying protocol or message structure,

The purpose of the analysis function is to reveal commonality of a group of similar messages and to decide which group an incoming request belongs to. To this end, the analysis function is required to: 1) cluster similar interactions into groups; 2) identify the static parts by comparing messages in a single cluster; 3) decide which cluster an incoming request belongs to. Correspondingly, the analysis function is performed in three steps: (i) %process interaction traces. The analysis function is to
clusters all recorded interactions between a client and the target service into {\em interaction clusters}, where the formats of all interactions in one cluster are more {\em similar} to each other than those in other clusters; (ii) formulates a single {\em cluster prototype} for the request messages of each {\em interaction cluster}; and (iii) selects a {\em cluster centre} for each {\em interaction cluster}. We define some notations in order to formalise the analysis function.

\textbf{Definition 6} (Interaction Cluster). An interaction cluster is composed of a set of interactions, denoted as $ {\InteractionCluster} = \{{\Interaction}_1,{\Interaction}_2,{\Interaction}_3,{\Interaction}_k\}
$ with ${\Interaction}_l \in \InteractionSet, 1 \leq l \leq k$, which is the subset of the interaction set {\InteractionSet} of a trace library.

Given a trace library ${\TraceLibrary} = <{\TraceSet}, {\InteractionSet}>$,

\begin{itemize}
    \item The interaction set $\InteractionSet$ is a union of interaction clusters,
        \begin{center}

         ${\InteractionSet}=\InteractionCluster_1 \bigcup \InteractionCluster_2 \bigcup \ldots \bigcup \InteractionCluster_n$

        \end{center}
        ,where $n$ is the total number of clusters in the interaction set;
    \item Each interaction can only exist in one interaction cluster, that is, $\forall i,j \in [1,n]$, ${\InteractionCluster}_i \cap {\InteractionCluster}_j = \emptyset$;
    \item Request and/or response message(s) are more similar to the request and/or response message(s) in the same interaction cluster, than those in other interaction clusters.

        $\forall {\Interaction}_i = {\Pair{\Request_i}{\Response_i}},{\Interaction}_j = {\Pair{\Request_j}{\Response_j}} \in {\InteractionCluster}_1, {\Interaction}_k = {\Pair{\Request_k}{\Response_k}} \in {\InteractionCluster}_2$:

        \begin{center}
         ${\SimiCalc}(\Request_i,\Request_j) > {\SimiCalc}(\Request_i,\Request_k)$ or
        \end{center}

        \begin{center}
         ${\SimiCalc}(\Response_i,\Response_j) > {\SimiCalc}(\Response_i,\Response_k)$
        \end{center}
        {\SimiCalc} is user-defined method for measuring similarity between two request messages or two response messages.
\end{itemize}

% to be the most ``similar'' interaction to the incoming request, which will be utilised to

\begin{figure}[th]
\centering
    \subfloat[Cluster 1 (Search Operations)\label{Chap4tab:searchcluster}]{
    \scriptsize
    \begin{tabular}{|c||l|l|}

        \hline
        Index & Request & Response \\ \hline\hline
        2 & \{id:2,op:S,sn:Du\} & \{id:2,op:SearchRsp,result:Ok,gn:Miao,sn:Du,mobile:5362634\} \\ \hline
        13 & \{id:13,op:S,sn:Versteeg\} & \{id:13,op:SearchRsp,result:Ok,gn:Steve,sn:Versteeg,mobile:9374723\} \\ \hline
        275& \{id:275,op:S,sn:Han\} & \{id:275,op:SearchRsp,result:Ok,gn:Jun,sn:Han,mobile:33333333\} \\ \hline
        490 & \{id:490,op:S,sn:Grundy\} &
        \{id:490,op:SearchRsp,result:Ok,gn:John,sn:Grundy,mobile:44444444\} \\ \hline
        2273 & \{id:2273,op:S,sn:Schneider\} & \{id:2273,op:SearchRsp,result:Ok,sn:Schneider,mobile:123456\} \\ \hline
    \end{tabular}
    }

    \subfloat[Cluster 2 (Add Operations)\label{Chap4tab:addcluster}]{
    \scriptsize
    \begin{tabular}{|c||l|l|}
    \hline
        Index & Request & Response \\ \hline\hline
        24 & \{id:24,op:A,sn:Schneider,mobile:123456\} & \{id:24,op:AddRsp,result:Ok\} \\ \hline
        2487 & \{id:2487,op:A,sn:Will\} & \{id:2487,op:AddRsp,result:Ok\} \\ \hline
        3106 & \{id:1106,op:A,sn:Hine,gn:Cam,postalCode:33589\} & \{Id:1106,Msg:AddRsp,result:Ok\} \\
    \hline
    %\parbox[p]{2cm}{\textbf{Cluster\\Centroid\\Interaction}} & {Id:3,Msg:A,Lastname:Schnei-der,Firstname:Jean-Guy, Telephone:123456} & {Id:3,Msg:AddRsp,Result:Ok} \\ \hline
    \end{tabular}
    }

    \subfloat[Cluster 3 (Bind Operations)\label{Chap4tab:bindcluster}]{
    \scriptsize
    \begin{tabular}{|c||l|l|}
    \hline
        Index & Request & Response \\ \hline\hline
        1 & \{id:1,op:B\} & \{id:1,op:BindRsp,result:Ok\} \\ \hline
    %\parbox[p]{2cm}{\textbf{Cluster\\Centroid\\Interaction}} & {Id:3,Msg:A,Lastname:Schnei-der,Firstname:Jean-Guy, Telephone:123456} & {Id:3,Msg:AddRsp,Result:Ok} \\ \hline
    \end{tabular}
    }

    \subfloat[Cluster 4 (Unbind Operations)\label{Chap4tab:unbindcluster}]{
    \scriptsize
    \begin{tabular}{|c||l|l|}
    \hline
        Index & Request & Response \\ \hline\hline
         3211 & \{id:3211,op:U\} & \{id:3211,op:UnbindRsp,result:Ok\} \\ \hline
    %\parbox[p]{2cm}{\textbf{Cluster\\Centroid\\Interaction}} & {Id:3,Msg:A,Lastname:Schnei-der,Firstname:Jean-Guy, Telephone:123456} & {Id:3,Msg:AddRsp,Result:Ok} \\ \hline
    \end{tabular}
    }
    \caption{Interaction Cluster Examples}
    \label{chap4tab:clusters}
\end{figure}

Consider the interaction trace example in Table~\ref{Chap4tab:t1}. After calculating response similarities with a similarity calculation method, we can cluster all interactions into four {\em interaction clusters}, requiring no knowledge regarding the underlying protocol. However, by observing interactions in each cluster, we know these clusters are the {\em search operation} cluster, the {\em add operation} cluster, the {\em bind operation} cluster and the {\em unbind operation} cluster, respectively, as shown in Figure~\ref{chap4tab:clusters}. These figures indicate each interaction appears in one and only one interaction cluster. None of these interactions can appear in more than one cluster.

\textbf{Definition 7} (Cluster centre). A cluster centre ${\Interaction}= {\Pair{{\Request}_{centre}}{{\Response}_{centre}}}$ is a recorded interaction that is selected for representing other interactions in this cluster.

A different cluster centre interaction is chosen by using a different kind of cluster centre selection approach. Let us consider interaction cluster examples in Figure~\ref{Chap4tab:addcluster}, \{id:24,op:A,sn:Schneider,mobile:123456\} is selected to be the cluster with the lowest sum of the absolute distances to other interactions in the cluster. The detailed description of this measurement can be referred to Section~\ref{chap6subsec:centers}.

\textbf{Definition 8} (Cluster Prototype). A {\em cluster prototype} is to describe common features of all request messages in an interaction cluster, denoted as {\ConsensusPrototype}.

A cluster prototype can be either an existing request message that has been seen before, such as {\em \{id:1,op:S,sn:Du\}} in the search operation cluster; or a non-existing message that is able to summarise the common features and accommodating variations among messages in a cluster. For example, {\em \{id:???,op:S,sn:???????\}}{\footnote{? is a `wildcard' symbol, allowing us to encode where there are high variability sections of the messages. Section~\ref{chap7:prototypederivation} introduces a detailed description of how we derive these cluster prototypes.}} is the cluster prototype of the search operation cluster in Figure~\ref{Chap4tab:searchcluster}, and we formulate a cluster prototype, {\em \{id:????,op:A,sn:??????????????????????\}}, for the add operation cluster in Figure~\ref{Chap4tab:addcluster}.

In our work, the aim of the analysis function is to derive a cluster prototype of the request messages of each cluster, assisting in searching for the most similar recorded interaction for an incoming request. Using the definitions introduced above, our analysis function {\AnalysisFunc} can thus be formalised as below.

{\hspace{0.25cm}}
${\AnalysisFunc} = ({\ClusteringStep}, {\FormulatingStep}, {\SelectingStep})$

The {\ClusteringStep} denotes the clustering step where we cluster the interaction set, for the purpose of organising interactions by message types0. The {\FormulatingStep} is the step that is performed to derive a cluster prototype for each message type cluster. The {\SelectingStep} is to select a cluster centre that is able to represent all other messages in the cluster.

Each step can be formalised as following equations. The right side of these equations shows inputs being taken by each step, and the left side shows each step's output.

{\hspace{0.25cm}}
$({\InteractionCluster}_1, {\InteractionCluster}_2, \ldots, {\InteractionCluster}_n) = \mathbf{\ClusteringStep} \ ({\InteractionSet})$

{\hspace{0.25cm}}
$({\ConsensusPrototype}_1, {\ConsensusPrototype}_2, \ldots, {\ConsensusPrototype}_n) = \mathbf{\FormulatingStep} \ ({\InteractionCluster}_1, {\InteractionCluster}_2, \ldots, {\InteractionCluster}_n)$

{\hspace{0.25cm}}
$({\Interaction}_1, {\Interaction}_2, \ldots, {\Interaction}_n) = \mathbf{\SelectingStep} \ ({\InteractionCluster}_1, {\InteractionCluster}_2, \ldots, {\InteractionCluster}_n)$

,where {\em {\ClusteringStep}}, {\em {\FormulatingStep}} and {\em {\SelectingStep}} denote user-defined approaches for each step of the analysis function. In Chapter~\ref{chap6:SoftMine} and Chapter~\ref{chap7:consensus}, we present implementations of these approaches. % denotes user-defined analysis function.
%
% The analysis function is the most important component at the preprocess stage. To establish a communication between two system elements, both of them must adhere to a particular
%  protocol specification, which implies that observable interaction traces contain protocol
%  knowledge. %, especially concerning message structures.
%  And furthermore, besides structural information, transmitted
%  messages often take variables consumed/produced by an application that use this protocol to exchange messages
%  with other application on other host, which are referred to as payloads.
%
%  With use of little or none of protocol knowledge, how to separate protocol-related information from application-related information is the core of interaction traces analysis.

\section{Matching Function}
\label{chap4sec:distance}

The analysis function formulates a cluster prototype for each cluster to summarise common features of all requests in this cluster. At runtime, the matching function is invoked in order to choose the most similar cluster prototype for a live request. An appropriate distance calculation approach, denoted as {\DistFunc}, is crucial at this stage, which is used to rank the distance between a request and every cluster prototype.

%For an unknown request, the most similar request in the transaction library can be identified.
Our proposed approach does not require require explicit knowledge of the protocols that the target service uses so that we cannot use protocol-specific metrics as our distance measurement. Instead, we use the minimum edit distance \cite{Ristad:1998} to describe the distance between a live request and a recorded request, or the distance between a live request and a cluster prototype. Given two messages $s_1$ and $s_2$, a number of edit operations can be performed to transform $s_1$ to $s_2$. Each edit operation has a cost. The minimum edit distance is a metric that measures the total cost of obtaining $s_2$ from $s_1$. One of the simplest sets of modification operations is defined by Levenshtein in 1966~\cite{levenshtein1966}, including three types of edit operations, that is, {\em insertion}, {\em deletion}, and {\em substitution}. Assuming $p_1$ and $p_2$ are positions in $s_1$ and $s_2$ respectively, these operations are defined as follows:

\begin{itemize}
  \item \textbf{{\em Insertion} of a message character}. A message character at position $p_2$ in $s_2$ is inserted into $s_1$ at position $p_1$. For example,
      \begin{center}
          \begin{tabular}{llcll}

            % after \\: \hline or \cline{col1-col2} \cline{col3-col4} ...
            $s_1$: & old & \multirow{2}{*}{$ \stackrel{insertion}{\longrightarrow} $}  & $s_1^{'}$: & \textit{g}old \\
            $s_2$: & gold &  & $s_2$: & gold \\
          \end{tabular}
      \end{center}
  \item \textbf{{\em Deletion} of a message character}. A message character is deleted from $s_1$ at position $p_1$. For example,

      \begin{center}
          \begin{tabular}{llcll}
            % after \\: \hline or \cline{col1-col2} \cline{col3-col4} ...
            $s_1$: & {\em g}old & \multirow{2}{*}{$ \stackrel{deletion}{\longrightarrow} $}  & $s_1^{'}$: & old \\
            $s_2$: & old &  & $s_2$: & old \\
          \end{tabular}
      \end{center}
  \item \textbf{{\em Substitution} of a message character}. A message character at position $p_2$ in $s_2$ is substituted for a character at position $p_1$ in $s_1$. For example,

      \begin{center}
          \begin{tabular}{llcll}
            % after \\: \hline or \cline{col1-col2} \cline{col3-col4} ...
            $s_1$: & gold & \multirow{2}{*}{$ \stackrel{substitution}{\longrightarrow} $}  & $s_1^{'}$: & gold \\
            $s_2$: & sold &  & $s_2$: & {\em g}old \\
          \end{tabular}
      \end{center}
\end{itemize}

The standard edit operations have associated costs which are application dependent, and are typically positive integers. For example, the costs of insertion, deletion, and substitution can all be set to 1. As substitution can be viewed as insertion and deletion, substitution is set to 2 in some contexts.

Considering the following two text messages:

\begin{minipage}[c]{\linewidth}
    \centering
	\texttt{\newline
         Where is my computer book?\quad\quad\quad\quad\quad
         \newline
         Where are your computer magazines?
         \newline}
\end{minipage}

To calculate their global dissimilarity, we set insertion and deletion to be 1, and substitution to be 2. The common subsequences are ``{\sf Where }'', ``{\sf \ computer }'', and
``{\sf ?}''. (Note the spaces in around ``{\sf \ computer }'' and ``{\sf Where }''.)  ``{\sf my}''
versus ``{\sf your}'', ``{\sf is}'' versus ``{\sf are}'', and ``{\sf book}'' versus ``{\sf magazines''} are the three
differing parts of the two sequences. %The edit distance calculation technique would align the character `{\sf y}' common to ``{\sf my}'' and ``{\sf your}'', although it probably makes more sense not to identify '{\sf y}' as a common subsequence, hence the need for a minimum length of common subsequence.
The fully aligned sequences will be as follows (we use the character `{\agap}' to denote required deletion and insertion operations):
%
% \vspace{-0.3cm}

\begin{minipage}[c]{\linewidth}
    \centering
	\texttt{\newline
    Where is{\agap}{\agap}{\agap} my{\agap}{\agap}{\agap} computer
    book{\agap}{\agap}{\agap}{\agap}{\agap}{\agap}{\agap}{\agap}{\agap}? \newline
    Where {\agap}{\agap}are {\agap}your computer
    {\agap}{\agap}{\agap}{\agap}magazines?
    \newline}
\end{minipage}

%%% NOTE JGS: I am not 100% sure whether we *really* use such a threshold
%%% in the implementation. I leave the text above as is for now, but we may
%%% need to address this in future drafts...

% \vspace{-0.3cm}
The distance between two sequences is defined by the number of gaps inserted
to both sequences in the alignment process -- 22 in the example above. In
order to allow for a better comparison of distances across multiple interactions
and/or protocol scenarios, we define the {\em dissimilarity ratio} as the ratio of the
``raw'' edit distance divided by the length ({\ie} number of elements) of both
sequences, \eg $22 / (26 + 33) = 0.37$ in the example given above. We require the {\em dissimilarity ratios} conform to the following two criteria: (i) the {\em dissimilarity ratios} is calculated from the absolute distance to normalise for cluster prototypes of different features, (ii) the {\em dissimilarity ratios} is in the range 0 to 1, inclusive, where 0 signifies the best possible match with the consensus, and 1 represents the furthest possible distance. %The dissimilarity ratio, as illustrated in this section, was used as the distance
%measure for the evaluation of our approach ({\cf}
%Section~\ref{sec:evaluation}). Two identical sequences will
%have a dissimilarity ratio of $0$ and the bigger the ratio, the more
%{\em dissimilar} two sequences are.

%
%By doing a comparison of two messages, the distance function ranks the similarity between an unknown message and a set of messages in the transaction library. For an unknown request, the most similar request in the transaction library can be identified using the distance function.

At runtime, when a live request ${\Request}_{in}$ being observed, the matching function adopts a distance calculation approach {\DistFunc} to calculate the {\em dissimilarity ratio} between this new request and every cluster prototype ${\ConsensusPrototype}_i$ for selecting a cluster prototype ${\ConsensusPrototype}_{centre}$ which is least distant to the new request. Once a cluster prototype is selected, the matching function outputs the cluster centre {\ClusterCentre} for the following processing. Using the definitions introduce above, the matching function can be formalised as below.%The similarity calculation between an unknown message, denoted as and a message in the transaction library, denoted

{\hspace{0.25cm}}
${\MatchingFunc} = (\DistFunc)$

where {\DistFunc} denotes user-defined {\em distance} function.

Assuming the analysis function outputs a set of cluster prototypes $({\ConsensusPrototype}_1, {\ConsensusPrototype}_2, \ldots, {\ConsensusPrototype}_n)$ for interaction clusters $({\InteractionCluster}_1, {\InteractionCluster}_2, \ldots, {\InteractionCluster}_n)$, the distance function {\DistFunc} can be further formalised as follows.

{\hspace{0.25cm}}
$ ({\Interaction}_i, {\ConsensusPrototype}_i) = \mathbf{\DistFunc} ({\Request}_{in}, ({\ConsensusPrototype}_1, {\ConsensusPrototype}_2, \ldots, {\ConsensusPrototype}_n))$

{\vspace{0.20cm}}
\noindent with % {\vspace{-0.4cm}}
\begin{itemize}
\item ${\ConsensusPrototype}_i$ and ${\Interaction}_i$ are the cluster prototype and the cluster centre of ${\InteractionCluster}_i$, where $i \in [1,n]$;

\item ${\Interaction}_i = {\Pair{{\Request}_{centre}}{{\Response}_{centre}}} \in {\InteractionSet}$;
and \\[-0.55cm]
\item $\forall j \in [1,n]$ and $i \neq j$ :
${\DistFunc}({\Request}_{in}, {\ConsensusPrototype}_i) \leq {\DistFunc}({\Request}_{in}, {\ConsensusPrototype}_{j})$
\end{itemize}

The right side of this equation shows the input, and the left side show the output. An implementation of the distance function is described in Section~\ref{chap5sec:distcal} in Chapter~\ref{chap5:qosa}.

\section{Translation Function}
\label{chap4sec:translation}

The cluster with the minimum distance to an incoming request is selected by the matching function as the matching cluster. The translation function is able to utilise the centroid interaction of this cluster for response generation. The translation function contains two key steps, that is, the {\em symmetric field identification} step and the {\em symmetric field substitution} step. The aim of the symmetric field identification step is to identify the {\em symmetric fields} existing in the selected centroid interaction. The symmetric field substitution is to utilise identified symmetric field knowledge for generating a response. We define some notations that are used to describe the translation function formally.

\textbf{Definition 9} (Symmetric Field). %{\SymmetricField} denotes a set of symmetric fields. A symmetric field information that is encoded in the request message of a given interaction, which is subsequently used in the corresponding responses. A single symmetric field $s \in {\SymmetricField}$ is a common subsequence%, of a length greater than a given minimum length threshold, which occur within both the request and responses of a given interaction.
Many protocols encode information in request messages that are subsequently used in the corresponding responses. %For example, application-level protocols such as LDAP add a unique message identifier to each request message. 
%The corresponding response message must contain the same information in order to be seen as a valid response. Therefore, A
Any approach that attempts to synthesize valid response(s) for a request must “copy” such information into the corresponding response message. %Similarly, information associated with a specific request operation (e.g., a search pattern for a search request) will often also be “copied” across from the request to its response.
We refer to such information as \textit{symmetric field}s, denoted as {\SymmetricField}.

For the same interaction, a number of message character substrings occur within both the request and responses. A message character substring is defined as a \textit{common subsequence} when it can meet the following two requirements: (1) having a length greater or equal to a given minimum threshold; (2) not wholly occurring within another longer symmetric field. A symmetric field is either a common subsequence, or contained by a common subsequence. Hence, if we can identify all common subsequences for one interaction, we are able to identify all symmetric fields.

Consider the following cluster centre:

\hspace{0.25cm}
\begin{tabular}{ll}

  % after \\: \hline or \cline{col1-col2} \cline{col3-col4} ...
  ${\Request}_{centre}$: &  Where\textbf{ is }m\textit{y computer book}? \\
  ${\Response}_{centre}$: &  M\textit{y computer book}\textbf{ is }on the desk. \\

\end{tabular}

In this example, \textit{computer book} is critical information that must be copied from the request message into corresponding response. To illustrate common subsequence identification method, we use 4 as the minimum length threshold for a common subsequence. By comparing these two messages, we can get two common subsequences, which are ``y computer book{\footnote {Lowercase letters and uppercase letters are represented by different ASCII codes so that `M' and 'm' are not included in the symmetric field in this example.}}'' and `` is ''. (Note spaces are around `` is '' so its length is 4, equivalent to the given threshold.) As `` computer '', ''book'', etc. occur wholely within a longer symmetric field (``y computer book''), we only take the longest one to be the common subsequence. By finding out these common sequences, symmetric fields ({\eg}\textit{computer book}) are identified.

Considering the following example, once symmetric fields within the request and responses messages of the selected interaction are identified, shown in bold and italic, the symmetric field substitution step is to utilise these knowledge to locate symmetric fields in the live request ${\Request}_{in}$. And then, it modifies responses of the cluster centre with new symmetric fields which are also highlighted in bold and italic, and produce a new response ${\Response}_{out}$.

\hspace{0.25cm}
    \begin{tabular}{ll}
      ${\Request}_{centre}$: & Where\textbf{\ is }m\textit{y computer book}? \\
      ${\Response}_{centre}$: & M\textit{y computer book}\textbf{ is }on the desk. \\
      ${\Request}_{in}$: & Where\textbf{ are }\textit{your computer magazines}? \\
      ${\Response}_{out}$: & M\textit{your computer book}\textbf{ are }on the desk.
    \end{tabular}

%\begin{itemize}
%    \item ${\Request}_{centre}$: \quad { Where\textbf{\ is }m\textit{y computer book}?}\\[-0.65cm]
%    \item ${\Response}_{centre}$: \quad { M\textit{y computer book}\textbf{ is }on the desk.}\\[-0.65cm]
%    \item ${\Request}_{in}$: \quad { Where \textbf{ are }\textit{your computer magazines}?}\\[-0.65cm]
%    \item ${\Response}_{out}$: \quad { M\textit{your computer book}\textbf{ are }{on the desk.}}\\[-0.65cm]
%\end{itemize}

The translation function {\TransFunc} can be formalised as below.

{\hspace{0.25cm}}
${\TransFunc} = ({\IdentifyingStep}, {\SubstitutingStep})$

The {\IdentifyingStep} is the step that is performed to identify symmetric fields occurring in the request and responses of a recorded interaction. Once the symmetric fields are identified, the {\SubstitutingStep} is to alter the responses in order to synthesize a response for an incoming request. We can formalise each step as follows. The right side of these equations shows inputs being taken by each step, and the left side shows each step's output.

{\hspace{0.25cm}}
${\SymmetricField} = \mathbf{\IdentifyingStep} \ {\Pair{{\Request}_{centre}}{{\Response}_{centre}}}$

{\hspace{0.25cm}}
${\Response}_{out} = \mathbf{\SubstitutingStep} \ ({\Request}_{in}, {\SymmetricField}, {\Response}_{centre})$

, where {\em {\IdentifyingStep}}, {\em {\SubstitutingStep}} denote user-defined approaches for each step of the translation function. Implementations of these two step are illustrated in Section~\ref{chap5subsec:compat} and Section~\ref{chap5subsec:fieldsubstitution} in Chapter~\ref{chap5:qosa}.

\section{Summary}
\label{chap4sec:summary}

In this chapter, we have presented an overview of our opaque response generation framework for automatic service virtualisation. Our framework includes three components: the {\em analysis function}, the {\em matching function} and the {\em translation function}. The steps for each component have been described and formalised while the implementation for each component will be presented in subsequent three chapters of this thesis. 
%introduce the alignment algorithm in details
\chapter{Opaque Response Generation Approach}
\label{chap5:qosa}

In chapter \ref{chap4:framework}, we describe an architecture of our automatic service virtualisation approach, which contains three main components, that is, the {\em analysis function}, the {\em matching function} and the {\em translation function}. %As the motivation behind our opaque response generation approach is to virtualise a service without explicit knowledge of the service protocol structure information and payload information ({\cf Chapter~\ref{chap3sec:conclusion} for definitions}), the {\em matching function} and the {\em translation function} are the major components of our approach for generating responses.
In this chapter, we introduce the {\em whole library} technique which implements our conceptual framework for automatic service virtualisation. In this implementation, the output of the analysis function is the whole trace library of captured interaction recordings. In other words, each interaction recording is considered as a separate cluster so that the number of clusters is equivalent to the number of interactions in the trace library. %The analysis function outputs the whole interaction recording takes the whole interaction recordings as its output these two components with technical details. %By using a genome sequencing alignment algorithm to calculate distance
A genome sequencing alignment algorithm and a field substitution algorithm are adopted to implement the matching function and the translation function, respectively. This technique is to generate responses directly from recorded interaction traces, without requiring explicit knowledge of the underlying protocol a system-under-test used to communicate with a target service.
%The genome sequencing alignment algorithm is executed by both the matching function and the translation function. Given an incoming request, at the matching stage, the sequencing alignment algorithm provides a way of ranking similarities between two messages and thereby searches for it a suitably “similar” request from the analysis function output. The substitution function 1) applies the sequencing alignment algorithm to indicate the common patterns between the recorded request and the recorded response, 2) uses the field substitution algorithm to generate a response with identified common patterns.

The organisation of this chapter is as follows. In Section~\ref{chap5sec:needlemanwunsch}, we introduce the genome sequencing alignment algorithm and its application. In Section~\ref{chap5sec:distcal}, we discuss the adoption of a genome sequencing alignment algorithm for implementing our conceptual {\em matching function}. %and
In Section~\ref{chap5sec:translation}, we describe the implementation of the {\em translation function}. Specifically, we introduce how we adapt a genome sequencing alignment algorithm for identifying common patterns between a request message and its corresponding response message(s) in Section~\ref{chap5subsec:compat}. In Section~\ref{chap5subsec:fieldsubstitution}, we present a novel field substitution algorithm for generating responses. %sequencing alignment algorithm for 1) ranking the similarity of two messages and 2) identifying their common patterns. The field substitution algorithm with its technical details is introduced in Section~\ref{chap5sec:substitution}.
Section~\ref{chap5sec:evaluation} demonstrates the experimental results. Finally, Section~\ref{chap5sec:summary} summarises this chapter.

This chapter is mainly based on our work presented in ~\cite{Du:2013}.
%------------------------------------------------------------------------ %

\section{Needleman-Wunsch Sequence Alignment}
\label{chap5sec:needlemanwunsch}
% Give a brief explanation of the genome sequencing alignment algorithm
%For the purpose of this work, we are using a modified version of the solution presented by Needleman and Wunsch \cite{needleman:1970} as our distance measure. This is because as the stochastic approach presented by Ristad and Yianilos \cite{Ristad:1998} relies on a suitably configured benchmark corpus, we may not always be able to generate this from the recorded interaction traces.

\begin{figure}[!htb]
\centering
    \subfloat[Alignment 1 (total score: 4)\label{chap5fig:alignNW}]{%
          \includegraphics[width=0.35\textwidth,height=0.075\textheight]{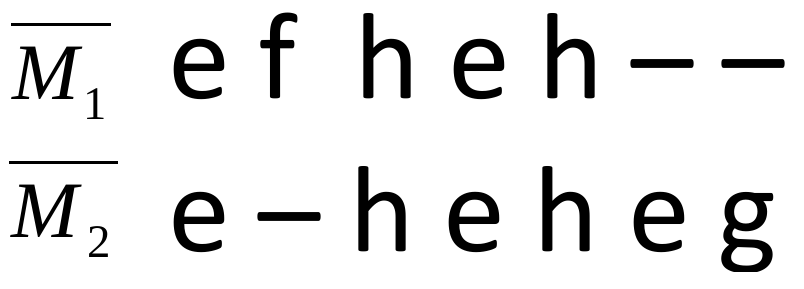}      }
     \quad
     \subfloat[Alignment 2 (total score: 1)\label{chap5fig:alignSW}]{%
          \includegraphics[width=0.35\textwidth,height=0.075\textheight]{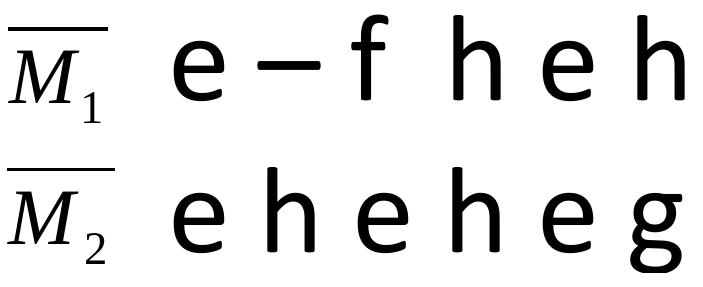}
          }
   \caption{Pairwise sequence alignments, match=1, mismatch=-1, gap=0. }
   \label{chap5fig:alignexample}
\end{figure}

%Bioinformatics is a discipline where various techniques from mathematics, statistics, and computer science are utilised to solve biological problems. The biological problems often include finding specific patterns in large sequence of complex data~\cite{SeqAlign}.
Sequence alignment is a way of arranging two or multiple sequences in order to identify a series of characters that are in the same order in all sequences. Approaches to aligning two sequences are called {\em pair-wise alignment}, and those to aligning three or more sequences are called {\em multiple sequence alignment}. When doing pair-wise sequence alignment, two sequences can be aligned by writing them into a table with two rows. %across a page in two rows.
Identical or similar characters are placed in the same column, and non-identical ones can either be placed in the same column as a mismatch or against a gap ({\em -} for example) in the other sequence. %Sequences that are aligned in this manner are said to be similar.
Sequence alignment result is useful for discovering functional, structural, and evolutionary information in biological sequences.
%Identified genetic patterns may be used to reflect evolutionary relationships between the sequences.
Figure \ref{chap5fig:alignexample} shows an example of aligning two message sequences $M_1=efheh$ and $M_2=eheheg$. %through a set of editing operations applied to one of them iteratively.

Given two messages, there are many variants of aligning them. To determine the best alignment between them, a very popular approach is to calculate the total score of each alignment, and then select one with the maximum total score. The total score of the alignment depends on each column of the alignment. For example, if the column has two identical characters, it will receive value +1 (a match); if the column has two different characters, it will receive value -1 (a mismatch); if the column has a gap, it will receive value 0 (Gap Penalty). For the alignment example in Figure~\ref{chap5fig:alignexample}, the best alignment will be the alignment 1 in Figure~\ref{chap5fig:alignNW} with the total score: $4 \times 1 + 0 \times (-1) + 3 \times (0) = 4$, while the Alignment 2 in Figure~\ref{chap5fig:alignSW} gives a total score: $3 \times 1 + 2 \times (-1) + 1 \times (0) = 1$. These parameters (match, mismatch and gap penalty) can be adjusted to different values according to the choice of sequences or experimental results.

One approach to computing similarity between two sequences is to generate all possible alignments and pick the best one. However, the number of possible alignments for two traces of length $l$ is $\approx (1+\sqrt{2})^{2l+1}l^{-1/2}$ \cite{BiologyIntroduction}, which grows exponentially with the length of the sequences. Therefore, it is infeasible to enumerate all possible alignments, and identify the best alignment. {\em Needleman-Wunsch algorithm} is proposed as a solution to produce an optimal alignment~\cite{BioinformaticsAlgorithms:2004}.

{\em Needleman-Wunsch algorithm} is a popular dynamic programming algorithm for finding the best alignment between two
message sequences under a given scoring matrix~\cite{needleman:1970}. It can be used to
compare two message sequences that have overall similarity and similar lengths. Moreover, it can also be used to indicate common patterns of two significantly different strings. Giving two message sequences Seq1 and Seq2 ({\cf Definition 2 in Chapter \ref{chap4sec:preliminaries}}) instead of determining the similarity between sequences as a whole, the Needleman-Wunsch algorithm builds up the solution by determining all similarities between arbitrary prefixes of the two sequences. The algorithm starts with shorter prefixes and uses previously computed results to solve the problem for larger prefixes. Finally, it outputs their optimal global alignment.

To better illustrate the Needleman-Wusnch algorithm, we symbolise the sequence alignment problem as follows. Given any two message sequences $M_1 = c_1c_2c_3\ldots c_h$, $M_2 = c_1c_2c_3\ldots c_l$ with $M_i \in {\MessageSet}$, a {\em pair-wise sequence alignment} is a pair of message character sequences $A_1$, $A_2$ with $A_1,A_2 \in ({\MsgCharSet} \cup {-})^+$:
\begin{itemize}
  \item {\em -} $\notin ${\MsgCharSet} is called {\em gap character};
  \item Removing all gaps from $A_1$ and $A_2$, we can obtain $M_1$ and $M_2$.
\end{itemize}

, where {\MsgCharSet} denotes a set of message character and {\MessageSet} is the set of every message on it. The algorithm contains three main steps:

\begin{enumerate}
  \item {\textbf{Initialization}

  This is a step that is to create a matrix {\em F} with $(h+1)$ rows and $(l+1)$ columns where $h$ and $l$ correspond to the size of the messages sequences to be aligned.}

  \item {\textbf{Matrix fill (scoring)}

  In the matrix filling step, the matrix {\em F} is filled, where the value F(i,j) is the score of the best alignment between the prefix ${M_1}^i$ of $M_1$ and the prefix ${M_2}^j$ of $M_2$. F(i,j) is constructed recursively by starting from the top left corner in the matrix, denoted as F(0,0), and then proceeding to fill the matrix from top left to bottom right using the following scoring scheme $f$.

    \begin{equation}
    \label{eq:score}
    f(c_i, c_j) = \left\{ \begin{array}{rl}
                          m &\mbox{ if $c_i=c_j \land \forall c_i, c_j \in {\MsgCharSet}$} \\
                          n &\mbox{ if $c_i \neq c_j \land \forall c_i, c_j \in {\MsgCharSet}$} \\
                          g &\mbox{ if $ c_i=c_j=-$ }

    \end{array}
    \right.
    \end{equation}

   There are three possible ways that the best score F(i,j) could be obtained: $M_1(i)$ could be aligned to $M_2(j)$, in which case $F(i,j)=F(i-1,j-1)+m/n$; or $M_1(i)$ is aligned to a gap, in which case $F(i,j)=F(i,j-1)+g$; or $M_2(j)$ is aligned to a gap, in which case $F(i,j)=F(i-1,j)+g$. The best score up to (i,j) will be the maximum of these three options. In other word, we have equation~\ref{eq:opticalscore}.

    \begin{equation}
    \label{eq:opticalscore}
    F_{i,j} = \left\{ \begin{array}{rl}
                                  F_{i-1,j-1}+m &\mbox{ if $c_i=c_j \land \forall c_i, c_j \in {\MsgCharSet}$} \\
                                  F_{i-1,j-1}+n &\mbox{ if $c_i \neq c_j \land \forall c_i, c_j \in {\MsgCharSet}$} \\
                                  F_{i,j-1}+g &\mbox{ if $- \in M_1$ }\\
                                  F_{i-1,j}+g &\mbox{ if $- \in M_2$ }

    \end{array}
    \right.
    \end{equation}
    }

   \item {\textbf{Traceback (alignment)}

       The traceback step begins at position F(h,l) in the matrix (the lower right hand corner), which is the position which has the best score for an alignment of $M_1$ and $M_2$. To find the alignment itself, we must find the path of choices from Equation~\ref{eq:opticalscore} that led to this best score. We move recursively from the current cell (i,j) to one of the cells (i-1,j-1), (i-1,j) or (i,j-1), from which the value F(i,j) was derived. At the end of this procedure, we will reach the start of the matrix, i=j=0.
   }
    \end{enumerate}

 We use a running example to illustrate each step of this algorithm. Considering the two given messages $M_1=efheh$ and $M_2=eheheg$. The length of the message 1 and message 2 are 5 and 6 respectively.
\begin{figure}%
    \centering
    \subfloat[Initialization\label{chap5fig:fmatrixini}]{%
     \centering\parbox{0.8\linewidth}{
          \includegraphics[width=0.8\textwidth,height=0.22\textheight]{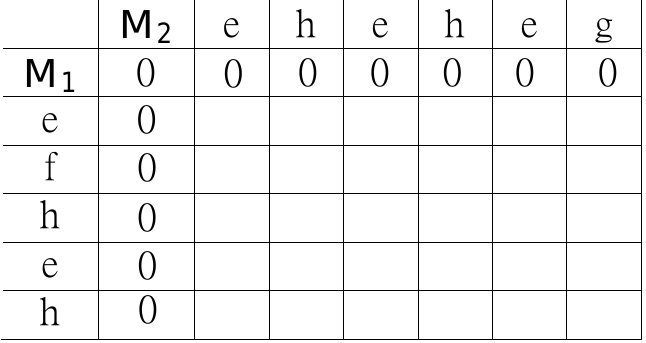}
          }
        }
     \\
     \subfloat[Matrix fill\label{chap5fig:fmatrixdistance}]{%
     \centering\parbox{0.8\linewidth}{
          \includegraphics[width=0.8\textwidth,height=0.22\textheight]{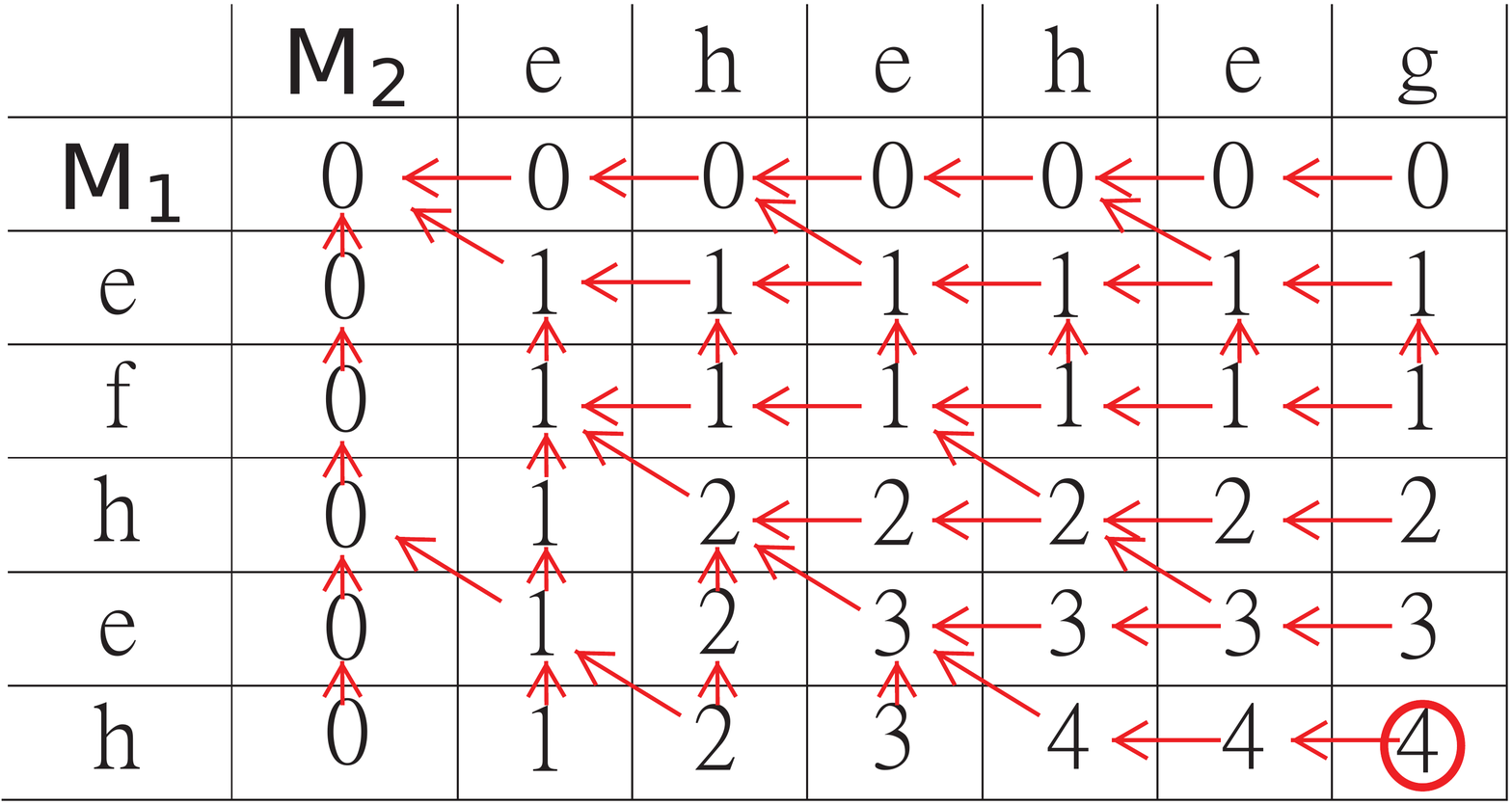}
          {m=1, n=-1, g=0\\}
          }
        }
     \\
     \subfloat[Traceback\label{chap5fig:tracebackdistance}]{%
     \centering\parbox{0.8\linewidth}{
          \includegraphics[width=0.8\textwidth,height=0.22\textheight]{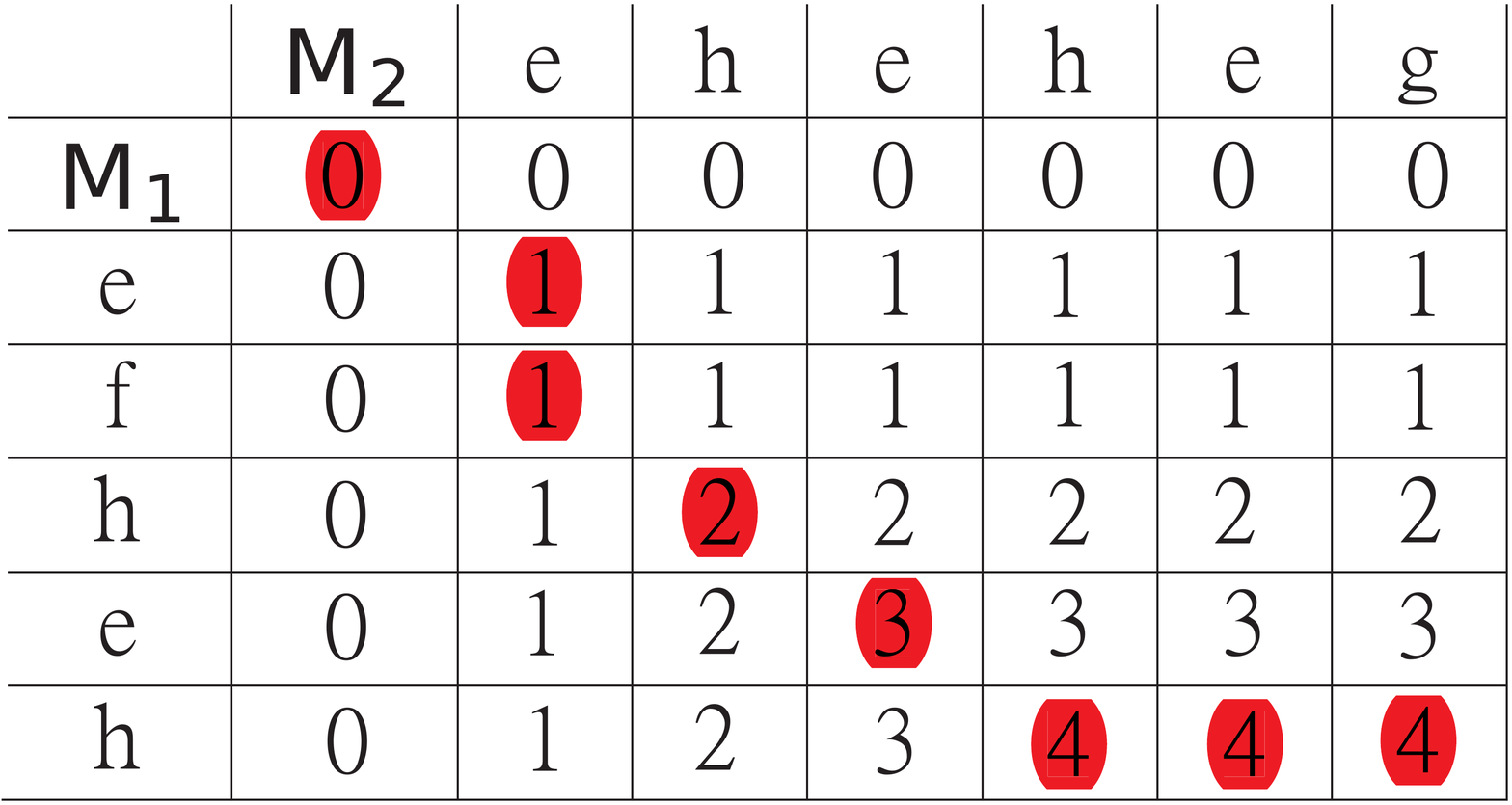}
          {$M_1^{'}$ = e f h e h -- --\\
          $M_2^{'}$ = e -- h e h e g }
          }
     }
   \caption{Needlman-Wunsch Algorithm. The matrix fill and the traceback computing an alignment between two traces using the scoring scheme, where m=1, n=-1, g=0. Their alignment can indicate different characters of these two messages. }
\label{chap5fig:distcalcexample}
\end{figure}

\begin{enumerate}

\item \textbf{Step 1: Initialization Step}

We can get a $6 \times 7$ matrix. Extra row and column are given so as to align with gaps. After creating the matrix, a scoring schema is introduced which can be defined with specific scores. For example, if two message characters at $i^{th}$ and $j^{th}$ positions are same, matching score is 1 $(f(c_i, c_j)=1)$ or if the two message characters at $i^{th}$ and $j^{th}$ positions are not same, mismatch score is assumed to be -1 ($f(c_i, c_j)=-1$). The gap score or gap penalty is assumed to be 0. Equation~\ref{eq:score1} shows a summary of this user defined scoring schema.

\begin{equation}
\label{eq:score1}
f(c_i, c_j) = \left\{ \begin{array}{rl}
                      1 &\mbox{ if $c_i=c_j \land \forall c_i, c_j \in {\MsgCharSet}$} \\
                      -1 &\mbox{ if $c_i \neq c_j \land \forall c_i, c_j \in {\MsgCharSet}$} \\
                      0 &\mbox{ if $ c_i=c_j=-$ }

\end{array}
\right.
\end{equation}

To initialise this matrix, we first put value 0 to the first position. Then, the gap score can be added to the previous cell of the row or column. As the gap value is assumed to be 0 in this example, first row and first column of the matrix can be initially filled with 0, as shown in  Figure~\ref{chap5fig:fmatrixini}.

\item \textbf{Step 2: Matrix fill}

The second step of the Needleman-Wunsch algorithm is to fill the matrix. One possible solution of the matrix filling step finds the maximum global alignment score by starting in the upper left corner in the matrix and finding the maximal score $F_{i,j}$ for each cell in the matrix. In order to find $F_{i,j}$ for any position (i,j), it is required to know the neighbouring scores for the matrix positions to the left, above and diagonal to the current position, that is, the scores for $F_{i-1,j}$, $F_{i,j-1}$, and $F_{i-1,j-1}$. By adding the match or mismatch scores to the diagonal value and adding the gap score to the left and above values, we can thus obtain three different values, where we can fill the position (i,j) with the maximum among them. Overall the equation can be shown as follows.

\hspace{0.25cm}
$F_{i,j} = Maximum[F_{i-1,j-1}+m/n, F_{i,j-1}+g, F_{i-1,j}+g]$

We can apply the above formula to score the matrix in Figure~\ref{chap5fig:fmatrixini}. Consider the current position (the first position $M_{1,1}$), the first message characters in the two sequences are ``{\em e}'' and ``{\em e}''. Since they are matching characters, the score would be 1.

\hspace{0.25cm}
$F_{1,1} = Maximum[F_{0,0}+m, F_{1,0}+g, F_{1,0}+g]$

\hspace{1cm}
= $Maximum[0+1, 0+0, 0+0]$

\hspace{1cm}
= $Maximum[1, 0, 0]$

\hspace{1cm}
= 1

The obtained score is 1, being placed in position (1,1) of the scoring matrix. Then, we can fill all the remaining rows and columns by using the above equation and method sequentially. Furthermore, we place back pointers to the cell from where the maximum score is obtained, seen as red arrows, which are predecessors of the current cell (Figure~\ref{chap5fig:fmatrixdistance}).

\item \textbf{Step 3: Traceback and alignment}

The final step in the algorithm is the trace back for the best alignment. Figure~\ref{chap5fig:fmatrixdistance} shows  the matrix filling with back pointers of the above mentioned example. We can see the bottom right corner score as 4. The current cell with value 4 has an immediate predecessor, where the maximum score is obtained from its left neighbour{\footnote {Note that if there are two or more values which points back, it suggests that there can be two or more possible alignments.}}. By continuing the trace back step with the above defined method, we can get a path to the $0^{th}$ row, $0^{th}$ column. When trace back is completed, the alignment of two sample messages can be obtained whereby, starting from the top left corner of this matrix, 1) if a diagonal score is on the path, we take characters of the two messages; 2) if a bottom score is on the path, we take only vertical axis character, and insert a gap character into current position of the horizontal message; 3) if a right score is on the path, we take only horizontal axis character, and insert a gap character into current position of the vertical message. An alignment of the above mentioned example is shown in Figure~\ref{chap5fig:tracebackdistance}. (Note if there are more than one optimal alignment, that is, two or more alignments have the same score, we select the first optimal alignment that we obtain to be our alignment result.)

\end{enumerate}

The Needleman-Wunsch algorithm is applied multiple times at different stages of our framework for different purposes.

\section{Message Distance Calculation for the Matching Function}
\label{chap5sec:distcal}
% This should be a good start for explaining why we choose this for distance calculation

As described in Section~\ref{chap5sec:needlemanwunsch}, the Needleman-Wunsch algorithm can perform global sequence alignment between two message sequences and indicate their common subsequences, which lends itself as a suitable edit distance calculation method. The {\em analysis function} outputs a whole library of interaction recordings for the matching function. At runtime, for a new request, the matching function 1) adopts the Needleman-Wunsch algorithm to calculate the edit distance between this new request and each recorded request message in the library, 2) calculates their {\em dissimilarity ratios} ({\cf} Section~\ref{chap4sec:distance}), and 3) selects a recorded interaction to be the interaction for the translation function, where its request message has smallest dissimilarity ratio.

The following example shows how our matching function selects a recorded interaction for the following translation function. Consider a request is received:

\hspace{0.25cm}
\textbf{Incoming Requeset ($Rq'$):}

\hspace{0.5cm}
\{id:75,op:S,sn:Hune\}

\begin{table}[t]
\footnotesize
\begin{center}
\begin{tabular}{|c||l||l|}
\hline

Index & Request Messages & Dissimilarity Ratio \\ \hline\hline
1 & \{id:1,op:B\} & 0.275 \\ \hline
2 & \{id:2,op:S,sn:Du\} & 0.125 \\ \hline
3 & \{id:13,op:S,sn:Versteeg\} & 0.1875 \\ \hline
4 & \{id:24,op:A,sn:Schneider,mobile:123456\} & 0.307 \\ \hline
\textbf{5} & \textbf{\{id:275,op:S,sn:Han\}} & 0.05 \\ \hline
6 & \{id:490,op:S,sn:Grundy\} & 0.152 \\ \hline
7 & \{id:2273,op:S,sn:Schneider\} & 0.185 \\ \hline
8 & \{id:2487,op:A,sn:Will\} & 0.182 \\ \hline
9 &\{id:3106,op:A,sn:Hine,gn:Cam,Postcode:33589\} & 0.318\\ \hline
10 & \{id:3211,op:U\} & 0.275 \\ \hline
11 & \{id:1,op:B\} & 0.275 \\ \hline
12 & \{id:12,op:S,sn:Hine\} & 0.075 \\ \hline
13 & \{id:34,op:A,sn:Lindsey,gn:Vanessa,PostalAddress1:83 Venton Road\} & 0.383 \\ \hline
14 & \{id:145,op:S,sn:Will\} & 0.143 \\ \hline
15 & \{id:1334,op:S,sn:Lindsey,gn:Vanessa,PostalAddress1:83\ Venton\ Road\} & 0.379  \\ \hline
16 & \{id:1500,op:U\} & 0.25 \\ \hline
\end{tabular}
\end{center}
\caption{Message Distance Calculation}
\label{chap5tab:distance}
\end{table}

Table~\ref{chap5tab:distance} shows the requests of all interactions in the trace library from Table~\ref{Chap1tab:tl}, and their comparison results against the above mentioned new request, using the Needleman-Wunsch algorithm (with the parameters: match = 1, mismatch = -1, gap = 0). As the similarity ratio between two messages is determined by the number of insert gaps ({\cf} Section~\ref{chap5sec:distcal}), different message characters are not able to be aligned at the same position. To this end, we assign bigger penalty value to mismatching characters than insert gaps. After performing the matching function, the 5th request ``\{id:275,op:S,sn:Han\}'' is determined to be the matching request. Hence, the centroid interaction ({\cf} Definition 7 in Section~\ref{chap4sec:distance}) is:

(\{id:275,op:S,sn:Han\},\{id:275,op:SearchRsp,result:Ok,gn:Jun,sn:Han,mobile:33333333\})

% ------------------------------------------------------------------------ %

\section{Implementation of the Translation Function}
\label{chap5sec:translation}

Given an incoming request ${\Request}'$ and a recorded interaction ${\Interaction}_i = {\Pair{{\Request}_{centre}}{{\Response}_{centre}}}$, the last step of the process is for the translation function to generate the final response. The translation function copies some information from the live request into the selected response for playback. As described in Section~\ref{chap4sec:translation}, it contains two steps. The first step is to discover symmetric fields by comparing the request message of this interaction and its corresponding response(s). Technical details of this step can be seen in Section~\ref{chap5subsec:compat}. The second step is to modify the recorded response(s) with corresponding information in the new request, which is illustrated in Section~\ref{chap5subsec:fieldsubstitution}.

\subsection{Symmetric Field Identification}
\label{chap5subsec:compat}

Many protocols encode information in request messages that are subsequently used in the corresponding responses. For example, application-level protocols such as LDAP add
a unique message identifier to each request message. The
corresponding response message must contain the same message identifier in order to be seen as a valid response. Such information is referred to as {\em symmetric fields} ({\cf} Definition 9 in Section~\ref{chap4sec:translation}). %They are defined as common subsequences, of a length greater than a given threshold, which occur within the same interaction for both the request and responses ({\cf} Definition 9 in Section~\ref{chap4sec:translation}).
The aim of the symmetric field identification method is to discover symmetric fields between one request message and its associated responses. %We use the Needleman-Wunsch algorithm described in Section~\ref{chap5sec:needlemanwunsch} in order to identify symmetric field: they are the common subsequences of a request and its associated response. However,

Symmetric fields may not appear in the request and responses in the same order and/or cardinality. Furthermore, the same symmetric field may occur multiple times within ${\Request}$ and/or ${\Response}$. Therefore, we record their positions within ${\Request}$ and ${\Response}$ respectively. In Figure~\ref{Agl:CommonStrings}, we present pseudocode of the symmetric field identification algorithm for finding symmetric fields within two messages. This algorithm takes two messages and a minimum length threshold as input, denoted as ${\Request}={\MessageCharacter}_{11}{\MessageCharacter}_{12}\ldots{\MessageCharacter}_{1m}$, ${\Response}={\MessageCharacter}_{21}{\MessageCharacter}_{22}\ldots{\MessageCharacter}_{2n}$, and $minLen$. It outputs a list of tuples. Each tuple represent a symmetric field in the format of {\em (match, rqpos, rsppos, length)}, where {\em match} is the sequence of message characters representing the symmetric field,  {\em rqpos} is a list (length 1 or more) of indices where the matching string starts in ${\Request}$, {\em rsppos} is a list (length 1 or more) of indices where the match starts in ${\Response}$, and $length$ is the number of message characters in the matching symmetric field.

\renewcommand{\algorithmicrequire}{\textbf{Input:}}
\renewcommand{\algorithmicensure}{\textbf{Output:}}

\begin{figure}[!t]
\algsetup{indent=1.8em}
\begin{algorithmic}[1]
    \REQUIRE
    {\Request}, {\Response}, minLen
    \ENSURE
    ${\SymmetricField}$ = a list of tuples (match,rqpos,rsppos,length)
    \item[]

    \COMMENT {// Generate Common Message Character Matrix}

    \FOR{$i=1$ \TO $n$}
        \FOR{$j=1$ \TO $m$}
            \IF{current character already exist in a common subsequence}
                \STATE {continue;}
            \ENDIF

            \COMMENT {// Length of Current Common Subsequence}

            \STATE length=0;

            \WHILE{$(i+length)<n \&\& (j+length)<m$}
                \IF{$c_{1(j+length)}=c_{2(i+length)}$}
                    \STATE length++;
                 \ENDIF
            \ENDWHILE
            \STATE skip all shorter common subsequences which are completely contained in current common subsequence
            \IF{$length>=minLen$}
                \STATE $rqpos \leftarrow j$;
            \ENDIF
        \ENDFOR
        \IF{$length>=minLen$}
        \STATE add (${\MessageCharacter}_{1(rqpos)}\ldots{\MessageCharacter}_{1(rqpos+length-1)}$, i,rqpos, length) to ${\SymmetricField}$
        \ENDIF
    \ENDFOR
    \RETURN ${\SymmetricField}$
\caption{Common subsequences identification algorithm}
\label{Agl:CommonStrings}
\end{algorithmic}
\end{figure}

For the above centroid interaction, there are two symmetric fields, which are highlighted in red and blue, respectively, as shown in Table~\ref{chap5tab:symmetricfields}. The values for {\em (match,rqpos,rsppos,length)} for both identified symmetric fields are shown in Table~\ref{chap5tab:symmetricfieldstuple}. The java source code of the implementation of the symmetric field and the symmetric field identification algorithm are listed in appendix~\ref{apx.symmetrixfield} and appendix~\ref{apx.symmetrixfieldidentify} respectively.

\begin{table}
\begin{center}
\begin{tabular}{|p{6cm}|p{7.5cm}|}
\hline
${\Request}_{center}$ & ${\Response}_{center}$ \\ \hline
 \color{blue}{\{id:275,op:S}\textbf{\color{red}{,sn:Han}}\} & \color{blue}{\{id:275,op:S}\color{black}{earchRsp,result:Ok,gn:Jun}\textbf{\color{red}{ ,sn:Han}}\color{black}{,mobile:33333333\}} \\ \hline
\end{tabular}
\end{center}
\caption{Common Subsequences Occurring in the Centroid Interaction}
\label{chap5tab:symmetricfields}
\end{table}

\begin{table}
\begin{center}
\begin{tabular}{|c|c|c|c|}
\hline
{\em match} & {\em rqpos} & {\em rsppos} & {\em length} \\ \hline
\{id:275,op:S & 0 & 0 & 12 \\ \hline
,sn:Han &  12 & 18 & 7 \\ \hline
\end{tabular}
\end{center}
\caption{Symmetric Field Tuples}
\label{chap5tab:symmetricfieldstuple}
\end{table}

\subsection{Field Substitution Method}
\label{chap5subsec:fieldsubstitution}

The field substitution method performs the symmetric field substitution to modify the response ${{\Response}_{centre}}$ for generating a response to an incoming request. It includes three steps. In the first step, we line up symmetric fields of the centroid request ${\Request}_{centre}$ with specific fields in an incoming request ${\Request}^{'}$, using the Needleman-Wunsch algorithm (with the parameters: match=1, mismatch=-1, gap=-1). By giving same penalty values to the mismatch characters and the insert gap, we can line up the recorded request with the incoming request with the least number of gaps being inserted. Table~\ref{chap5tab:alignnewrequest} shows an alignment between a new request and the recorded request, where ${\Request}^{'}$ represents an incoming request and ${\Request}$ is the request message of the recorded interaction.

\begin{table}
\footnotesize
\begin{tabular}{l*{21}{c}r}
&0&1&2&3&4&5&6&7&8&9&10&11&12&13&14&15&16&17&18&19&20\\ \hline
Rq' &\{ & i & d & : & {\agap} & 7 & 5 & , & o & p & : & S & , & s & n & : & H & u & n & e &\} \\
\hline
Rq &\{&i&d&:&2&7&5&,&o&p&:&S&,&s&n&:&H&a&n&{\agap}&\} \\
\hline
\end{tabular}
\caption{A New Request and the Centroid Request Alignment}
\label{chap5tab:alignnewrequest}
\end{table}

In the second step, we update the {\em rqpos} position indices for all symmetric fields, in order to compensate for any gaps which may have been inserted as part of the recorded request and the incoming request alignment process.  Table~\ref{chap5tab:symmetricfieldsupdate} shows the modified positions ${rqpos}^{'}$.  If gaps are inserted within the symmetric field, then the length of the symmetric field will need to be modified correspondingly; let ${length}^{’}$ be the modified length.  The match string should also be modified to contain the aligned bytes or characters in the symmetric field position for ${\Request}^{'}$. Let ${match}^{'}$ be the modified symmetric field value.

The third step is to generate a response ${\Response}^{'}$ by modifying the response ${\Response}_{centre}$. For each symmetric field, we copy the symmetric field value ${match}^{'}$ to overwrite the response message characters at position ${rsppos}^{'}$. If ${length}^{'}$ is different to the original symmetric field length, then extra characters will need to be inserted or deleted into the response at the symmetric field positions, to compensate.  Let us call the response with the substituted symmetric fields ${\Response}^{’}$. Consider the above example, after substituting the symmetric field strings copied from ${\Request}^{'}$, copied to the symmetric field position in ${\Response}_{centre}$, the modified response (${\Response}^{'}$) becomes:

\color{blue}{\{id:75,op:S}\color{black}{earchRsp,result:Ok,gn:Jun}\textbf{\color{red}{,sn:Hune}}\color{black}{,mobile:33333333\}}

The java source code of the implementation of modifying the centroid response is listed in the appendix~\ref{apx.substitution}.

\begin{table}
\begin{center}
\begin{tabular}{|c|c|c|c|}
\hline
{\em match'} & {\em rqpos'} & {\em rsppos} & {\em length'} \\ \hline
\{id:\color{red}{75}\color{black}{,op:S} & 0 & 0 & \color{red}{11} \\ \hline
,sn:\color{red}{Hune} &  12 & \color{red}{19} & \color{red}{8} \\ \hline
\end{tabular}
\end{center}
\caption{Update Symmetric Field Properties}
\label{chap5tab:symmetricfieldsupdate}
\end{table}

%\subsection{Implementation}
%
%We have developed a proof of concept realisation of our framework, including
%the sequence alignment, the symmetric field identification and substitution
%algorithms, as well as the underlying modified Needleman-Wunsch algorithm.
%This was proof of concept prototype was implemented in the Java programming
%language. Both Wireshark and LISA were used to capture network traffic and
%exported into a format suitable for input into our Java implementation. At the
%time of writing, the implementation was not specifically optimized (both from
%a performance and memory consumption perspective).  This is a task we intend
%to cover in future work.

%%%%%%%%%%%%%%%%%%%%%%%%%%%%%%%%%%%%%%%%%%%%%%%%%%%%%%%%%%%%%%%%%%%%%%%%%%%%%

\section{Evaluation}
\label{chap5sec:evaluation}

In this section, we present a set of experiments that we conducted to evaluate the
effectiveness of the approach presented in the previous section and discuss
the results of our experiments. More specifically, we introduce our
experimental setup as well as our evaluation criteria in sections
\ref{subsec:experiment} and \ref{chap5subsec:criteria}, respectively. In Section
\ref{subsec:results}, we present the results of our cross-validation and
illustrate the accuracy of the synthesized responses. Finally, we discuss
limitations of our current approach and identify possible areas of future
improvements in Section \ref{sec:discussion}.

% ------------------------------------------------------------------------- %

\subsection{Experimental Setup}
\label{subsec:experiment}

Although one of the aims of our work is to enable emulation for unknown or
ill-specified protocols, for evaluation purposes, we used two protocols where
the precise message structures as well as the corresponding temporal
properties are known: the Simple Object Access Protocol (SOAP) \cite{SOAPv1.1}
and the Lightweight Directory Access Protocol (LDAP) \cite{Sermersheim2006}.
Both are commonly used application-layer protocols and hence lend themselves
as case studies for our evaluation. SOAP is a light-weight protocol % that is
designed for exchanging structured information in a decentralised, distributed
environments whereas LDAP is widely used in large enterprises for maintaining
and managing directory information. Our aim is to evaluate the effectiveness of our approach for generating valid responses. We assume all recorded interactions comfort to the message format specification and the protocol specification. There are no error interaction recordings in these two trace samples.

\smallskip

The interaction trace for SOAP used for our evaluation was generated based on
a recording of a banking example using the service virtualisation tool \cite{Michelsen:11}. Refer to Appendix~\ref{apx.soapexample} for the SOAP interaction examples which were used in our experiment. The
protocol consists of $7$ different request types, each with a varying number
of parameters, encoding ``typical'' transactions one would expect from a
banking service. From a pre-defined set of account attributes ({\eg} account ids, account names etc). We
then randomly generated an interaction trace containing $1,000$
request/response pairs. Amongst those, we had 548 unique requests (with only
22 requests occurring multiple times), 714 uniqe responses (the replicated ones
are predominantly due to the fact that the {\tt deleteTokenResponse} message
only had true or false as possible return values), and $23$ duplicated
request/response pairs. For the purpose of our evaluation, we considered this
a sufficiently diverse ``population'' of messages to work with.

% \smallskip

LDAP is a binary protocol that uses an ASN.1~\cite{ASN.1} encoding to encode and decode
text-based message information to and from its binary representation,
respectively. For the purpose of our study, we used a ASN.1 decoder to translate recorded LDAP messages into a text format and an encoder to
check whether the synthesized responses were well-formed ({\cf}
Section~\ref{chap5subsec:criteria}). LDAP message examples can be found in Appendix~\ref{apx.ldapexample}. In future work, we plan to investigate whether
we can omit the encoding/decoding steps and directly manipulate the binary representations.

The LDAP interaction trace used for the evaluation consisted of $498$ unique
interactions containing the core LDAP operations, such as \textit{adding},
\textit{searching}, \textit{modifying} etc. applied to CA's {\em DemoCorp}
sample directory \cite{ca:democorp}. The trace did not contain any duplicated
requests or responses, and the search responses contained a varying number of
matching entries, ranging from zero to $12$.
% ------------------------------------------------------------------------- %

\subsection{Cross-Validation Approach and Evaluation Criteria}
\label{chap5subsec:criteria}

 \begin{figure}[h]
 \centering
 \includegraphics[width=.8\textwidth]{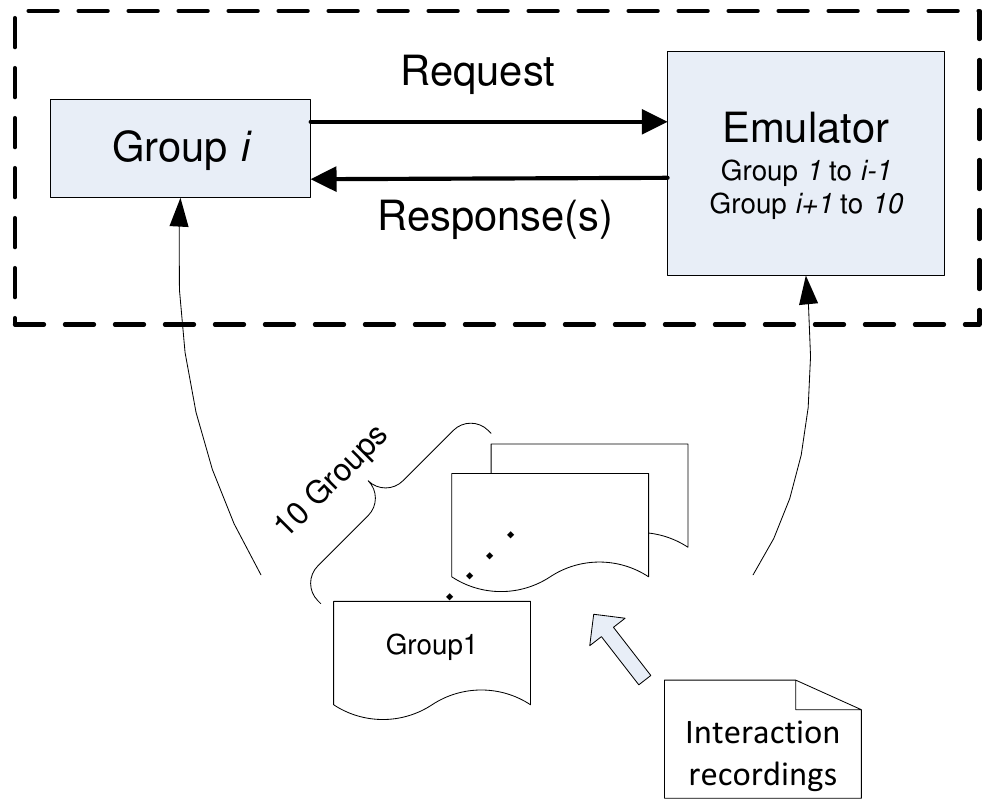}
 \caption{10-fold Cross Validation Approach}
 \label{fig:evaluationprocedure}
 \end{figure}

A {\em cross-validation approach} \cite{devijver:1982} is one of the most
popular methods for assessing how the results of a statistical analysis will
generalise to an independent data set. For the purpose of our evaluation, we
applied the commonly used 10-fold cross-validation approach
\cite{mclachlan:2004} to both the recorded SOAP and LDAP messages,
respectively.

As shown in Figure \ref{fig:evaluationprocedure}, we randomly partitioned the original
interactions' data set into 10 groups. Of the 10 groups,
group $i$ ({\cf} top-left rectangle in Figure \ref{fig:evaluationprocedure})
is considered to be the {\em evaluation group} for testing our approach, and
the remaining 9 groups constitute the {\em training set}. The cross-validation
process is then repeated 10 times (the same as the number of groups), so that
each of the 10 groups will be used as the evaluation group once.

\smallskip

In order to investigate the applicability and effectiveness of our approach,
for each message in the evaluation group, we compared the resulting
synthesized response with the corresponding recorded response. % To do so, we
We defined four criteria to evaluate the ``validity'' of synthesized
responses. These four criteria are explained as follows using an example shown in Table~\ref{tab:criteriaexample}. Consider the \emph{incoming
  request} {\texttt{\footnotesize{\{id:37,op:A,sn:Durand\}}}} with the
associated response {\texttt{\footnotesize{\{id:37,op:AddRsp,result:Ok\}}}}
in the transaction library. The emulated response is considered to be:

\begin{enumerate}
\item \textbf{identical} if its character sequence is identical to the
  recorded (or expected) response ({\cf} Example (i) in
  Table~\ref{tab:criteriaexample});
 % the synthesized response is identical to the recorded response
%     this criterion requires all characters in the generated response to be the
%     same compared to the corresponding recorded, real service response for the
%     incoming request.
%
%\item \textbf{consistent} if it is of the expected operation type and
%  has the critical fields in the payload replicated ({\cf} Example (ii) in
%  Table~\ref{tab:criteriaexample} where {\texttt{\footnotesize{{id}}}} is
%  identical, but some of the other payload differs);

\item \textbf{protocol conformant} if its operation type corresponds to the
  expected response, but it differs in some payload information ({\cf} Example (ii) and (iii)
  in Table~\ref{tab:criteriaexample} where both the
  {\texttt{\footnotesize{{id}}}} and {\texttt{\footnotesize{{result}}}}
  tags may differ);

\item \textbf{well-formed} if it is structured correctly (that is, it
  corresponds to one of the valid response messages), but has the wrong
  operation type ({\cf} Example (iv) in Table~\ref{tab:criteriaexample} where
  the generated response is of a valid structure, but its operation type
  {\texttt{\footnotesize{{op:SearchRsp}}}} does not match the expected
  operation type {\texttt{\footnotesize{{op:AddRsp}}}}); and

  \item \textbf{malformed} if it does not meet any of the above criteria ({\cf} Example (v)
  in Table~\ref{tab:criteriaexample} where operation type
  {\texttt{\footnotesize{{op:AearchRsp}}}} in the generated response is invalid).

\end{enumerate}

We further consider a generated response to be {\em valid} if it meets one of
the first two criteria, that is, {\em identical} or {\em
  protocol conformant}. Otherwise, a generated response is considered to be
{\em invalid}.

% Table~\ref{tab:criteriaexample} illustrates the five criteria using an
% example. Consider the following \emph{incoming request}:
%
%     \hspace{0.2cm} \{id:37,op:A,sn:Durand\}
%
% Table~\ref{tab:criteriaexample} lists its {\em expected response} and
% different {\em generated responses}, each of which meets a particular
% criterion.

\begin{table}[t]
\begin{center}
\begin{tabular}{|l|c||c|l|}
\hline
\label{identical}
{Identical} & \multirow{2}{*}{(i)} & Expected & \{id:37,op:AddRsp,result:Ok\} \\ \cline{3-4}
&& Generated & \{id:37,op:AddRsp,result:Ok\} \\ \hline
\multirow{4}{2cm}{Protocol conformant} & \multirow{2}{*}{(ii)} & Expected & \{id:\textbf{\emph{37}},op:AddRsp,result:\emph{Ok}\} \\\cline{3-4}
& & Generated & \{id:\textbf{\emph{37}},op:AddRsp,result:\textbf{\emph{AlreadyExists}}\} \\ \cline{2-4}
& \multirow{2}{*}{(iii)} & Expected & \{id:37,op:AddRsp,result:Ok\} \\\cline{3-4}
& & Generated & \{id:\textbf{\emph{15}},op:AddRsp,result:\textbf{\emph{AlreadyExists}}\} \\ \hline
Well-formed & \multirow{2}{*}{(iv)} & Expected & \{id:37,op:\emph{AddRsp},result:Ok\} \\\cline{3-4}
& & Generated & \{id:15,op:\textbf{\emph{SearchRsp}},result:Ok,gn:Miao,sn:Du\} \\ \hline
Malformed & \multirow{2}{*}{(v)}& Expected & \{id:37,op:\emph{AddRsp},result:Ok\} \\\cline{3-4}
&& Generated & \{id:15,op:\textbf{\emph{AearchRsp}},result:Ok,gn:Miao,sn:Du\} \\ \hline
\end{tabular}
\end{center}
\caption{Examples for accuracy criteria: (i) Identical, (ii)(iii) Protocol conformant, (iv) Well-formed, (v) Malformed.}
\label{tab:criteriaexample}
\end{table}

For the purpose of our evaluation, we used a weaker notion of {\em protocol
  conformance} as the order in which the requests are selected from the
evaluation set is {\em random} and, as a consequence, unlikely to conform to a
sequence of protocol conformant requests. Therefore, we consider a synthesized
response to be protocol conformant if it conforms to the temporal properties
at some point of time.

If a synthesized response is {\em identical}, then the other two properties
({\em well-formed} and {\em protocol conformant}) are implied. We can
guarantee this under the assumption that the recorded interaction traces we
use are considered to be valid and conform to the temporal interaction
properties of the protocol. However, it is very well possible that the
response generation process synthesises a well-formed response that is {\em
  not} protocol conformant (as we will further discuss in
Section~\ref{sec:discussion}).

For the purpose of emulation, protocol conformance is the most important
property a synthesized response needs to exhibit. The aim of an emulatable
endpoint model is not necessarily to reproduce the behaviour of a real
endpoint to 100\% - as long as the responses an emulated endpoint provides are conformant to corresponding protocol structure specification, this will be sufficient for many quality assurance
activities \cite{hine:thesis}.

% ------------------------------------------------------------------------- %

%%!!%% FORMATTING %%!!%%
% \bigskip

\subsection{Evaluation Results}
\label{subsec:results}

To benchmark the effectiveness of our approach for synthesizing responses, we
used a random selection strategy as baseline where for an ``incoming''
request, the corresponding response is randomly selected from the responses
contained in the training set. All generated responses for both, the {\em whole library} approach
as well as the random selection strategy, were categorised according to the criteria introduced in Section~\ref{chap5subsec:criteria}.

Table~\ref{tab:results} summarizes the result of our experiments. Besides the
number of responses falling in each of the four categories, it lists the
number of {\em valid} response messages, that is, the sum of identical and
protocol-conformant messages. Please note that the column {\em Well-form.}
does {\em not} include the number of valid messages, that is, only those
well-formed responses that are not protocol conformant are listed.
Furthermore, for all {\em non-identical} responses, Table~\ref{tab:results}
also lists the mean as well as the maximum dissimilarity ratios,
respectively.

\begin{table}
\footnotesize
\begin{center}
\begin{tabular}{|m{1.5cm}||r|r|c|c|c|c||m{1cm}|m{1cm}|}
\hline
Experiment & No. & Valid & Ident. & Conf. & \parbox[p]{1cm}{Well\\-form.} & \parbox[p]{1cm}{Ill\\-form.} &
             Mean dsim.$*$ & Max dsim.$*$ \\
\hline\hline
\parbox[p]{1cm}{SOAP\\Random} & 1,000 &    33 & 33 &   0 & 967 &  0 & 0.046 & 0.259 \\\hline
SOAP Whole Library    & 1,000 & 1,000 & 93 & 907 &   0 &  0 & 0.020 & 0.046 \\
\hline
LDAP Random &   498 & 438 &   2  & 436 &  39 &  0 & 0.067 & 0.873 \\\hline
LDAP Whole Library  &   498 & 484 & 466  &  18 &   9 &  5 & 0.200 & 0.775 \\
\hline
\end{tabular}
\end{center}
\caption{Summary of Evaluation Results.}
\label{tab:results}
\end{table}

\subsubsection{Evaluation results for SOAP.}

Table~\ref{tab:results} compares the different outcomes
of the random response strategy and our whole library approach. Most importantly, no
malformed SOAP responses were generated by either the baseline approach or
the whole library approach. However, our approach
outperformed the random selection strategy in a number of
aspects. Specifically, (i) {\em all} $1,000$ synthesized responses using our
approach were protocol conformant, compared to only $33$ of the randomly
selected responses, and (ii) $9.3$\% of the generated responses were identical
to the recorded responses in our approach, compared to $3.3$\% in the random
selection strategy.

Analysing the non-identical responses in more detail, we observed that the
worst dissimilarity ratio of the whole library approach is
$0.046$ (all other dissimilarity ratios are smaller). With an average response
length of $239$ characters, this gives us a maximum edit distance of $24$
between the synthesized response and the ``expected'' response ({\ie} the
response associated with the most similar request). This shows that for the
SOAP case study used, our approach was able to synthesize responses
significantly more accurately than the random strategy.

\subsubsection{LDAP results}

A summary of the result of the LDAP experiments are also given in
Table~\ref{tab:results}. For the whole library approach, $466$
(out of $498$) generated response messages were identical to the corresponding
recorded responses ($89.9$\%), and an additional $18$ of the generated
responses met the protocol conformant criterion ($3.6$\%). Therefore, a total
of $487$ (or $97.8$\%) of all generated responses were considered to be
valid. Of the remaining $14$ responses, $9$ were well-formed, but had the
wrong message type, and $5$ responses were {\em malformed}. Both aspects will
be discussed further in the following section.

In case of the random selection strategy, all responses were well-formed (as
expected), but as many as $438$ responses were valid ($87.5$\%), which is not
much worse than the whole library approach. This rather surprising result can be
explained by the fact that about $90$\% of all recorded requests are {\tt
  searchRequest} messages (with different search criteria), and hence the
likelihood of randomly choosing another {\tt searchRequest} as the ``best''
match is rather high. This also explains the rather low number of only
well-formed messages.

With regards to the rather high maximum dissimilarity ratio, there are a
number of very similar search requests in our data set, some of them
resulting in responses with zero or one search result entries only, others
with a large number of entries. Therefore, if a response with a small number
of entries is used as the basis to synthesize a response for a request that
expects a large number of entries (or vice versa), then the edit difference
between the synthesized and expected responses is rather large and,
consequently, the dissimilarity ratio as well. However, for the purpose of our
overall goal of being able to generate valid responses, this is not a problem
as despite a high dissimilarity, a valid response is generated as long as all
symmetric fields are replaced correctly.

% ------------------------------------------------------------------------- %

\subsection{Discussion and Limitations}
\label{sec:discussion}

Based on the investigation of both the SOAP and LDAP experimental results, we can
see that our approach is able to automatically generate valid responses in
most situations. However, as illustrated in the results for LDAP, a small
proportion of protocol non-conformant or even malformed responses were
synthesized. In order to better illustrate the underlying reasons, consider
the following example where a protocol non-conformant response was
synthesized. The following request

\vspace{-0.15cm}
\small\begin{verbatim}
    Message ID: 171
    ProtocolOp: addRequest
      ObjectName: cn=Miao DU,ou=Finance,
                  ou=Corporate,o=DEMOCORP,c=AU
      Scope: 0 ( baseObject )
  \end{verbatim}
\normalsize

\vspace{-0.35cm} \noindent resulted in the generation of the following
response:

\vspace{-0.15cm}
\small\begin{verbatim}
    Message ID: 171
    ProtocolOp: modifyResponse
      resultCode: success
  \end{verbatim}
\normalsize

\vspace{-0.3cm} The response is well-formed and the {\tt Message} {\tt Id}
field has been substituted properly. However, according to the LDAP protocol
specification, an {\tt addRequest} adding an extra node to an LDAP directory,
must result in an {\tt addResponse}, and not in an {\tt modifyResponse} as
given in the example above. The reason for this unexpected response can be
explained by the fact that the test set contains a {\tt modifyReqest} with
precisely the same {\tt ObjectName} and {\tt Scope} as the {\tt addRequest}
above and a {\tt Message} {\tt ID} of 151. Our distance measure identified
this {\tt modifyRequest} as the most similar match and hence, the associated
modify response was used as the basis for synthesizing the response.

Most application-level protocols define message structures containing some
form of {\em operation} or service name in their requests, followed by a
{\em payload} on what data this service is expected to operate upon
\cite{hine:thesis}. In the example above, the fact that {\tt addRequest} and
{\tt modfiyRequest} denote % such operation names and identify
different operations was not taken into consideration when the most similar
request was chosen. In future work we intend to devise suitable heuristics
allowing us to (semi-)automatically identify which part(s) of a request
message most likely correspond to a service name, use this information to
divide the set of interaction traces into clusters containing a single
service type only, and restrict the search for the most similar request to one
cluster only. This should also improve the run-time performance of our
approach.

%% Above is an well-formed response example. If we only take the response into
%% consideration, it has precise message structure. Moveover, the
%% \textbf{Message Id} field has been substituted properly. However, the field
%% of \textbf{protocolOp} with (\textit{modifyResponse}) does not conform to
%% the LDAP protocol, which is expected to be a {\textit{addResponse}} in
%% practice.

The following example indicates an malformed LDAP response. It is worth
noting that the {\tt Message} {\tt Id} and {\tt ObjectName} fields have been
properly substituted from the corresponding request. However, the {\tt
  protocolOp} values of {\tt addResEntry} and {\tt addResDone} are invalid
LDAP operation names and were flagged as such by the LDAP encoder used.

\vspace{-0.15cm}
\small
\begin{verbatim}
    Message ID: 154
    ProtocolOp: addResEntry
      ObjectName: cn=Miao DU,ou=Legal,
                  ou=Corporate,o=DEMOCORP,c=AU
      Scope: 0 ( baseObject )
    Message ID: 154
    ProtocolOp: addResDone
      resultCode: success
\end{verbatim}
\normalsize

\vspace{-0.1cm} Similar to the previous example, there is a mismatch in the
operation name of the most similar request: whereas the request message denotes
an {\tt addRequest}, the test set contains a {\tt searchRequest} with a very
similar message id and an identical {\tt ObjectName}. The message id was
substituted correctly, but all occurrences of {\tt search} in the response were
substituted to {\tt add}, resulting in an malformed LDAP response. Again,
clustering the set of interactions according to the service/operation name
would have most likely prevented the selection of a {\tt searchRequest} as
the most similar request to an {\tt addRequest}.

\smallskip

Comparing the dissimilarity measures of our LDAP and SOAP results ({\cf} the
corresponding values in Table~\ref{tab:results}), we noticed that non-zero
SOAP dissimilarities are generally significantly lower than the non-zero LDAP
results, indicating that our non-exact matching SOAP responses are typically
less dissimilar to the real responses than their LDAP counterparts. This can
be attributed to the fact that SOAP messages contain a significant amount of
structural information which is easily duplicated in the generated
responses. This makes the generated and real SOAP responses \emph{similar}
even when there are, perhaps significant, differences in the payload.

This is not a major issue for our approach in general. However, it implies
that comparing the effectiveness of various distance and translation functions
{\em across} protocols needs to be done carefully as low(er) dissimilarity
ratios in one protocol may be more due to the amount of common {\em
  structural} information than the properties of the distance and translation
functions used. Similar to the abovementioned clustering approaches, we intend
to use heuristics to (semi-)automatically separate payload and structure in
messages and devise similarity measures that give payload information a higher
weighting than structural information in order to improve the cross-protocol
comparisons.

\smallskip

In our tests we have examined text-based messages with SOAP being a text-based
protocol and for LDAP, we used a text representation. Future work will attempt
to synthesise responses directly for binary protocols. This will bring extra
challenges. In order to give one example, binary packets often contain the
packet length as part of the encoding. Our field substitution method could
change the length of packets and, therefore easily produce an malformed
response. In order to address this issue, without using explicit knowledge of
the message structure, we will need to devise methods to automatically
identify fields such as the packet length.
\section{Summary}
\label{chap5sec:summary}

In this chapter, we introduce a whole library technique that is to automatically build executable interactive models of software service behaviour by generating responses from recorded interaction traces, without requiring explicit knowledge of the internals of the target service or of the protocols the service uses to communicate. We adopt a genome sequence alignment algorithm and a field substitution algorithm for implementing the matching function and the translation function of our proposed framework. We evaluate our method against two common application-layer protocols: LDAP and SOAP. The experimental results have shown a greater than 98\% accuracy for the two protocols tested. %first introduce the concept of sequence alignment, and then discuss one of the most popular alignment algorithm, {\em the Needleman-Wunsch algorithm}. Then we describe how the matching function utilises the Needleman-Wunsch algorithm in order to find the most similar recorded interaction. In addition, we also introduce a novel method for modifying the recorded response in order to generate a response. After introducing the experiments that we conducted to evaluate this approach, we discuss the results of our experiments and identify some limitations of it, which motives us to conduct the following work. 
\chapter{Interaction Trace Analysis}
\label{chap6:SoftMine}

% introduce last
%In Chapter~\ref{chap4:framework}, we introduced an architecture for automatic service virtualisation, which contains three main functions: the {\em analysis function}, the \emph{matching function} and the \emph{translation function}.
In Chapter~\ref{chap5:qosa}, we described the adoption of a genome sequencing alignment algorithm, a symmetric field identification algorithm and a field substitution algorithm to implement the service virtualisation architecture. %Specifically, given an incoming request, the genome sequencing alignment algorithm is applied by the matching function to search for the most similar request in the previously recorded interaction traces by comparing their distances. Then the field substitution algorithm and the field substitution method are performed to synthesize a valid response.
However, as this implementation has to process the whole recorded interaction collection - which may become extremely large, it becomes very inefficient in practice. % our experimental results show an approach which has to process the whole recorded interaction collection %at runtime - which may become extremely large - becomes very inefficient in practice.
In this chapter, we introduce a new implementation of the analysis function, called the \emph{cluster centroid} technique, to improve runtime efficiency of our previous approach. This technique is performed to condense the whole interaction collection into a small number of selected representative interactions at the offline stage so that it can accelerate response inference time at runtime.%that is able to select and provide representative interaction data to the response synthesis approach, thereby accelerating response inference time at runtime. Given a collection of interaction traces, our new technique (i) calculates the distance between pairwise interactions and builds a distance matrix; (ii) clusters interactions; and (iii) exports the clusters and infers the cluster centres for use later in the process.

The organisation of this chapter is as follows. Section \ref{chap6sec:approach} describes the design of the cluster centroid technique. % for pre-processing interaction traces to assist the response synthesis task.
In Section \ref{chap6sec:evaluation},
we present the results of our evaluation and discuss the
relevant findings as well as identified limitations. Finally, we
summarises this chapter in Section \ref{chap6sec:summary}.

This chapter is mainly based on our work presented in~\cite{Du:2013SoftMine}

\section{Approach}
\label{chap6sec:approach}

\begin{figure*}[ht]
     \centering
     \includegraphics[width=\textwidth,height=0.4\textheight]{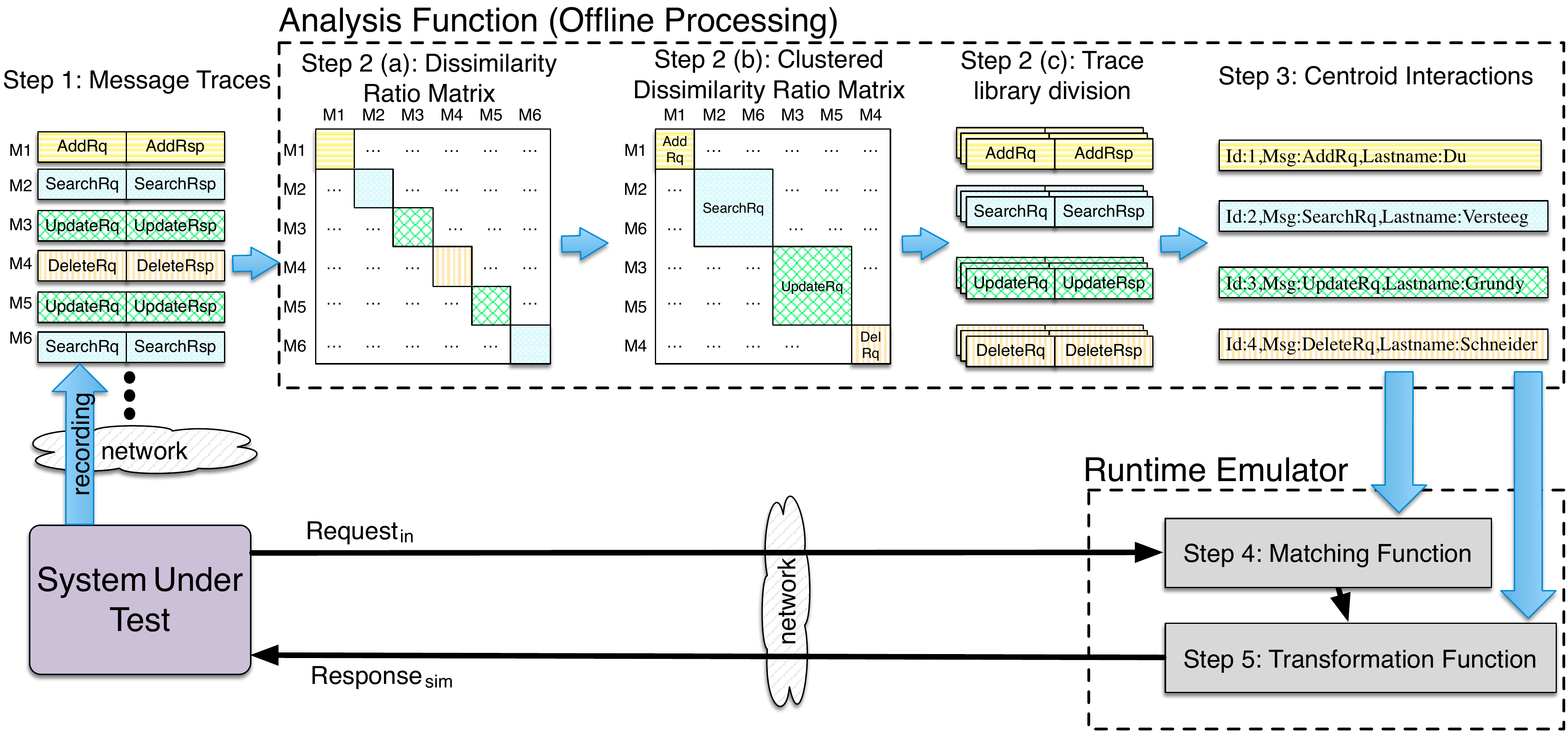}
     \caption{System overview}
     \label{chap6fig:system_overview}
\end{figure*}

The target of our new approach is to reduce the number of interactions that need to be searched in finding a matching interaction. To achieve this, our approach is to classify trace library into groups of similar interactions to assist in searching for the most similar recorded message for an incoming request. %Since we use the sequence alignment algorithm to calculate distance of pairwise messages, This algorithm can be considered as a natural selection of clustering and sequence alignment techniques for inferring protocol structure information.
The motivation behind our approach is that if request/response pairs
are less distant to some other requests/responses, then these less
distant requests/responses are more likely to have the same structure and
information. Hence, computing the distance between each pair of messages
should give us an indication of how to classify recorded interactions.
These classified interactions then enable us to generate responses conforming better
to the expected responses. %of the enterprise system under test.

Figure~\ref{chap6fig:system_overview} shows a schematic overview of our approach. Step 2 and 3 are the implementation of the {\em analysis function}, which are performed offline to provide cluster centres for the {\em matching function}. As these two steps are the keys of this chapter, we will illustrate them  with technical details in the following sections. Step 4 and 5 are performed to generate a response for the live request. We adopt previous techniques to achieve this task ({\cf} Section~\ref{chap5sec:distcal} and Section~\ref{chap5sec:translation}). Our solution works as follows:

\begin{enumerate}

  \item {We collect network traffic of communications between a client and the real target service, and transform the traffic into a suitable format. Tools such as Wireshark \cite{lamping:12a} have the functionality to filter network traffic and record
      interactions in a suitable textual format for further processing. Appendix~\ref{apx.ldapexample} shows how Wireshark converts binary LDAP messages into their corresponding textual format. }

  \item {We then divide trace library into groups of similar interactions by:}

      \begin{enumerate}

        \item {Calculating dissimilarity ratios for all pairs of request messages or response messages in that trace library, and building up a dissimilarity ratio matrix;}

        \item {Transforming the origin dissimilarity ratio matrix to clustered dissimilarity ratio matrix where items with similar dissimilarity ratio are closer to each other, and outputting the reordered matrix as grayscale image;}

        \item {Visualising the reordered matrix as a grayscale image and dividing the trace library into separate clusters of similar interactions.}

      \end{enumerate}

  \item {We then select a sample interaction from each cluster to be the centroid interaction of that cluster.}

  \item {At runtime, the {\em matching function} adopts the Needleman-Wunsch algorithm to compare incoming requests to centroid interactions for calculating dissimilarity ratios. The centroid interaction with the minimum dissimilarity ratio is selected for the {\em translation function}.}

  \item {At the final step, the {\em translation function} adopts previous symmetric field identification and field substitution algorithms ({\cf} Section~\ref{chap5sec:translation}) to generate a response to be returned to the system under test.}

\end{enumerate}

%
% ------------------------------------------------------------------------ %
%
%
%\subsection{Interaction trace collection and translation (Step 1)}
%
%We assume that for a given protocol
%under investigation, we are able to record a sufficiently large number of
%interactions between a system under test and a target service system.%two (or more) deployed software endpoints (components or services).
%These recordings are assumed
%to be ``valid'', that is, that the sequence of
%recorded interactions are (i) correct with regards to the temporal properties
%of the underlying protocol and that (ii) each request and response message is
%well-formed. Without loss of generality,
%we further assume that each request is always followed by a single response. If a
%request does not generate a response, we insert a dedicated ``no-response'' message into
%the recorded interaction traces. If, on
%the other hand, a request that leads to multiple responses, these are concatenated
%into a single response.
%%
%Given these assumptions, {\Pair{\Request}{\Response}} denotes a single
%interaction, where {\Request} represents the request,
%the corresponding response of {\Request} is defined by
%{\Response}. Both {\Request} and {\Response} are a sequence of
%characters describing the message structure and payload. An {\em interaction trace} is defined as a finite, non-empty sequence of interactions,
%which is denoted by ({\Pair{\Request_1}{\Response_1}}, {\Pair{\Request_2}{\Response_2}} ... {\Pair{\Request_n}{\Response_n}}).

\subsection{Building Request/Response Dissimilarity Ratio Matrix}
\label{chap6subsec:distance}

% We give a small sample set of request and response messages for the directory service in Table~\ref{chap6tab:interactionlibraryexample}. In the following sections we use this sample interaction library to illustrate our approach. %use this sample trace library as a running example to illustrate how our method works.

\begin{table}[h]
    \begin{center}
    \begin{tabular}{|c||m{5cm}|m{7cm}|}
    \hline
    Index & Request & Response \\ \hline\hline
    1 & \{id:1,op:S,sn:Du\} & \{id:1,op:SearchRsp,result:Ok, gn:Miao,sn:Du,mobile:5362634\} \\ \hline
    2 & \{id:13,op:S,sn:Versteeg\} & \{id:13,op:SearchRsp,result:Ok, gn:Steve,sn:Versteeg,mobile:9374723\} \\ \hline
    3 & \{id:24,op:A,sn:Schneider, mobile:123456\} & \{id:24,op:AddRsp,result:Ok\} \\ \hline
    4 & \{id:275,op:S,sn:Han\} & \{id:275,op:SearchRsp,result:Ok, gn:Jun,sn:Han,mobile:33333333\} \\ \hline
    5 & \{id:490,op:S,sn:Grundy\} &
    \{id:490,op:SearchRsp,result:Ok, gn:John,sn:Grundy,mobile:44444444\} \\ \hline
    6 & \{id:2273,op:S,sn:Schneider\} & \{id:2273,op:SearchRsp,result:Ok, sn:Schneider,mobile:123456\} \\ \hline
    7 & \{id:2487,op:A,sn:Will\} & \{id:2487,op:AddRsp,result:Ok\} \\ \hline
    8 & \{id:3106,op:A,sn:Hine, gn:Cam,Postcode:33589\} & \{id:3106,op:AddRsp,result:Ok\} \\
\hline
    \end{tabular}
    \end{center}
    \caption{Interaction Library Example}
    \label{chap6tab:interactionlibraryexample}
\end{table}

% ------------------------------------------------------------------------ %

Once the interaction traces collection and translation step has been done, we have a collection of a sufficiently large number of interactions between a service under test and a target services. Table~\ref{chap6tab:interactionlibraryexample} demonstrates a small sample set of request and response messages for the directory service. This is from a fictional protocol that has some similarities to the widely used LDAP protocol, but is simplified to make our running example easier to follow.
The next step (Step 2(a)) is to build a $N \times N$ matrix for describing dissimilarity ratios for all pairs of request messages or response messages in the trace library, where $N$ is the number of interactions. All dissimilarity ratios are measured by edit distances that are calculated using the Needleman-Wunsch algorithm ({\cf} Section~\ref{chap5sec:needlemanwunsch}). Consider the above example, its request dissimilarity ratio matrix and response dissimilarity ratio matrix are shown as $DM_{req}$ and $DM_{res}$ respectively.

    $DM_{req} = \kbordermatrix{
    ~ & 1 & 2 & 3 & 4 & 5 & 6 & 7 & 8 \\
     1 & 0 & 0.1875 & 0.3333 & 0.1500 & 0.1739 & 0.2407 & 0.2045 & 0.36 \\
     2 & 0.1875 & 0 & 0.2949 & 0.2083 & 0.1875 & 0.1852 & 0.2292 & 0.3100 \\
     3 & 0.3333 & 0.2949 & 0 & 0.3077 & 0.2949 & 0.2051 & 0.2692 & 0.25  \\
     4 & 0.1500 & 0.2083 & 0.3077 & 0 & 0.1739 & 0.1852 & 0.1591 & 0.3400 \\
     5 & 0.1739 & 0.1875 & 0.2949 & 0.1739 & 0 & 0.2037 & 0.1956 & 0.3300 \\
     6 & 0.2407 & 0.1852 & 0.2051 & 0.1852 & 0.2037 & 0 & 0.2037 & 0.3200 \\
     7 & 0.2045 & 0.2292 & 0.2692 & 0.1591 & 0.1957 & 0.2037 & 0 & 0.3400 \\
     8 & 0.36 & 0.31 & 0.25 & 0.3400 & 0.3300 & 0.3200 & 0.3400 & 0 \\ }$

    $DM_{res} = \kbordermatrix{
    ~ & 1 & 2 & 3 & 4 & 5 & 6 & 7 & 8 \\
     1 & 0 & 0.1364 & 0.3103 & 0.1230 & 0.1493 & 0.1742 & 0.3103 & 0.3017 \\
     2 & 0.1364 & 0 & 0.3333 & 0.1515 & 0.1567 & 0.1515 & 0.3333 & 0.3258 \\
     3 & 0.3103 & 0.3333 & 0 & 0.3115 & 0.3284 & 0.3258 & 0.0345 &0.0690 \\
     4 & 0.1230 & 0.1515 & 0.3115 & 0 & 0.1493 & 0.1667 & 0.3033 & 0.3197 \\
     5 & 0.1493 & 0.1567 & 0.3284 & 0.1493 & 0 & 0.1716 & 0.3284 & 0.3284 \\
     6 & 0.1742 & 0.1515 & 0.3258 & 0.1667 & 0.1716 & 0 & 0.3182 & 0.3258 \\
     7 & 0.3103 & 0.3333 & 0.0345 & 0.3033 & 0.3284 & 0.3182 & 0 & 0.0690 \\
     8 & 0.3017 & 0.3258 & 0.0690 & 0.3197 & 0.3284 & 0.3258 & 0.0690 & 0 \\
    }$
%Both
%the optimal alignment for pairwise messages and the distance (\ref{equa:dist}) can be
%computed efficiently using the Needleman-Wunsch algorithm \cite{needleman:1970}.

% ------------------------------------------------------------------------ %

\subsection{Clustering dissimilarity ratio matrix and exporting trace library division}

% we need some dissimilarity based matrix clustering
% bea is one option, summarise its process, cite dong's paper
% vat is another option, visual assessment of cluster tendency
% only summarise these two methods' characteristics
% explain reorder results in the evaluation section
Once the dissimilarity ratio matrix $DM$ has been constructed, the next step is to operate clustering algorithms on this matrix for grouping interactions into clusters (Step 2(b)). A cluster is a group of interactions which are similar among them and are dissimilar to the interactions belonging to other clusters. As the target of our approach is to virtualise services without explicit knowledge of the service protocol, we cannot define the number of clusters to seek prior to clustering. Therefore, most of clustering techniques that require pre-defined threshold values, {\eg the number of clusters}, are not applicable in our work. We require approaches that can analyse the trace library to assist our opaque service virtualisation technique, capable of determining how many types of interaction can be identified and if the trace library can be partitioned.%{\em cluster tendency} prior to actual clustering to determine if clusters exist.

VAT~\cite{VAT}\cite{VATIMPL} is a technique that exists in data mining for the visual assessment of cluster tendency prior to actual clustering to determine if clusters exist. This approach is able to present the cluster tendency in a way that the human eye can easily distinguish the clusters and recognise the number of operation type clusters. %justification of manual interaction
Users only need to know that they are presented with a cluster tendency. With minimal manual interaction, interactions of the same operation type can be grouped into the same cluster without requiring any protocol-specific knowledge.  %can analyse cluster tendency using reordered distance matrix.
The VAT technique is performed in two steps. At the first step, it reorders an dissimilarity matrix and produces a reordered {\em DM'}. At the second step, VAT produces an intensity representation of the matrix showing clusters as dark block along the diagonal. % and visual analysis of cluster tendency (VAT)~\cite{VAT} algorithm. These two algorithms are performed to rearrange columns and rows of a matrix. Items with similar values are grouped together so that the clustered matrix is able to indicate how many interaction clusters to seek.

\subsubsection{Reorder a dissimilarity matrix}
\label{chap6subsubsection:reorder}

To reorder a dissimilarity matrix, a matrix permutation technique is required. This technique takes an initial dissimilarity ratio matrix {\em DM} as the input {\em DM} and produces a reordered {\em DM'} where similar dissimilarity values $d_{ij}$ are places together in the matrix. This can be performed using a number of methods. We considered two alternative techniques -- modified Prim's minimal spanning tree algorithm~\cite{PrimSpanningTree} and the bond energy algorithm (BEA)~\cite{BEA}.

\begin{itemize}
\item{\textbf{Modified Prim's minimal spanning tree algorithm}}

Modified Prim's minimal spanning tree algorithm is the original VAT reordering method. It is a greedy approach that starts with the most distant interaction and continuously places next most similar in the next position of the matrix until all values are reordered. To assist with the reordering, we define {\em M}, {\em N} and {\em P} to be the set of sorted matrix indices, the set of unsorted matrix indices and the new position of {\em $d_{ij}$} respectively. Consider the response dissimilarity ratio matrix $DM_{res}$, we can reorder this matrix using Prim's minimal spanning tree algorithm as follows.

Initially, {\em M}, {\em N}, {\em P} are as follow.

\hspace{0.3cm}
$M=\o$, $N=\{1,2,3,4,5,6,7,8\}$ and $P=(0,0,0,0,0,0,0,0)$

The first step of the reordering is to select
the most dissimilar message in the trace library, chosen as the column or row with the largest dissimilarity ratio in {\em DM}. The index of this interaction is placed as the first position in {\em P}, added to the sorted indices {\em M} and removed from the unsorted indices {\em N}. The highest dissimilarity value in $DM_{res}$ is 0.3333, one of which in position $d_{2,3}$. We choose this as the most dissimilar value, and update $M$, $N$ and $P$ as follows.

\hspace{0.3cm} $M=\{2\}$, $N=\{1,3,4,5,6,7,8\}$ and $P=(2,0,0,0,0,0,0,0)$

The next step is to choose an interaction in
$N$ that is most similar to the interactions in $M$. Searching for the smallest $d_{mn}$ where $m \in \{2\}, n \in \{1,3,4,5,6,7,8\}$, the position $d_{2,1}$ has been identified with the smallest value 0.1364. After this
interaction is found, the index is placed into the next position in $P$, the index is added to $M$ and removed from $N$.

\hspace{0.3cm} $M=\{1,2\}$, $N=\{3,4,5,6,7,8\}$ and $P=(2,1,0,0,0,0,0,0)$

This step is repeated until all the indices have been reordered.

\hspace{0.3cm} $M=\{1,2,4\}$, $N=\{3,5,6,7,8\}$ and $P=(2,1,4,0,0,0,0,0)$

\hspace{0.3cm} $M=\{1,2,4,5\}$, $N=\{3,6,7,8\}$ and $P=(2,1,4,5,0,0,0,0)$

\hspace{0.3cm} $M=\{1,2,4,5,6\}$, $N=\{3,7,8\}$ and $P=(2,1,4,5,6,0,0,0)$

\hspace{0.3cm} $M=\{1,2,4,5,6,8\}$, $N=\{3,7\}$ and $P=(2,1,4,5,6,8,0,0)$

\hspace{0.3cm} $M=\{1,2,3,4,5,6,8\}$, $N=\{7\}$ and $P=(2,1,4,5,6,8,3,0)$

\hspace{0.3cm} $M=\{1,2,3,4,5,6,7,8\}$, $N=\o$ and $P=(2,1,4,5,6,8,3,7)$

Finally, using the reordered
positions stored in P, we can get a reordered matrix $DM'_{res}$.

$DM'_{res} = \kbordermatrix{
    ~ & 2 & 1 & 4 & 5 & 6 & 8 & 3 & 7 \\
     2 & 0 & 0.1364 & 0.1515 & 0.1567 & 0.1515 & 0.3258 & 0.3333 & 0.3333 \\
     1 & 0.1364 & 0 & 0.1230 & 0.1493 & 0.1742 & 0.3017 & 0.3103 & 0.3103 \\
     4 & 0.1515 & 0.1230 & 0 & 0.1493 & 0.1667 & 0.3197 & 0.3115 & 0.3033 \\
     5 & 0.1567 & 0.1493 & 0.1493 & 0 & 0.1716 & 0.3284 & 0.3284 & 0.3284 \\
     6 & 0.1515 & 0.1742 & 0.1667 & 0.1716 & 0 & 0.3258 & 0.3258 & 0.3182 \\
     8 & 0.3258 & 0.3017 & 0.3197 & 0.3284 & 0.3258 & 0 & 0.0690 & 0.0690 \\
     3 & 0.3333 & 0.3103 & 0.3115 & 0.3284 & 0.3258 & 0.0690 & 0 & 0.0345 \\
     7 & 0.3333 & 0.3103 & 0.3032 & 0.3284 & 0.3182 & 0.0690 & 0.0345 & 0 \\}$

\item{\textbf{Bond Energy Algorithm (BEA)}}

The Bond Energy Algorithm (BEA) is a permutation algorithm, which has been widely utilised in distributed database systems for the vertical partition of large tables~\cite{BEA}. In our work, it takes the dissimilarity ratio matrix ({\em DM}) as input, reorders rows and columns of this matrix and generates a clustered dissimilarity ratio matrix {\em DM'}. In {\em DM'}, similar interactions are grouped together along the matrix diagonal, that is, large values with other large values and small values with other small values. To transform {\em DM} to {\em DM'}, the permutation is done in a way to maximise the following global measure equation (\ref{chap6equa:bea}), denoted by $GM$,
\begin{equation}
\label{chap6equa:bea}
GM={\sum_{i=1}^n}{\sum_{j=1}^n}(1-d_{ij})(2-d_{i(j-1)}-d_{i(j+1)})
\end{equation}
where $d_{ij}$ denotes the dissimilarity ratio between $msg_i$ and $msg_j$, $d_{0j}=d_{i0}=d_{n+1j}=d_{in+1}=0$. The algorithm is done in three steps:

\begin{enumerate}

    \item
    {\em Initialisation}. Place the first two columns of $DM_{res}$ into $DM'_{res}$.

    \item
    {\em Iteration}. For the remaining $n-i$ columns,  we try to place them in the remaining $i+1$ positions in the matrix, where $i$ is the number of columns already placed. The placement is to make the greatest {\em contribution} to the global measure described in Equation~\ref{chap6equa:bea}. The {\em contribution} to the global measure of placing interaction $I_k$ between $I_i$ and $I_j$ is

    \begin{equation}
    \label{chap6equa:contribution}
        cont(I_i,I_k,I_j) = 2bond(I_i,I_k) +2bond(I_k,I_j)-2bond(I_i,I_j)
    \end{equation}
    where $bond(I_x,I_y)={\sum_{z=1}^n}(1-d_{zx})(1-d_{zy})$.
    This step is repeated until no more columns remain to be placed.

    \item
    {\em Row ordering}. Once the columns are reordered, the rows should also be changed so that their relative positions match the relative positions of the columns.

\end{enumerate}

The details of this algorithm
can be found in \cite{DistributionDatabase}. Consider the response dissimilarity ratio matrix $DM_{res}$, the BEA algorithm is performed to reorder the matrix as follows.

Initially, Column 1 and 2 are chosen as below.

     $DM'_{res} = \kbordermatrix{
    ~ & 1 & 2 & & & & & & & & & & &&&&&&&&& & & & & &\\
     1 & 0 & 0.1364 &       &       &       &       & \\
     2 & 0.1364 & 0 &       &       &       &       & \\
     3 & 0.3103 & 0.3333 & & & & &\\
     4 & 0.1230 & 0.1515 & & & & &\\
     5 & 0.1493 & 0.1567 & & & & &\\
     6 & 0.1742 & 0.1515 & & & & &\\
     7 & 0.3103 & 0.3333 & & & & &\\
     8 & 0.3017 & 0.3258 & & & & &\\
    }$

For column 3, there are three alternative places where column 3 can be placed: to the left of column 1, resulting in the ordering (3-1-2), in between column 1 and 2, having (1-3-2), and to the right of column 2, resulting in (1-2-3). The following show the contributions to the global measure of each alternative.

\textbf{Ordering (0-3-1):}

\begin{quote}
$cont(I_0,I_3,I_1)= 2bond(I_0,I_3)+2bond(I_3,I_1)-2bond(I_0,I_1)$

$bond(I_0,I_1)=bond(I_0,I_3)=0$

$bond(I_3,I_1)=1 \times 0.6897 + 0.8636 \times 0.6667 + 0.6897 \times 1 + 0.8770 \times 0.6885 + 0.8615 \times 0.6769 + 0.8644 \times  0.6949 + 0.6897 \times 0.9655 + 0.6983 \times 0.9310 = 5.058$
\end{quote}

Thus

\hspace{5cm} $cont(I_0,I_3,I_1)=10.116$

\textbf{Ordering (1-3-2):}

\begin{quote}
 $cont(I_1,I_3,I_2)=2bond(I_1,I_3)+2bond(I_3,I_2)-2bond(I_1,I_2)$\\
 $bond(I_1,I_3)=bond(I_3,I_1)=5.058$ \\
 $bond(I_3,I_2)=4.949$\\
 $bond(I_1,I_2)=5.326$
\end{quote}

Thus

\hspace{5cm} $cont(I_1,I_3,I_2)=9.362$

\textbf{Ordering (2-3-4):}

\begin{quote}
 $cont(I_2,I_3,I_4)=2bond(I_2,I_3)+2bond(I_3,I_4)-2bond(I_2,I_4)$\\
 $bond(I_2,I_3)=bond(I_3,I_2)=4.949$ \\
 $bond(I_3,I_4)=5.035$ \\
 $bond(I_2,I_4)=5.298$
\end{quote}

Thus

\hspace{5cm} $cont(I_2,I_3,I_4)=9.372$

Since the contribution of the ordering (0-3-1) is the largest, $I_3$ is placed to the left of $I_1$.

$DM'_{res} = \kbordermatrix{
    ~ & 3 & 1 & 2 & & & & & & & & & & &&&&&&&&& &    \\
     1 & 0.3103 & 0 & 0.1364 &       &       &       &       & \\
     2 & 0.3333 & 0.1364 & 0 &       &       &       &       & \\
     3 & 0 & 0.3103 & 0.3333 & & & & &\\
     4 & 0.3115 & 0.1230 & 0.1515 & & & & &\\
     5 & 0.3284 & 0.1385 & 0.1515 & & & & &\\
     6 & 0.3258 & 0.1356 & 0.1515 & & & & &\\
     7 & 0.0345 & 0.3103 & 0.3333 & & & & &\\
     8 & 0.0690 & 0.3017 & 0.3258 & & & & &\\
    }$

For column 4, there are four alternative places where column 4 can be placed, which results in the orderings (0-4-3), (3-4-1), (1-4-2) and (1-2-4) respectively.

\begin{quote}
$cont(I_0,I_4,I_3)=2bond(I_0,I_4)+2bond(I_4,I_3)-2bond(I_0,I_3)
    = 0 + 2 \times 5.035 - 0 = 10.07$

 $cont(I_3,I_4,I_1)=2bond(I_3,I_4)+2bond(I_1,I_4)-2bond(I_1,I_3)
    = 2 \times 5.035 + 2 \times 5.403 - 2 \times 5.059 = 10.758$

$cont(I_1,I_4,I_2)=2bond(I_1,I_4)+2bond(I_4,I_2)-2bond(I_1,I_2)
    = 2 \times 5.403 + 2 \times 5.298 - 2 \times 5.326 = 10.75$

$cont(I_1,I_2,I_4)=2bond(I_1,I_2)+2bond(I_2,I_4)-2bond(I_1,I_4)
    = 2 \times 5.326 + 2 \times 5.298 - 2 \times 5.403 = 10.44$
\end{quote}

The contribution of the ordering (3-4-1) is the largest, enabling $I_4$ is placed to the right of $I_3$. The $DM'_{res}$ becomes as follows.

$DM'_{res} = \kbordermatrix{
    ~ & 3 & 4 & 1 & 2 & & & & & & & & & &&&&&&  \\
     1 & 0.3103 & 0.1230 & 0 & 0.1364 &       &       &       &       & \\
     2 & 0.3333 & 0.1515 & 0.1364 & 0 &       &       &       &       & \\
     3 & 0 & 0.3115 & 0.3103 & 0.3333 & & & & &\\
     4 & 0.3115 & 0& 0.1230 & 0.1515 & & & & &\\
     5 & 0.3284 & 0.1493 & 0.1385 & 0.1567 & & & & &\\
     6 & 0.3258 & 0.1742 & 0.1356 & 0.1515 & & & & &\\
     7 & 0.0345 & 0.3103 & 0.3103 & 0.3333 & & & & &\\
     8 & 0.0690 & 0.3017 & 0.3017 & 0.3258 & & & & &\\
    }$

For column 5, five alternative places can be placed, which results in the orderings (0-5-3), (3-5-4), (4-5-1), (1-5-2) and (1-2-5) respectively.

\begin{quote}

    $cont(I_0,I_5,I_3)=2bond(I_0,I_5)+2bond(I_5,I_3)-2bond(I_0,I_3)
        = 0 + 2 \times 4.968 - 0 = 9.936$

    $cont(I_3,I_5,I_4)=2bond(I_3,I_5)+2bond(I_5,I_4)-2bond(I_3,I_4)
        = 2 \times 4.968 + 2 \times 5.312 - 2 \times 5.035 = 10.49$

    $cont(I_4,I_5,I_1)=2bond(I_4,I_5)+2bond(I_5,I_1)-2bond(I_4,I_1)
        = 2 \times 5.312 + 2 \times 5.336 - 2 \times 5.403 = 10.49$

    $cont(I_1,I_5,I_2)=2bond(I_1,I_5)+2bond(I_5,I_2)-2bond(I_1,I_2)
        = 2 \times 5.336 + 2 \times 5.236 - 2 \times 5.326 = 10.492$

    $cont(I_1,I_2,I_5)=2bond(I_1,I_2)+2bond(I_2,I_5)-2bond(I_1,I_5)
        = 2 \times 5.326 + 2 \times 5.403 - 2 \times 5.336 = 10.786$
\end{quote}

The contribution of the ordering (1-2-5) is the largest. Therefore, column 5 is placed at the right of column 2.

$DM'_{res} = \kbordermatrix{
    ~ & 3 & 4 & 1 & 2 & 5 & & & & & & & & &&&  \\
     1 & 0.3103 & 0.1230 & 0 & 0.1364 & 0.1493 &       &       &       & \\
     2 & 0.3333 & 0.1515 & 0.1364 & 0 & 0.1567 &       &       &       & \\
     3 & 0 & 0.3115 & 0.3103 & 0.3333 & 0.3284 & & & &\\
     4 & 0.3115 & 0& 0.1230 & 0.1515 & 0.1493 & & & &\\
     5 & 0.3284 & 0.1493 & 0.1385 & 0.1567 & 0 &  & & & &\\
     6 & 0.3258 & 0.1742 & 0.1356 & 0.1515 & 0.1716 & & & & &\\
     7 & 0.0345 & 0.3103 & 0.3103 & 0.3333 & 0.3284 & & & & &\\
     8 & 0.0690 & 0.3017 & 0.3017 & 0.3258 & 0.3284 & & & & &\\
    }$

The step is repeated until all columns are inserted into the matrix. Finally, we can get a reordered matrix as follows.

$DM'_{res} = \kbordermatrix{
    ~ & 8 & 3 & 7 & 4 & 1 & 2 & 5 & 6 \\
     1 & 0.3017 & 0.3103 & 0.3103 & 0.1230 & 0 & 0.1364 & 0.1493 & 0.1742 \\
     2 & 0.3258 & 0.3333 & 0.3333 & 0.1515 & 0.1364 & 0 & 0.1567 & 0.1515 \\
     3 & 0.0690 & 0 & 0.0345 & 0.3115 & 0.3103 & 0.3333 & 0.3284 & 0.3258 \\
     4 & 0.3197 & 0.3115 & 0.3033 & 0 & 0.1230 & 0.1515 & 0.1493 & 0.1667 \\
     5 & 0.3284 & 0.3284 & 0.3284 & 0.1493 & 0.1493 & 0.1567 & 0 & 0.1716 \\
     6 & 0.3258 & 0.3258 & 0.3182 & 0.1667 & 0.1742 & 0.1515 & 0.1716 & 0\\
     7 & 0.0690 & 0.0345 & 0 & 0.3033 & 0.3103 & 0.3333 & 0.3284 & 0.3182 \\
     8 & 0 & 0.0690 & 0.0690 & 0.3197 & 0.3017 & 0.3258 & 0.3284 & 0.3258 \\
    }$

The last step is to reorder rows, enabling their relative positions match the relative positions of the columns. The final matrix is as following.

$DM'_{res} = \kbordermatrix{
    ~ & 8 & 3 & 7 & 4 & 1 & 2 & 5 & 6 \\
    8 & 0 & 0.1364 & 0.1515 & 0.1567 & 0.1515 & 0.3258 & 0.3333 & 0.3333 \\
    3 & 0.1364 & 0 & 0.1230 & 0.1493 & 0.1742 & 0.3017 & 0.3103 & 0.3103 \\
    7 & 0.1515 & 0.1230 & 0 & 0.1493 & 0.1667 & 0.3197 & 0.3115 & 0.3033 \\
    4 & 0.1567 & 0.1493 & 0.1493 & 0 & 0.1716 & 0.3284 & 0.3284 & 0.3284 \\
    1 & 0.1515 & 0.1742 & 0.1667 & 0.1716 & 0 & 0.3258 & 0.3258 & 0.3182 \\
    2 & 0.3258 & 0.3017 & 0.3197 & 0.3284 & 0.3258 & 0 & 0.0690 & 0.0690 \\
    5 & 0.3333 & 0.3103 & 0.3115 & 0.3284 & 0.3258 & 0.0690 & 0 & 0.0345 \\
    6 & 0.3333 & 0.3103 & 0.3033 & 0.3284 & 0.3182 & 0.0690 & 0.0345 & 0 \\}$
\end{itemize}

By observing reordered matrix consequences, we find both of these permutation algorithms gather similar messages together, along the main diagonal.

\subsubsection{Dissimilarity matrix image}

Once the initial dissimilarity ratio matrix has been reordered using a matrix reordering algorithm, the second step is to display the reordered matrix as a gray-scale image. The step is performed to display the dissimilarity matrix as an intensity image, which is called a {\em dissimilarity image} in~\cite{VAT}. The intensity or gray scale of pixel (i,j) depends on the value of $d_{ij}$. The value $d_{ij}=0$ corresponds to pure black; the value $d_{ij}=d_{max}$, where $d_{max}$ denotes the largest similarity value in $DM$, gives pure white. Intermediate values of $d_{ij}$ correspond pixels with intermediate levels of grey.

In Table~\ref{Chap6:VATImage}, we list the reordered dissimilarity matrix that is produced by using the modified Prim's minimal spanning tree algorithm, as well as its corresponding dissimilarity image. The 0 values on the main diagonal of $DM$ generate main diagonal pixels that are black. The larger dissimilarity values give lighter grey pixels in the dissimilarity image. This image shows a clear interaction cluster tendency. There are two solid dark squares along the diagonal. Each of them represents interactions of the same message type that can be separated as an individual partition.

\begin{table}[ht]
        \begin{tabular}{ c }
        $DM'_{res} = \begin{bmatrix}
              0 & 0.1364 & 0.1515 & 0.1567 & 0.1515 & 0.3258 & 0.3333 & 0.3333 \\
             0.1364 & 0 & 0.1230 & 0.1493 & 0.1742 & 0.3017 & 0.3103 & 0.3103 \\
             0.1515 & 0.1230 & 0 & 0.1493 & 0.1667 & 0.3197 & 0.3115 & 0.3033 \\
             0.1567 & 0.1493 & 0.1493 & 0 & 0.1716 & 0.3284 & 0.3284 & 0.3284 \\
             0.1515 & 0.1742 & 0.1667 & 0.1716 & 0 & 0.3258 & 0.3258 & 0.3182 \\
             0.3258 & 0.3017 & 0.3197 & 0.3284 & 0.3258 & 0 & 0.0690 & 0.0690 \\
             0.3333 & 0.3103 & 0.3115 & 0.3284 & 0.3258 & 0.0690 & 0 & 0.0345 \\
             0.3333 & 0.3103 & 0.3033 & 0.3284 & 0.3182 & 0.0690 & 0.0345 & 0
        \end{bmatrix}$ \\
            \Huge $\Downarrow$ \normalsize \\

        \includegraphics[width=0.7\textwidth]{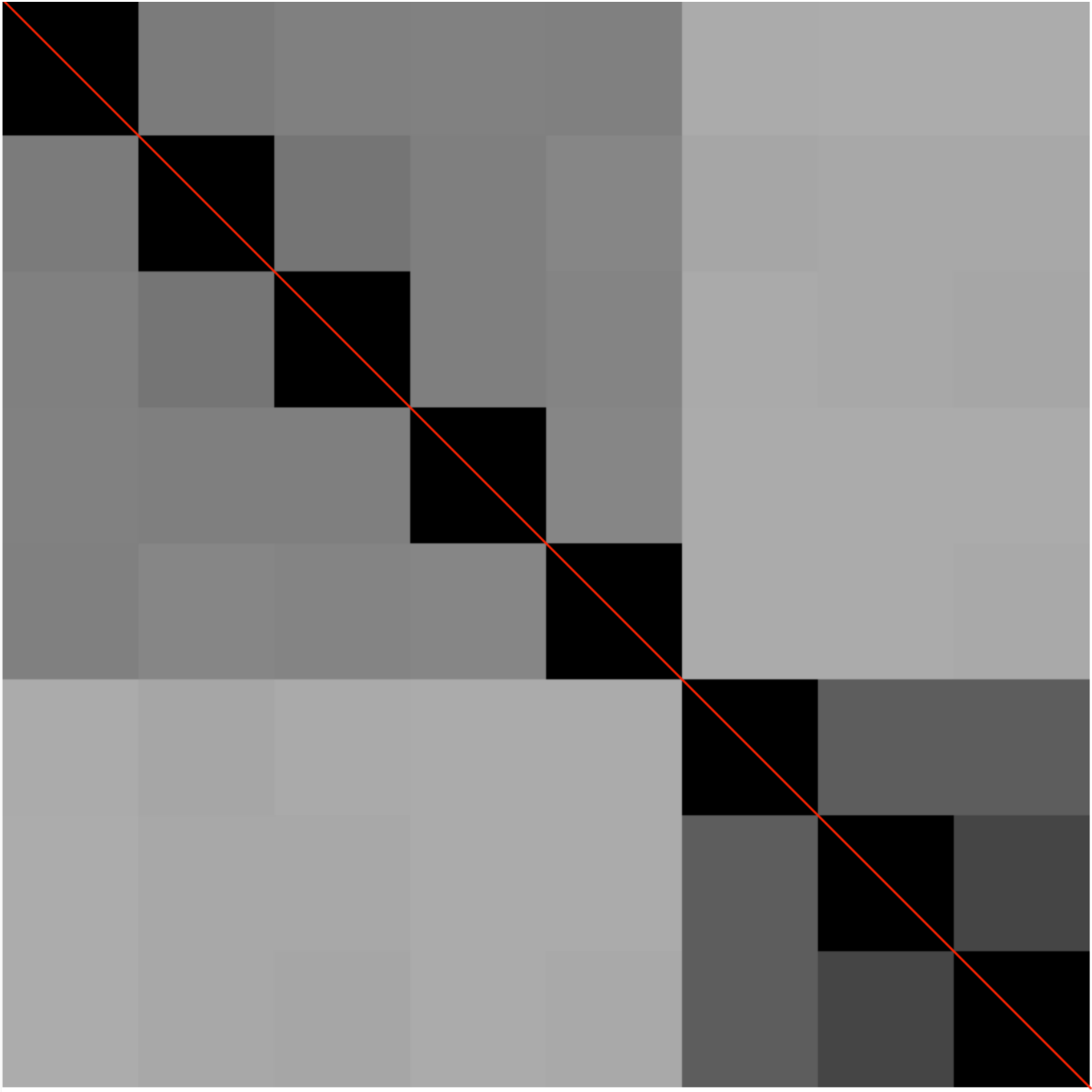}\\
    \end{tabular}
    \caption{Reordered dissimilarity matrix by using Prim's minimal spanning tree algorithm and its corresponding dissimilarity image}\label{Chap6:VATImage}
\end{table}
%Figure~\ref{chap6:BEAImage} shows the grey-scale image of the reordered dissimilarity matrix using the BEA algorithm.
%
%
%\begin{figure}[h]
%  \centering
%  % Requires \usepackage{graphicx}
%  \includegraphics[width=0.8\textwidth]{BEAImage}\\
%  \caption{Reordered Matrix using the BEA algorithm}
%  \label{chap6:BEAImage}
%\end{figure}

%add two graphs here, for displaying bea reording request and vat reording results
%highlight, with the use of image, non-expert is able to decide which messages should be grouped
%together, they can manually drop a point and drag in order to mark a cluster. By selecting
%exporting, a number of clusters are exported.

\subsubsection{Partition the reordered dissimilarity image and export divisions}

\begin{figure}[ht]
  \centering
  \includegraphics[width=0.8\textwidth]{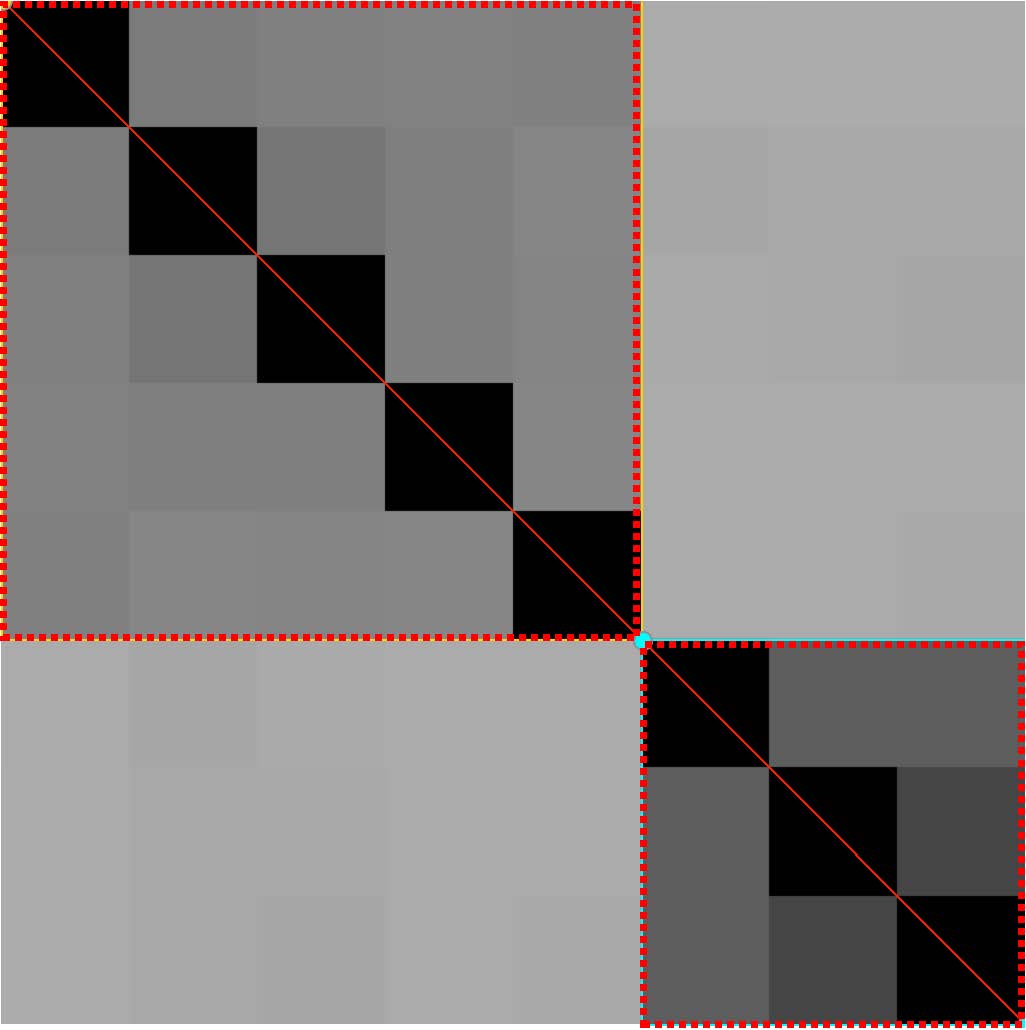}\\
  \caption{Partitioning on the Trace Library by dropping and dragging a rectangle}
  \label{Chap6:ClusterPartition}
\end{figure}

A reordered dissimilarity matrix image indicates cluster tendency in the data by dark {\em blocks} of pixels along the main diagonal. Partitioning on the trace library and exporting trace library division with minimal manual effect become straightforward (Step 2(c)). %Justification of manual interaction within our assumption
At this stage, users only need to know dark blocks show them a cluster tendency. They need to partition interaction trace library by dropping and dragging rectangles to select interactions of the same operation type. This step requires limited manual interaction, which is users' visual knowledge rather than explicit protocol knowledge. The manual interaction is required by our current approach. We intend to apply other techniques to automate this step, which will be discussed in Chapter~\ref{chap8sec:futurework}. Considering the matrix image in Figure~\ref{Chap6:ClusterPartition}, selected solid dark squares are highlighted with the red dash line. Then, we export and preserve partition consequence as separate groups of different operation types. These groups will be used for following steps.

% ------------------------------------------------------------------------ %

\subsection{Selecting a cluster centre from each group (Step 3)}
\label{chap6subsec:centers}
% define a formular to explain current method to select the center message for
% each cluster

% ------------------------------------------------------------------------ %

Applying the cluster tendency analysis technique to the trace library yields interaction groups. Consider the trace library example in Table~\ref{chap6tab:interactionlibraryexample}, we can export two interaction clusters after the cluster tendency analysis, as shown in tables~\ref{chap6tab:searchcluster} and ~\ref{chap6tab:addcluster}, corresponding to search interactions and add interactions, respectively.

\begin{table}[ht]
\begin{center}
\begin{tabular}{|c||m{4.8cm}|m{7cm}|}
\hline
Index & Request & Response \\ \hline\hline
1 & \{id:1,op:S,sn:Du\} & \{id:1,op:SearchRsp,result:Ok, gn:Miao,sn:Du,mobile:5362634\} \\ \hline
2 & \{id:13,op:S,sn:Versteeg\} & \{id:13,op:SearchRsp,result:Ok, gn:Steve,sn:Versteeg,mobile:9374723\} \\ \hline
3 & \{id:275,op:S,sn:Han\} & \{id:275,op:SearchRsp,result:Ok, gn:Jun,sn:Han,mobile:33333333\} \\ \hline
4 & \{id:490,op:S,sn:Grundy\} &
\{id:490,op:SearchRsp,result:Ok, gn:John,sn:Grundy,mobile:44444444\} \\ \hline
5 & \{id:2273,op:S,sn:Schneider\} & \{id:2273,op:SearchRsp,result:Ok, sn:Schneider,mobile:123456\} \\ \hline\hline
\parbox[p]{2cm}{\textbf{Centroid\\Interaction}}& \{id:275,op:S,sn:Han\} & \{id:275,op:SearchRsp,result:Ok, gn:Jun,sn:Han,mobile:33333333\} \\
\hline
\end{tabular}
\end{center}
\caption{Search Request Cluster}
\label{chap6tab:searchcluster}
\end{table}

\begin{table}[ht]
\begin{center}
\begin{tabular}{|c||m{5.2cm}|m{6.5cm}|}
\hline
Index & Request & Response \\ \hline\hline
1 & \{id:24,op:A,sn:Schneider, mobile:123456\} & \{id:24,op:AddRsp,result:Ok\} \\ \hline
2 & \{id:2487,op:A,sn:Will\} & \{id:2487,op:AddRsp,result:Ok\} \\ \hline
3 & \{id:1106,op:A,sn:Hine,gn:Cam, postalCode:33589\} & \{Id:1106,Msg:AddRsp,result:Ok\} \\ \hline\hline
\parbox[p]{2cm}{\textbf{Centroid\\Interaction}} & \{id:24,op:A,sn:Schneider, mobile:123456\} & \{id:24,op:AddRsp,result:Ok\} \\ \hline
\end{tabular}
\end{center}
\caption{Add Request Cluster}
\label{chap6tab:addcluster}
\end{table}

Having divided our trace library, the next step is to select a cluster centre from each cluster. %At runtime, for a new request, the {l\em matching function} is able to search its best matching interaction from the set of centroid interactions, thereby reducing time consumption for matching.
In our approach, the cluster centre is chosen by finding the request of one interaction with the lowest sum of the distances to requests of all other interactions in the cluster. Hence, a cluster centre interaction is selected in a way to minimize the following equation~\ref{chap6equa:dist},

\begin{equation}
\label{chap6equa:dist}
\sum_{k=1}^n{dist({\Request}_i,{\Request}_k)}, i \in [1,...,n]
\end{equation}

where ${\Request}_{i,k}$ denotes request messages of any interactions in the cluster, $n$ denotes the total number of interactions
in the cluster.

Consider add interactions in Table~\ref{chap6tab:addcluster}, the centroid interaction is selected as follows.

For ${\Interaction}_1$,
\begin{quote}
   $ \sum_{k=1}^3{dist({\Request}_1,{\Request}_k)} = 0 + 0.2692 + 0.25 = 0.5192$
\end{quote}

For ${\Interaction}_2$,
\begin{quote}
   $ \sum_{k=1}^3{dist({\Request}_2,{\Request}_k)} = 0.2692 + 0 + 0.34 = 0.6092$
\end{quote}

For ${\Interaction}_3$,
\begin{quote}
    $\sum_{k=1}^3{dist({\Request}_3,{\Request}_k)} = 0.25 + 0.34 + 0 = 0.59$
\end{quote}

Having these calculations, ${\Interaction}_2$ has the lowest sum of the distances. Hence, interaction {\em \{id:24,op:A,sn:Schneider, mobile:123456\} \& \{id:24,op:AddRsp,result:Ok\}} is selected to be the centroid interaction of the add cluster. Interaction {\em \{id:275,op:S,sn:Han\} \& \{id:275,op: SearchRsp,result:Ok, gn:Jun,sn:Han,mobile:33333333\}} is selected to be the centroid interaction of the search cluster using the same measure.

%In Section \ref{sec:evaluation}, we describe evaluations of both methods and discuss our experimental results.

%\subsection{Improving cluster homogeneity}
%2.	Cluster messages into operations
%    a.	We get better clusters by clustering by response.MSA is sensitive to cluster noise so cluster homogeneity is critical to cluster consensus inference.
%
%    b.	Use entropy weighted Needleman-Wunsch distance function

%\subsection{Implementation}
%
%
%We have developed a proof-of-concept realisation of this trace analysis
%approach, including
%the Needleman-Wunsch algorithm and using two clustering algorithms, BEA and VAT.
%We have integrated this implementation with our prior interaction analysis and response generation prototype \cite{Du:2013}.
%This prototype interaction analysis and response generation
%environment was implemented in the Java programming
%language. We used the Wireshark tool to capture network
%traffic and exported this into a format suitable for input into our Java implementation.
%However, the format transformation requirement significantly limits the
%applicability of our approach for building diverse interaction models.
%This is a task we intend to cover in future work.

%%%%%%%%%%%%%%%%%%%%%%%%%%%%%%%%%%%%%%%%%%%%%%%%%%%%%%%%%%%%%%%%%%%%%%%%%%%%%
% Section about the evaluation of the approach

%%%%%%%%%%%%%%%%%%%%%%%%%%%%%%%%%%%%%%%%%%%%%%%%%%%%%%%%%%%%%%%%%%%%%%%%%%%%%

\section{Evaluation}
\label{chap6sec:evaluation}

We evaluate our new opaque response generation approach for two aspects: {\em effectiveness} and {\em efficiency}, aiming to answer following questions:

\begin{enumerate}
    \item {\textbf{Question 1 (Effectiveness)}}: After applying our new approach to condense the trace library, how is the accuracy of generated responses? Compared with our previous approach in Chapter~\ref{chap5:qosa}, what is the impact of this condensate to the accuracy of generated responses?

    \item {\textbf{Question 2 (Efficiency)}}: Is our new approach efficient to generate responses in real time for large trace library? Compared with our previous approach in Chapter~\ref{chap5:qosa}, how is the impact of our new approach to the runtime efficiency of generated responses?

\end{enumerate}

To answer these two questions, we applied our new approach to two application-layer protocols, the Simple Object Access Protocol (SOAP)~\cite{SOAPv1.1} and the Lightweight Directory Access Protocol (LDAP)~\cite{Sermersheim2006}. We chose these two protocols because: (i) they are commonly used in enterprise environment; (2) they were used as case study examples to evaluate our prior work ({\cf} Section~\ref{chap5sec:evaluation}). Message examples for each test case can be referred to Appendix~\ref{apx.traceexamples}.
%

%--------------------------------------------------------------------------
\subsection{Cross-Validation Approach and Evaluation Criteria}
\label{chap6subsec:criteria}

{\em Cross-validation} \cite{devijver:1982} is one of the most
popular methods for assessing how the results of a statistical analysis will
generalise to an independent data set. For the purpose of our evaluation, we
applied the commonly used 10-fold cross-validation approach
({\cf}~\ref{chap5subsec:criteria}) to both SOAP and LDAP trace libraries.

We investigated both the effectiveness and efficiency of our proposed approach for response
synthesis. We utilise four criteria (identical, protocol-conformant, well-formed, malformed) to evaluate the accuracy of synthesized responses, as defined in Section~\ref{chap5subsec:criteria}.

%\begin{enumerate}
%  \item \textbf{Identical}:
%  the synthesized response is identical to the recorded response if all characters in
%  the synthesized response are the same as those in the recorded response
%  \item \textbf{Protocol Conformant}\footnote {We use a weaker notion of {\em protocol comformant} as the order in which the requests are selected from the evaluation set is {\em random} and, as a consequence, unlikely to conform to the temporal sequence of request-response pairs.}:
%  this criterion indicates %the synthesized response are not identical to the recorded
%%  response. However, it requires
%  that the responses conform to the temporal interaction
%  properties of the given protocol, that is, the temporal consistency between request
%  and response is preserved.
%  \item \textbf{Well-Formed}:
%  this criterion requires that the synthesized responses correspond to the structure
%  required for responses as defined by the underlying protocol.
%  \item \textbf{Ill-Formed}:
%  Synthesized responses do not meet above criteria.
%\end{enumerate}
%

For the efficiency investigation, for each message in the evaluation group,
we record how much time was taken to synthesize a response for an
incoming request.

% ------------------------------------------------------------------------- %

\subsection{Evaluation Results}
\label{chap6subsec:results}

In Chapter~\ref{chap5:qosa}, we introduced the {\em whole library} approach, the primary implementation of our opaque response generation approach. This approach is to search every interaction in a large trace library for the closest matching one. For the purpose of evaluation, we use this approach to benchmark both the effectiveness and efficiency of
the proposed cluster-based approach for synthesizing responses. In our experiment, in the
pre-processing stage, we apply the BEA reordering algorithm and Prim's minimal spanning tree reordering algorithm to reorder interaction distance matrix respectively.

At the runtime stage, the matching function and the translation function generate response %build executable interaction
%models of software service behaviours (\cf \cite{Du:2013})
with two methods: 1)
{\em cluster centroid} and 2) {\em whole cluster}. Specifically, given an incoming request,
the {\em cluster centroid} method perform transformation on the selected cluster centroid interaction, % synthesize the response by comparing the incoming request with the selected cluster center only,
%while
while {\em whole cluster} method requires the \emph{matching function} to further compare this incoming request against all recorded request messages in the selected cluster for searching the most similar, which may be the centroid interaction, but also may be someone else.
%synthesis is based on comparison results of the incoming request with all interaction elements in the cluster.At the runtime stage, we further use the
%\emph {Cluster Centroid} method and \emph {Whole Cluster} method ({\cf} Section \ref{chap6:CenterOnly}) to synthesize responses.
The reasons of running this experiment is to assess if the centroid interaction is able to represent all other interactions in the cluster.

In Table \ref{chap6tab:SOAP} and Table \ref{chap6tab:LDAP}, we summarise the number of generated SOAP responses and LDAP
responses, which were categorised according to the criteria introduced in Section
\ref{chap6subsec:criteria}. %As both the identical and protocol-conformant responses are \emph{valid}
%responses, we
In addition, we also list the number of valid responses,
that is, the sum of responses falling in the \emph{Identical} and \emph{Protocol-conformant} categories,
for both protocol examples.

Figure \ref{chap6fig:soaptimewhole} and Figure \ref{chap6fig:ldaptimewhole} outline the amount of time that was taken to
generate responses for incoming requests of different lengths, \ie the number of characters per incoming
request, by using the \emph{whole library} method, the \emph{cluster centroid} method
and the \emph{whole cluster} method. For both the \emph{cluster centroid} method and the \emph{whole cluster} method, we
further compare the response generation times of using the Prim's minimal spanning tree algorithm and BEA algorithm respectively,
as shown in Figure \ref{chap6fig:soaptimedetail} and Figure \ref{chap6fig:ldaptimedetail}.
\\

\subsubsection{Evaluation results for SOAP}

\begin{figure}[ht]%
    \centering
     \subfloat[]{
         \includegraphics[width=0.9\textwidth]{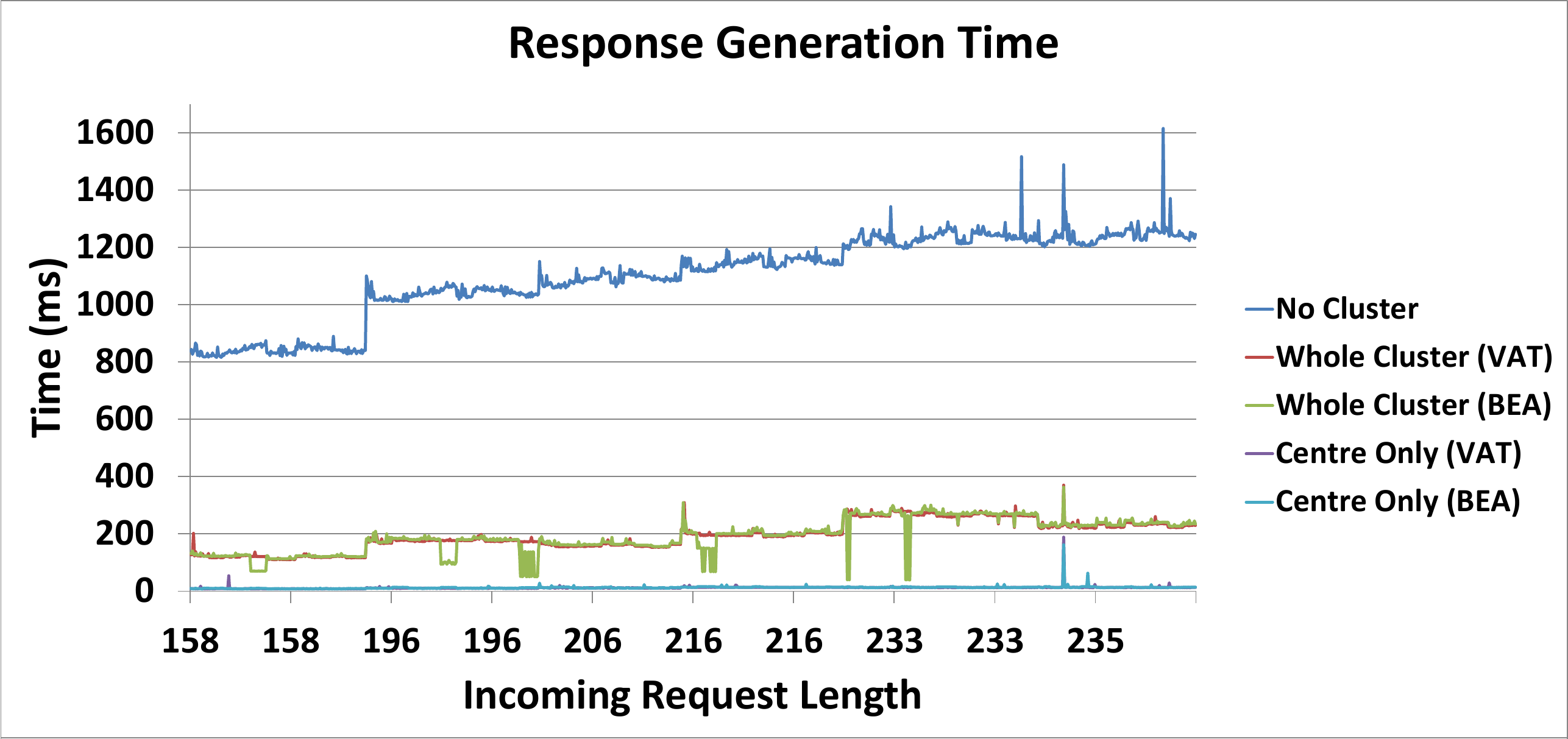}
         \label{chap6fig:soaptimewhole}}

     \subfloat[]{%
          \includegraphics[width=0.9\textwidth]{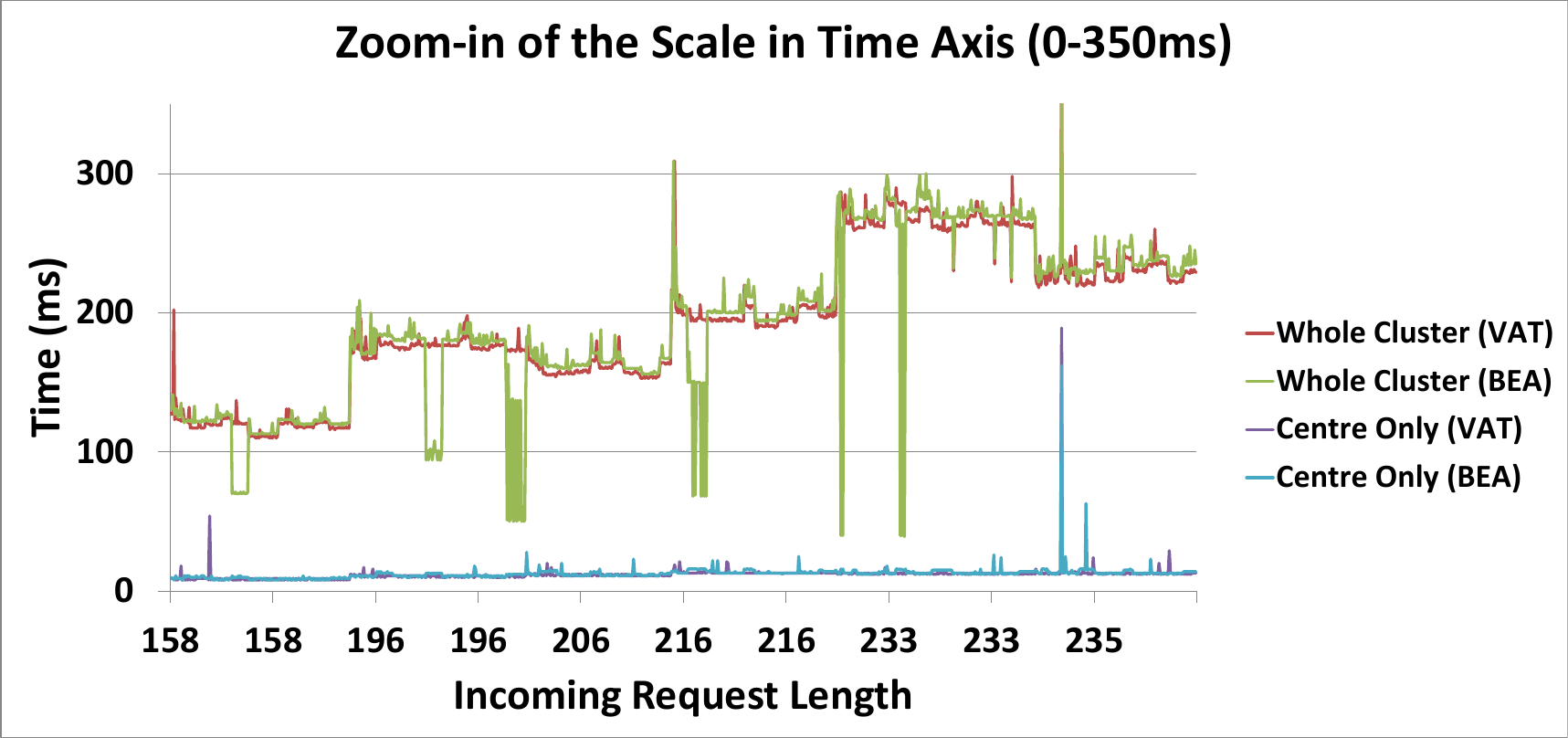}
          \label{chap6fig:soaptimedetail}}
   \caption{SOAP Responses Generation Time}
   \label{chap6fig:soaptime}
\end{figure}

Table~\ref{chap6tab:SOAP} and Figure~\ref{chap6fig:soaptime} compare the different outcomes
of the {\emph whole library} method and cluster-based approaches.
From Table~\ref{chap6tab:SOAP} we can see that neither the {\em cluster centroid}
method nor the {\em whole cluster} method generate malformed SOAP
responses, which produce the same outcomes
as using the {\em whole library} method. This shows that for the
SOAP case study used, our cluster-based approach has the same effectiveness
as the {\em whole library} method in synthesizing accurate responses.

\begin{table}[ht]

\begin{center}
\begin{tabular}{|c||c|c|c|c|c|c|c|}
\hline
\parbox[p]{1cm}{Exp.} &\parbox[p]{1.2cm}{Cluster Method} &No. & Valid & Ident. & Conf. \footnotemark[1]& \parbox[p]{1.2cm}{Well-form.~ \footnotemark[2]}& \parbox[p]{1cm}{Malformed}%&Mean dsim.$*$ & Max dsim.$*$
\\

\hline\hline
\parbox[p]{2cm}{No Cluster}& None &1,000 &  1,000 & 85 & 915 & 0 &  0 %& 0.021 &0.046 \\
\\
\hline
\multirow{2}{*}{\parbox[p]{1.5cm}{Whole Cluster}}& Prim &1,000 & 1,000 & 90 & 910 &   0 &  0 %&0.021& 0.048
\\
                   \cline{2-8}&BEA& 1,000 & 1,000 & 82 & 918 & 0 & 0 %&0.021&0.048
                   \\
\hline
\multirow{2}{*}{\parbox[p]{1.5cm}{Centre Only}}& Prim &1,000 & 1,000 & 87 & 913 &   0 &  0 %&0.021& 0.046
\\
                 \cline{2-8}&BEA& 1,000 & 1,000 & 92 & 908 & 0 & 0 %&0.021&0.044
                 \\
\hline

\end{tabular}
\end{center}
\caption{Validity of Synthesized SOAP Responses}
\label{chap6tab:SOAP}
\end{table}

From the Figure \ref{chap6fig:soaptime}, we can see that whatever approach we use
to generate a response (\ie { either \emph{No Cluster} method or cluster-based
methods}), more time is consumed to synthesize responses for longer incoming requests. However, a significant improvement of the generation time is observed by
using cluster--based approaches. Specifically, response generation time of using \emph {Whole Cluster}
methods is at least 5 times quicker than using the \emph {No Cluster} method; using \emph {Cluster Centroid}
is able to further accelerate response message synthesis by approximately 120 times than that of using the \emph
{No Cluster} method.

By further investigating cluster-based approaches in Figure \ref{chap6fig:soaptimedetail},
we can see that although there is more fluctuation in the response generation time when using the BEA algorithm
to cluster messages, the selection of clustering approaches does not impact the response generation time too much.
Based on these observations, we conclude that for the SOAP protocol, the \emph {Cluster Centroid} method is able to provide a good response message synthesis.

\subsubsection{Evaluation results for LDAP}

\begin{table}[ht]
\begin{center}
\begin{tabular}{|m{1.5cm}||c|c|c|c|c|c|c|}
\hline

\parbox[p]{1cm}{Exp.} &\parbox[p]{1.2cm}{Cluster Method} &No. & Valid & Ident. & Conf. \footnotemark[1]& \parbox[p]{1.2cm}{Well-form.~ \footnotemark[2]}& \parbox[p]{1cm}{Malformed}\\

\hline\hline
\parbox[p]{1.5cm}{Whole Library} & None &1,000 &  908 & 451 & 457 & 89 &  3 \\%&0.066&0.453 \\
\hline
\multirow{2}{*}{\parbox[p]{1.5cm}{Whole Cluster} }& Prim &1,000 & 751 & 360 & 391 &   246 &  3\\%&0.092&0.843 \\
\cline{2-8}&BEA& 1,000 & 753 & 383 & 370 & 241 & 6\\%&0.096& 0.843\\
\hline
\multirow{2}{*}{\parbox[p]{1.5cm}{Cluster Centroid}}& Prim &1,000 & 751 & 296 & 455 &  235 &  14\\%&0.094&0.842  \\
\cline{2-8}&BEA& 1,000 & 761 & 330 & 431 & 232 & 7 \\%&0.097&0.843 \\
\hline

\end{tabular}
\end{center}
\caption{Validity of Synthesized LDAP Responses}
\label{chap6tab:LDAP}
\end{table}

Table~\ref{chap6tab:LDAP} gives a summary of the result of our LDAP experiments. %are also given in
For the {\em No Cluster} method, $451$
(out of $1,000$) generated response messages are identical to the
corresponding recorded responses ($45.1$\%), and an additional $457$ of the
generated responses met the protocol conformant criterion ($45.7$\%).
Therefore, a total of $908$ (or $90.8$\%) of all generated responses were
considered to be valid. In contrast, Cluster-based approaches generate less valid responses, the number
of which decrease by 14.7\% to around 761 (out of 1000). By observing valid responses of cluster-based
approaches, we identify that the Prim's minimal spanning tree algorithm performs better than the BEA
algorithm. However, there is no distinguishable difference between results of
the {Cluster Centroid} method and the {Whole Cluster} method.

%

%\begin{subfigures}[!h]%
%    \begin{figure}\centering
%     \includegraphics[width=3.6in]{LDAPGenerationTimeStatistics}
%     \label{chap6fig:ldaptimewhole}
%    \end{figure}
%     \begin{figure}\centering
%     \includegraphics[width=3.6in]{LDAPZoomInGenerationTimeStatistics}
%     \label{chap6fig:ldaptimedetail}
%    \end{figure}
%
%\label{chap6fig:ldaptime}
%\end{subfigures}

\begin{figure}[!ht]%
    \centering
     \subfloat[]{
         \includegraphics[width=0.8\textwidth]{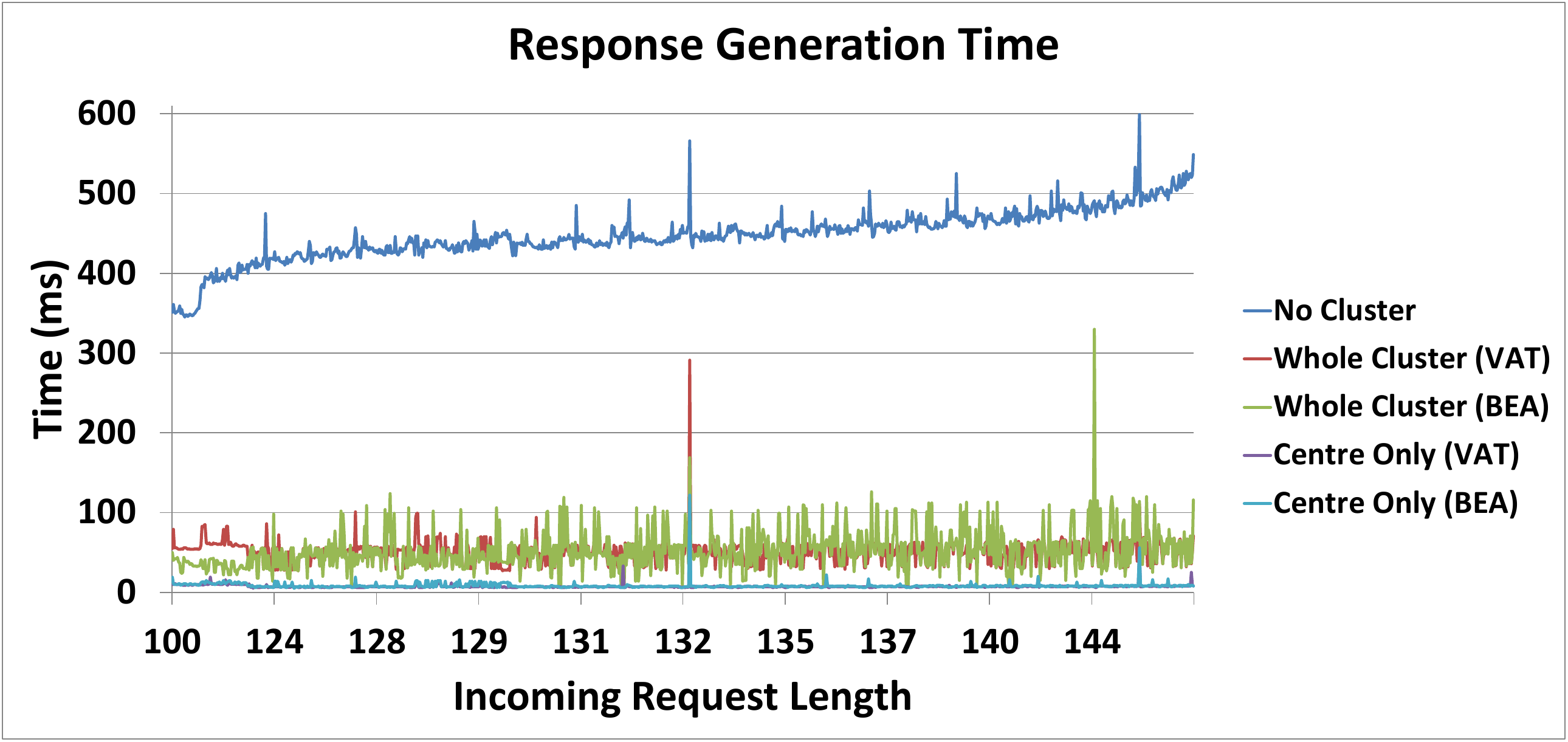}
         \label{chap6fig:ldaptimewhole}}

     \subfloat[]{%
          \includegraphics[width=0.8\textwidth]{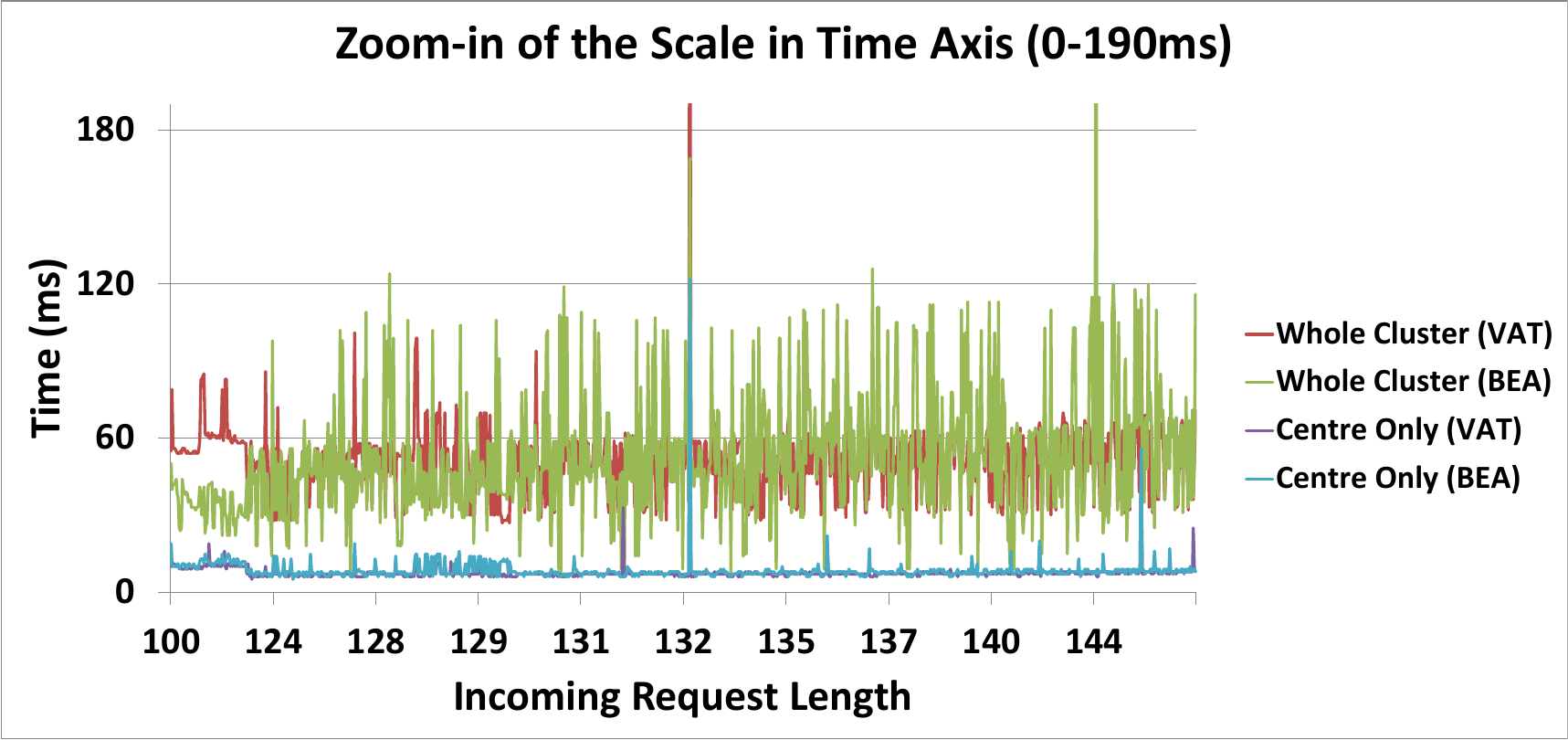}
          \label{chap6fig:ldaptimedetail}}
   \caption{LDAP Response Generation Time}
   \label{chap6fig:ldaptime}
\end{figure}

Figure \ref{chap6fig:ldaptime} also shows an increasing trend of response
generation time when the length of incoming requests becomes longer. As the length of
the majority of incoming requests is shorter than the SOAP incoming requests, the LDAP
response generation time is shorter than SOAP response generation time, shown in Figure~\ref{chap6fig:soaptime}. Specifically,
compared with the average response generation time of using \emph{No Cluster} method (about 53.28ms),
using \emph{Whole Cluster} method is able to produce responses about 9 times faster (about 5.46ms),
while using \emph{Centre Only} method can further improve the generation time to around
0.79ms. As we can see from the Figure \ref{chap6fig:ldaptimedetail}, the response generation time of using \emph{Whole cluster}
method fluctuates significantly. This is because the sizes of the clusters generated by the clustering algorithms
are different. Therefore, the time taken when using them to generate responses are accordingly different. In contrast, as
the number of clusters is stable, the response generation time of using the \emph {Centre Only} method is
observed to have only a slight fluctuation.

\subsection{Discussion}
\label{chap6subsec:discussion}
Based on summaries of both the SOAP and LDAP experimental results in Table~\ref{chap6tab:summary}, we can see that
the cluster centroid approach is able to generate valid responses much more efficiently than
searching the entire trace library.
However, as illustrated in the results for LDAP, the cluster centroid approaches generate fewer valid
responses. This is attributed to differences between the SOAP and LDAP protocols. Most application-level protocols
define message structures containing some form of {\em operation} or service name in their request,
followed by a {\em payload}, containing the data the service is expected to operate upon \cite{hine:thesis}.
In LDAP, some messages contain much more {\em payload} information than {\em operation} information.  Two LDAP
messages of different operation types, but with a similar payload, may be found to be the closest matching
messages.  In such a case, a response of the wrong operation type may be sent back, which would not be a valid protocol conformant response.

\begin{table}[ht]
\begin{center}
\begin{tabular}{|c|l|c|c|c|}
\hline
Experiment & \parbox{1.4cm}{Cluster\\Method\\(VAT)} & \parbox{0.6in}{Response\\Time(ms)} & \parbox{0.4in}{Dist.\\Calcs.} & Accuracy\\
\hline \hline

\multirow{3}{*}{\parbox{0.7in}{SOAP}}&No Cluster & 128.6 &  899 & 100\% \\ \cline{2-5} & Whole & 22.92 &156&100\% \\ \cline{2-5} & Centre &0.90 & 6 & 100\% \\

\hline
\multirow{3}{*}{\parbox{0.7in}{LDAP}}& No Cluster & 53.28 &  899 & 90.8\% \\ \cline{2-5} & Whole & 5.46 & 96 &75.1\% \\ \cline{2-5} & Centre & 0.79 & 11 & 75.1\% \\ \hline

\end{tabular}
\end{center}
\caption{Average Response Times, Number of Distance Calculations and Accuracy}
\label{chap6tab:summary}
\end{table}

In Table~\ref{chap6tab:summary}, we can see that the time cost of generating
responses depends on the number of distance calculations. In our prototype,
we use the Needleman-Wunsch algorithm as the distance function, which has a relatively high time complexity of
{\em O(MN)}, where {\em M} and {\em N} denote the length of pairwise messages.
%distance function, the Needleman-Wunsch algorithm is th
% needle-wunch algorithm is time consuming, the time complexity of which is {\em O(MN)}, where
% M, N stand for the length of aligned sequences. Our implementation needs to be further
%optimized.

\section{Summary}
\label{chap6sec:summary}

In this chapter, we have utilised data mining techniques in a large enterprise software emulation environment to efficiently generate system responses. Specifically, we utilized two popular clustering algorithms, the BEA algorithm and the Prim's minimal spanning tree algorithm, to pre-process recorded interaction traces. Using these clustered results, our new approach is able to mimic service interaction behaviours in a runtime fashion. As our proposed approach does not require explicit knowledge of the protocols that the target service uses to communicate, it eliminates the human effort of manually specifying interaction models. Moreover, by utilizing data mining techniques, the efficiency of response generation in the emulation environment has been greatly improved. However the increased efficiency came at
the expense of decreased accuracy. %Experimental results conducted on LDAP and SOAP protocols demonstrate that the response generation time has been reduced by 99\% on average compared to our prior approach, while the ac- curacy of response generation (the valid response rate) was 100\% for SOAP and 75\% for LDAP.
In next chapter, we will introduce a novel method that can automatically summarize the most common characters of all messages within a cluster, which will achieve both efficiency and accuracy for opaque response generation,.
%%%%%%%%%%%%%%%%%%%%%%%%%%%%%%%%%%%%%%%%%%%%%%%%%%%%%%%%%%%%%%%%%%%%%%%%%%%%% 
\chapter{Cluster Consensus Prototype Inference}
\label{chap7:consensus}
%motivation
%limitations of QoSA technique and SoftMine technique, should refer to respective chapters
%QoSA limitation: 1) time-consuming; 2) found recorded interaction of different operation type; 3) relying on transformation function of the traffic capture tool, only work for interactions recorded in textual format
The whole library technique, proposed in Chapter~\ref{chap5:qosa}, can be used to create virtual services with high accuracy. However, this approach has three main limitations. Firstly, it is very time-consuming as it has to iterate over the whole interaction collection to search for the most similar interaction recording for an incoming request. Hence, it is not suitable to be used in the runtime environment. Secondly, this technique is prone to some errors where the incoming request has similar payload information to a recorded request of a different operation type. This is because payload sections are usually longer than operation type sections, so have more weight during the distance calculation. Finally, it can only be applied to textual formats and relies on a binary-to-text transformation tool for binary formats.

%limitation of SoftMine technique: (payload versus operation type) 1) interactions of different types are clustered into the same group, which is attributed to distance function that is not able to distinguish type section and payload section; 2) the incoming request is matched with a wrong cluster consensus, which is attributed to i) distance calculation that is not able to distinguish type section and payload section; ii) chosen cluster consensus is not %representative
The cluster centroid technique, described in Chapter~\ref{chap6:SoftMine}, can be used to emulate services and provide real time response generation. Clustering interaction messages can reduce the number of interactions to be searched from the entire trace library to just comparing to one interaction for each cluster, thereby speeding up the matching procedure. However, as this technique condenses all messages in a cluster into only one message, this discards a lot of important information. New requests are more often matched with the wrong message compared to the whole library technique, which leads to lower accuracy of generated responses. A new approach is needed which is able to reduce the number of comparisons but retains the important information.

In this chapter, we present a novel consensus prototype technique that can: 1) produce a representative consensus sequence which robustly captures most of the important information contained in the cluster messages; 2) introduce a refined distance function, which is able to assign different weights to different sections of the message according to whether it is payload or operation type; and 3) apply the refined distance function to cluster either request messages or response messages to achieve homogeneous clustering. Our evaluation results show that this results in a more accurate and run-time efficient request matching and response generation approach.

\section{Motivating example}

The clustered centroid method ({\cf} Section~\ref{chap6:SoftMine}) extends the whole library method by making it more efficient, such that responses can be generated in real time, even for large interaction libraries. This method implements the analysis function with clustering techniques at the preprocessing stage to assist in searching for the most similar recorded interaction for an incoming request. The analysis function: 1) adopts a clustering technique that classifies interactions into different groups; and 2) selects a message from each cluster, which is the closest neighbour to the others, to be the representative of all messages in the cluster. Instead of searching all recordings of the interaction library, the clustered interaction library method only needs to search the number of cluster representatives. It hence can generate responses very efficiently in practice. However, this method produces less accurate responses. The reasons for the low accuracy of this technique are:

\begin{enumerate}
\item The clustering process is not 100\% accurate. Interactions of different types are sometimes clustered into the same group. This is caused by the distance function not being able to distinguish the operation type section and the payload section of the messages;
\item Incoming requests are sometimes matched with the wrong cluster centroid. This is because the cluster centroid is not representative of all messages in the cluster, but simply one sample message in the cluster. During the matching process, the incoming request is likely to match with the wrong cluster centroid, if the payload information in the incoming request is similar to that of the cluster centroid.
\end{enumerate}

We illustrate this point with a running example as follows.

  %  \begin{table}[h]
%    \begin{center}
%    \begin{tabular}{|c||m{5cm}|m{7cm}|}
%    \hline
%    Index & Request & Response \\ \hline\hline
%    1 & \{id:1,op:S,sn:Du\} & \{id:1,op:SearchRsp,result:Ok, gn:Miao,sn:Du,mobile:5362634\} \\ \hline
%    13 & \{id:13,op:S,sn:Versteeg\} & \{id:13,op:SearchRsp,result:Ok, gn:Steve,sn:Versteeg,mobile:9374723\} \\ \hline
%    24 & \{id:24,op:A,sn:Schneider, mobile:123456\} & \{id:24,op:AddRsp,result:Ok\} \\ \hline
%    275& \{id:275,op:S,sn:Han\} & \{id:275,op:SearchRsp,result:Ok, gn:Jun,sn:Han,mobile:33333333\} \\ \hline
%    490 & \{id:490,op:S,sn:Grundy\} &
%    \{id:490,op:SearchRsp,result:Ok, gn:John,sn:Grundy,mobile:44444444\} \\ \hline
%    2273 & \{id:2273,op:S,sn:Schneider\} & \{id:2273,op:SearchRsp,result:Ok, sn:Schneider,mobile:123456\} \\ \hline
%    2487 & \{id:2487,op:A,sn:Will\} & \{id:2487,op:AddRsp,result:Ok\} \\ \hline
%    3106 & \{id:3106,op:A,sn:Hine, gn:Cam,Postcode:33589\} & \{id:3106,op:AddRsp,result:Ok\} \\
%\hline
%    \end{tabular}
%    \end{center}
%    \caption{Interaction Library Example}
%    \label{chap7tab:interactionlibraryexample}
%    \end{table}

\begin{table}[ht!]
\begin{center}
\begin{tabular}{|c||m{4.8cm}|m{7cm}|}
\hline
Index & Request & Response \\ \hline\hline
1 & \{id:1,op:S,sn:Du\} & \{id:1,op:SearchRsp,result:Ok, gn:Miao,sn:Du,mobile:5362634\} \\ \hline
13 & \{id:13,op:S,sn:Versteeg\} & \{id:13,op:SearchRsp,result:Ok, gn:Steve,sn:Versteeg,mobile:9374723\} \\ \hline
275& \{id:275,op:S,sn:Han\} & \{id:275,op:SearchRsp,result:Ok, gn:Jun,sn:Han,mobile:33333333\} \\ \hline
490 & \{id:490,op:S,sn:Grundy\} &
\{id:490,op:SearchRsp,result:Ok, gn:John,sn:Grundy,mobile:44444444\} \\ \hline
2273 & \{id:2273,op:S,sn:Schneider\} & \{id:2273,op:SearchRsp,result:Ok, sn:Schneider,mobile:123456\} \\ \hline\hline
\parbox[p]{2cm}{\textbf{Cluster\\Centre}}& \{id:275,op:S,sn:Han\} & \{id:275,op:SearchRsp,result:Ok, gn:Jun,sn:Han,mobile:33333333\} \\
\hline
\end{tabular}
\end{center}
\caption{Cluster 1 (Search Cluster)}
\label{chap7tab:searchcluster}
\end{table}

\begin{table}[ht!]
\begin{center}
\begin{tabular}{|c||m{6.2cm}|m{5.5cm}|}
\hline
Index & Request & Response \\ \hline\hline
24 & \{id:24,op:A,sn:Schneider,mobile:123456\} & \{id:24,op:AddRsp,result:Ok\} \\ \hline
2487 & \{id:2487,op:A,sn:Will\} & \{id:2487,op:AddRsp,result:Ok\} \\ \hline
3106 & \{id:1106,op:A,sn:Hine,gn:Cam, postalCode:33589\} & \{Id:1106,Msg:AddRsp,result:Ok\} \\ \hline\hline
\parbox[p]{2cm}{\textbf{Cluster\\Centre}} & \{id:24,op:A,sn:Schneider,mobile:123456\} & \{id:24,op:AddRsp,result:Ok\} \\ \hline
\end{tabular}
\end{center}
\caption{Cluster 2 (Add Cluster)}
\label{chap7tab:addcluster}
\end{table}

Given a trace library ({\cf} Definition 5 in Chapter~\ref{chap4sec:preliminaries}), shown as Table~\ref{chap6tab:interactionlibraryexample}, a data clustering method (like the one used in Chapter~\ref{chap6:SoftMine}) is used to group interactions into clusters of the same operation type, which are the \textit{Search Cluster} in Table~\ref{chap7tab:searchcluster} and the \textit{Add Cluster} in Table~\ref{chap7tab:addcluster}, respectively. Then, a recorded interaction{\footnote{The request message of this selected interaction is the closest neighbour to request messages of all the other interactions in this cluster.}} is selected from each cluster to be the \textit{cluster centre} ({\cf} Definition 7 in Chapter~\ref{chap4sec:analysis}), while the request message of this interaction is the \textit{cluster prototype} ({\cf} Definition 8 in Chapter~\ref{chap4sec:analysis}) for this cluster, seen in Table~\ref{chap7tab:searchcluster} and Table~\ref{chap7tab:addcluster}.

Now considering the following incoming add request:

    \hspace{1cm} \textbf{Incoming Request:}

    \hspace{2cm}\{id:37,op:A,sn:Durand\}

By calculating the matching distance between this incoming request against all consensus prototypes, this incoming request is much closer to the cluster prototype of the cluster 1 (search cluster) as they have similar lengths. It is, therefore, assigned to the wrong cluster. %, our clustered interaction library technique matches it with the cluster of %different operation type.
The translation function operates on the centroid interaction of the search cluster as follows. Two symmetric field are identified, which are highlighted in red and blue, respectively.

\begin{tabular}{ll}
\textit{${\Request}_{centre}$:} & \texttt{\footnotesize{\boxed{\color{red}{\{id:275,op:S}}\boxed{\color{blue}{,sn:Han}}\color{black}{\}}}} \\
\textit{${\Response}_{centre}$:} & \texttt{\footnotesize{\boxed{\color{red}{\{id:275,op:S}}earchRsp,result:Ok,gn:Jun\boxed{\color{blue}{,sn:Han}}\color{black}{,mobile:33333333\}}}} \\
\end{tabular}

After aligning the incoming request with the request of the centroid interaction, we can locate symmetric fields in the incoming request as follows.

\begin{tabular}{ll}
\textit{Incoming request:} & \texttt{\footnotesize{\color{red}{\{id:275,op:S}\color{blue}{,sn:{\agap}{\agap}Han{\agap}}\color{black}{\}}}} \\
\textit{${\Request}_{centre}$:} & \texttt{\footnotesize{\color{red}{\{id:37{\agap},op:A}\color{blue}{,sn:Durand}\color{black}{\}}}} \\
\end{tabular}

Finally, the translation function generates a response by modifying symmetric field information in the recorded response with the new information in the new request.

\hspace{0.25cm}
\textit{Generated response:}

\hspace{0.25cm}
\texttt{\footnotesize{\boxed{\color{red}{\{id:37,op:A}}\color{black}{earchRsp,result:Ok,gn:Jun}\boxed{\color{blue}{,sn:Durand}}\color{black}{, mobile:33333333\}}}}

However, the generated response is invalid because its operation type field value does not conform to the protocol specification (\textit{op:AerchRsp} versus \textit{op:AddRsp}). To address this problem, we need a better consensus prototype inference approach in order to summarise common features of requests in a cluster. Moreover, we also need a better matching approach to match an incoming request with a consensus prototype with the right operation type. %is not consistent with that in the incoming request ()hence not conformant to the protocol specification. The generated generated response and the expected one will be as follows:
%
%    \hspace{0.5cm} \textbf{Generated Response:}
%
%    \hspace{1.5cm} { \{Id:8,Msg:AearchRsp,Result:Ok,Firstname:Miao,Lastname:    \hspace{1.5cm}Durand,Telephone:12345678\}}
%
%    \hspace{0.5cm} \textbf{Expected Response:}
%
%    \hspace{1.6cm}\{Id:8,Msg:AddRsp,Result:Ok\}
%

%

\section{Improved Approach}
\label{sec:approach}

\begin{figure*}[ht]
     \centering
     \includegraphics[width=\textwidth]{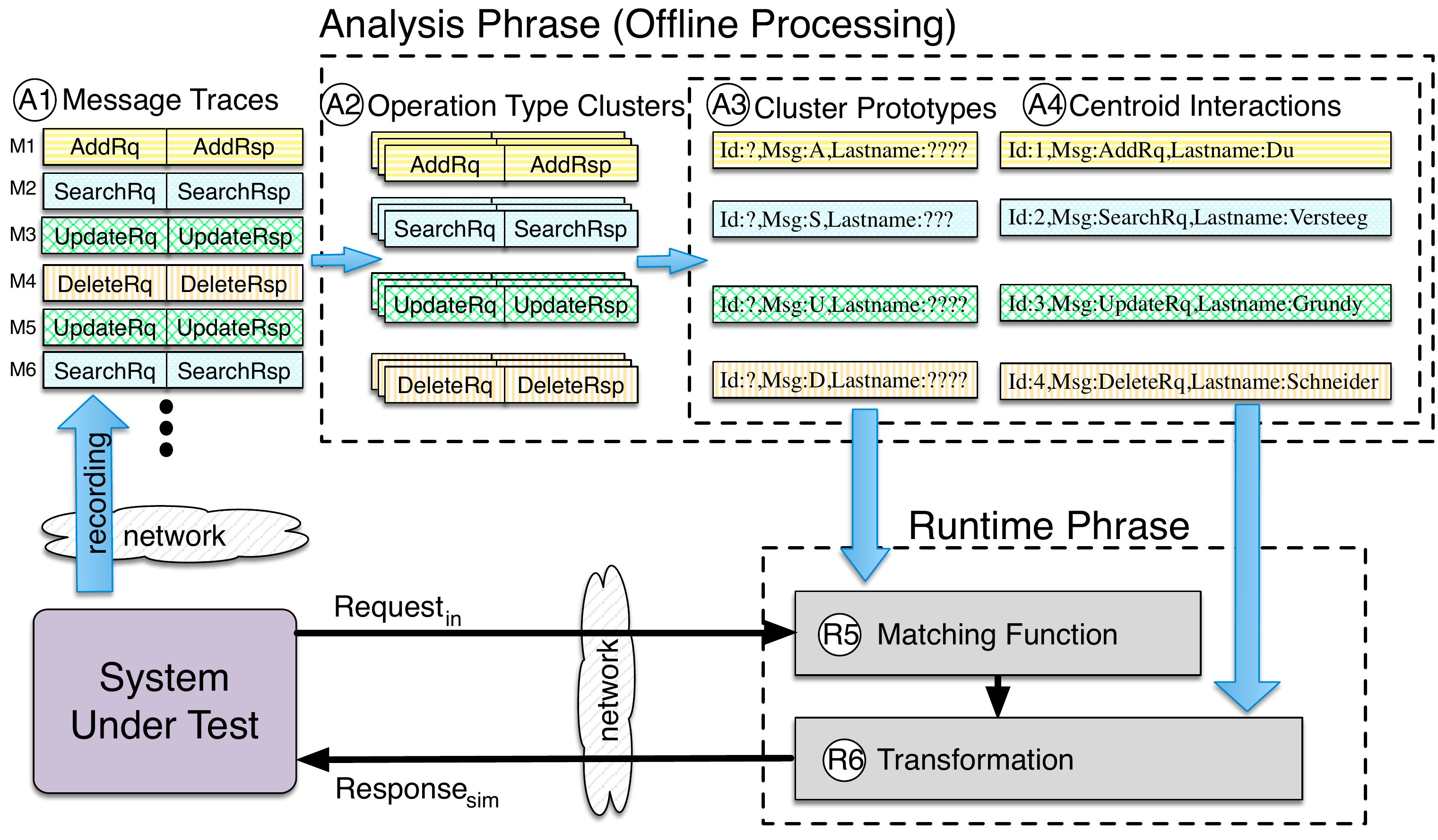}
     \caption{System overview}
     \label{fig:system_overview}
\end{figure*}

Our goal is to produce a system which can produce a model of a virtual service
which can be invoked by a system under test from message trace recordings.
Our improved approach is as follows:

\begin{enumerate}
\item We collect message traces of communications between a client and the real target service and store them in a {\em trace library} with each {\em interaction} being a request-response pair.
\item We then cluster the trace library, with the goal of grouping interactions by operation type. We do not consider the state of service under which the requests are issued. (This may give lower accuracy but is still useful in many cases as discussed in Section~\ref{chap7sec:evaluation})
\item We then derive a \emph{request consensus prototype} (a message pattern) for each operation type cluster by:
\begin{enumerate}
    \item Aligning all request messages of each cluster to reveal their common features for each operation type by inserting gap characters;
	\item Extracting a cluster prototype ({\cf} Definition 8) from the aligned request messages by selecting or deleting a character at each character position;
    \item Calculating positional weightings to prioritise different sections of each cluster prototype.
\end{enumerate}
\item At runtime, for an incoming request message, the {\em matching function} ({\cf} Section~\ref{chap4sec:distance} in Chapter~\ref{chap4:framework}) adopts a \emph{runtime matching distance calculation} technique to select the nearest matching cluster prototype.
\item At the final step, the {\em {translation function}} ({\cf} Section~\ref{chap4sec:translation} in Chapter~\ref{chap4:framework}) is performed on a specifically chosen response message from the identified operation type cluster to generate a response message to be sent back to the enterprise system under test.
\end{enumerate}

Steps 1 to 3 are performed offline to prepare the consensus sequence prototypes.
Steps 4 and 5 are performed at runtime to generate live responses sent to the
system under test.
A schematic overview is depicted in figure~\ref{fig:system_overview}. Each step with technical details is illustrated as follows.

%\subsection{Preliminaries}
%
%The aim of the approach is to work for any protocol, irrespective of
%whether it is textual or binary.
%Our approach can be applied to any protoco
%bytes characters used interchangeably.

% could reuse qosa definition and then extend
\subsection{Collecting and Clustering Interaction Messages (Step 1 and 2)}
\label{sec:clustering}

The first step is to record real message exchanges between a client (such as a previous version of the system under test) and the service we aim to emulate. An interaction is a request-response pair of the communications, where the service may respond with zero or more response messages to a request message from the client. We record requests and responses at the network level (using a tool such as Wireshark), recording the bytes in TCP packet payloads. This makes no assumptions about the message format of the service application. Table~\ref{chap6tab:interactionlibraryexample} shows a sample trace library. This library is from a fictional
protocol, but its key properties similar to LDAP, including that the operation type is a single character.

Having recorded our trace library, the next step is to group the transactions by operation type without assuming any knowledge of the message format. To achieve this we use a distance function-based clustering technique. In Chapter~\ref{chap6:SoftMine} we discussed multiple cluster distance functions and found that the response similarity (as measured by edit distance, calculated using the Needleman-Wunsch algorithm ({\cf} Section~\ref{chap5sec:needlemanwunsch} in Chapter~\ref{chap5:qosa}) was the most effective method for grouping interactions of the same operation type. We therefore use the same technique in this work, grouping interactions by the similarity of their response messages. The clustering algorithm used was VAT~\cite{VAT}, one among many alternatives, as we found it to be effective in our previous work introduced in Chapter~\ref{chap6:SoftMine}.
Applying our clustering technique to the example trace library yields two clusters, as shown in Table~\ref{chap7tab:searchcluster} and Table~\ref{chap7tab:addcluster}, corresponding to search operations and add operations, respectively.

\subsection{Deriving a request consensus prototype for each cluster (Step 3)}
\label{chap7:prototypederivation}
Once clustering interaction messages has been done, the next step is to formulate a representative cluster prototype for each cluster. We first align the request messages in a cluster to determine the common features of these request messages. Then we extract the common features while accommodating variations.

\subsubsection{Aligning Request Message (Step 3a)}
\label{chap7subsec:alignrequestmsg}

In aligning the requests of a cluster, we considered two alternative approaches from other work in pattern matching sequences of data. One is the $n-gram$ approach~\cite{gu2008ngram} and the other is \textit{multiple sequence alignment} (MSA) approach~\cite{wang2012semantics}.  %to derive a consensus message.
 %what the n-gram is and reference{ngram:wang2012semantics}
 A $n-gram$ is a contiguous sequence of n items from a given sequence message $m$ of text or binary. Given a set of text or binary messages, the $n-gram$ approach is way of extracting a number of n-grams by moving a sliding window of length $n$ over every message. At each position, a substring of length $n$ is considered and the frequency of its occurrence is counted. The $n-gram$ based technique is able to exhibit a frequency distribution of the set of n-grams, which can be used to infer commonly occurring patterns in both text and binary protocols. The multiple sequence alignment is a technique to align three or more biological sequences in order to revealing their structural commonalities~\cite{biologicalsequenceanalysis}.% In a multiple sequence alignment, homologous residues among a set of sequences are aligned together in columns, which can reveal structural commonalities of sequences. Therefore, multiple sequence alignment algorithms are widely used for reverse engineering the message %structure~\cite{Comparetti:2009}\cite{PI:2004}.

 %explain why we use msa to reveal commonalities of interaction messages
 After considering both approaches, we decided to adopt the multiple sequence alignment approach in our work. This is because: 1) for some binary protocols, {\eg} LDAP, the message type field is only 1 byte in length. If we use n-grams to identify this field, we have to set n to be 1, which is a degenerate case equivalent to just counting byte frequencies. In contrast, the multiple sequence alignment approach is able to indicate a 1-byte critical field if the operation type section of all messages is correctly aligned; 2) n-gram requires {\em n} to be specified while the multiple sequence alignment approach does not require a predefined threshold; 3) For a new request sequence, multiple sequence alignment can be used as basis to detect common features between it and existing groups of sequences.

 %dynamic programming
 There are several techniques that have been proposed to produce a multiple sequence alignment. A direct method uses a dynamic programming technique to identify the globally optimal alignment solution. Essentially, this method extends the basic Needleman-Wunsch algorithm~\cite{needleman:1970} to handle multiple dimensions by lining up $N$ sequences along $N$ 1-dimensional edges of an $N$-dimensional hypercube. Assuming $L$ represents the length of the final sequence after alignment, the time complexity of scoring function calculation is $O(2^NL^N)$ and the memory complexity is $O(L^N)$~\cite{biologicalsequenceanalysis}. Computing an optimal multiple sequence alignment is a NP complete problem that can only be solved for a small number of sequences~\cite{MSASurvey}. Since the dynamic programming technique is too expensive in both time and space, it is rarely used for aligning more than 3 or 4 sequences. The practical algorithms use heuristics rather than global optimisation.

 %progressive alignment
 %The most popular approach to multiple sequence alignment is the
 {\em Progressive alignment} is the most widely used heuristic-based approach. The basic idea behind a progressive alignment is to align the two most close sequences first and then progressively align the next most close related sequences until all the sequences are aligned. Examples of the progressive alignment method include: ClustalW~\cite{clustalw}, T-coffee~\cite{tcoffee} and Probcons~\cite{probcons}. %  the two It works by constructing a succession of pairwise alignments iteratively. It is a heuristic as 1) it does not separate the process of scoring an alignment from the optimisation algorithms and 2) it does not directly optimise any global scoring function of alignment correctness. It is, therefore, more fast and efficient. The most important heuristic of progressive alignment is to align the most similar pairs of sequences first as these are the most reliable alignments. %Initially, two sequences are chosen and aligned by standard pairwise alignment. Then, a third sequence is chosen and aligned to the previous alignment, and this process is repeated until all sequences have been aligned.  which requires two stages: . There are plenty of algorithms for progressive alignment.
% Procedure of the progressive alignment requires two stages: 1) at the first stage, similarities between the sequences are calculated and are organised into a tree, called a {\it guide tree}; 2) at the second stage, a multiple sequence alignment (MSA) is built by first aligning sequence messages that are the most similar to each other and then adding the messages sequentially to the growing MSA according to the {\it guide tree}.
%
%
% A number of authors propose methods for multiple sequence alignment, using sequence-sequence, sequence-profile, and profile-profile alignment. A profile is a table that lists the frequencies of each character at each position of the group's sequence alignment, which can reveal position-specific information which are somehow typical for this group. The idea behind profile alignment methods is to penalize mismatches more strongly at highly conserved positions than in variable positions. They are, therefore, able to produce a high quality alignment while introducing the least number of gaps.
In our work, we apply the most popular progressive alignment algorithm, $ClustalW$~\cite{clustalw}, to perform the multiple sequence alignment. ClustalW incorporates a novel position-specific scoring scheme and a weighting scheme for downweighting overrespresented sequence groups, with the ``W'' representing ``weights''. An overview of the $ClustalW$ algorithm is as follows:

 \begin{enumerate}
   \item The Needleman-Wunsch algorithm (Section.~\ref{chap5sec:needlemanwunsch}) is used to compute pairwise scores for all pairs of sequence messages. Then, a $N \times N$ distance matrix is built for describing the $N(N-1)/2$ distance calculation results, where $N$ is the number of sequences. Table~\ref{chap7tab:distancematrix} shows a distance matrix of search request messages within the search request cluster in Table~\ref{chap7tab:searchcluster}.

       \begin{table}[ht!]
        \begin{center}
        \begin{tabular}{|p{1.5cm}|p{1.5cm}|p{1.5cm}|p{1.5cm}|p{1.5cm}|p{1.5cm}|}
        \hline
         & $S_1$ & $S_2$ & $S_3$ & $S_4$ &$S_5$ \\ \hline
        $S_1$ & 0.0 & 0.375 & 0.300 & 0.348 & 0.481 \\ \hline
        $S_2$ & & 0.0 & 0.437 & 0.396 & 0.389 \\ \hline
        $S_3$ & & & 0.0 & 0.348 & 0.370 \\ \hline
        $S_4$ & & & & 0.0 & 0.407 \\ \hline
        $S_5$ & & & & & 0.0 \\ \hline
        \end{tabular}
        \end{center}
        \caption{Distance Matrix}
        \label{chap7tab:distancematrix}
        \end{table}

   \item A {\it guide tree} is constructed from the distance matrix by applying a neighbour-joining clustering algorithm~\cite{neighborjoining}, which ensures similar sequences are closer in the tree. Figure~\ref{chap7fig:guidTree} shows a guide tree for the above request messages.

       \begin{figure}[ht]
        \centering
         \includegraphics[width=0.6\textwidth]{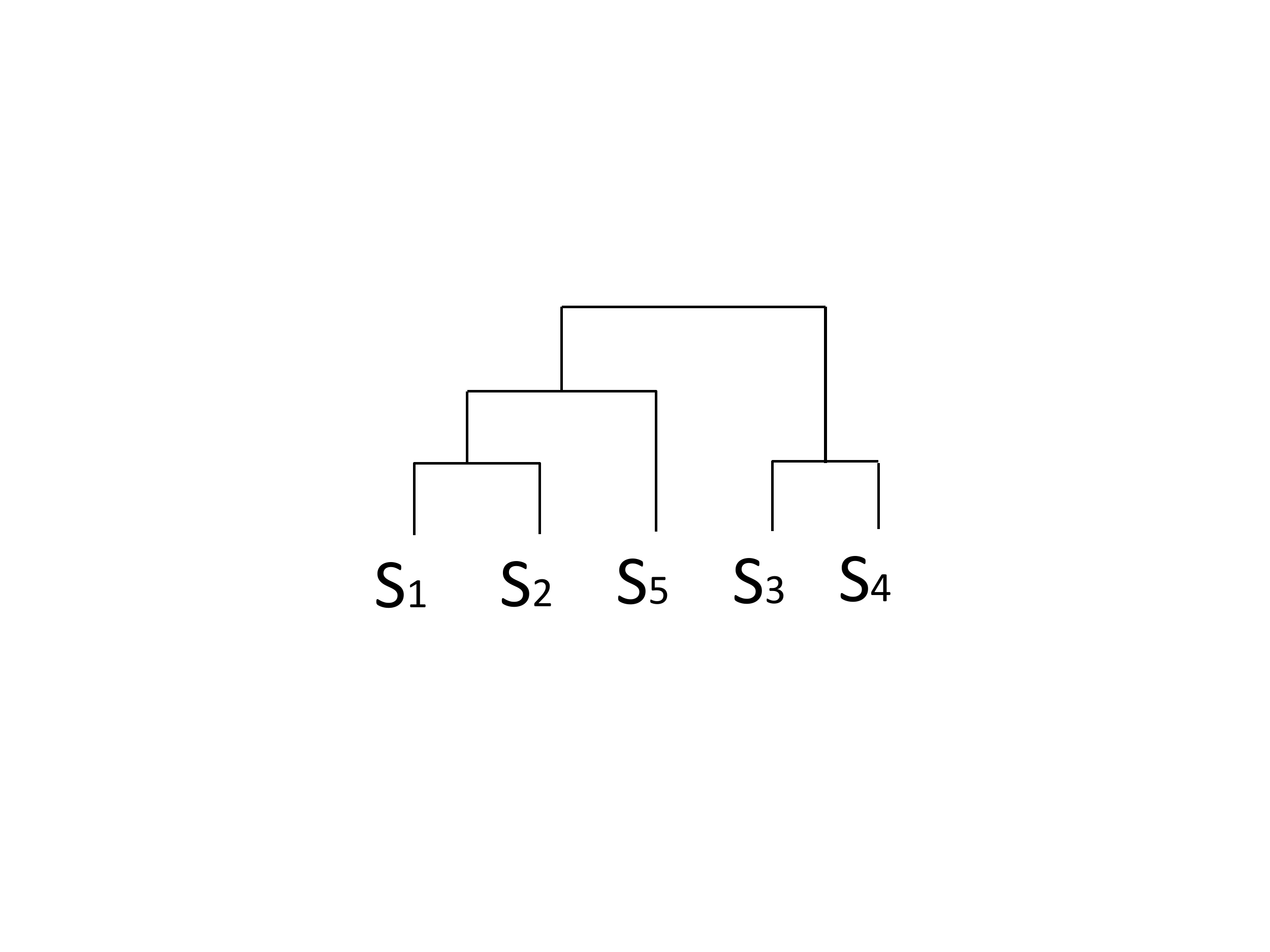}
        \caption{Guide Tree}
        \label{chap7fig:guidTree}
       \end{figure}

   \item The {\it guide tree} is used to guide a progressive alignment of sequences from the leaves to the root of the tree. Aligning $S_1$ and $S_2$, we get

	\hspace{1cm}$S_1$=\texttt{	\{id:1{\agap},op:S,sn:{\agap}{\agap}{\agap}{\agap}{\agap}{\agap}Du\}}

    \hspace{1cm}$S_2$=\texttt{
	\{id:13,op:S,sn:Versteeg\}}

    Aligning $S_3$ and $S_4$, we get

    \hspace{1cm}$S_3$=\texttt{ \{id:275,op:S,sn:{\agap}Han{\agap}{\agap}\}}

    \hspace{1cm}$S_4$=\texttt{
	\{id:490,op:S,sn:Grundy\}}

    Aligning ($S_1$, $S_2$) with $S_5$, we get

    \hspace{1cm}$S_1$=\texttt{	\{id:1{\agap},op:S,sn:{\agap}{\agap}{\agap}{\agap}{\agap}{\agap}Du\}}

    \hspace{1cm}$S_2$=\texttt{
	\{id:13,op:S,sn:Versteeg\}
	}

    \hspace{1cm}$S_5$=\texttt{
    \{id:2273,op:S,sn:Schneider\}
    }

    Aligning ($S_1$, $S_2$, $S_5$) with ($S_3$, $S_4$), we get

    \hspace{1cm}$S_1$=\texttt{\{id:\agap{\agap}1{\agap},op:S,sn:{\agap}{\agap}{\agap}{\agap}{\agap}{\agap}{\agap}Du\}}

    \hspace{1cm}$S_2$=\texttt{\{id:{\agap}{\agap}13,op:S,sn:{\agap}Versteeg\}}

    \hspace{1cm}$S_5$=\texttt{\{id:2273,op:S,sn:Schneider\}}

    \hspace{1cm}$S_3$=\texttt{\{id:275{\agap},op:S,sn:{\agap}Han{\agap}{\agap}{\agap}{\agap}{\agap}\}}

    \hspace{1cm}$S_4$=\texttt{\{id:490{\agap},op:S,sn:Grundy{\agap}{\agap}{\agap}\}}
	%\{id:13,op:S,sn:Versteeg\}

 \end{enumerate}
 %This algorithm can be illustrated as the following example.

    Figure~\ref{chap7fig:alignment} shows the multiple sequence alignment results of applying the ClustalW algorithm to the example clusters from tables~\ref{chap7tab:searchcluster} and \ref{chap7tab:addcluster}. The MSA results are known as profiles. Gaps which were inserted during the alignment process are denoted with the ‘{\agap}’ symbol. Note that the common sequences for the requests in each cluster have now been aligned.

    \newcommand{\alignedSrqs}{
	\begin{minipage}[c]{\linewidth}
	\centering \texttt{\{id:\agap{\agap}1{\agap},op:S,sn:{\agap}{\agap}{\agap}{\agap}{\agap}{\agap}{\agap}Du\} \newline
	\{id:{\agap}{\agap}13,op:S,sn:{\agap}Versteeg\}\newline
	\{id:2273,op:S,sn:Schneider\} \newline
	\{id:275{\agap},op:S,sn:{\agap}Han{\agap}{\agap}{\agap}{\agap}{\agap}\} \newline
	\{id:490{\agap},op:S,sn:Grundy{\agap}{\agap}{\agap}\} \newline
	}
\end{minipage}
}

\newcommand{\alignedArqs}{
	\begin{minipage}[c]{\linewidth}
\centering
\texttt{\{id:24{\agap}{\agap},op:A,sn:Schne{\agap}{\agap}{\agap}{\agap}{\agap}ider{\agap}{\agap},mo{\agap}bil{\agap}{\agap}{\agap}e:123456\} \newline \{id:2487,op:A,sn:W{\agap}{\agap}{\agap}{\agap}{\agap}{\agap}{\agap}{\agap}{\agap}{\agap}{\agap}{\agap}{\agap}{\agap}{\agap}{\agap}{\agap}{\agap}{\agap}{\agap}il{\agap}{\agap}{\agap}l{\agap}{\agap}{\agap}{\agap}{\agap}{\agap}{\agap}\}
\newline	\{id:3106,op:A,sn:Hi{\agap}ne,gn:Cameron,postalCode:3{\agap}3589\}
\newline
	}
	\end{minipage}
}

\begin{figure}[ht]%
	\subfloat[Cluster 1 alignment \label{chap7fig:reqalignment}]{\alignedSrqs}%
     \qquad \newline
	%\hline
	\subfloat[Cluster 2 alignment \label{chap7fig:reqalignment}]{\alignedArqs}
\caption{MSA results of the requests in tables \ref{chap7tab:searchcluster} and \ref{chap7tab:addcluster}}%
\label{chap7fig:alignment}%
\end{figure}

\subsubsection{Formulating the Request Consensus Prototype (Step 3b)}
\label{subsec:consensusseqcal}

Having derived the MSA profile for the request messages of each cluster, the next step is to extract the common features from the MSA profile into a single character sequence, which we call the {\em request consensus prototype}, to facilitate efficient runtime comparison with an incoming request message from the system under test.

\begin{figure}[ht]
     \centering
     \includegraphics[width=\textwidth]{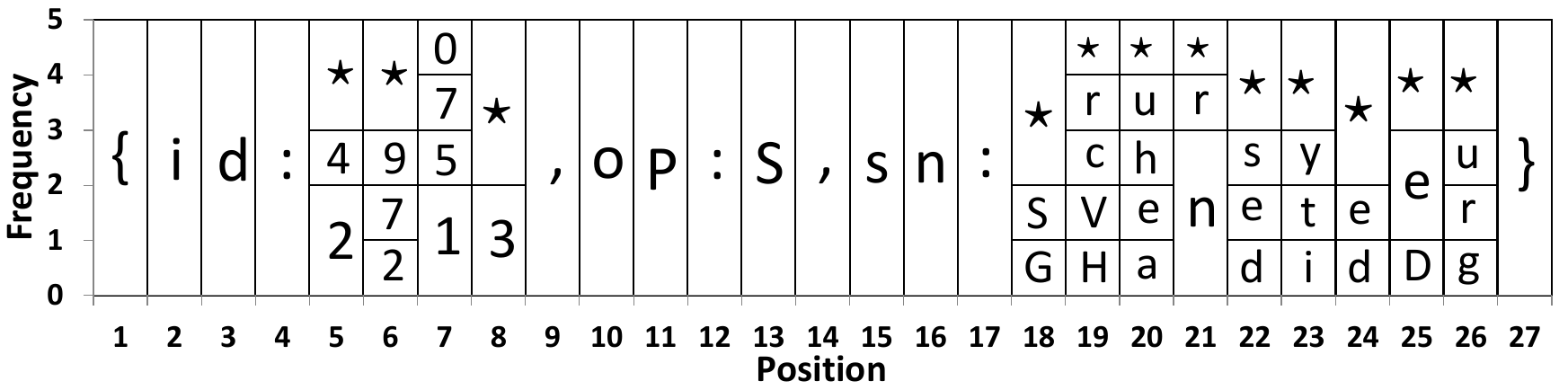}
     \caption{Character Frequencies in the Alignment Result of Search Requests.}
     \label{chap7tab:charfreq}
\end{figure}

From all the aligned request messages in a cluster, we derive a message character occurrence count table. Having all messages in a cluster aligned, we can derive a byte/character occurrence count table from the alignment result. Figure~\ref{chap7tab:charfreq} graphically depicts byte frequencies at each position for the example alignment in Figure~\ref{chap7fig:alignment}. Each column represents a position in the alignment result. The frequencies of the different bytes which occur at each position are displayed as a stacked bar graph.

Based on the byte occurrence table, we formulate the \emph{consensus request message prototype} by extending the concept of a \emph{consensus sequence}~\cite{ConsensusComparison} commonly used in summarising a MSA profile.  A consensus sequence can be viewed as a sequence of \emph{consensus symbols}, where the consensus symbol $c_i$ is the most commonly occurring byte at the position $i$. In our extension, a consensus request message prototype, $\mathbf{p}$, is calculated by iterating each byte position of the MSA profile, to calculate a \emph{prototype symbol}, $p_i$, at each position, according to equation~\ref{eq:consensus}.

\begin{equation}
\label{eq:consensus}
    p_i = \left\{ \begin{array}{rl}
                 c_i &\mbox{ if $q(c_i) \geq f \land c_i \neq \agap$} \\
                 \perp &\mbox{ if $q(c_i) \geq \frac{1}{2} \land c_i = \agap$} \\
                 {\wildcard} &\mbox{ otherwise }
                 %{\wildcard} &\mbox{ if $q(c_i) < f \land \lnot (c_i = \gap \land q(c_i) \geq \frac{1}{2}) $ }
    \end{array}
        \right.
\end{equation}

Where $q(c_i)$ denotes the relative frequency at position $i$ of the consensus symbol $c_i$, $f$ is the relative frequency threshold, `$\gap$' denotes a gap, `$\wildcard$' is the `wildcard' symbol and `$\perp$' represents a truncation. After calculating the prototype symbol for each position, any truncation symbols are then deleted from the consensus request message prototype.

Introducing wildcards and truncations into the message prototypes allows us to distinguish between gaps and where there is no consensus.  If the relative frequency $q(c_i)$ is at or above the threshold $f$, then we insert the consensus symbol into our prototype (unless the consensus symbol is a gap.)  If the relative frequency is below threshold, then we insert a wildcard.  If the consensus symbol is a gap and it is in the majority, then we leave that position as empty (i.e., deleted).  Wildcards allow us to encode where there are high variability sections of the message.  Without truncations, consensus sequences become very long as there tend to be many gaps.  By truncating the gaps, the lengths of the prototypes become similar to the typical lengths of messages in the cluster. The consensus prototype for a cluster of request messages can differentiate stable positions from variant positions. Moreover, it can identify consensus symbols that can be utilised for matching.

Applying our consensus request message prototype method, using a frequency threshold $f=0.8$, to the example clusters in tables \ref{chap7tab:searchcluster} and \ref{chap7tab:addcluster} yields the following results:

    \hspace{-0.2cm} \textbf{Consensus prototype for the search request cluster:}

    \hspace{0.2cm}\texttt{\footnotesize{\{id:${\wildcard}{\wildcard}{\wildcard}$,op:S,sn:${\wildcard}{\wildcard}{\wildcard}{\wildcard}{\wildcard}{\wildcard}{\wildcard}$\}}}

    \hspace{-0.2cm} \textbf{Consensus prototype for the add request cluster:}

    \hspace{0.2cm}\texttt{\footnotesize{\{id:${\wildcard}{\wildcard}{\wildcard}{\wildcard}$,op:A,sn:${\wildcard}{\wildcard}{\wildcard}{\wildcard}{\wildcard}{\wildcard}{\wildcard}{\wildcard}{\wildcard}{\wildcard}{\wildcard}{\wildcard}{\wildcard}{\wildcard}$l${\wildcard}{\wildcard}{\wildcard}{\wildcard}{\wildcard}{\wildcard}{\wildcard}$\}}}

Note that the add prototype contains an `l' from coincidentally aligning `l's from `mobile', `Will' and `postalCode'.

\subsubsection{Deriving Entropy-Based Positional Weightings (Step 3c)}
\label{ss:weightings}

The final step of the offline analysis is to calculate weightings for each byte position
in the consensus prototype to support the runtime distance matching process for request messages.
In our service emulator, generating a response message of the correct operation type is more critical
than the contents of the message payload. Thus we give a higher weighting to
the sections of the message that likely relate to the operation type.
To do this we make use of the observation that structure information (such as operation type)
is more stable than payload information.  We use entropy as a measurement of variability,
and use it as the basis to calculate a weighting for each byte position of the
consensus prototype.

Using the MSA profile for a cluster (from section~\ref{chap7subsec:alignrequestmsg}), we calculate the entropy
for each column using the Shannon Index \cite{shannon:1948}, as given by equation~\ref{eq:shannon}.
\begin{equation}
\label{eq:shannon}
E_i = -\sum^{R}_{j=1} q_{ij} \log q_{ij}
\end{equation}
Where $E_i$ is the Shannon Index for the $i$th column, $q_{ij}$ is the relative frequency of the $j$th character in the character set at column $i$ and $R$ is the total number of characters in the character set.

Since we wish to give a high weighting to stable parts of the message,
for each column we invert the entropy by applying a scaling function of the form given
in equation~\ref{eq:scaling}
\begin{equation}
\label{eq:scaling}
w_i = \frac{1}{(1 + bE_i)^c}
\end{equation}
Where $w_i$ is the weighting for the $i$th column, $E_i$ is the entropy of the $i$th column, and $b$ and $c$ are positive constants.  The higher the values of $b$ and $c$, the higher entropy columns are deweighted.  In our experiments we found the best results were obtained with $b=1$ and $c=10$.  This allows structural information to strongly dominate in the matching process.  Payload similarity is only used as a `tie-breaker'.

Columns that correspond to gaps removed from the consensus prototype are also dropped
from the weightings array.
Table~\ref{tab:weighting} gives an example weightings array for the search request consensus prototype.

\newcolumntype{C}[1]{>{\centering}m{#1}}
\newcolumntype{L}[1]{>{\raggedleft}m{#1}}

\begin{table}
\centering
%\begin{tabular}{|l|c|c|c|c|C{0.25cm}|C{0.25cm}|C{0.4cm}|c|c|c|c|c|c|c|c|c|c|c|C{0.4cm}|C{0.4cm}|C{0.4cm}|C{0.4cm}|C{0.4cm}|c|}
\begin{tabular}{|l|c|c|c|c|c|c|c|c|c|c|c|c|}
\hline
Idx. & 0 & 1 & 2 & 3 & 4 & 5 & 6 & 7 & 8 & 9 & 10 & 11  \\ \hline
$\mathbf{p}$ & \{ & i & d & : & ${\wildcard}$ & $\wildcard$ & $\wildcard$ & , & o & p & : & S \\
\hline
$\mathbf{E}$ & 0  & 0 & 0 & 0 & 1.05           & 1.33            & 1.33            & 0 & 0 & 0 & 0 & 0  \\ \hline
$\mathbf{w}$ & 1  & 1 & 1 & 1 &$\frac{1}{1342}$& $\frac{1}{4760}$&$\frac{1}{4760}$ & 1 & 1 & 1 & 1 & 1  \\[0.12cm]
\hline \hline
Idx. & 13 & 14 & 15 & 16 & 17 & 18 & 19 & 20 & 21 & 22 & 23 & 24 \\ \hline
$\mathbf{p}$ & , & s & n & : & $\wildcard$ & $\wildcard$ & $\wildcard$ & $\wildcard$ & $\wildcard$ & $\wildcard$ & $\wildcard$ & \} \\
\hline
$\mathbf{E}$ & 0 & 0 & 0 & 0 & 1.61             & 1.61             & 0.95           & 1.33            & 1.33            & 1.05            & 1.33            & 0 \\ \hline
$\mathbf{w}$ & 1 & 1 & 1 & 1 &$\frac{1}{14638}$ &$\frac{1}{14638}$ &$\frac{1}{796}$ &$\frac{1}{4760}$ &$\frac{1}{4760}$ &$\frac{1}{1342}$ &$\frac{1}{4760}$ & 1 \\[0.12cm]
\hline
\end{tabular}
\caption{Example weightings for search consensus prototype calculated from the MSA profile in Figure~\ref{chap7tab:charfreq} (using equation~\ref{eq:scaling} constants $b = 1, c = 10$.)}
\label{tab:weighting}
\end{table}

\subsection{Entropy Weighted Runtime Matching for Request Messages (Step 4)}
\label{ss:runtimematching}

At runtime, the formulated consensus request message prototypes are used to match
incoming requests from the system under test.
We extend the Needleman-Wunsch algorithm to calculate the
matching distance between an incoming request and the consensus request prototype for each operation type (or cluster).
The modifications include matching the wildcards and applying the entropy weights calculated
as described in Section~\ref{ss:weightings}.

When a character from the incoming request
is aligned with a wildcard character in the prototype,  the alignment is given a score that is different to an identical match or a non-match (\ie a difference), as given in equation~\ref{eq:wildcardnw}.

\begin{equation}
\label{eq:wildcardnw}
S({p}_i,{r}_j) = \left\{ \begin{array}{rl}
		 w_i M &\mbox{ if ${p}_i = {r}_j \land {p}_i \neq \wildcard $} \\
		 w_i D &\mbox{ if ${p}_i \neq {r}_j \land {p}_i \neq \wildcard $} \\
		 w_i X &\mbox{ if ${p}_i = \wildcard $}
    \end{array}
	\right.
\end{equation}
Where ${p}_i$ is the $i$th character in the consensus prototype,
${r}_j$ is the $j$th character in the incoming message,
$w_i$ is the weighting for the $i$th column,
$M$ and $D$ are constants denoting the Needleman-Wunsch identical score and difference penalty, respectively,
and $X$ is the wildcard matching constant.

A global alignment between the consensus prototype and incoming request is
made by applying the modified scoring equation~\ref{eq:wildcardnw} to the Needleman-Wunsch
algorithm, giving an absolute alignment score, $s$.

The relative distance, denoted as $d_{\mathrm{rel}}$, is calculated from the absolute alignment score to
normalise for consensus prototypes of different lengths, different entropy weights
and different numbers of wildcards.
The relative distance is in the range 0 to 1, inclusive, where 0 signifies the best
possible match with the consensus, and 1 represents the furthest possible distance.
It is calculated according to equation~\ref{eq:reldist}:

\begin{equation}
\label{eq:reldist}
d_{\mathrm{rel}}(\mathbf{p}, \mathbf{r}) =
	1 - \frac{s(\mathbf{p}, \mathbf{r}) - s_{\mathrm{min}}(\mathbf{p}) }
		     {s_{\mathrm{max}}(\mathbf{p})}
\end{equation}
Where $s_{\mathrm{max}}$ is the maximum possible alignment score for the given consensus prototype and
$s_{\mathrm{min}}$ is the minimum possible alignment score for the given consensus prototype.  These
are calculated in equations \ref{eq:maxscore} and \ref{eq:minscore}, respectively.

\begin{equation}
\label{eq:maxscore}
{s_{\mathrm{max}}(\mathbf{p})} = \sum_{i=0}^{|\mathbf{p}|-1} S(\mathbf{p}_i,\mathbf{p}_i)
\end{equation}

\begin{equation}
\label{eq:minscore}
{s_{\mathrm{min}}(\mathbf{p})} = \sum_{i=0}^{|\mathbf{p}|-1} S(\mathbf{p}_i,\varnothing)
\end{equation}
Where $\varnothing$ is a symbol different to all of the characters in
the consensus prototype.

The consensus prototype that gives the least distance to the incoming message is identified as the matching prototype, therefore identifying the matching transaction cluster.

As an example, suppose we receive an incoming add request with the byte sequence

\texttt{
\{id:37,op:A,sn:Durand\}}.
Comparing the request against the search consensus prototype yields the alignment:

\begin{tabular}{ll}
\emph{request:} & \texttt{\footnotesize{\{id:3*7,op:A,sn:*Durand\}}} \\
\emph{prototype:}        & \texttt{\footnotesize{\{id:???,op:S,sn:???????\}}} \\
\end{tabular}

\smallskip \noindent
Using equation~\ref{eq:reldist}, the weighted relative distance is calculated to be 0.0715.
Comparing against the add consensus prototype produces the alignment below with a relative distance of 0.068.

\begin{tabular}{ll}
\emph{request:} & \texttt{\footnotesize{\{id:37**,op:A,sn:********D*u*r*an*d***\}}} \\
\emph{prototype:}        & \texttt{\footnotesize{\{id:????,op:A,sn:?????????????l???????\}}} \\
\end{tabular}

\smallskip \noindent
Consequently, the add prototype is selected as the matching prototype.

\subsection{Response Transformation through Dynamic Substitution (Step 5)}

The final step in our approach is to send a customised response
for the incoming request by performing some dynamic substitutions on a response message from the matched cluster. Here we
use the symmetric field technique described in Section~\ref{chap5sec:translation} in Chapter~\ref{chap5:qosa}, where
character sequences which occur in both the request and response
messages of the chosen transaction are substituted with the corresponding characters from the live request in the generated response. We take the response from the centroid interaction ({\cf} Section~\ref{chap6subsec:centers} in Chapter~\ref{chap6:SoftMine}) of the selected cluster and apply the symmetric field substitution.

In our example, the centroid interaction from the add cluster is given below. There is one symmetric field (boxed).

\begin{tabular}{ll}
\textit{request:} & \texttt{\footnotesize{\boxed{\texttt{\{id:24,op:A}},sn:Schneider,mobile:123456\}}} \\
\textit{response:} & \texttt{\footnotesize{\boxed{\texttt{\{id:24,op:A}}ddRsp,result:Ok\}}} \\
\end{tabular}

\smallskip \noindent
After performing the symmetric field substitution the final generated response
is:

\texttt{\footnotesize{\{id:37,op:AddRsp,result:Ok\}}}

\section{Evaluation}
\label{chap7sec:evaluation}

In our evaluation, we assess our technique in three aspects: \emph{accuracy}, \emph{efficiency} and \emph{robustness}, which will answer following questions:

\begin{enumerate}
  \item \textbf{Q1} (\textbf{Accuracy}): Having cluster \emph{cluster prototypes} formulated at the preprocessing stage as described in Section.~\ref{chap7subsec:alignrequestmsg} and Section.~\ref{subsec:consensusseqcal}, is our approach able to generate accurate, protocol-conformant responses? At runtime, can \emph{positional weighting based runtime matching} (as discussed in Section.~\ref{ss:weightings} and Section.\ref{ss:runtimematching}) further improve the accuracy of our approach for generating such responses?
  \item \textbf{Q2} (\textbf{Efficiency}): Is our technique efficient enough to generate timely responses, even for large interaction libraries?
  \item \textbf{Q3} (\textbf{Robustness}): Given non-homogeneous clusters, which contain some messages of operation types different to the majority of messages, is our technique robust to generate accurate responses?
\end{enumerate}
To answer these questions, we apply our technique on six case study protocols. In the following, we first describe how we collect data of six case study protocols in Section.~\ref{ss:casestudies}. And then, we introduce how we perform to evaluate our technique in Section.~\ref{ss:crossvalidation}. After that, we present our results in Section.~\ref{ss:results}.

\subsection{Case Study Protocols}
\label{ss:casestudies}

\begin{table}[h]
\centering
\begin{tabular}{|l|c|c|c|c|}
  \hline
  Protocol & Binary/Text & Fields & \#Ops. & \#Transactions \\
  \hline \hline
  IMS & binary & fixed length & 5 & 800 \\ \hline
  LDAP & binary & length-encoded & 10 & 2177 \\ \hline
  LDAP text (1) & text & delimited & 10 & 2177\\ \hline
  LDAP text (2) & text & delimited & 6 & 1000 \\ \hline
  SOAP & text & delimited & 5 & 1000 \\ \hline
  Twitter (REST) & text & delimited & 6 & 1825 \\ \hline
\end{tabular}
\caption{Sample protocol message trace datasets}
\label{tab:sampleset}
\end{table}

We applied our techniques to four real-world protocols,
\emph{IMS}~\cite{imsredbook} (a binary mainframe protocol),
\emph{LDAP}~\cite{Sermersheim2006} (a binary directory service protocol),
\emph{SOAP}~\cite{SOAPv1.1} (a textual protocol, with an Enterprise Resource Planning (ERP) system messaging system services),
and \emph{RESTful Twitter}~\cite{twitter} (a JSON protocol for the Twitter social media service).

We have used one message trace dataset for each of these protocols.  In addition, \emph{LDAP} has two additional datasets: a dataset with textual representation converted from the binary dataset (denoted by
\emph{LDAP text (1)}), and another textual dataset that was used in
our prior work \cite{Du:2013SoftMine} (\emph{LDAP text (2)}).

We chose these protocols because: (i) they are widely used in enterprise
environments; (ii) some of them (LDAP text (2) and SOAP) were used in the
evaluation of our prior work \cite{Du:2013,Du:2013SoftMine}; (iii) they
represent a good mix of text-based protocols (SOAP and RESTful Twitter) and
binary protocols (IMS and LDAP); (iv) they use either fixed length, length encoding or
delimiters to structure protocol messages;\footnote{Given a protocol message,
  length fields or delimiters are used to convert its structure into a
  sequence of bytes that can be transmitted over the wire. Specifically, a
  length field is a number of bytes that show the length of another field,
  while a delimiter is a byte (or byte sequence) with a
  known value that indicates the end of a field.} and (v) each of them includes a diverse
number of operation types, as indicated by the \emph{Ops} column. The number
of request-response interactions for each test case is shown as column
\emph{Transactions} in Table~\ref{tab:sampleset}. Message examples for each test case can be referred to Appendix~\ref{apx.traceexamples}.

\subsection{10-fold Cross-Validation Approach and Evaluation Criteria}
\label{ss:crossvalidation}

%{\em Cross-validation} \cite{devijver:1982} is a popular model validation method for assessing how accurately a predictive model will perform in practice.
For the purpose of our evaluation, we
applied the commonly used 10-fold cross-validation approach ({\cf} Section~\ref{chap5subsec:criteria} in Chapter~\ref{chap5:qosa})
to data sets of all six case study protocols. Having generated a response for each incoming request, we utilised
% the following
five criteria (identical, consistent, protocol-conformant, well-formed and malformed) to determine its accuracy, thereby evaluating the ability of our
approach to generate protocol-conformant responses. To obtain deep insight into generated response, we further divide previous protocol-conformant criterion into the consistent criterion and the protocol-conformant criterion. A generated response is considered to be consistent if it is of the required
operation type and has the critical fields {\footnote{Critical fields are some payload information that are required to occur in both a request message and its corresponding responses. In this example, the \footnotesize{id} field is the critical field.}} in the payload replicated ({\cf} Example (ii) in Table~\ref{tab:criteriaexample} where {\footnotesize{id}} is identical, but some the other payload information differs). A generate response is considered to be protocol conformant if its operation type corresponds to the expected response, but it differs in some payload information, which may include the critical fields ({\cf} Example (iii) in Table~\ref{tab:criteriaexample} where both the {\footnotesize{id}} and {\footnotesize{result}} tags differs).

\subsection{Evaluation Results}
\label{ss:results}

In our prior work, we have proposed the {\em Whole Library} approach in Chapter~\ref{chap5:qosa} and the {\em Cluster Centroid} approach in Chapter~\ref{chap6:SoftMine}. For synthesizing the response(s) for an incoming request, the {\em Whole Library} approach can achieve high accuracy by searching the entire collection of previously recorded interaction traces, while the \textit{Cluster Centroid} approach performs efficiently at runtime by reducing the number of searching to the number of selected interaction representatives. We use these two approaches as baseline to evaluate both effectiveness (Section~\ref{ssb:effectiveness}) and efficiency (Section~\ref{ssb:efficiency}) of the proposed technique for synthesizing responses. Moreover, we present the robustness evaluation results of our approach in Section~\ref{ssb:robustness}.

\subsubsection{\textbf{Effectiveness Evaluation(Q1)}}
\label{ssb:effectiveness}

The accuracy evaluation is conducted to assess the capability of our approach for generating accurate responses. As discussed in Section~\ref{sec:approach}, our approach has two important features aimed at enhancing response accuracy, the \emph{cluster prototypes} combined with \emph{entropy-weighted distance calculation} at runtime. To measure the impact of these techniques we ran two separate sets of experiments, which are referred to as \emph{Consensus Only} and \emph{Consensus+Weighting}, respectively.
In addition, for both sets of experiments we tested for the best pre-defined
frequency threshold $f$, trying three different values.\footnotemark

% to be able to produce better results~.}, which are $0.5$, $0.8$ and $1$.}%A number of investigations have been done toIn \cite{PI:2004}, authors have indicated that the threshold $1$ only allow the most prevalent character to be preserved in consensus sequences which are very likely to lose important information. The experiment of threshold $1$ is hence used as the baseline in our evaluation. The frequency thresholds $0.5$, $0.8$ are empirical, which are found.

Table~\ref{tab:accuracy} summarises the evaluation results of \emph{Consensus
  Only}, \emph{Consensus+Weighting}, and our prior work ({\ie}\emph{Whole
  Library} and \emph{Cluster Centroid}) experiments for the six test
datasets. The \textbf{Accuracy Ratio} column is calculated by dividing the
number of valid generated responses by the total number of interactions
tested. The last five columns give a more detailed breakdown of the different
categories of valid and invalid responses generated.

%%% moved the footnote here (JGS) %%%
\footnotetext{As illustrated in Equation~\ref{eq:consensus}, a pre-defined
  frequency threshold is required to calculate the consensus sequence
  prototype. In bioinformatics, a number of investigations have been done for
  identifying the best threshold \cite{ConsensusComparison}. In our
  experiments, we selected the most popular 3 values, that is, $0.5$, $0.8$
  and $1$.}

Table~\ref{tab:accuracy} shows that the combined {\em Consensus+Weighting} approach achieves the highest accuracy overall for the datasets tested.  The combined approach achieves 100\% accuracy for four of the datasets, and 99.95\% and 99.34\% for the remaining two (LDAP binary and Twitter, respectively).  Twitter is the only case where the {\em Whole Library} approach is marginally better.

With respect to the impact for the frequency threshold $f$, the results show that allowing some tolerance (\ie $f < 1$) in the multiple sequence alignment can yield better results.  For the LDAP (binary) dataset the thresholds of $f=0.5$ and $f=0.8$ produced significantly higher accuracy than $f=1$.  For the other datasets the threshold had no impact on the accuracy.  The general conclusion appears that the results are not very sensitive to the value of the threshold for most scenarios.

An interesting result is that the {\em Consensus+Weighting} approach has a higher accuracy than the \emph{Whole Library} approach, even though the latter uses all the available data points from the trace library (for three datasets {\em Consensus+Weighting} is significantly more accurate, for two it has the same accuracy, for one it is slightly lower). The reason for the higher accuracy is that the {\em Consensus+Weighting} abstracts away the message payload information sections (using wildcards), so is less susceptible to matching a request to the wrong operation type but with the right payload information, whereas the \emph{Whole Library} approach is susceptible to this type of error (note the well-formed but invalid responses for the \emph{Whole Library} approach in Table~\ref{tab:accuracy}).

The impact of the entropy weightings can only be observed for the LDAP binary dataset.  For this test, the weightings significantly improve the accuracy results.  For the other datasets, no impact from the weightings can be observed, as the consensus sequence prototype by itself (\emph{Consensus Only}) already produces 99-100\% accuracy.

\begin{table}
\footnotesize
\centering
%\begin{tabular}{|C{0.4cm}|C{0.4cm}|C{0.4cm}|C{0.4cm}|C{0.4cm}|C{0.4cm}C{0.4cm}C{0.4cm}|C{0.4cm}C{0.4cm}|}
\begin{tabular}{|c|c|c|c|c|ccc|cc|}
  \hline
  % after \\: \hline or \cline{col1-col2} \cline{col3-col4} ...
  \multirow{2}{1cm}{EXP.} & \multicolumn{2}{c|}{\multirow{2}{*}{Method}} & \multirow{2}{*}{AR} & \multirow{2}{*}{No.} & \multicolumn{3}{c|} {Valid} & \multicolumn{2}{c|} {Invalid} \\ \cline{6-10}
&\multicolumn{2}{c|}{}& & &Ident. & Cons. & Conf. & Wf. & Mf. \\ \hline\hline

  \multirow{8}{1cm}{\centering IMS (binary) } & \multicolumn{2}{c|}{Whole Library} & 75.25\% & \multirow{8}{*}{800} &  400 & 202 & 0 & 198 & 0 \\ \cline{2-4} \cline{6-10} &\multicolumn{2}{c|}{Cluster Centroid} & 97.88\% & & 400 & 383 & 0 & 17 & 0 \\ \cline{2-4} \cline{6-10}
    & \multicolumn{1}{c|}{\multirow{3}{1.5cm}{\centering Consensus Only}} & f=0.5 & \textbf{\textit{100\%}}&& 400 & 400 & 0 & 0 & 0 \\ \cline{3-4}\cline{6-10}
  & & f=0.8 & \textbf{\textit{100\%}}&& 400 & 400 & 0 & 0 & 0 \\\cline{3-4} \cline{6-10}
  && f=1 & \textbf{\textit{ 100\%}}&& 400 & 400 & 0 & 0 & 0 \\ \cline{2-4}\cline{6-10}
  & \multicolumn{1}{c|}{\multirow{3}{1.5cm}{\centering Consensus + Weighting}} & f=0.5 & \textbf{\textit{100\%}}&& 400 & 400 & 0 & 0 & 0 \\ \cline{3-4}\cline{6-10}
  & & f=0.8 & \textbf{\textit{100\%}}&& 400 & 400 & 0 & 0 & 0 \\\cline{3-4} \cline{6-10}
  && f=1 & \textbf{\textit{100\%}}&& 400 & 400 & 0 & 0 & 0 \\ \hline

   \multirow{8}{1cm}{\centering LDAP (binary) } & \multicolumn{2}{c|}{Whole Library} & 94.12\% & \multirow{8}{*}{2177}  & 248 & 17 & 1784 & 36 & 92 \\ \cline{2-4} \cline{6-10}
   & \multicolumn{2}{c|}{Cluster Centroid} & 91.59\% && 263 & 17 & 1714 & 183 & 0 \\ \cline{2-4} \cline{6-10}
  & \multicolumn{1}{c|}{\multirow{3}{1.5cm}{\centering Consensus Only}} & f=0.5 & \textit{87.74\%} &&268 & 14 & 1628 &267 & 0 \\ \cline{3-4}\cline{6-10}
  & & f=0.8 & \textit{84.66\%}&& 264 & 14 & 1565 & 334 & 0 \\\cline{3-4} \cline{6-10}
  && f=1 & \textit{79.97\%}&& 259 & 14 & 1468 & 436 & 0 \\\cline{2-4}\cline{6-10}
  & \multicolumn{1}{c|}{\multirow{3}{1.5cm}{\centering Consensus + Weighting}} & f=0.5 & \textit{98.71\%} &&278 & 18 & 1853 &28 & 0 \\ \cline{3-4}\cline{6-10}
  & & f=0.8 & \textbf{\textit{99.95\%}}&& 278 & 18 & 1880 & 1 & 0 \\\cline{3-4} \cline{6-10}
  && f=1 & \textit{86.90\%}&& 267 & 16 & 1609 & 285 & 0 \\ \hline

   \multirow{8}{1cm}{\centering LDAP text (1) (text)} & \multicolumn{2}{c|}{Whole Library} & 100\% & \multirow{8}{*}{2177} &  1648 & 415 & 114 & 0 & 0 \\ \cline{2-4} \cline{6-10}
    & \multicolumn{2}{c|}{Cluster Centroid} & 100\% & &  811 & 1325 & 41 & 0 & 0 \\ \cline{2-4} \cline{6-10}
  & \multicolumn{1}{c|}{\multirow{3}{1.5cm}{\centering Consensus Only}} & f=0.5 & \textbf{\textit{100\%}}&& 1555 & 622 & 0 & 0 & 0 \\ \cline{3-4}\cline{6-10}
  & & f=0.8 & \textbf{\textit{100\%}}&& 1555 & 622 & 0 & 0 & 0 \\\cline{3-4} \cline{6-10}
  && f=1 & \textbf{\textit{100\%}}&& 1527 & 650 & 0 & 0 & 0 \\\cline{2-4} \cline{6-10}
  & \multicolumn{1}{c|}{\multirow{3}{1.5cm}{\centering Consensus + Weighting}} & f=0.5 & \textbf{\textit{100\%}}&& 1559 & 618 & 0 & 0 & 0 \\ \cline{3-4}\cline{6-10}
  & & f=0.8 &\textbf{\textit{100\%}}&& 1559 & 618 & 0 & 0 & 0 \\\cline{3-4} \cline{6-10}
  && f=1 & \textbf{\textit{100\%}}&& 1559 & 618 & 0 & 0 & 0 \\ \hline

   \multirow{8}{1cm}{\centering LDAP text (2) (text)} & \multicolumn{2}{c|}{Whole Library} &92.9\% & \multirow{8}{*}{1000}  &  927 & 2 & 0 & 71 & 0 \\ \cline{2-4} \cline{6-10}
   & \multicolumn{2}{c|}{Cluster Centroid} &73.4\% & &  509 & 225 & 0 & 252 & 14 \\ \cline{2-4} \cline{6-10}
  & \multicolumn{1}{c|}{\multirow{3}{1.5cm}{\centering Consensus Only}} & f=0.5 & \textbf{\textit{100\%}} && 808 & 192 & 0 & 0 & 0 \\ \cline{3-4}\cline{6-10}
  & & f=0.8 & \textbf{\textit{100\%}}&& 808 & 192 & 0 & 0 & 0 \\\cline{3-4} \cline{6-10}
  && f=1 & \textbf{\textit{100\%}}&& 808 & 192 & 0 & 0 & 0 \\ \cline{2-4} \cline{6-10}
   & \multicolumn{1}{c|}{\multirow{3}{1.5cm}{\centering Consensus + Weighting}} & f=0.5 & \textbf{\textit{100\%}} && 808 & 192 & 0 & 0 & 0 \\ \cline{3-4}\cline{6-10}
  & & f=0.8 & \textbf{\textit{100\%}}&& 808 & 192 & 0 & 0 & 0 \\\cline{3-4} \cline{6-10}
  && f=1 & \textbf{\textit{100\%}}&& 808 & 192 & 0 & 0 & 0 \\ \hline

  \multirow{8}{1cm}{\centering SOAP (text)}  & \multicolumn{2}{c|}{Whole Library} & 100\% & \multirow{8}{*}{1000} &  77 & 923 & 0 & 0 & 0 \\ \cline{2-4} \cline{6-10}
    & \multicolumn{2}{c|}{Cluster Centroid} & 100\% &&  98 & 902 & 0 & 0 & 0 \\ \cline{2-4} \cline{6-10}
  & \multicolumn{1}{c|}{\multirow{3}{1.5cm}{\centering Consensus Only}} & f=0.5 & \textbf{\textit{100\%}} && 96 & 904 & 0 & 0 & 0 \\ \cline{3-4}\cline{6-10}
  & & f=0.8 & \textbf{\textit{100\%}}&& 96 & 904 & 0 & 0 & 0 \\\cline{3-4} \cline{6-10}
  && f=1 & \textbf{\textit{100\%}}&& 96 & 904 & 0 & 0 & 0 \\ \cline{2-4} \cline{6-10}
  & \multicolumn{1}{c|}{\multirow{3}{1.5cm}{\centering Consensus + Weighting}} & f=0.5 & \textbf{\textit{100\%}} && 96 & 904 & 0 & 0 & 0 \\ \cline{3-4}\cline{6-10}
  & & f=0.8 & \textbf{\textit{100\%}}&& 96 & 904 & 0 & 0 & 0 \\\cline{3-4} \cline{6-10}
  && f=1 & \textbf{\textit{100\%}}&& 96 & 904 & 0 & 0 & 0 \\ \hline

  \multirow{8}{1cm}{\centering Twitter (REST) (text)}& \multicolumn{2}{c|}{Whole Library} & \textbf{99.56\%} & \multirow{8}{*}{1825} &  150 & 994 & 673 & 7 & 1  \\ \cline{2-4} \cline{6-10}
  & \multicolumn{2}{c|}{Cluster Centroid} &99.34\% & &  0 & 896 & 917 & 11 & 1  \\ \cline{2-4} \cline{6-10}
  & \multicolumn{1}{c|}{\multirow{3}{1.5cm}{\centering Consensus Only}} & f=0.5 & \textit{99.34\%}&& 0 & 893 & 920 & 11 & 1 \\ \cline{3-4}\cline{6-10}
  & & f=0.8 & \textit{99.34\%}&& 0 & 893 & 920 & 11 & 1 \\\cline{3-4} \cline{6-10}
  && f=1 & \textit{99.34\%}&& 0 & 893 & 920 & 11 & 1 \\ \cline{2-4} \cline{6-10}
    & \multicolumn{1}{c|}{\multirow{3}{1.5cm}{\centering Consensus+ Weighting}} & f=0.5 & \textit{99.34\%}&& 0 & 893 & 920 & 11 & 1 \\ \cline{3-4}\cline{6-10}
  & & f=0.8 & \textit{99.34\%}&& 0 & 893 & 920 & 11 & 1 \\\cline{3-4} \cline{6-10}
  && f=1 & \textit{99.34\%}&& 0 & 893 & 920 & 11 & 1 \\ \hline
\end{tabular}
\caption{Accuracy Comparison Results.}
\label{tab:accuracy}
\end{table}

\medskip

\subsubsection{\textbf{Efficiency Assessment(Q2)}}
\label{ssb:efficiency}

\newcommand{\resgentime}{
   \begin{tabular}{|>{\centering\arraybackslash}m{1.5cm} |c|>{\centering\arraybackslash}m{1.5cm} |>{\centering\arraybackslash}m{1.5cm} |>{\centering\arraybackslash}m{1.5cm} |>{\centering\arraybackslash}m{1.5cm}|}
        \hline
      % after \\: \hline or \cline{col1-col2} \cline{col3-col4} ...
       &No.& Whole Library  & Cluster Centroid & Consensus +Weighting & Real System \\\hline\hline
      IMS  &800& 470.99 & 4.94 & 3.24& 518\\ \hline
      LDAP &2177& 835.91 & 2.77 & 2.88 & 28  \\ \hline
      LDAP text(1) &2177 & 1434.70 & 5.69 & 7.30 & 28 \\ \hline
      LDAP text(2) &1000& 266.30 & 2.44 & 1.63 & 28 \\ \hline
      SOAP &1000& 380.24 & 2.97 & 3.35 & 65 \\ \hline
      Twitter&1825& 464.09 & 32.86 & 36.62 & 417 \\ \hline
    \end{tabular}
}
\newcommand{\matchingtime}{

    \begin{tabular}{|c|c|c|c|c|c|c|c|}

   % \begin{tabular}{|c|c|c|c|c|c|c|c|}
        \hline
      % after \\: \hline or \cline{col1-col2} \cline{col3-col4} ...
       \multirow{2}{0.5cm}{}&\multirow{2}{0.5cm}{\centering No.}& \multicolumn{2}{c|}{\multirow{1}{*}{Whole Library}}  & \multicolumn{2}{c|}{\multirow{1}{*}{Cluster Centroid}} & \multicolumn{2}{c|}{Consensus+W} \\\cline{3-8}
       &&M&S&M&S&M&S \\\hline\hline
      \multirow{1}{0.5cm}{IMS} &800& 460.78&10.2 & 3.67 &1.27 & 2.68&0.56 \\ \hline
      LDAP &2177& 828.95 & 6.96 & 2.38 &0.39& 2.60&0.28 \\ \hline
      LDAP text(1) &2177 & 1425.23 &9.47& 4.38&1.31 & 5.67&1.63 \\ \hline
      LDAP text(2) &1000& 257.92 &8.38& 1.35 &1.09& 1.14 &0.49\\ \hline
      SOAP &1000& 372.58 &7.66& 1.92 &1.05 & 2.45 &0.9\\ \hline
      Twitter &1825& 412.98 &51.11& 1.47 &31.39& 1.67 &34.95 \\ \hline
    \end{tabular}

}

\begin{table}[t]
\centering
\subfloat[Average Total Response Generation Time (ms)\label{tab:resgen}]{\resgentime}%
     \qquad \newline
\subfloat[Average Matching Time and Average Substitution Time. M represents the matching time, and S represents the substitution time. (ms)\label{tab:matching}]{\matchingtime}
    \qquad \newline
%\subfloat[Average Transformation Time (ms)\label{tab:transformation}]{\transfmtime}
\caption{Approach Efficiency Evaluation Results}
\label{tab:efficiency}
\end{table}

Table \ref{tab:resgen} compares the average response generation time of the
{\em Consensus+Weighting} approach with the \emph{Whole Library} and
\emph{Cluster Centroid} approaches. Tests were run on an
Intel Xeon E5440 2.83GHz CPUs with 24GB of main memory available.
The times represent the average times
spent generating requests, for all the requests in the datasets sets. In
addition, Table \ref{tab:resgen} lists the average response times of the real
services from which the original traces were recorded. In order to get a better
insight of the runtime performance of our approach, we separately measured
matching time and substitution time, results of which are presented in Table
\ref{tab:matching}.

The results show that the \emph{Consensus+Weighting} approach is very
efficient at generating responses, indeed much faster than the real services
being emulated. The response generation time is comparable to the
\emph{Cluster Centroid} approach, being faster for some datasets, slower for
others. Both of these approaches are about two orders of magnitude faster
than the \emph{Whole Library} approach. However, whereas the \emph{Cluster
  Centroid} approach trades off accuracy for speed, the
\emph{Consensus+Weighting} has a high accuracy at a similar speed.

% Comment about response time distributions

Comparing the matching time versus the substitution time (shown in Table~\ref{tab:matching}) we can observe that the \emph{Whole Library} approach consumes most of its time during the matching process (because a Needleman-Wunsch alignment is made with every request in the transaction library).  \emph{Consensus+Weighting} and \emph{Cluster Centroid} have greatly reduced matching times.  Twitter has unusually long substitution times, such that for the fast approaches, most time is spent performing the substitution.  This is due to the Twitter responses being very long, causing the symmetric field identification (common substring search) to become time consuming.

The \emph{Consensus+Weighting} generates responses faster than the real services being emulated.  This is crucial for supporting testing  of an enterprise system under test under realistic performance conditions (delays can be added to slow down the emulated response, but not the other way around). A major limitation of the \emph{Whole Library} approach is that it cannot generate responses in a time which matches the real services for fast services (such as LDAP).

\subsubsection{\textbf{Robustness to Noisy Clustering(Q3)}}
\label{ssb:robustness}

Our final test is to evaluate whether our \emph{Consensus+Prototype} approach
is robust in generating accurate responses when the clustering process (from Section~\ref{sec:clustering}) is
imperfect.  For this test we deliberately inject noise into our clusters, \ie
to create clusters where a fraction of the interaction messages are of
different operation types. The noise ratios tested were 5\%, 10\% and 20\%.
We repeated the experiments with different frequency thresholds ({\ie} 0.5,
0.8, and 1).

Table~\ref{tab:accuracynoisycluster} summarises the experimental results. The
results show that having a frequency threshold below 1 has a very big impact
on preserving the accuracy when the clustering is noisy.  A threshold of $f =
0.5$ gives the best accuracy.  When using this threshold, the accuracy stays
above 97\% for all datasets, when the noise ratio is 5\%.  As the noise ratio
increases to 20\%, the accuracy decreases significantly for binary LDAP, but
stays high for the other datasets.

Overall this is a good result.  For our approach to work best the clusters produced should be relatively clean, but there is tolerance for a small amount of noise.  A noise ratio of 20\% is considered very high.  Our actual clustering process produced perfect separation ({\ie} 0\% noise) of interaction messages by operation type for the six datasets tested.

\begin{table}[h]
\centering
\begin{tabular}{|c|c|c|c|c|}
  \hline
    \multicolumn{1}{|c|}{}&\multirow{2}{*}{Noise Ratio}& \multicolumn{3}{c|}{Consensus Prototype} \\ \cline{3-5}
  && f = 0.5 &
  f = 0.8 & f = 1 \\
  \hline \hline
  % after \\: \hline or \cline{col1-col2} \cline{col3-col4} ...
    IMS & 5\% & \textit{99.38\%} & \textit{99.75\%}  & \textit{72.25\%} \\ \cline{2-5}
    &10\% &  \textit{95.75\%} & \textit{98.63\%}  & \textit{68.5\%} \\ \cline{2-5}
    &20\% & \textit{ 99.75\%} & \textit{96.5\%} & \textit{68.13\%}
    \\ \hline \hline
    LDAP & 5\% & \textit{97.01\%} & \textit{97.29\%}  & \textit{33.49\% } \\ \cline{2-5}
     &10\% & \textit{87.09\%} & \textit{79.83\%} &\textit{49.93\%} \\ \cline{2-5}
     &20\% & \textit{73.81\%} & \textit{63.67\%} & \textit{42.03\%} \\ \hline \hline
   LDAP text (1)  & 5\% & \textit{100\%} & \textit{100\%}& \textit{83.74\%} \\ \cline{2-5}
   &10\% & \textit{100\%} & \textit{100\%} & \textit{84.20\%}  \\ \cline{2-5}
   &20\% & \textit{100\%} & \textit{100\%} & \textit{86.50\%}  \\ \hline \hline
  LDAP text (2) & 5\% &\textit{100\%}&\textit{100\%}&\textit{100\%} \\ \cline{2-5}
  &10\% &\textit{100\%}&\textit{100\%}&\textit{100\%} \\ \cline{2-5}
  &20\% &\textit{100\%} & \textit{100\%} & \textit{100\%} \\ \hline \hline
  SOAP & 5\% &\textit{98.2\%}&\textit{98.2\%}&\textit{91.5\%}\\ \cline{2-5}
  &10\% &\textit{100\%} &\textit{100\%}&\textit{98.2\%}\\ \cline{2-5}
  &20\% & \textit{100\%} & \textit{100\%} & \textit{97.7\%} \\ \hline \hline
  Twitter (REST)& 5\% & \textit{99.34\%} & \textit{99.34\%}& \textit{22.41\%} \\ \cline{2-5}
  &10\% & \textit{97.59\%}& \textit{95.01\%}& \textit{32.60\%} \\ \cline{2-5}
  &20\% & \textit{96.71\%}& \textit{94.13\%}& \textit{29.00\%}\\ \hline
\end{tabular}
\caption{Response Accuracy for Clusters with Noisy Data}
\label{tab:accuracynoisycluster}
\end{table}

%Table~\ref{tab:accuracynoisycluster} summarises experimental results, and also demonstrates comparison results of our approach with the \emph{Cluster Centroid} approach. From this table, we can see that 1) our approach are more robust to textual protocols than binary protocols. 2) Compared with the \emph{cluster centroid} approach, noisy data has more impacts on the effectiveness of our approach for generating correct responses. Specifically, our approach can generate 100\% or close to 100\% accurate results for textual protocol experiments, while the accuracy ratio of binary protocol experiments decrease when the number of noisy data in clusters increases. 3) For binary protocols, when clusters only contain few noisy data (5\%), different frequency thresholds rarely influence accuracy ratio of generated response. However, as the number of noisy data increases, the greater frequency thresholds of our approach ({\ie} 0.8 and 1) leads to decreasing number of accurate responses. with the increasing  The robustness of our approach is not influenced.

%\subsection{LISA integration}

\subsection{Industry Validation}

The opaque service emulation technique has been integrated into CA Technologies' commercial product: CA Service Virtualization~\cite{theaustralian}. An earlier version of the technique was released as a new feature in version 8.0 of the product and has been sold to customers. Opaque service emulation has been used at customer sites to successfully emulate services of protocols not otherwise supported by the product. The present technique is on the backlog for version 9.0.

\subsection{Threats to Validity}

We have identified some threats to validity which should be taken into consideration when generalising our experimental results:
\begin{itemize}
\item Our evaluation was performed on six datasets from four protocols.  Given the great diversity in message protocols, further testing should be performed on other message protocols.
\item The datasets were obtained by randomly generating client requests for services of different protocols.  Some real system interactions are likely to be more complicated than those of our datasets.  Further testing on real system interactions are warranted.
\end{itemize}

\section{Discussion}

We have developed an approach for automatically generating service responses from message traces which requires no prior knowledge of message structure or decoders or message schemas.
Our approach of using multiple sequence alignment to automatically generate consensus prototypes for the purpose of matching request messages is shown to be accurate, efficient and robust.  Wildcards in the prototypes allow the stable and unstable parts of the request messages for the various operation types to be separated.  Rather than using the prototypes directly for strict matching (such as using it as a regular expression) we instead calculate matching distance through a modified Needleman-Wunsch alignment algorithm.  Since we look for the closest matching prototype, the method is robust even if the prototypes are imperfect.  Moreover, this process can match requests which are slightly different to the prototypes, or are of different length to the prototypes.  This allows the system to handle requests which are outside of the cases directly observed in the trace recordings.  Weighting sections of the prototype with different importance based on the entropy, further improves the matching accuracy.

Our experimental results using the 6 message trace datasets demonstrate that our approach is able to automatically generate accurate responses in real time for most cases. Moreover, we can also see that our approach can also generate accurate responses from imperfect message clusters that contain a small number of messages of different operation types.

\section{Summary}

In this chapter, we discuss a new technique for automatically generating realistic response messages from network traces for enterprise system emulation environments that outperforms current approaches. We use the bioinformatics-inspired multiple sequence alignment algorithm to derive message prototypes, adding wildcards for high variability sections of messages. A modified Needleman-Wunsch algorithm is used to calculate message distance and entropy weightings are used in distance calculations for increased accuracy. Our technique is able to automatically separate the payload and structural information in complex enterprise system messages, making it highly robust. We have shown in a set of experiments with four enterprise system messaging protocols a greater than 99\% accuracy for the four protocols tested. Additionally they show efficient emulated service response performance enabling scaling within an emulated deployment environment. We successfully demonstrate that our opaque response generation approach is able to automatically create virtual service models in the absence of experts or expert knowledge, thereby achieving our goals of accuracy, efficiency and robustness.

\chapter{Conclusion and Future Work}
\label{chap8:conclusion}

In this chapter, we summarise the whole thesis. Section~\ref{chap8sec:summary} summaries the contents of the whole thesis. Then we suggest topics for future work in Section~\ref{chap8sec:futurework}.

\section{Summary of This Thesis}
\label{chap8sec:summary}

In this thesis we have demonstrated that:

\begin{center}
\begin{tabular}{p{0.85\textwidth}}
   {\em The opaque response generation is an approach to automatically generating the messages of an application-layer protocol from its network trace, enabling corresponding virtual service models to be created in the absence of experts or expert knowledge. This approach is able to mimic server-side interaction behaviours by returning client-side system-under-test accurate approximation of the ``real'' responses at runtime.}
\end{tabular}
\end{center}

% why devops? what is devops? what is the relationship between devops and service virtualisation? Service virtulisation allows continuous access
In a large enterprise software environment, an enterprise software system interacts with many other systems performing its functionalities. Once such a system gets upgraded, assessing its run-time properties is a task best undertaken in an alternative environment where this system under test is able to operate in the same manner that it would in the real environment. Given the scale and complexity of typical enterprise deployment environments, producing such an approximate environment is a significant engineering challenge. Service virtualisation has been proposed as an approach to providing a reactive production-like environment. By creating virtual service models and simultaneously executing a number of those virtual service models, the service virtualisation technique provides an interactive representation of the enterprise software environment, capable of achieving the required level of complexity and interactivity.

Being able to generate virtual service models is pivotal to the service virtualisation technique. By providing responses for the external system under test, a virtual service model can mimic interaction behaviours of server-side systems in the alternative environment. A virtual service can be conceptualised as a series of steps to be executed when a request is received. We developed the opaque response generation approach to automating those steps in a manner of synthesizing responses from network traffic recordings. A key property of our opaque response generation approach is that it operates in a protocol-independent fashion: it does not require any specifics about the particular service it mimics. It allows virtual service models to be created automatically, without requiring any knowledge of the internals of the service or of the protocols the service uses to communicate. This approach can also assist quality assurance teams in testing against poorly documented protocols. %summarise how this approach works,

By studying some popularly used application-level protocols, we specified three requirements which our opaque response generation approach needs to meet. The three requirements include,
 
\begin{itemize}
    \item {processing and analyzing network traffic for extracting protocol structure information;}
    \item {creating virtual service models on the basis of extracted information;}
    \item {replacing the real target service for quality assurance purpose.}
\end{itemize}
 
We proposed a highlevel framework to discuss how our approach meets these requirements. Our framework contains three components, which are the analysis function, the matching function and the translation function. Collectively, these functions tackle these three requirements. Specifically, the purpose of the analysis function is to learn protocol knowledge from the network traffic recordings. The matching function is to look for appropriate protocol knowledge for an incoming request. The translation function is to utilise protocol knowledge to generate responses.

As a proof of concept realisation of opaque response generation, we described the \textit{whole library} approach, for automatically generating responses from the network traces. This approach is the first implementation of our proposed framework. It is also the foundation of this thesis. The approach adopted a genome sequence alignment algorithm and a field substitution algorithm to implement the matching function and the translation function. We evaluated this method against two common application-layer protocols: LDAP and SOAP. The experimental results showed a greater than 98\% accuracy for the two protocols tested. We have demonstrated that our opaque response generation approach can synthesize accurate responses in the absence of protocol experts or protocol knowledge from network traces, enabling the creation of interactive virtual service models automatically.

The whole library approach can provide accurate responses to a system under test. However, it takes a long time to generate a response. In practice, a slow response time may lead to a service failure, which will terminate the rest of the operation. To accelerate response generation time, and meet our goal of efficiency, we invented the \textit{cluster centroid} approach. This approach adopted clustering techniques and data mining techniques to analyse the trace library. The clustering techniques were to group interactions in a trace library into clusters of interactions of the same type. The data mining techniques were used to select representative interactions, which were utilised by the matching function to accelerate response inference time at runtime. However the increased efficiency came at the expense of decreased accuracy.

To achieve both efficiency and accuracy for opaque response generation, we proposed a new technique: the \textit{consensus protototype} approach. This is a further refinement of the whole library approach. This approach utilised the bioinformatics-inspired multiple sequence alignment algorithm to derive consensus prototypes of message groups of the same operation type. The consensus prototypes capture the common patterns for each operation. Moreover, this approach adopted a modified Needleman-Wunsch algorithm to calculate the message distance. Experimental results showed the consensus prototype approach was efficient and robust to generate accurate responses for service virtualisation.

In summary, we conclude that:

\begin{center}
\begin{tabular}{p{0.85\textwidth}}
  {\em The opaque response generation is an approach to automatically generating the messages of an application-layer protocol from its network trace, enabling corresponding virtual service models to be created in the absence of experts or expert knowledge. This approach is able to mimic server-side interaction behaviours by returning client-side system-under-test accurate approximation of the ``real'' responses at runtime. It is both effective and efficient when using our consensus prototype approach.}
\end{tabular}
\end{center}

\section{Future Work}
\label{chap8sec:futurework}

In the software engineering area, opaque response generation approach is a novel technique to create virtual service models for service virtualisation, enabling the provision of interaction representation of an enterprise software environment. To create better service models, there are many possible aspects for improving our current work. We discuss some of these aspects here for our future work.

\begin{enumerate}

    \item {\textbf{Discovering stateful virtual service models}}

        %Protocols defines some rules of formatting messages that specify how data is packaged into message sent and received. These rules are called protocol syntax.
        We have discussed an application-layer protocol can be either stateless or stateful in Chapter~\ref{chap3:requirements}. For a stateless protocol, the server system responds client requests without retaining any client information. On the other hand, a stateful protocol defines temporal rules expressing data dependencies among exchanged messages. Let us consider LDAP~\cite{Sermersheim2006} as an example. The LDAP server remembers that which directory a LDAP client bound with, and allows the client to get access to data on that directory. Once the server received an unbind request, it must unbind with the same entry.

        Our current work synthesized responses solely depending on the incoming request and the recorded interaction traces, but not the service state history. Hence, it can only create stateless virtual service models since it does not consider protocol temporal properties in formulating responses. In practice a stateless model is sufficient in many cases. For example: (i) when the emulation target service is stateless, or (ii) when the testing scenario does not contain two equivalent requests, requiring different state affected responses, or (iii) where the testing scenario does not require highly accurate responses (e.g. for performance and scaling testing.) To further improve the accuracy of generated responses, we need to take protocol temporal properties into consideration. An avenue of future exploration is to process mine~\cite{processmining} the operation sequences to discover stateful models.

        %1. consider protocol temporal properties, reverse engineer protocol behavioural models, generate responses conforming to protocol temporal constrains
%        2. further improve the efficiency
%           possible solution: refine algorithm implementations (Jean-Guy's new paper)
%        3. improve the robustness of our approach to noisy clusters
%
%        To further improve the robustness of our approach to noisy clustering, we will utilise outlier detection techniques \cite{outlierdetection} to remove outliers of clusters before applying the alignment method.

    \item {\textbf{Improving diversity of the generated responses}}

    Our most sophisticated technique - the consensus prototype approach - lacks diversity in the responses generated. We are working on an approach to identify common patterns of all responses in a cluster. A possible solution is to apply multiple sequence alignment to response messages, for distinguish stable positions from variable positions. The variable parts of responses could then be stochastically generated. % Accordingly, the symmetric field technique \cite{Du:2013} will be refined, where only character sequences which occur in both the recorded request and {\em variable positions} of the recorded response will be substituted with characters from the live request.
    %This will also improve the efficiency of the substitution method for long messages.

%scv: not clear how making the symmetric field substitution more accurate improves the diversity. You may wish to mention injecting random strings into the unstable parts of the response, which are not part of a symmetric field substitution.

\end{enumerate}

In the long term, the following directions are interesting and important.

\begin{enumerate}

\item {\textbf{Extending opaque response generation approach to other network application paradigms}}

    The research in this thesis works for the client-server paradigm. Our approach is invoked by receiving a request from an external system under test, not in the opposite direction. In reality, however, there are two other network application paradigms, that is, the peer-to-peer architecture and the middleware architecture. For applications using these two paradigms, the virtual service models are required to switch roles, acting as both communication initiator and responder. Process mining techniques can be studied, assisting in determining the role of a virtual service model.

\item {\textbf{Visualising virtual service models}}

In our current approach, the generated opaque service models are a black box to the engineers using them. However, sometimes engineers want to inspect and modify generated responses. Presenting virtual service models to quality assurance engineers in appropriate forms can facilitate comprehension of application-layer protocols. Moreover, it will greatly increase the flexibility and usability of the targeted virtual deployment environment. Engineers will have the ability to tune and adjust the automatically generated service models. Model transformation and visualisation techniques can be investigated to transform the opaque models to human readable models.
\end{enumerate}

\appendix
\chapter{Java Code}
\label{apd.javacode}

\section{Symmetric Field}
\label{apx.symmetrixfield}

Listing~\ref{lst:cs} contains the Java code for describing the symmetric field information.

\begin{lstlisting}[language=Java,caption={Symmetric Field},
label=lst:cs]
package com.ca.calabs.bilby.substitution;

/**
 * Describes the matching positions and lengths of a symmetric field, relative
 * to a request and response.
 * @author scv, Miao Du
 *
 */

public class SymmetricField
{
    /** The start position of the symmetric field in the response */
    public final int rsppos;
    /** The start position of the symmetric field in the request */
    public final int rqpos;
    /** The length (n chars) of the symmetric field */
    public final int length;
    /** The length (n chars) of the symmetric field */
    public final String substring;
    
    public SymmetricField(int rsppos, int rqpos, int len)
    {
        this.rsppos = rsppos;
        this.rqpos = rqpos;
        this.length = len;
        this.substring = null;
    }
    
    public SymmetricField(int rsppos, int rqpos, int len, String substring)
    {
        this.rsppos = rsppos;
        this.rqpos = rqpos;
        this.length = len;
        this.substring = substring;
    }
    
    public int getRspPos() { return rsppos; }
    public int getRqPos() { return rqpos; }
    public int getLength() { return length; }
    
    public String toString()
    {
        if (substring == null)
        	return "(rspos:"+rsppos+" ,rqpos:"+rqpos+", len:"+length+")";
        else
        	return "(rspos:"+rsppos+" ,rqpos:"+rqpos+", len:"+length+","+substring+")";
    }
    
    @Override
    public boolean equals(final Object o) {
        if (this == o) return true;
        if (o == null || getClass() != o.getClass()) return false;
   
        final SymmetricField that = (SymmetricField) o;
   
        if (length != that.length) return false;
        if (rqpos != that.rqpos) return false;
        if (rsppos != that.rsppos) return false;
    
        return true;
    }
    
    @Override
    public int hashCode() {
        int result = rsppos;
        result = 31 * result + rqpos;
        result = 31 * result + length;
        return result;
    }
}
\end{lstlisting}

\section{Symmetric Field Identification}
\label{apx.symmetrixfieldidentify}

Listing~\ref{lst:fastcsfinder} contains the Java code for the
symmetric field identification class which can be used to find symmetric fields within a request and its corresponding responses.

\begin{lstlisting}[language=Java,caption={Symmetric Field Identification},
label=lst:fastcsfinder]
package com.ca.calabs.bilby.substitution;

import java.util.ArrayList;
import java.util.List;
import java.util.Map;
import java.util.HashMap;
import java.util.HashSet;

/**
 * Provides method for finding common substrings between a request and response.
 * @author scv, Miao Du, Jean-Guy Schneider
 *
 */

public class FastCSFinder
{
    private static final boolean debug=true;
    private final char[] rq;
    private final char[] rsp;
    private final int rqLen;
    private final int rspLen;
    /** Minimum length of a matching string */
    private final int minLen;
    /** Exclude strings made up of this char */
    private boolean doexclude;
    private final HashSet<Character> excludeChars;
    private final boolean removeconflicts;
    
    private final List<SymmetricField> matches;
    
	public FastCSFinder(char[] rq, char[] rsp, int minLen, boolean doexclude, char[] exclude, boolean removeconflicts)
    {
        this.rq = rq;
        this.rsp = rsp;
        this.rqLen = rq.length;
        this.rspLen = rsp.length;
        this.minLen = minLen;
        this.doexclude = doexclude && (exclude != null);
        if (doexclude && exclude != null)
        {
        	excludeChars = new HashSet<Character>(exclude.length);
            for (int i = 0; i < exclude.length; i++)
                excludeChars.add(exclude[i]);
        }
        else
        	excludeChars = null;
        this.removeconflicts = removeconflicts;
        matches = findCommonStrings2();
        if (this.removeconflicts)
            removeConflicts(matches);
    }
    
	public FastCSFinder(char[] rq, char[] rsp, int minLen, boolean doexclude, char[] exclude)
	{
		this(rq, rsp, minLen, doexclude, exclude, true);
	}
    
	public FastCSFinder(char[] rq, char[] rsp, int minLen, char exclude[])
	{
		this(rq, rsp, minLen, true, exclude);
	}
    
	public FastCSFinder(char[] rq, char[] rsp, int minLen)
	{
		this(rq, rsp, minLen, null);
        doexclude = false;
	}
    
    /**
     * Response centric symmetric field matching.
     */
    private List<SymmetricField> findCommonStrings2()
    {
        List<SymmetricField> matches = new ArrayList<SymmetricField>();
        boolean[][] skip = new boolean [rspLen][rqLen];
    	for (int i = 0; i < rspLen; i++)
    	{
            boolean[] skipi = skip[i];
            int maxmatch = minLen - 1;
            int rqpos = -1;
        	for (int j = 0; j < rqLen; j++)
        	{
                if (skipi[j])
                    continue;
                int matchlen = 0;
        		while ((i+matchlen) < rspLen && (j+matchlen) < rqLen && rsp[i+matchlen] == rq[j+matchlen])
                    matchlen++;
                // Update skip table to avoid finding shorter substrings
                for (int k = 1; k < matchlen; k++)
                {
                    skip[i+k][j+k] = true;
                }
                // Save match if its the longest so far
                if (matchlen >= maxmatch &&
                    (!doexclude || acceptString(rsp, i, matchlen, excludeChars)))
                {
                    rqpos = j;
                    maxmatch = matchlen;

                }
        	}
            if (maxmatch >= minLen)
            {
                if (!debug)
                    matches.add(new SymmetricField(i, rqpos, maxmatch));
                else
                {
                    char[] substring = new char[maxmatch];
                    for (int z = 0; z < maxmatch; z++)
                        substring[z] = rq[rqpos+z];
                    matches.add(new SymmetricField(i, rqpos, maxmatch, new String(substring)));
                }
            }
    	}
        return matches;
    }
    
    /**
     * If two symmetric fields overlap, take the longest. If they are
     * the same length, take the left one.
     */
    private static void removeConflicts(List<SymmetricField> matches)
    {
        if (matches == null || matches.size() < 2)
            // nothing to do
            return;
        SymmetricField prev = matches.get(0);
        SymmetricField curr;
    	for (int i = 1; i < matches.size(); i++)
    	{
            curr = matches.get(i);
            if ((prev.rsppos+prev.length) > curr.rsppos)
            {
                // take longest
                if (prev.length >= curr.length)
                {
                	matches.remove(i);
                    --i;
                }
                else
                {
                	matches.remove(i-1);
                    --i;
                }
            }
            prev = curr;
    	}
    }
    
    /**
     * Creates a list of symmetric field objects describing the matches found.
     * @return
     */
    public List<SymmetricField> getMatches()
    {
    	// convert internal rqMap and rspMap structures to symmetric field descs
        return matches;
    }
    
    private static final boolean acceptString(char[] str, int st, int len, HashSet<Character> excludeChars)
    {
    	for (int i = st; i < (st+len); i++)
    	{
            if (!excludeChars.contains(str[i]))
                return true;
    	}
        return false;
    }
}
\end{lstlisting}

\section{Field Substitution Method}
\label{apx.substitution}

Listing~\ref{lst:substitution} contains the Java code for the field substitution method that performs symmetric field substitution to modify a recorded response for generating a response.

\begin{lstlisting}[language=Java,caption={Field Substitution Method},
label=lst:substitution]
package com.ca.calabs.bilby.substitution;

import java.util.List;
import java.util.ArrayList;
import java.util.Map;

import org.slf4j.Logger;
import org.slf4j.LoggerFactory;

import com.ca.calabs.bilby.Transformation;
import com.ca.calabs.bilby.Interaction;
import com.ca.calabs.bilby.Message;
import com.ca.calabs.bilby.Alignment;
import com.ca.calabs.bilby.nw.FixedAlignment;
import com.ca.calabs.bilby.AlignmentUtil;
import com.ca.calabs.bilby.AlignmentResult;
import com.ca.calabs.bilby.util.StringUtil;

/**
 * Provides method for dynamically substituting symmetric fields (aka
 * magic strings).
 * @author scv, miao, jean-guy
 *
 */

public class Substitution
    implements Transformation
{
    private static final Logger log = LoggerFactory.getLogger(Substitution.class);
    
    private final int minLen;
    private final boolean doAdjustSf;
    private final boolean doexclude;
    private final Alignment alignmeth;
    private final char[] excludeChars;
    
    /**
     * @param minLen minimum matching length
     * @param excludeChars characters to exclude from common subsequence matching, 
     *         only the first character is excluded (for now)
     * @param alignmeth
     */
    public Substitution(int minLen, char[] excludeChars, Alignment alignmeth, boolean adjustSf)    
    {
    	this.minLen = minLen;
    	doexclude = (excludeChars != null && excludeChars.length > 0);
        this.excludeChars = excludeChars;
        this.alignmeth = alignmeth;
        this.doAdjustSf = adjustSf;
    }

    public Substitution(int minLen, char[] excludeChars, Alignment alignmeth)    
    {
    	this(minLen, excludeChars, alignmeth, false);
    }    
    
	public List<Message> translate (Interaction anInteraction, Message newRequest)
	{
		List<Message> rsps = anInteraction.getResponses();
		List<Message> newrsps = new ArrayList<Message>(1);
        char[] newrqchars = newRequest.getData();
        char[] mrqchars = anInteraction.getRequest().getData();
        int rspcount = 0;
        boolean isfixed = (alignmeth instanceof FixedAlignment);
        for (Message rsp: rsps)
        {
            char[] rspchars = rsp.getData();
            // identify common substrings between request and response
        	FastCSFinder csf = new FastCSFinder(mrqchars, rspchars, minLen, doexclude, excludeChars);
            // matches: (rsppos, rqpos, len)*
            List<SymmetricField> matches = csf.getMatches();
            // align old rq and new rq
            AlignmentResult alignr = alignmeth.align(mrqchars, newrqchars);
            // calculate copy operations
            List<CopyOperation> copyops = calcCopyOps(alignr, matches, isfixed, newrqchars.length);
            
            // new step method(copyops. alignr, originalreq, originalres) 
            // check boundary, modify boundary, stretch the length
            if (!isfixed && doAdjustSf)
                copyops = adjustCopyOps(copyops, matches, alignr, mrqchars, rspchars, newrqchars);
            
            char[] newrspchars = doSubst(newrqchars, rspchars, copyops);
            //Map<Integer, Integer> seqmap = alignr.getMapping();
            //char[] newrspchars = doSubst(newrqchars, rspchars, seqmap, matches);
            Message nrsp = new Message(newRequest.getDestination(), newRequest.getSource(), new String(newrspchars));
            newrsps.add(nrsp);
            if (log.isDebugEnabled())
                logSubstitution(rspcount, anInteraction.getRequest(), rsp, newRequest, nrsp, matches, alignr);
            rspcount++;
        }
        return newrsps;
	}
    
	/**
	 *  Adjust boundaries of magic fields 
	 * @param CopyOperation substitution instruction
	 * @param alignr alignment result for the matching request and the incoming request
	 * @param mrqchars original matched request
	 * @param rspchars original matched response
	 * @return
	 */
	protected List<CopyOperation> adjustCopyOps(List<CopyOperation> copyops, List<SymmetricField> matches,
			AlignmentResult alignr, char[] mrqchars, char[] rspchars, char[] newrqchars)
	{
        if (log.isDebugEnabled())
        {
            AlignmentUtil ar = new AlignmentUtil(alignr, mrqchars, newrqchars);
            log.debug(ar.printAlignment());
        }
		List<CopyOperation> copyops2 = new ArrayList<CopyOperation>();
		Map<Integer, Integer> ralignMap = alignr.getMapping();
		for (int i = 0; i < matches.size(); i++)
		{
			SymmetricField sf = matches.get(i);
			int mrqidx = sf.getRqPos();
			int nrqidx = ralignMap.get(mrqidx);
			CopyOperation op2 = copyops.get(i);
            
			// Step 0: test for no overlap

			// Step 1: find true start in new request
			int nstart = nrqidx;
			if(alignr.isSeq1PrevGap(mrqidx))
			{
				// Case 1: gaps at start of old request
				if(mrqidx>0)
				{
					// This pattern is not at the beginning
					if(mrqchars[mrqidx-1]
								==newrqchars[ralignMap.get(mrqidx-1)])
					{
						nstart = ralignMap.get(mrqidx-1) + 1;
					}
				}
				// if this pattern is at the start of sequence alignment, just extend symmetric field's length
				else
				{
					nstart = 0;
				}
			}
			else
			{
				// Case 2: gaps at start of symmetric field in new request
				int nongapstart = alignr.getNextSeq2NonGap(mrqidx);
				nstart = ralignMap.get(nongapstart);
			}

			
			// Step 2: find true end in new request
			int nend = ralignMap.get(mrqidx + sf.length);
			if(alignr.isSeq1NextGap(mrqidx))
			{
				// Case 1: gaps at start of old request
				if(mrqidx + sf.length < ralignMap.size() && mrqidx + sf.length < mrqchars.length)
				{
					// This pattern is not at the end
					int temp_idx = Math.min(mrqidx + sf.length, mrqchars.length-1);
					if(mrqchars[temp_idx]
								==newrqchars[ralignMap.get(temp_idx)])
					{
						nend = ralignMap.get(temp_idx);
					}
				}
				// if this pattern is at the start of sequence alignment, just extend symmetric field's length
				else
				{
					nend = ralignMap.get(ralignMap.size()-1);
				}
			}
			else
			{
				// TODO
				// Case 2: gaps not at end of new request
				if(nend < ralignMap.size())
				{
					if(alignr.isSeq1PrevGap(mrqidx + sf.length)){
						
						// Case 3: gaps after symmetric fields of the old request
						nend = ralignMap.get(mrqidx + sf.length-1) + 1;
					}
				}
			
			}			
			// Step 3: calculate length and modify copy op
			int nlen = nend - nstart;
			op2 = new CopyOperation(nstart, sf.getRspPos(), nlen, sf.length);

			copyops2.add(op2);
		}
		return copyops2;
	}

	protected static final List<CopyOperation> calcCopyOps(AlignmentResult alignr, List<SymmetricField> matches, boolean fixed, int newrqlen)
	{
		List<CopyOperation> copyops = new ArrayList<CopyOperation>();
        Map<Integer, Integer> seqmap = alignr.getMapping();
        for (SymmetricField m : matches)
        {
        	int rqi = m.getRqPos();
            int rqi2 = seqmap.get(rqi);
        	int rpj = m.getRspPos();
            int len = m.getLength();
            int len2;
            if (rqi2+len > newrqlen)
            {
            	if (!fixed)
                    len2 = newrqlen - rqi2;
            	else
                    // set flag to drop this symmetric field
                    len2 = -1;
            }
            else
            	len2 = len;
            if (len2 >= 0) 
                copyops.add(new CopyOperation(rqi2, rpj, len, len2));
        }
        return copyops;
	}
    
 	protected static final char[] doSubst(char[] newrq, char[] oldrsp, List<CopyOperation> copyops)
	{
		try {
    		// oversize buffer then resize (inefficient)
    		char[] buf = new char [oldrsp.length + newrq.length];
            int nptr = 0;	// new response pointer
            int optr = 0;	// old response pointer
            for (CopyOperation op : copyops)
            {
                // copy non-symmetric field bytes into buffer
                int cpylen = op.getRspPos()-optr;
                if (cpylen > 0)
                {
                    buf = safeCopy(oldrsp, optr, buf, nptr, cpylen);
                }
                nptr += cpylen;
                optr += cpylen;
                // copy symmetric field bytes into buffer
                cpylen = op.getRqLen();
                int i0 = op.getRqPos();
                buf = safeCopy(newrq, i0, buf, nptr, cpylen);
                nptr += cpylen;
                optr += op.getRspLen();
            }
            // copy the residual
            int cpylen = oldrsp.length - optr;
            buf = safeCopy(oldrsp, optr, buf, nptr, cpylen);
            nptr += cpylen;
            // downsize buffer
            char[] newrsp = new char [nptr];
            System.arraycopy(buf, 0, newrsp, 0, nptr);
            return newrsp;
		}
        catch (RuntimeException e) {
			e.printStackTrace();
            log.error("Exception in substitution of newrq: \""+
			StringUtil.toPrintableString(String.valueOf(newrq))+
			"\" into oldrsp: \""+
			StringUtil.toPrintableString(String.valueOf(oldrsp))+
			"\". "+"Reverting to oldrsp");
        	// revert to original response
            return oldrsp;
        }
	}

	private static final char[] safeCopy(char[] src, int src_idx, char[] buf, int buf_idx, int cpylen)
	{
		if (buf_idx + cpylen >= buf.length)
			buf = resize(buf, buf_idx, cpylen);
		cpylen = Math.min(cpylen, src.length - src_idx);
        if (cpylen > 0)
    		System.arraycopy(src, src_idx, buf, buf_idx, cpylen);
		return buf;
	}
	
	protected static final char[] doSubst(char[] newrq, char[] oldrsp, Map<Integer, Integer>seqmap, List<SymmetricField> matches)
	{
		try {
    		// oversize buffer then resize (inefficient)
    		char[] buf = new char [oldrsp.length + newrq.length];
            int nptr = 0;
            int optr = 0;
            for (SymmetricField m : matches)
            {
            	int i0 = m.getRqPos();
                int j0 = seqmap.get(i0);
                int i1 = i0 + m.length;
                int j1 = seqmap.get(i1);
                // copy non-symmetric field bytes into buffer
                int cpylen = m.getRspPos()-optr;
                if (cpylen > 0)
                {
                    if (nptr + cpylen > buf.length)
                        buf = resize(buf, nptr, cpylen);
                    System.arraycopy(oldrsp, optr, buf, nptr, cpylen);
                }
                nptr += cpylen;
                optr += cpylen;
                // copy symmetric field bytes into buffer
                cpylen = j1 - j0;
                if (nptr + cpylen > buf.length)
                    buf = resize(buf, nptr, cpylen);
                System.arraycopy(newrq, j0, buf, nptr, cpylen);
                nptr += cpylen;
                optr += (i1 - i0);
            }
            // copy the residual
            int cpylen = oldrsp.length - optr;
            if (nptr + cpylen > buf.length)
                buf = resize(buf, nptr, cpylen);
            System.arraycopy(oldrsp, optr, buf, nptr, cpylen);
            nptr += cpylen;
            // downsize buffer
            char[] newrsp = new char [nptr];
            System.arraycopy(buf, 0, newrsp, 0, nptr);
            return newrsp;
		}
        catch (RuntimeException e) {
            e.printStackTrace();
            log.error("Exception in substitution of newrq: \""+
                String.valueOf(newrq)+"\" into oldrsp: \""+String.valueOf(oldrsp)+"\". "+
                "Reverting to oldrsp");
        	// revert to original response
            return oldrsp;
        }
	}
    
	private static final char[] resize(char[] buf, int nptr, int cpylen)
	{
        int newsize = Math.max(buf.length+cpylen, nptr+cpylen+1);
        char[] tmpbuf = new char [newsize];
        System.arraycopy(buf, 0, tmpbuf, 0, nptr);
        return tmpbuf;
	}
    
    private static final void logSubstitution(int rspno, Message oldRq, Message oldRsp, Message newRq, Message newRsp, List<SymmetricField> matches, AlignmentResult alignr)
	{
		log.debug("Reponse["+rspno+"]");
        log.debug("newRq =      "+newRq.printableBody());
        log.debug("oldRq =      "+oldRq.printableBody());
        // Show aligned new request
        char[] oldRqChars = oldRq.getData();
        char[] newRqChars = newRq.getData();
        StringBuffer sb = new StringBuffer();
        Map<Integer, Integer> seqmap = alignr.getMapping();
        for (int i = 0; i < oldRqChars.length; i++)
        {
            int idx2 = seqmap.get(i);
            if (idx2 < newRqChars.length)
                sb.append(newRqChars[idx2]);
            else if (idx2 > newRqChars.length)
                log.warn("(seqmap("+i+")="+idx2+") 
				> (newRqChars.length="+newRqChars.length+")");
        }
        log.debug("alignedNewRq="
		+StringUtil.toPrintableString(sb.toString()));
        sb = new StringBuffer();
        for (int i = 0; i < oldRqChars.length; i++)
        {
            sb.append(""+i+":"+seqmap.get(i));
            if (i < oldRqChars.length-1)
                sb.append(", ");
        }
        log.debug("mapping={"+sb+"}");
        // Show responses
        log.debug("oldRsp =     "+oldRsp.printableBody() +" [len="+oldRsp.getDataAsString().length()+"]");
        log.debug("newRsp =     "+newRsp.printableBody() +" [len="+newRsp.getDataAsString().length()+"]");
        for (SymmetricField m : matches)
        {
        	log.debug("symmetricField: "+m+"\""
			+oldRsp.getDataAsString().substring(m.rsppos, m.rsppos+m.length)+"\"");
        }
	}
    
}
\end{lstlisting} 
\chapter{Interaction Trace Examples}
\label{apx.traceexamples}

\section{Simple Object Access Protocol (SOAP)}
\label{apx.soapexample}

% The following is one of the requests recorded in this interaction trace:
The SOAP{\footnote{We removed any newlines,
        whitespaces etc. introduced for presentation purposes during processing.}} trace example contains six different types of messages, which are to {\em get new token}, to {\em get account}, to {\em deposit money}, to {\em get transaction}, to {\em withdraw money}, and to {\em delete token}. For each type, we present an example as follows.

\begin{enumerate}

\item {\textbf{Get new token}}

    The following is one of the recorded getNewToken request:

    \vspace{-0.1cm}
    \small\begin{verbatim}
    <?xml version="1.0"?>
     <S:Envelope xmlns:S="http://schemas.xmlsoap.org/soap/envelope/">
      <S:Body>
       <ns2:getNewToken xmlns:ns2="http://bank/"/>
      </S:Body>
     </S:Envelope>
    \end{verbatim}
    \normalsize

    \vspace{-0.1cm} \noindent with the following the corresponding response:

    % \textbf{SOAP GetUser Response}\\
    \vspace{-0.1cm}
    \small\begin{verbatim}
    <?xml version="1.0"?>
     <S:Envelope
      xmlns:S="http://schemas.xmlsoap.org/soap/envelope/">
      <S:Body>
       <ns2:getNewTokenResponse xmlns:ns2="http://bank/">
        <return>
         458897
        </return>
       </ns2:getNewTokenResponse>
      </S:Body>
     </S:Envelope>
    \end{verbatim}

\item{\textbf{Get account}}

    The following is one of the recorded getAccount requests:

    % \Textbf{SOAP GetUser Request}\\
    \vspace{-0.1cm}
    \small\begin{verbatim}
    <?xml version="1.0"?>
     <S:Envelope
      xmlns:S="http://schemas.xmlsoap.org/soap/envelope/">
      <S:Body>
       <ns2:getAccount xmlns:ns2="http://bank/">
        <accountId>867-957-31</accountId></ns2:getAccount>
      </S:Body>
     </S:Envelope>
    \end{verbatim}
    \normalsize

    \vspace{-0.1cm} \noindent with the following the corresponding response:

    % \textbf{SOAP GetUser Response}\\
    \vspace{-0.1cm}
    \small\begin{verbatim}
    <?xml version="1.0"?>
     <S:Envelope
      xmlns:S="http://schemas.xmlsoap.org/soap/envelope/">
      <S:Body>
       <ns2:getAccountResponse xmlns:ns2="http://bank/">
        <return>
         <accountId>867-957-31</accountId>
         <fname>Steve</fname>
         <lname>Hine</lname>
        </return>
       </ns2:getAccountResponse>
      </S:Body>
     </S:Envelope>
    \end{verbatim}
    \normalsize

\item{\textbf{Deposit money}}

 The following is one of the recorded depositMoney requests:

    % \Textbf{SOAP GetUser Request}\\
    \vspace{-0.1cm}
    \small\begin{verbatim}
    <?xml version="1.0"?>
     <S:Envelope
      xmlns:S="http://schemas.xmlsoap.org/soap/envelope/">
      <S:Body>
       <ns2:depositMoney xmlns:ns2="http://bank/">
        <accountId>761-964-73</accountId>
        <amount>133.74</amount>
       </ns2:depositMoney>
      </S:Body>
     </S:Envelope>
    \end{verbatim}
    \normalsize

    \vspace{-0.1cm} \noindent with the following the corresponding response:

    % \textbf{SOAP GetUser Response}\\
    \vspace{-0.1cm}
    \small\begin{verbatim}
    <?xml version="1.0"?>
     <S:Envelope
      xmlns:S="http://schemas.xmlsoap.org/soap/envelope/">
      <S:Body>
       <ns2:depositMoneyResponse xmlns:ns2="http://bank/">
        <return>1263.46</return>
       </ns2:depositMoneyResponse>
      </S:Body>
     </S:Envelope>
    \end{verbatim}
    \normalsize

\item{\textbf{Get transaction}}

 The following is one of the recorded getTransaction requests:

    % \Textbf{SOAP GetUser Request}\\
    \vspace{-0.1cm}
    \small\begin{verbatim}
    <?xml version="1.0"?>
     <S:Envelope
      xmlns:S="http://schemas.xmlsoap.org/soap/envelope/">
      <S:Body>
       <ns2:getTransactions xmlns:ns2="http://bank/">
        <accountId>601-667-86</accountId>
       </ns2:getTransactions>
      </S:Body>
     </S:Envelope>
    \end{verbatim}
    \normalsize

    \vspace{-0.1cm} \noindent with the following corresponding response:

    % \textbf{SOAP GetUser Response}\\
    \vspace{-0.1cm}
    \small\begin{verbatim}
    <?xml version="1.0"?>
     <S:Envelope
      xmlns:S="http://schemas.xmlsoap.org/soap/envelope/">
      <S:Body>
       <ns2:getTransactionsResponse xmlns:ns2="http://bank/">
        <return>
         <transID>12345</transID>
         <fromAccount>601-667-86</fromAccount>
         <toAccount>761-964-73</toAccount>
         <amount>472.73</amount>
        </return>
       </ns2:getTransactionsResponse>
      </S:Body>
     </S:Envelope>
    \end{verbatim}
    \normalsize

\item{\textbf{Withdraw money}}

 The following is one of the recorded withdrawMoney requests:

    % \Textbf{SOAP GetUser Request}\\
    \vspace{-0.1cm}
    \small\begin{verbatim}
    <?xml version="1.0"?>
     <S:Envelope
      xmlns:S="http://schemas.xmlsoap.org/soap/envelope/">
      <S:Body>
       <ns2:withdrawMoney xmlns:ns2="http://bank/">
        <accountId>802-706-19</accountId>
        <amount>713.25</amount>
       </ns2:withdrawMoney>
      </S:Body>
     </S:Envelope>
    \end{verbatim}
    \normalsize

    \vspace{-0.1cm} \noindent with the following the corresponding response:

    % \textbf{SOAP GetUser Response}\\
    \vspace{-0.1cm}
    \small\begin{verbatim}
    <?xml version="1.0"?>
     <S:Envelope
      xmlns:S="http://schemas.xmlsoap.org/soap/envelope/">
      <S:Body>
       <ns2:withdrawMoneyResponse xmlns:ns2="http://bank/">
        <return>3517.88</return>
       </ns2:withdrawMoneyResponse>
      </S:Body>
     </S:Envelope>
    \end{verbatim}
    \normalsize

\item {\textbf{Delete token}}

    The following is one of the recorded deleteToken request:

    \vspace{-0.1cm}
    \small\begin{verbatim}
    <?xml version="1.0"?>
     <S:Envelope xmlns:S="http://schemas.xmlsoap.org/soap/envelope/">
      <S:Body>
       <ns2:deleteToken xmlns:ns2="http://bank/">
        <token>240294</token>
       </ns2:deleteToken>
      </S:Body>
     </S:Envelope>
    \end{verbatim}
    \normalsize

    \vspace{-0.1cm} \noindent with the following the corresponding response:

    % \textbf{SOAP GetUser Response}\\
    \vspace{-0.1cm}
    \small\begin{verbatim}
    <?xml version="1.0"?>
     <S:Envelope
      xmlns:S="http://schemas.xmlsoap.org/soap/envelope/">
      <S:Body>
       <ns2:deleteTokenResponse xmlns:ns2="http://bank/">
        <return>false</return>
       </ns2:deleteTokenResponse>
      </S:Body>
     </S:Envelope>
    \end{verbatim}

\end{enumerate}
\vspace{-0.1cm}

\section{Lightweight Directory Access Protocol (LDAP)}
\label{apx.ldapexample}

Some typical LDAP operations{\footnote{For each LDAP operation, we demonstrate its raw data, as well as the textual representation, which is transformed by Wireshark.}} are shown as follows.

\begin{enumerate}
\item {\textbf{Search operation: }}

LDAP Search Request

\begin{itemize}
	\item {Binary format}
	
30 7c 02 01 1a 63 5a 04 3a 63 6e 3d 42 65 6c 69 6e 64 61 20 4c 49 4e 44 53 45 59 2c 6f 75 3d 46 69 6e 61 6e 63 65 2c 6f 75 3d 43 6f 72 70 6f 72 61 74 65 2c 6f 3d 44 45 4d 4f 43 4f 52 50 2c 63 3d 41 55 0a 01 00 0a 01 02 02 01 00 02 01 00 01 01 00 87 0b 6f 62 6a 65 63 74 43 6c 61 73 73 30 00 a0 1b 30 19 04 17 32 2e 31 36 2e 38 34 30 2e 31 2e 31 31 33 37 33 30 2e 33 2e 34 2e 32
	
	\item {Textual format}

    \small\begin{verbatim}
192.168.92.1          192.168.92.128        LDAP
searchRequest(26) "cn=Belinda LINDSEY,ou=Finance,
ou=Corporate,o=DEMOCORP,c=AU" baseObject
      \end{verbatim}
    \normalsize

%\begin{lstlisting}
%192.168.92.1          192.168.92.128        LDAP     searchRequest(26) "cn=Belinda LINDSEY,ou=Finance,ou=Corporate,o=DEMOCORP,c=AU" baseObject
%\end{lstlisting}

\end{itemize}

LDAP Search Response

LDAP search response consists of the
merge of a {\em search result entry} and a {\em search result done} message:

\begin{itemize}
\item {Binary format}

30 82 01 62 02 01 1a 64 82 01 5b 04 3a 63 6e 3d 42 65 6c 69 6e 64 61 20 4c 49 4e 44 53 45 59 2c 6f 75 3d 46 69 6e 61 6e 63 65 2c 6f 75 3d 43 6f 72 70 6f 72 61 74 65 2c 6f 3d 44 45 4d 4f 43 4f 52 50 2c 63 3d 41 55 30 82 01 1b 30 17 04 02 63 6e 31 11 04 0f 42 65 6c 69 6e 64 61 20 4c 49 4e 44 53 45 59 30 1e 04 0b 6f 62 6a 65 63 74 43 6c 61 73 73 31 0f 04 0d 69 6e 65 74 4f 72 67 50 65 72 73 6f 6e 30 0f 04 02 73 6e 31 09 04 07 4c 49 4e 44 53 45 59 30 22 04 05 74 69 74 6c 65 31 19 04 17 45 6c 65 63 74 72 69 63 61 6c 20 43 6f 2d 6f 72 64 69 6e 61 74 6f 72 30 1d 04 0f 74 65 6c 65 70 68 6f 6e 65 4e 75 6d 62 65 72 31 0a 04 08 32 31 32 20 33 30 34 38 30 18 04 0b 64 65 73 63 72 69 70 74 69 6f 6e 31 09 04 07 54 61 72 69 66 66 73 30 26 04 04 6d 61 69 6c 31 1e 04 1c 42 65 6c 69 6e 64 61 2e 4c 49 4e 44 53 45 59 40 44 45 4d 4f 43 4f 52 50 2e 63 6f 6d 30 34 04 0d 70 6f 73 74 61 6c 41 64 64 72 65 73 73 31 23 04 21 33 38 37 20 52 75 73 68 64 61 6c 65 20 50 6c 61 63 65 24 43 61 72 72 69 6e 67 74 6f 6e 20 4e 53 57 30 14 04 0a 70 6f 73 74 61 6c 43 6f 64 65 31 06 04 04 32 32 39 34 30 0c 02 01 1a 65 07 0a 01 00 04 00 04 00

\item {Textual format}

\small\begin{verbatim}
192.168.92.128        192.168.92.1          LDAP
searchResEntry(26) "cn=Belinda LINDSEY,ou=Finance,
ou=Corporate,o=DEMOCORP,c=AU"  | searchResDone(26)
\end{verbatim}
    \normalsize

\end{itemize}

\item{Add operation}

{\textbf{LDAP Add Request:}}

\begin{itemize}
	\item {Binary format}
	
30 81 c0 02 01 24 68 81 9d 04 39 63 6e 3d 4d 69 61 6f 20 44 55 2c 6f 75 3d 41 64 6d 69 6e 69 73 74 72 61 74 69 6f 6e 2c 6f 75 3d 43 6f 72 70 6f 72 61 74 65 2c 6f 3d 44 45 4d 4f 43 4f 52 50 2c 63 3d 41 55 30 60 30 0f 04 02 63 6e 31 09 04 07 4d 69 61 6f 20 44 55 30 41 04 0b 6f 62 6a 65 63 74 43 6c 61 73 73 31 32 04 0d 69 6e 65 74 4f 72 67 50 65 72 73 6f 6e 04 14 6f 72 67 61 6e 69 7a 61 74 69 6f 6e 61 6c 50 65 72 73 6f 6e 04 06 70 65 72 73 6f 6e 04 03 74 6f 70 30 0a 04 02 73 6e 31 04 04 02 44 55 a0 1b 30 19 04 17 32 2e 31 36 2e 38 34 30 2e 31 2e 31 31 33 37 33 30 2e 33 2e 34 2e 32
	
	\item {Textual format}

\small\begin{verbatim}
192.168.92.1          192.168.92.128        LDAP
addRequest(36) "cn=Miao DU,ou=Administration,
ou=Corporate,o=DEMOCORP,c=AU"
\end{verbatim}
    \normalsize
\end{itemize}

{\textbf{LDAP Add Response:}}

\begin{itemize}

	\item {Binary format}
	
	30 0c 02 01 24 69 07 0a 01 00 04 00 04 00
	
	\item{Textual format}
	
\small\begin{verbatim}
192.168.92.128        192.168.92.1          LDAP
addResponse(36) success
\end{verbatim}
    \normalsize

\end{itemize}

\item{\em Modify operation}

{\textbf{LDAP Modify Request:}}

\begin{itemize}

	\item {Binary format}

30 7f 02 01 20 66 5d 04 3c 63 6e 3d 43 72 61 69 67 20 4c 49 4e 4b 2c 6f 75 3d 41 64 6d 69 6e 69 73 74 72 61 74 69 6f 6e 2c 6f 75 3d 43 6f 72 70 6f 72 61 74 65 2c 6f 3d 44 45 4d 4f 43 4f 52 50 2c 63 3d 41 55 30 1d 30 1b 0a 01 02 30 16 04 0c 75 73 65 72 50 61 73 73 77 6f 72 64 31 06 04 04 31 32 33 34 a0 1b 30 19 04 17 32 2e 31 36 2e 38 34 30 2e 31 2e 31 31 33 37 33 30 2e 33 2e 34 2e 32
	
	\item {Textual format}

\small\begin{verbatim}
192.168.92.1          192.168.92.128        LDAP
modifyRequest(32) "cn=Craig LINK,ou=Administration,
ou=Corporate,o=DEMOCORP,c=AU"
\end{verbatim}
    \normalsize

\end{itemize}

{\textbf{LDAP Modify Response:}}

\begin{itemize}

\item {Binary format}

30 0c 02 01 20 67 07 0a 01 00 04 00 04 00

\item{Textual format}

\small\begin{verbatim}
192.168.92.128        192.168.92.1          LDAP
modifyResponse(32) success
\end{verbatim}
    \normalsize

\end{itemize}

\item{\em Delete operation}

{\textbf{LDAP Delete Request:}}

\begin{itemize}
	\item {Binary format}
	
30 5b 02 01 27 4a 39 63 6e 3d 4d 69 61 6f 20 44 55 2c 6f 75 3d 41 64 6d 69 6e 69 73 74 	 72 61 74 69 6f 6e 2c 6f 75 3d 43 6f 72 70 6f 72 61 74 65 2c 6f 3d 44 45 4d 4f 43 4f 52 50 2c 63 3d 41 55 a0 1b 30 19 04 17 32 2e 31 36 2e 38 34 30 2e 31 2e 31 31 33 37 33 30 2e 33 2e 34 2e 32
	
	\item {Textual format}

        \small\begin{verbatim}
192.168.92.1          192.168.92.128        LDAP
delRequest(39) "cn=Miao DU,ou=Administration,
ou=Corporate,o=DEMOCORP,c=AU"
        \end{verbatim}
            \normalsize

\end{itemize}

{\textbf{LDAP Delete Response:}}

\begin{itemize}

 \item {Binary format}

 30 0c 02 01 27 6b 07 0a 01 00 04 00 04 00

 \item {Textual format}

 \small\begin{verbatim}
192.168.92.128        192.168.92.1          LDAP
delResponse(39) success
\end{verbatim}
 \normalsize

\end{itemize}

\end{enumerate}

\section{Twitter REST API}
\label{apx.twitterexample}

Twitter is not only a useful tool within the social media space, it also provides Web application developers a number of services to enable automation of Twitter functionality. One of those services (and perhaps the most popular) is the REST API. REST is an acronym for Representational State Transfer that enables developers to access information and resources using a simple HTTP invocation.
Twitter REST API Resources

The Twitter REST API provides flexible access to any available tweets. Moreover, nearly all common Twitter functionality can be programmatically accessed, e.g., posting new tweets, retweeting, following a user, searching, etc. Basically, the Twitter REST API provides operations that can be classified into 16 categories, which are:

\begin{itemize}
    \item	Timelines: Timelines are collections of Tweets, ordered with the most recent first.
    \item	Tweets: Tweets are the atomic building blocks of Twitter, 140-character status updates with additional associated metadata. People tweet for a variety of reasons about a multitude of topics.
    \item	Search: Find relevant Tweets based on queries performed by your users.
    \item Streaming: Give developers low latency access to Tweet data and other events have occured
    \item Direct Messages: Direct Messages are short, non-public messages sent between two users.
    \item Friends \& Followers: Users follow their interests on Twitter through both one-way and mutual following relationships.
    \item Users: Users are at the centre of everything Twitter: they follow, they favourite, and tweet and retweet.
    \item Suggested Users: Categorical organization of users that others may be interested to follow.
    \item Favourites: Users favourite tweets to give recognition to awesome tweets, to curate the best of Twitter, to save for reading later, and a variety of other reasons. Likewise, developers make use of "favs" in many different ways.
    \item Lists: Lists are collections of tweets, culled from a curated list of Twitter users. List timeline methods include tweets by all members of a list.
    \item Saved Searches: Allows users to save references to search criteria for reuse later.
    \item Places \& Geo: Users tweet from all over the world. These methods allow you to attach location data to tweets and discover tweets \& locations.
    \item Trends: With so many tweets from so many users, themes are bound to arise from the zeitgeist. The Trends methods allow you to explore what's trending on Twitter.
    \item Spam Reporting: These methods are used to report user accounts as spam accounts.
    \item OAuth: Twitter uses OAuth for authentication.
    \item Help: These methods assist you in working \& debugging with the Twitter API.
\end{itemize}

Most of Twitter API operations are used to provide the twitter data for 3rd party Twitter applications. In order to control the use of these operations, Twitter sets them a limit on how many times they can be used in 15 minute window, called rate limit. Operations  that are intuitively used most often by twitter users and their corresponding categories will be introduced in details as follows.
\begin{enumerate}
    \item Timelines
    \begin{itemize}
        \item GET statuses\/mentions\_timeline
        \begin{enumerate}
            \item Returns the 20(default) most recent mentions for the authenticating user.
            \item The timeline returned is the equivalent of the one seen when you view your mentions on twitter.com.
            \item The maximum number of mentions that this method can return is 800 tweets.
            \item Rate limit: 15/user
        \end{enumerate}

        \item  GET statuses/user\_timeline
        \begin{enumerate}
            \item Returns a collection of the most recent Tweets posted by the user indicated by the screen\_name or user\_idparameters.
            \item User timelines belonging to protected users may only be requested when the authenticated user either "owns" the timeline or is an approved follower of the owner.
            \item The timeline returned is the equivalent of the one seen when you view a user's profile on twitter.com
            \item This method can only return up to 3,200 of a user's most recent Tweets. Native retweets of other statuses by the user is included in this total, regardless of whether include\_rts is set to false when requesting this resource.
            \item Rate limit: 180/user
        \end{enumerate}

        \item   GET statuses/home\_timeline
        \begin{enumerate}
            \item Returns a collection of the most recent Tweets and retweets posted by the authenticating user and the users they follow. The home timeline is central to how most users interact with the Twitter service.
            \item Up to 800 Tweets are obtainable on the home timeline. It is more volatile for users that follow many users or follow users who tweet frequently.
            \item Rate limit: 15/user
        \end{enumerate}

        \item  GET statuses/retweets\_of\_me
        \begin{enumerate}
            \item Returns the most recent tweets authored by the authenticating user that have been retweeted by others.
            \item This timeline is a subset of the user's GET statuses/user\_timeline.
            \item Rate limit: 15/user
        \end{enumerate}
    \end{itemize}

    \item Tweets
    \begin{itemize}
        \item GET statuses/retweets/:id
        \begin{enumerate}
            \item Returns a collection of the 100 most recent retweets of the tweet specified by the unique tweet\_id parameter.
            \item Rate limit: 15/user
        \end{enumerate}

        \item GET statuses/show/:id
        \begin{enumerate}
            \item Returns a single Tweet, specified by the id parameter.
            \item The Tweet's author will also be embedded within the tweet.
            \item Rate limit: 180/user
        \end{enumerate}

        \item POST statuses/destroy/:id
        \begin{enumerate}
            \item Destroys the status specified by the required ID parameter.
            \item The authenticating user must be the author of the specified status.
            \item Returns the destroyed status if successful.
            \item Rate limit: No
        \end{enumerate}

        \item POST statuses/update
        \begin{enumerate}
            \item Updates the authenticating user's current status, also known as tweeting.
            \item A user cannot submit the same status twice in a row. Otherwise, a 403 error will be sent back.
            \item Rate limit: No
        \end{enumerate}

        \item POST statuses/retweet/:id
        \begin{enumerate}
            \item Retweets a tweet.
            \item Returns the original tweet with retweet details embedded.
            \item Rate limit: No
        \end{enumerate}
    \end{itemize}

    \item Search
    \begin{itemize}
        \item GET search/tweets
        \begin{enumerate}
            \item Returns a collection of relevant Tweets matching a specified query.
            \item Rate limit: 180/user
        \end{enumerate}
    \end{itemize}

    \item Direct Message
    \begin{itemize}
        \item GET direct\_messages/sent
        \begin{enumerate}
            \item Returns the most recent direct messages sent to the authenticating user, the default number  is 20
            \item Request includes detailed information about the sender and recipient user
            \item The maximum number of direct messages for per call is 200
            \item The maximum number of incoming direct messages that one can request is 800
            \item Rate limit: 15/user
        \end{enumerate}

        \item GET direct\_messages/show
        \begin{enumerate}
            \item Returns a single direct message, specified by an id parameter
            \item Request includes detailed information about the sender and recipient user
            \item Rate limit: 15/user
        \end{enumerate}

        \item POST direct\_messages/destroy
        \begin{enumerate}
            \item Destroys the direct message specified in the required ID parameter
            \item The authenticating user must be the recipient of the specified direct message
            \item Rate limit: No
        \end{enumerate}

        \item POST direct\_messages/new
        \begin{enumerate}
            \item Sends a new direct message to the specified user from the authenticating user
            \item Requires both the user and text parameters and must be a POST
            \item Rate limit: No
        \end{enumerate}
    \end{itemize}

    \item Friend \& Follower
    \begin{itemize}
        \item GET friendships/show
        \begin{enumerate}
            \item Returns detailed information about the relationship between two arbitrary users
            \item Rate limit: 180/user
        \end{enumerate}

        \item GET friendships/incoming
        \begin{enumerate}
            \item Returns a collection of numeric IDs for every user who has a pending request to follow the authenticating user
            \item Rate limit: 15/user
        \end{enumerate}

        \item POST friendships/create
        \begin{enumerate}
            \item Allows the authenticating users to follow the user specified in the ID parameter
            \item Different returns:
            \begin{enumerate}
                \item Returns the befriended user in the requested format when successful.
                \item Returns a string describing the failure condition when unsuccessful. If you are already friends with the user a HTTP 403 may be returned, though for performance reasons you may get a 200 OK message even if the friendship already exists
            \end{enumerate}
            \item Rate limit: No
        \end{enumerate}
    \end{itemize}
\end{enumerate}

\subsection{Network Traffic Generation}

To collect network traffic, we need to build a twitter client that is able to use the Twitter API. In fact, it is easy to make a twitter client in java using a third-part library for the Twitter API, which is called Twitter4j .  Although there are plenty of Java library options, the reasons why I choose Twitter4j are because 1) it covers more complete Twitter API; 2) its releases are more frequent; 3) its documentation is more readable and 4) it provides a number of examples.
The current client program uses 4 resources: reading user timeline, reading a tweet, searching tweets and showing friendship, which are highlighted in the previous section with the red colour. The reasons to select these resources are 1) intuitively twitter users use selected 5 resource categories the most frequently; 2) selected resources from these categories allow the most calls every 15 minutes, which enables us to collect interactions as fast as possible; 3) these resources are the most simple to use; 4) as this is the first try of using Twitter API, it would be better to start with reading twitter data, rather than changing/updating data.  Although the current client program seems very simple, I can easily extend the client program by adding various resources. We can discuss and decide which resources would be worth adding in the client program and hence we could make the Twitter traffic more complex.
%\vspace{-0.4cm} \noindent Consider these examples, LDAP requests contain a (unique) message
%identifier ({\tt 26, 36, 32, 39}) and a specific object name ({\tt ObjectName:}
%$\ldots$) as the root node for the search to be used. The corresponding
%responses use the same message identifier (to indicate the request they
%are in response to) and the {\tt searchResEntry} message refers to the same
%object name as the request. For our approach to synthesize correct LDAP
%responses, the corresponding information needs to be copied across from the
%incoming request to the most similar response to be modified.

%\listoffigures
%\listoftables

%\mainheadings
%

\bibliographystyle{abbrv}
\bibliography{thesis,protocols,icse2015,Reportbib}

\end{document}